\documentclass{aa} 
\usepackage{graphicx}
\usepackage{txfonts}
\usepackage{longtable}
\usepackage{subfig}
\usepackage{booktabs}
\usepackage{natbib}
\bibpunct{(}{)}{;}{a}{}{,} 
\usepackage{amssymb}
\usepackage{pifont}
\usepackage{ gensymb }

\begin{document}

   \title{The X-ray variability of Seyfert 1.8/1.9 galaxies}

   \subtitle{}

 \author{Hern\'{a}ndez-Garc\'{i}a, L.\inst{1}$^,$\inst{2}; Masegosa, J.\inst{1};
   Gonz\'{a}lez-Mart\'{i}n, O.\inst{3}; M\'{a}rquez, I. \inst{1}; Guainazzi, M.\inst{4}; Panessa, F.\inst{2} }

   \institute{Instituto de Astrof\'{i}sica de Andaluc\'{i}a, CSIC,
     Glorieta de la Astronom\'{i}a, s/n, 18008 Granada,
     Spain \and INAF - Istituto di Astrofisica e Planetologia Spaziali di Roma (IAPS-INAF), Via del Fosso del Cavaliere 100, I-00133 Roma, Italy\\ \email{lorena.hernandez@iaps.inaf.it}  \and Instituto de radioastronom\'{i}a
     y Astrof\'{i}sica (IRyA-UNAM), 3-72 (Xangari), 8701, Morelia,
     Mexico \and European Space Research and Technology Centre (ESA/ESTEC), Kepleriaan 1, 2201 AZ Noordwijk, The Netherlands\\}

   \date{Received XXX / Accepted XXX}

\authorrunning{Hern\'{a}ndez-Garc\'{i}a et al.}
\titlerunning{X-ray variability in Seyfert 1.8/1.9}

 
  \abstract
{Seyfert 1.8/1.9 are sources showing weak broad $H_{\alpha}$ components in their optical spectra. According to unification schemes, they are seen with an edge on inclination, similarly to type 2 Seyfert galaxies, but with slightly lower inclination angles.}
   {We aim at testing whether Seyfert 1.8/1.9 have similar properties at UV and X-ray wavelengths. }
   {We use the 15 Seyfert 1.8/1.9 in the V\'{e}ron Cetty and V\'{e}ron catalogue with public data available from the \emph{Chandra} and/or \emph{XMM--Newton} archives at different dates, with timescales between observations ranging from days to years. All the spectra of the same source were simultaneously fitted with the same model and different parameters were left free to vary in order to select the variable parameter(s). Whenever possible, short-term variations from the analysis of the X-ray light curves and long-term UV variations from the optical monitor onboard \emph{XMM--Newton} were studied. Our results are homogeneously compared with a previous work using the same methodology applied to a sample of Seyfert 2 \citep{lore2015}.}
   {X-ray variability is found in all 15 nuclei over the aforementioned ranges of timescales. The main variability pattern is related to intrinsic changes in the sources, which are observed in ten nuclei. Changes in the column density are also frequent, as they are observed in six nuclei,  and variations at soft energies, possibly related to scattered nuclear emission, are detected in six sources.
X-ray intra-day variations are detected in six out of the eight studied sources. Variations at UV frequencies are detected in seven out of nine sources.}
   {A comparison between the samples of Seyfert 1.8/1.9 and 2 shows that, even if the main variability pattern is due to intrinsic changes of the sources in the two families, these nuclei exhibit different variability properties in the UV and X-ray domains. In particular, variations in the broad X-ray band on short time-scales (days/weeks), and variations in the soft X-rays and UV on long time-scales (months/years) are detected in Seyfert 1.8/1.9 but not in Seyfert 2. Overall, we suggest that optically classified Seyfert 1.8/1.9 should be kept separated from Seyfert 2 galaxies in UV/X-ray studies of the obscured AGN population because their intrinsic properties might be different. }

   \keywords{ Galaxies: active -- X-rays: galaxies -- Ultraviolet: galaxies
               }

   \maketitle
%

\newcommand{\xmark}{\ding{55}}%
\newcommand{\cmark}{\ding{51}}%

\begin{table*}
\begin{center}
\caption{\label{properties} General properties of the sample galaxies.}
\begin{tabular}{lccccccccccc} \hline
\hline
Name    & RA & DEC &  Dist.$^1$  & N$_{Gal}$ & $m_V$ & Morph. & Seyfert & log$M_{BH}$    \\
 & (J2000)  & (J2000) & (Mpc) & ($10^{20}$ cm$^{-2}$) & & type & type & $M_{\odot}$ \\
(1) & (2) & (3) & (4) & (5) & (6) & (7) & (8) & (9)    \\  \hline
ESO\,540-G01 & 00 34 13.8 & -21 26 20 &      110.5 & 1.62 &    13.7  & SBc & 1.8 & - \\
ESO\,195-IG21 & 01  00 36.5 & -47 52  03 &     201.8 & 1.65 &    16.7  & - & 1.8 & - \\
ESO\,113-G10 & 01  05 17.0 & -58 26 13 &      104.1 & 2.95 &    14.6  & SBa & 1.8 & 6.85 \\
NGC\,526A  &  01 23 54.4 & -35  03 56 &  77.8 &      2.19 &     14.6 & S0 & 1.9 & 7.90 \\
MARK\,609 & 03 25 25.4 & -06  08 39 &          141.1 & 4.42 &    14.1 & S0 & 1.8 & - \\
NGC\,1365 & 03 33 36.4 & -36  08 24 &         18.0 & 1.35 &     13.0 & Sb & 1.8 & 7.54 \\
NGC\,2617 & 08 35 38.8 & -04  05 19 &          56.1 & 3.65 &     14.0  & SBc & 1.8 & 7.60 \\
MARK\,1218 & 08 38 11.1 & 24 53 45 &         116.6 & 3.54 &    14.1 & Sb & 1.8 & - \\
NGC\,2992  & 09 45 42.0 & -14 19 35 & 30.5 & 4.99 & 	      13.8 & Sa & 1.9 & 7.73 \\
POX\,52 & 12  02 56.8 & -20 56  03 &          87.3 & 4.03 &     17.2  & - & 1.8 & 5.14 \\
NGC\,4138  & 12  09 29.9 & 43 41  06 & 16.0 &     1.36 &        12.2 & S0-a & 1.9 & 7.30 \\
NGC\,4395 & 12 25 48.9 & 33 32 48 &         4.5 & 1.35 &      10.3 & Sm & 1.8 & 4.82 \\
NGC\,4565  & 12 36 20.6 & 25 59 11 &        12.1 & 1.30 &     12.4 & Sb & 1.9 & 6.30 \\
MARK\,883  & 16 29 52.8 & 24 26 39 &        155.7 & 3.97 &    14.4 & I &1.9 & 7.28  \\
IRAS\,20051-1117 & 20  07 51.4 & -11  08 35 & 128.9 & 6.57 &    14.0  & - & 1.9 & 7.11 \\
\hline
\end{tabular}
\caption*{  (Col. 1) Name, (Col. 2) right ascension, (Col. 3) declination, (Col. 4) distance, (Col. 5) galactic absorption, (Col. 6) aparent magnitude in the Johnson filter V from \cite{veron2010}, (Col. 7) galaxy morphological type from Hyperleda, (Col. 8) AGN type as in \cite{veron2010}, and (Col. 9) black-hole mass on
  logarithmical scale, determined using the correlation between
  stellar velocity dispersion (from HyperLeda) and black-hole mass
  \citep{tremaine2002}, or obtained from the literature otherwise
  (ESO\,113-G10 from \cite{cackett2013}, NGC\,526A from \cite{vasudevanfabian2009}, NGC\,2617 from \cite{shappee2014}, MARK\,883 from \cite{benitez2013}, and IRAS\,20051-1117 from \cite{wangzhang2007}.)}
\end{center}
\vspace*{-0.5cm}
\footnote*{All distances are taken from the NED and correspond to the average redshift-independent distance estimates.}
\end{table*}

\section{\label{intro}Introduction}

Active galactic nuclei (AGN) are though to be powered by accretion of matter onto the supermassive black hole (SMBH) that resides in the center of the galaxies \citep{rees1984}. Historically, these nuclei have been classified as type 1 when broad Balmer permitted lines (full-width at half maximum (FWHM)$\sim$1000-20000 km/s) are detected in their optical spectra, while they are classified as type 2 when detecting only narrow lines (FWHM$\sim$300-1000 km/s). Using the relative intensity of broad and narrow lines, the nuclei can also be classified as type 1.2, 1.5, 1.8, or 1.9 AGN (intermediate Seyferts), the latter having the weaker broad component \citep[e.g.,][]{osterbrock1977,osterbrock1993}. In particular, the optical spectra of Seyfert 1.8 are characterized by strong narrow emission lines combined with weak broad $H_{\alpha}$ and $H_{\beta}$ emission lines, whereas Seyfert 1.9 present the narrow lines but only a weak broad $H_{\alpha}$ emission line  \citep{osterbrock1981}.

The detection of broad components in polarized light of type 2 sources set the unified model of AGN \citep{lawrence1987,antonucci1993,urrypadovani1995,moran2000}. Under this scenario, the different properties observed in AGN can be explained by orientation effects, i.e., they are the same kind of source observed at different angles. The cornerstone of this model is a dusty structure (often simplified as a torus) that surrounds the SMBH, which plays a fundamental role as it is responsible for obscuring the broad line region (BLR) where the broad lines are created.
In support of this model, X-ray observations have shown that type 2 sources are more obscured than type 1s, whereas type 1.8 and 1.9 AGN are less absorbed than strictly type 2s \citep{risaliti1999}.

X-rays are indeed a powerful tool for the comprehension of AGN as they are capable to reach closer to the SMBH than other wavelengths. At these energies the absorbing column density, $N_H$, is used to classify sources as unobscured (type 1) when $N_H$ is below $\sim 10^{22} cm^{-2}$ and obscured (type 2) sources for larger values. For $N_H$ values larger than $1.5 \times 10^{24} cm^{-2}$, the sources are classified as Compton-thick \citep{maiolino1998}. Sometimes transitions from Compton-thick to Compton-thin (or vice versa) have been observed; these are known as changing-look sources according to the original nomenclature by \cite{matt2003b}.

Variability is one of the properties characterizing AGN, a highly valuable tool for the comprehension of their physical structure \citep{bradley1997,netzer2013}.
The first systematic studies of AGN showed that
short-term X-ray variability (from hours to days) is common in type 1s, but not in type 2s, while long-term
(from months to years) variations are common in both \citep[e.g.,][]{nandra1997,turner1997,vaughan2005}. Nowadays we believe that the X-ray variations might be related to intrinsic changes
of the nuclear source \citep[e.g.,][]{uttley2005a,uttley2007,parker2015}, or to absorbing clouds that intersect the line of sight to the observer \citep[e.g.,][]{risaliti2007}. These changes can be studied by modelling the X-ray spectrum of
AGN, whose continuum is dominated by a power-law component extending up to a cut-off at energies $\ge$100~keV \citep[e.g.,][]{zdziarski1995,guainazzi2005b,fabian2015}. Changes in the power law might indicate a change in the accretion disk or the 
X-ray corona, while changes in the absorption may be related to clouds in our line of sight, more likely in the BLR, the torus, or the boundary between them \citep{risaliti2002b, risaliti2005b, risaliti2011,braito2013,markowitz2014}.

Due to their similar optical and X-ray spectra, it is usually assumed that optically classified Seyfert 1.2 and 1.5 behave more likely type 1s, whereas types 1.8 and 1.9 behave as type 2s. Indeed, many studies  aiming at analyzing the properties of type 2 sources have included Seyfert 1.8/1.9 in their samples \citep[e.g.,][]{guainazzi2001,risaliti2002,akylas2009}.

However, it is not clear whether the properties of Seyfert 1.8/1.9 are directly related to differences in the nuclear continuum or to an obscurer in our line of sight, since weaker broad lines may be produced by a lower ionizing continuum flux or by reddening from the BLR or the host galaxy \citep{osterbrock1981, goodrich1995, trippe2011}.
Through the analysis of variability, we are able to differentiate between changes in the accretion state and the configuration of the clouds. The main purpose of the present work is to 
homogeneously compare the variability properties of optically classified Seyfert 1.8/1.9 and Seyfert 2. The ultimate goal of our study is to understand the physical origin of the phenomenological differences between Seyfert 1.8/1.9 and Seyfert 2 in the optical, UV and X-ray. We employ X-ray variability as gauge in this paper.
This study is part of a systematic analysis of the variability properties of nearby AGN; by now we have analyzed the properties of a sample of optically classified low ionization nuclear emission line regions \citep[LINERs,][]{lore2013,lore2014}, and a sample of Seyfert 2 \citep{lore2015}. A comparison between the properties of LINERs and Seyfert 2 was carried out in \cite{lore2016}.

The paper is organized as follows. The sample selection is presented in Sect. \ref{sample}. The data reduction and the methodology are explained in Sect. \ref{reduction} and \ref{method}. The results of the analysis are presented in Sect. \ref{results}, which are discussed in Sect. \ref{discusion}. Finally, the conclusions of this study are summarized in Sect. \ref{conclusion}.


\section{\label{sample}Sample and data}

We used the 13$^{th}$ edition of the V\'{e}ron-Cetty and V\'{e}ron catalogue \citep{veron2010}, which contains quasars and AGN. We selected nearby sources located at redshifts below 0.05\footnote{The redshift of 0.05 corresponds to a distance of d=214.3 Mpc (using $H_0 = 70 \hspace*{0.1cm} km \hspace*{0.1cm} s^{-1}$). The limit on distance was chosen to be the same as in \cite{lore2015} for Seyfert 2.} that were classified as Seyfert type 1.8 and 1.9. In this way we selected 142 Seyfert 1.8 and 189 Seyfert 1.9.

We used the HEASARC database\footnote{http://heasarc.gsfc.nasa.gov/} to search for public data in the \emph{Chandra} and/or \emph{XMM--Newton} archives of these sources. In order to study X-ray variability, we selected those sources with more than one observation with these satellites. This included 12 Seyfert 1.8 and another 12 Seyfert 1.9. 

We restricted further our sample to sources whose spectra have a minimum of 400 number counts in the 0.5--10 keV energy band (to use $\chi^2$-statistics) and to not be affected by a pileup fraction larger than 10\%. This leaves us with nine Seyfert 1.8 and seven Seyfert 1.9. We removed MARK\,1018 from the sample because, although being classified as a Seyfert 1.9 by \cite{osterbrock1981} using optical data, \cite{cohen1986} reported variations from Seyfert 1.9 to Seyfert 1 also using optical data, and remained as a Seyfert 1 at least up to 2007 \citep{trippe2010}.

Therefore, the final sample includes nine Seyfert 1.8 and six Seyfert 1.9. The sample properties are presented in Table \ref{properties}.

It is worth to notice that a caveat in the analysis could be related to the non simultaneity of the X-ray data with the optical spectroscopic data used for the optical classification of the sources. 
Unfortunately, the only case in our sample where the X-ray and optical data were obtained at close epochs is NGC\,2617, where the X-ray data were taken in 2013, while it was reclassified as a Seyfert 1 using optical spectroscopy gathered in 2014, confirming that variability might be an important issue.

\section{\label{reduction}Data reduction}

\subsection{Chandra data}

\emph{Chandra} observations were obtained from the ACIS instrument
\citep{garmire2003}. Data reduction and analysis were carried out in a
systematic, uniform way using CXC Chandra Interactive Analysis of
Observations (CIAO\footnote{http://cxc.harvard.edu/ciao4.4/}), version
4.6. Level 2 event data were extracted by using the task { \sc
  acis-process-events}.  Background flares were cleaned using the task
{ \sc
  lc\_clean.sl}\footnote{http://cxc.harvard.edu/ciao/ahelp/lc\_clean. html},
which calculates a mean rate from which it deduces a minimum and
maximum valid count rate and creates a good time intervals file.

Nuclear spectra were extracted from a circular region centered on the
positions given by NED\footnote{http://ned.ipac.caltech.edu/}. We
chose circular radii, aiming to include all possible photons, while
excluding other sources or background effects. The radii are in the
range between 2-4$\arcsec$ (see
Table~\ref{obsSey}). The background was extracted from circular
regions in the same chip that are free of sources and close to the
object.

For the source and background spectral extractions, the {\sc
  dmextract} task was used. The response matrix file (RMF) and
ancillary reference file (ARF) were generated for each source region
using the {\sc mkacisrmf} and {\sc mkwarf} tasks,
respectively. Finally, the spectra were binned to have a minimum of 20
counts per spectral bin using the {\sc grppha} task (included in {\sc
  ftools}), to be able to use the $\chi^2$ statistics, as customary in X-ray spectroscopy.

\subsection{\label{xmm}XMM-Newton data}

\emph{XMM}-Newton observations were obtained with the EPIC pn camera
\citep{struder2001}. The data were reduced in a systematic, uniform
way using the Science Analysis Software
(SAS \footnote{http://xmm.esa.int/sas/}), version 14.0.0. First,
Good-Time Intervals (GTIs) were selected using a method that maximizes the
signal-to-noise ratio of the net source spectrum by applying a
different constant count rate threshold on the single events, E $>$ 10
keV field-of-view background light curve.  We extracted the spectra of
the nuclei from circles of 20--35$\arcsec$ radius
centered on the positions given by NED, while the background spectra were
extracted from circular regions using an algorithm that automatically
selects the best area - and the closest to the source - that is free of
sources. This selection was manually checked to ensure the best
selection for the backgrounds.

Source and background spectra were extracted with the {\sc
  evselect} task. The response matrix files (RMF) and the ancillary
response files (ARF) were generated using the {\sc rmfgen} and {\sc
  arfgen} tasks, respectively. To be able to use the $\chi^2$
statistics, the spectra were binned to obtain at least 20 counts per
spectral bin using the {\sc grppha} task.

\subsection{ \label{sigma} Light curves}

Light curves in three energy bands (0.5--2.0 keV, 2.0--10.0 keV, and
0.5--10 keV) for the source and background regions as defined above
were extracted using the {\sc dmextract} task (for \emph{XMM--Newton})
and {\sc evselect} task (for \emph{Chandra}) with a 1000s bin. To be
able to compare the variability amplitudes in different light curves
of the same object, only those observations with a net exposure time
longer than 30 ksec were taken into account. For observations longer than 40 ksec,
the light curves were divided into segments of 40 ksec, so in some
cases more than one segment of the same light curve can be extracted.
Our light curves are occasionally affected by high particle background events (“flares”), whose flux dominates the observed count rates. We decided to remove these intervals from the source background-subtracted light curves due to their poor signal-to-noise ratio that could affect the estimate of the normalized effect variance (cf. Sect. \ref{short}). High particle background flux intervals were identified using the same algorithm described in Sect. \ref{xmm}. As the fraction of high particle background intervals is small, our procedure does not significantly affect the results discussed in this paper \citep{vaughan2003}.
We notice that after excluding these events, the
exposure time of the light curve could be shorter, thus we recall that
only observations with a net exposure time longer than 30 ksec were
used for the analysis. The light curves are shown in Appendix
\ref{lightcurves}. We recall
that the values of the continuum (median value of the count rate) and dashed (1$\sigma$ standard deviation) lines are used only for visual inspection of the data and
not as estimators of the variability (as in \citealt{lore2014}).

\section{\label{method}Methodology}

The methodology used in this work is presented in \cite{lore2013}. Here we review the most important aspects but we refer the reader to this paper for details on the analysis. 

\subsection{\label{indiv}Individual spectral analysis}

The first step is to select a model to fit all the data of the same source simultaneously. For that purpose, we used five different models that were fitted to each spectrum individually. We notice that more complex models were also tested but they were not required by the data. The models are as follows:

\begin{itemize}
\item[$\bullet$] 
PL: A single power law representing the continuum of
  a non-stellar source. The empirical model is

$e^{N_{Gal} \sigma (E)} \cdot e^{N_{H} \sigma (E(1+z))}[N_{H}] \cdot Norm e^{-\Gamma}[\Gamma, Norm].$
\vspace*{0.2cm}
\item[$\bullet$] ME: The emission is dominated by hot diffuse gas, i.e., a thermal plasma. A MEKAL (in XSPEC) model is used to fit the spectrum. The model is

$e^{N_{Gal} \sigma (E)} \cdot e^{N_{H} \sigma (E(1+z))}[N_{H}] \cdot MEKAL[kT, Norm].$
\vspace*{0.2cm}

\item[$\bullet$] 2PL: In this model the primary continuum is an
  absorbed power law representing the non stellar source, while the
  soft energies are due to a scattering component that is represented
  by another power law. Mathematically the model is explained as

$e^{N_{Gal} \sigma (E)} \big( e^{N_{H1} \sigma (E(1+z))}[N_{H1}] \cdot Norm_1 e^{-\Gamma}[\Gamma, Norm_1] + e^{N_{H2} \sigma (E(1+z))}[N_{H2}] \cdot Norm_2 e^{-\Gamma}[\Gamma, Norm_2]\big)$.
\vspace*{0.2cm}

\item[$\bullet$] MEPL: The primary continuum is represented by an
  absorbed power law, but at soft energies a thermal plasma dominates
  the spectrum. Empirically it can be described as

$e^{N_{Gal} \sigma (E)} \big(e^{N_{H1} \sigma (E(1+z))}[N_{H1}] \cdot MEKAL[kT, Norm_1] + e^{N_{H2} \sigma (E(1+z))}[N_{H2}] \cdot Norm_2 e^{-\Gamma}[\Gamma, Norm_2]\big)$.
\vspace*{0.2cm}

\item[$\bullet$] ME2PL: This is same model as MEPL, but an additional
  power law is required to explain the scattered component at soft
  energies, so mathematically it is

$e^{N_{Gal} \sigma (E)} \big( e^{N_{H1} \sigma (E(1+z))}[N_{H1}] \cdot Norm_1 e^{-\Gamma}[\Gamma, Norm_1] + MEKAL[kT] + e^{N_{H2} \sigma (E(1+z))}[N_{H2}] \cdot Norm_2 e^{-\Gamma}[\Gamma, Norm_2]\big)$.
\vspace*{0.2cm}

\end{itemize}

\noindent In the equations above, $\sigma (E)$ is the photo-electric
cross-section, $z$ is the redshift, and $Norm_i$ are the
normalizations of the power law and/or the thermal component. For
each model, the parameters that vary are written in brackets.  The
Galactic absoption, $N_{Gal}$, is included in each model and fixed to
the predicted value (Col. 5 in Table \ref{properties}) using the tool
{\sc nh} within {\sc ftools} \citep{dickeylockman1990, kalberla2005}.
Even if not included in the mathematical expressions above, all the models
include three narrow Gaussian lines to take the iron lines at 6.4 keV
(FeK$\alpha$), 6.7 keV (FeXXV), and 6.95 keV (FeXXVI) into account.

The $\chi^2/d.o.f$ and F-test were used to select the simplest model
that represents the data best. We considered an improvement of the spectral fit significant when the F-test results in a value lower than $10^{-5}$.

\subsection{\label{simult} Simultaneous spectral analysis}

We determined the best-fit model for each individual observation using the procedure described in Sect. \ref{indiv}.
As a baseline model we used the one corresponding to the individual observation with the largest count number, and we checked that it matches the best fit model of the remaining spectra of the same source\footnote{Note that for NGC\,4138 we used the PL model because the \emph{Chandra} spectrum did not have counts below 2 KeV, therefore the analysis was performed in the 2--10 keV band.}. This model was applied to all the observations of the same source simultaneously with its parameters linked amongst them -- note that the values of the parameters are able to change from the initial values given in the baseline model. If this fit (SMF0) resulted in a good fit (see below), we considered the source as non-variable.

When SMF0 did not give a good resut, the next step was to let different parameters in the model vary one-by-one (SMF1). These parameters are the column densities at soft ($N_{H1}$) and hard ($N_{H2}$) energies, the temperature ($kT$), the spectral index ($\Gamma$), and the normalizations at soft ($Norm_1$) and hard ($Norm_2$) energies. 

When SMF1 failed to be a good fit, we also tested to vary two parameters at the same time (SMF2), and also three parameters (SMF3) were needed in one case. 

Each `next step' (e.g., SMF1 versus SMF0) was always tested in order to confirm an improvement of the spectral fit. A $\chi^2_r$ in the range between 0.9--1.5 --and as close as possible to the unity and an F-test value lower than $10^{-5}$ were the criteria to accept a new step. If different models at a given step yielded a significant improvement with respect to the previous step, we chose the model corresponding to the lowest $\chi^2_r$.

Whenever possible, this analysis was applied to observations of the same satellite. However, in some cases there was only one observation per instrument available. In order to compare the data extracted from different apertures, we fit the extranuclear emission in the annular region in the \emph{Chandra} image between the \emph{Chandra} aperture around the nucleus and the \emph{XMM--Newton} aperture (see Table \ref{obsSey}) using the same procedure described in Sect. \ref{indiv}. This allowed us to define the best-fit model of the \emph{Chandra} extranuclear emission. This model was included in the spectral analysis of the \emph{XMM--Newton} data, when comparing with \emph{Chandra} data. This procedure was applied whenever \emph{XMM--Newton} and \emph{Chandra} data were available.

\subsection{Flux variability}

The luminosities in the soft and hard X-ray energy bands were computed
using XSPEC for both the fits of the individual observations, as well as for the simultaneous fit of all the observations together. The distances were taken from NED, corresponding to
the average redshift-independent distance estimate for each object,
when available, or to the redshift-estimated distance otherwise;
distances are listed in Table \ref{properties}.

When data from the optical monitor (OM) onboard \emph{XMM}--Newton
were available, UV luminosities (simultaneously to X-ray data) were
estimated in the available filters.  We recall that UVW2 is centered
at 1894$\AA$ (1805-2454) $\AA$, UVM2 at 2205$\AA$ (1970-2675) $\AA$,
and UVW1 at 2675$\AA$ (2410-3565) $\AA$.  We used the OM observation
FITS source lists
(OBSMLI)\footnote{ftp://xmm2.esac.esa.int/pub/odf/data/docs/XMM-SOC-GEN-ICD-0024.pdf}
to obtain the photometry. When OM data were not available, we searched
for UV information in the literature. We note that in this case, the
X-ray and UV data might not be simultaneous (see Appendix
\ref{indivnotes}).

We assumed an object to be variable when the square root of the
squared errors was at least three times smaller than the dynamical range covered by the luminosities \citep[see][for details]{lore2014}.

\subsection{\label{short}Short-term variability}

Initially, we assumed a constant count rate for segments of 30-40 ksec
of the observation in each energy band and calculated
$\rm{\chi^2/d.o.f}$. We considered the
source as a variable candidate if the count rate differed from
the average by more than 3$\rm{\sigma}$ (or 99.7\% probability).

Secondly, we calculated the normalized excess
variance, $\rm{\sigma_{NXS}^2}$, for each light curve segment with
30-40 ksec following prescriptions in \cite{vaughan2003} \citep[see
  also][]{omaira2011a,lore2014}. We recall that $\rm{\sigma_{NXS}^2}$
is related to the area below the power spectral density (PSD) shape.

When $\rm{\sigma_{NXS}^2}$ was negative or compatible with zero within
the errors, we estimated the 90\% upper limits using Table 1 in
\cite{vaughan2003}. We assumed a PSD slope of -1, the upper limit from
\cite{vaughan2003}, and we added the value of
1.282err($\rm{\sigma_{NXS}^2}$) to the limit to account for Poisson
noise.  For a number of segments, N, obtained from an individual light
curve, an upper limit for the normalized excess variance was
calculated. When N segments were obtained for the same light curve and
at least one was consistent with being variable, we calculated the
normalized weighted mean and its error as the weighted variance.

We considered short-term variations for $\rm{\sigma_{NXS}^2}$
detections above 3$\sigma$ of the confidence level.

\subsection{\label{thick}Compton thickness}

We tested the possibility of some sources being so heavily absorbed that their spectra can be completely reflected below 10 keV, i.e., Compton-thick sources. Since the Compton-thick column densities cannot be directly measured at the energies analyzed here, the following indirect indicators (using X-ray and [O III] data) are taken to classify these sources: $\Gamma < 1$. 
EW(FeK$\alpha)>$ 500 eV, and
$F(2-10 keV)/F_{[O III]}$ $<$ 1 \citep{ghisellini1994, bassani1999, panessabassani2002}. Where $\Gamma$ and EW(FeK$\alpha)$ were obtained from individual spectral fits in the 3-10 keV energy band using the PL model, the extinction-corrected [O III] fluxes were obtained from the literature (and corrected when needed following \citealt{bassani1999}), and the hard X-ray luminosities, $L(2-10 keV)$, from the individual fits were used (see Table \ref{lumincorrSey}) for the calculation.

We considered that a source is a \emph{\emph{Compton}}-thick candidate
when at least two of the three criteria above were met. Otherwise, the
source is considered to be a \emph{\emph{Compton}}-thin
candidate. When different observations of the same source result in
different classifications, the object was considered to be a
changing-look candidate.

{\tiny
\begin{table*}
\begin{center}
\caption{\label{variab} Results of the variability analysis. } 
\begin{tabular}{lccccccccc} \hline \hline
Name  & log ($L_{soft}$) & log ($L_{hard}$) &  log ($R_{Edd}$) &  \multicolumn{3}{c}{Long-term variability}  & $\Delta$T$_{max}$ & Short & UV  \\ \cline{5-7} 
 &  (0.5-2 keV) & (2-10 keV) & & SMF0 &  SMF1 & SMF2/3 & (Years) & term & Variab. \\ 
(1) & (2) & (3)          & (4) & (5) & (6) & (7) & (8) & (9) & (10)   \\ \hline
ESO\,540-G01 (X,C) &   41.53$\pm$0.13  & 41.72$\pm$0.13 &  - & MEPL & $Norm_2$ & & 1 & - & - \\
                   & 40\% & 38\% &   &      &   74\% \\ 
ESO\,195-IG21 (X,C)  &   42.54$\pm$0.37  & 43.03$\pm$0.37  & - & MEPL & $Norm_2$ & $Norm_1$ & 4 & - & - \\
                   & 98\% & 97\% &   &      &   91\% &   98\% \\ 
ESO\,113-G10 (X) &  43.07$\pm$0.05  & 42.70$\pm$0.05 &  -0.74 & ME2PL & $Norm_1$  &  & 4 & TSH & W2 \\
                   & 17\% & 17\% &       &  &   40\% \\                    
NGC\,526A (X) &  42.91$\pm$0.10  & 43.32$\pm$0.09   & -1.16 & 2PL & $Norm_2$ & - & 11 & TSH & W1 \\
                   & 48\% & 46\% &       &  &   48\% \\ 
MARK\,609 (X) &    42.55$\pm$0.04 & 42.69$\pm$0.04 &  - & 2PL & $Norm_1$ & & 5 & - & - \\
                   & 13\% & 13\% &       &  &   22\% \\
NGC\,1365 (X)$^{CL}$ &   41.15$\pm$0.35  & 42.18$\pm$0.42 &  -1.95 & ME2PL & $N_{H2}$ & $N2$/$N1$* & 10  & TSH & W1, M2 \\
                   & 81\% & 24\% &       &  &   68\% &   33/35\%  \\ 
\hspace*{1.2cm}   (X,C) &  41.80$\pm$0.38 & 41.73$\pm$0.27 &  & ME2PL & $N_{H2}$ & $Norm_2$ & 2 & - & - \\
                   & 99\% & 77\% &   &      &   37\% &   30\% \\ 
NGC\,2617 (X) &   43.24$\pm$0.15  & 43.25$\pm$0.15 &  -0.94 & 2PL & $Norm_1$ & $N_{H2}$ & 0.1 & TS & W1 \\
                   & 46\% & 45\% &   &      &   59\% &   30\% \\
MARK\,1218 (X)  &   41.87$\pm$0.22  & 42.56$\pm$0.21 &  - & PL & $Norm$ & & 0.08 & - & No \\
                   & 64\% & 64\% &   &      &   63\% \\ 
NGC\,2992 (X)$^{CL}$ &   41.57$\pm$0.23  & 42.03$\pm$0.22 &  -2.30 & 2PL & $N_{H2}$ & $Norm_2$ & 3 & TH & M2 \\
                   & 19\% & 19\% &       &  &   5\% &   21\% \\ 
POX\,52 (X,C) &    41.89$\pm$0.01  & 41.75$\pm$0.02 &  0.18 & ME2PL & $N_{H2}$ & $N_{H1}$ & 1 & No & - \\
                   & 2\% & 7\% &   &      &   44\% & 100\% \\ 
NGC\,4138 (X,C) &   - & 41.53$\pm$0.07 &  -2.42 & PL** & $Norm$ & & 2 & - & - \\
                   &  & 21\% &       &  &   98\% \\ 
NGC\,4395 (C) &   39.50  & 39.94 &  -1.15 & ME2PL & $N_{H2}$ & & 0.003 & - & W1 \\
                   & 0\% & 0\% &   &     &   31\% \\ 
\hspace*{1.5cm}   (X) &  39.69$\pm$0.06  & 40.28$\pm$0.40 &    & ME2PL & $N_{H2}$ & $Norm_2$ & 12 & TSH & -  \\
                   & 15\% & 13\% &   &      &   20\% &   88\% \\ 
\hspace*{1.2cm}   (X,C)  & 39.78$\pm$0.21 & 40.21$\pm$0.22 &  & ME2PL & $Norm_2$ & & 2 & - & - \\
                   & 61\% & 65\% &   &      &   93\% \\              
NGC\,4565 (X,C) &  39.51$\pm$0.03  &39.65$\pm$0.07  &  -4.63 & PL & $N_H$ & & 2 & No & - \\
                   & 11\% & 21\% &       &  &   48\% \\ 
MARK\,883 (X) &  42.42$\pm$0.09  & 42.71$\pm$0.08 &  -1.15 & PL & $Norm$ & & 4 & - & W1, W2 \\
                   & 28\% & 28\% &       &  &   28\% \\ 
IRAS\,20051-1117 (X) &  42.39$\pm$0.09  & 42.53$\pm$0.09 &  -1.16 & PL & $Norm$ & & 0.5 & - & No \\
                   & 29\% & 29\% &   &      &   29\% \\ 
\hline
\end{tabular}
\caption*{{\bf Notes.} (Col. 1) Name, and the
  instrument (C: \emph{Chandra} and/or X: \emph{XMM}--Newton) in
  parenthesis ($^{CL}$ refer to changing-look candidates); (Cols. 2 and 3)
  logarithm of the soft (0.5--2 keV) and hard (2--10 keV) X-ray
  luminosities, where the mean was calculated for variable objects,
  and percentages in flux variations;
  (Col. 4) Eddington ratio, $L_{bol}/  L_{Edd}$, calculated from
  \cite{eracleous2010b} using $L_{bol}=33L_{2-10 keV}$; (Col. 5) best
  fit for SMF0; (Col. 6) parameter varying in SMF1, with the
  percentage of variation; (Col. 7) parameter varying in SMF2 and SMF3 (for NGC\,1365, *N1=$Norm_1$, and N2=$Norm_2$), with
  the percentage of variation; (Col. 8) the sampling timescale,
  corresponding to the difference between the first and the last
  observation. The percentages correspond to this $\Delta$T$_{max}$; (Col. 9) short-term variations in the total (T), soft (S), and/or hard (H) energy bands; and (Col. 10) filters where variations are detected at UV frequencies with the OM. A `-' means that data were not available, while `No' means that variations were not detected.
  
    ** Note that the \emph{XMM--Newton} data of NGC\,4138 is best fitted by the ME2PL model, but \emph{Chandra} data does not have counts below 2 keV, thus the PL model is used for the simultaneous fit.}

\end{center}
\end{table*} }
\normalsize

\begin{figure*}
\centering
{\includegraphics[width=0.30\textwidth]{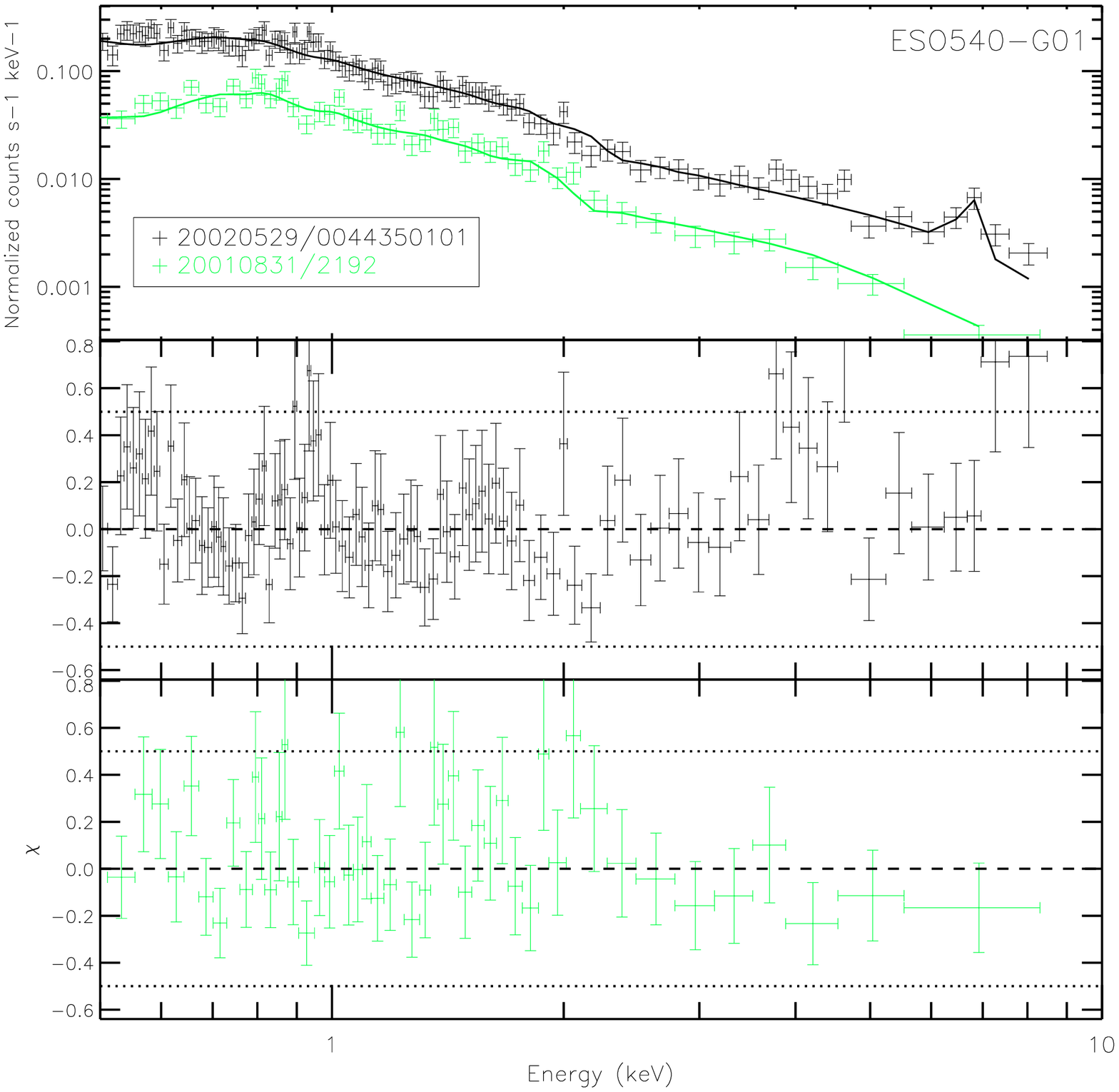}}
{\includegraphics[width=0.30\textwidth]{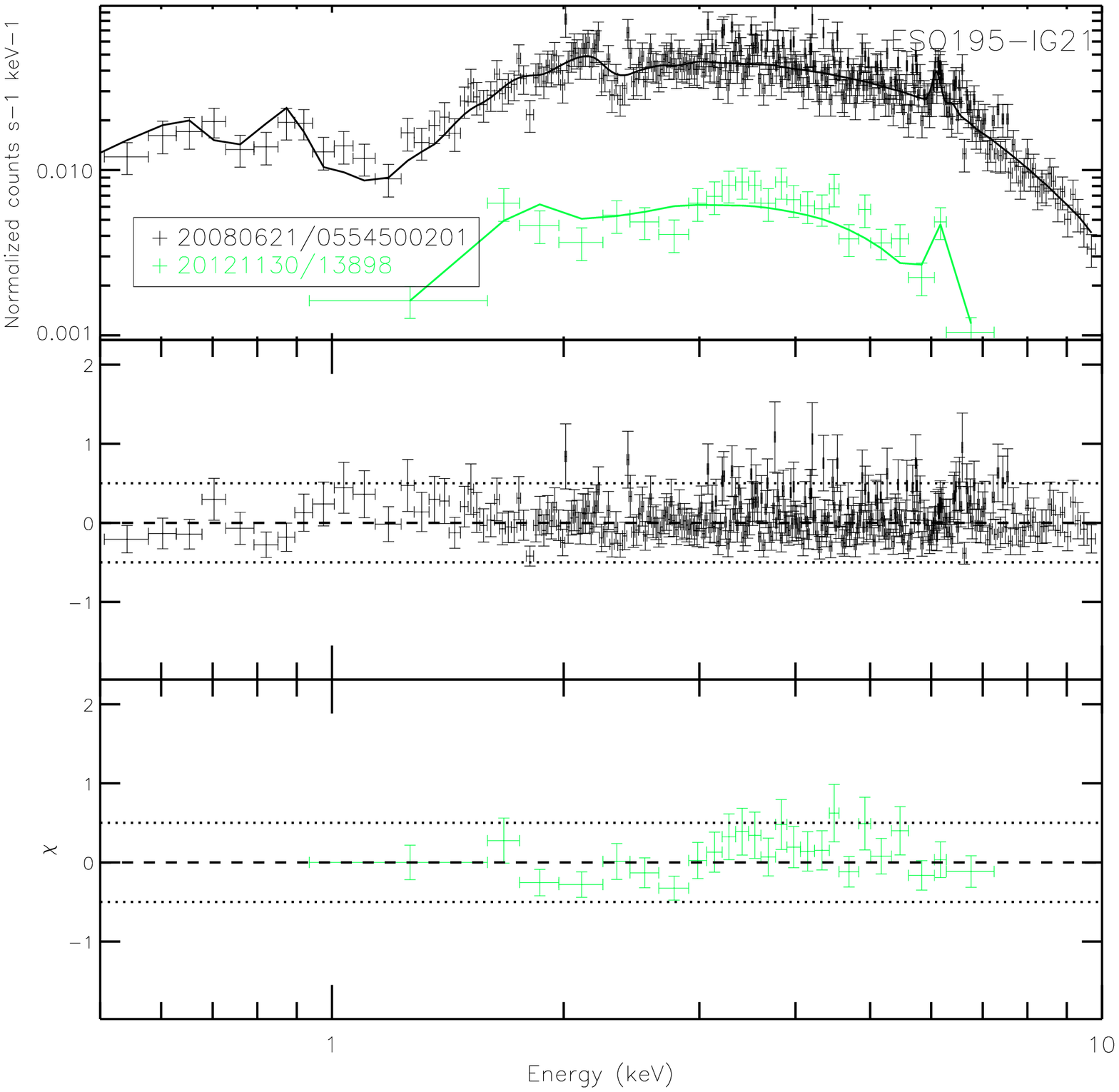}}
{\includegraphics[width=0.30\textwidth]{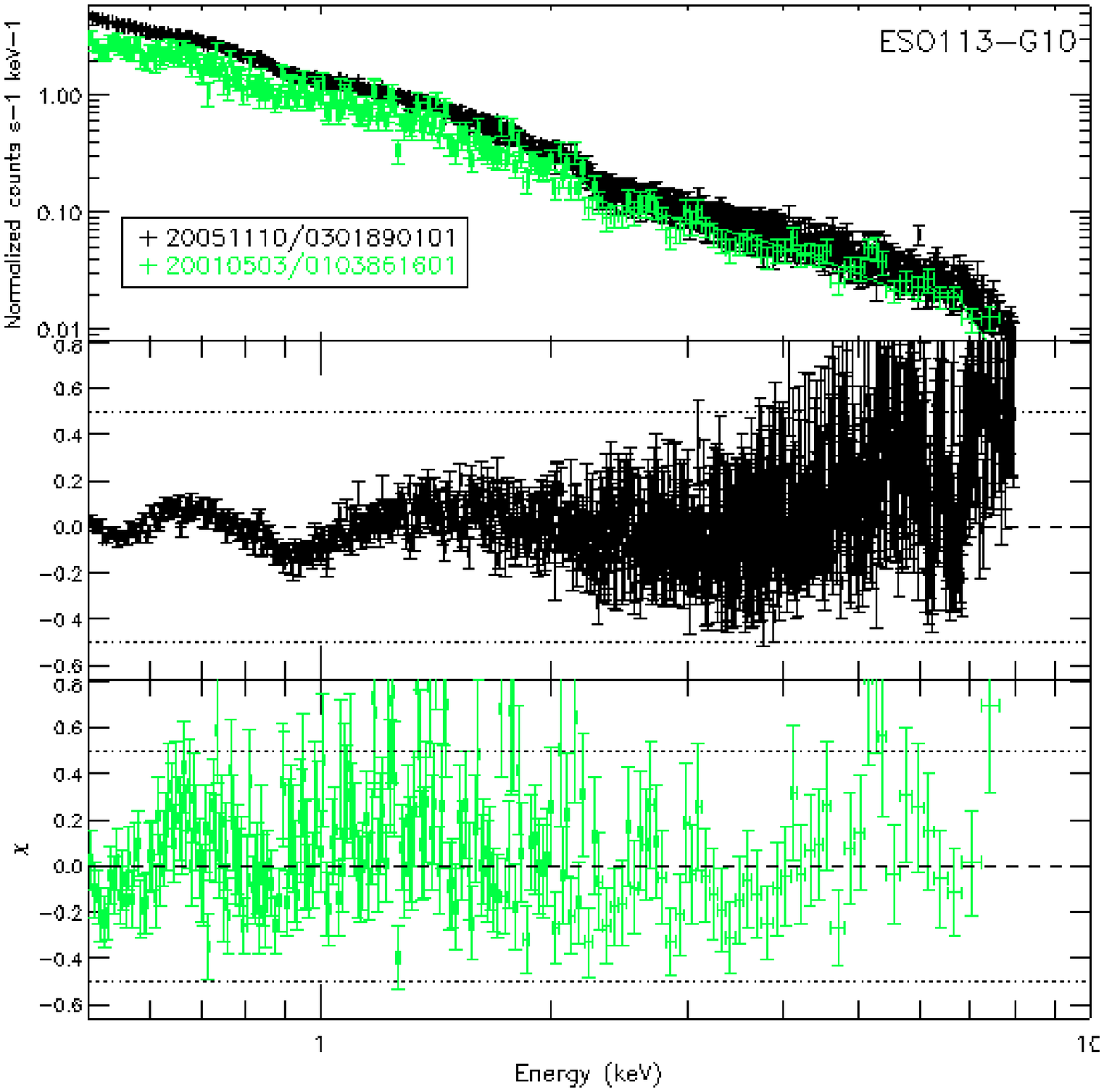}}

{\includegraphics[width=0.30\textwidth]{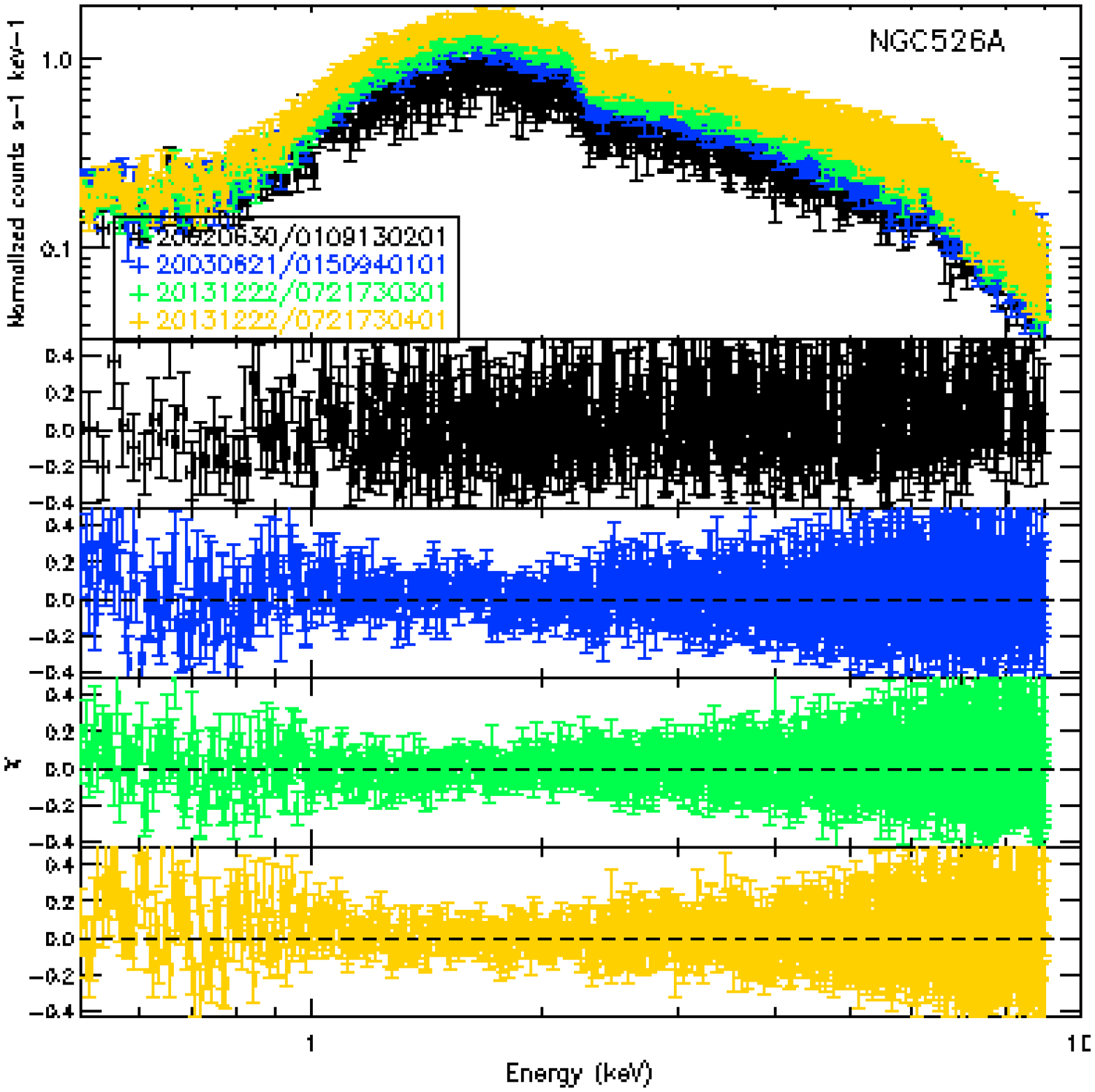}}
{\includegraphics[width=0.30\textwidth]{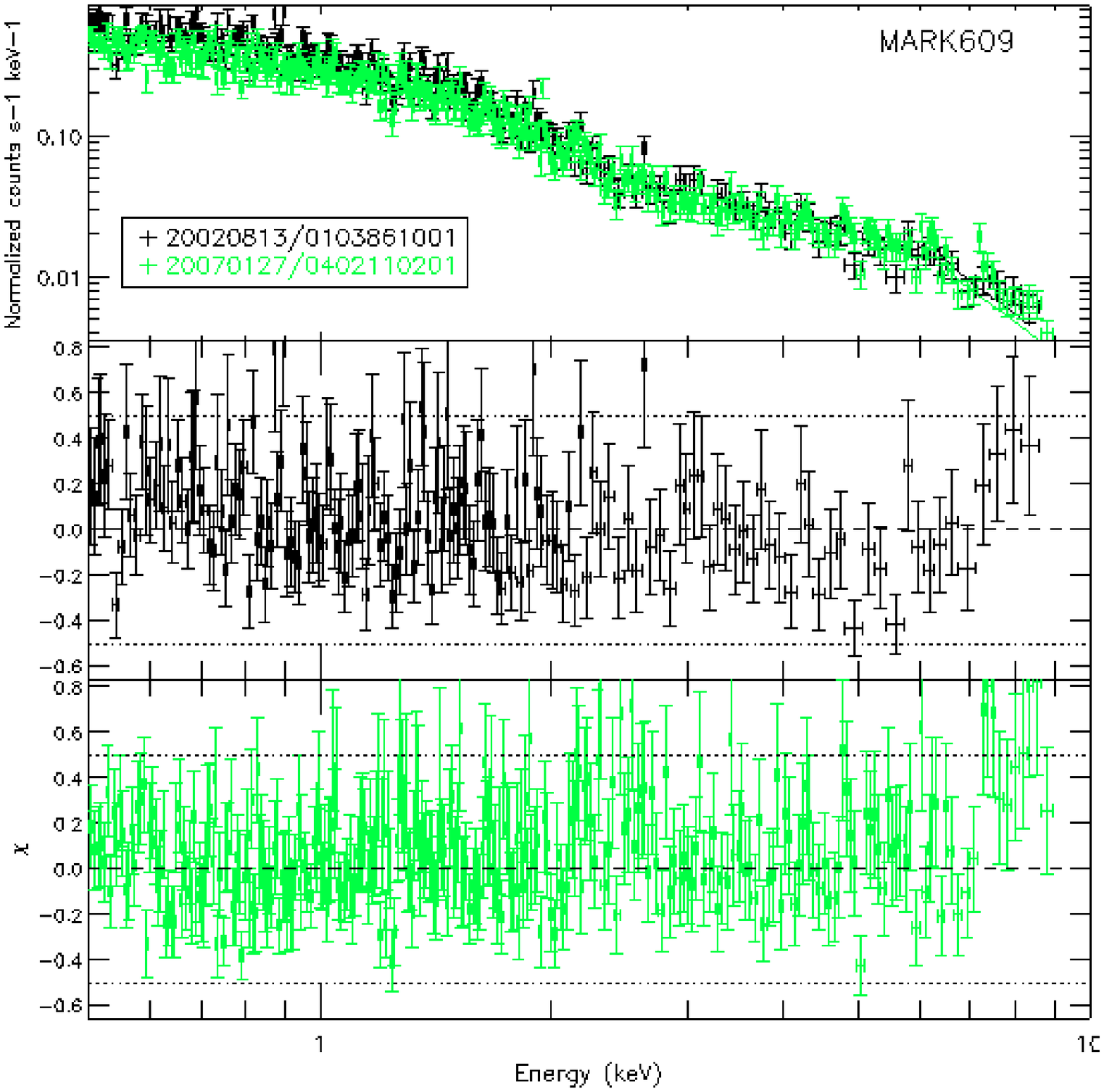}}
{\includegraphics[width=0.30\textwidth]{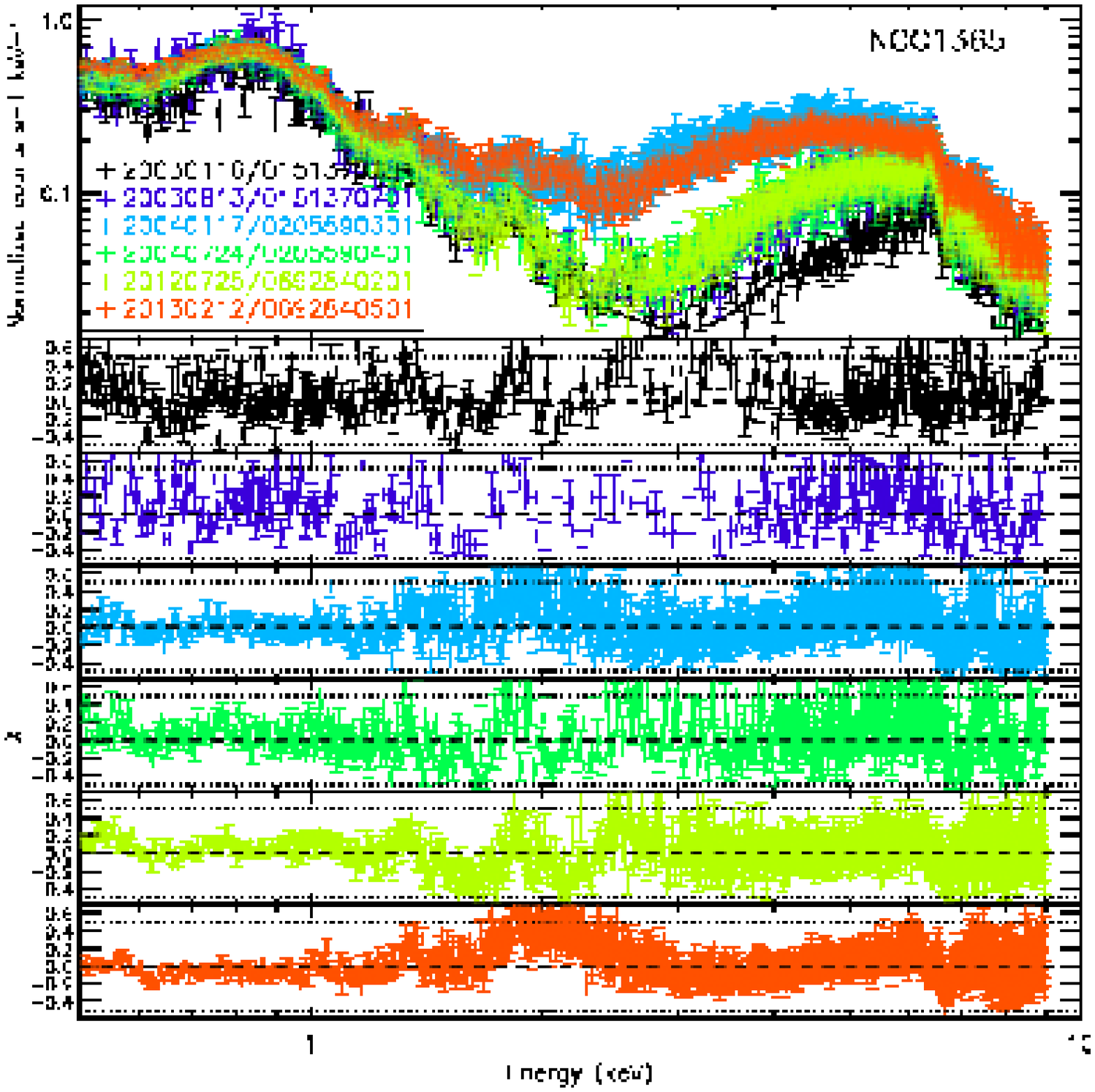}}

{\includegraphics[width=0.30\textwidth]{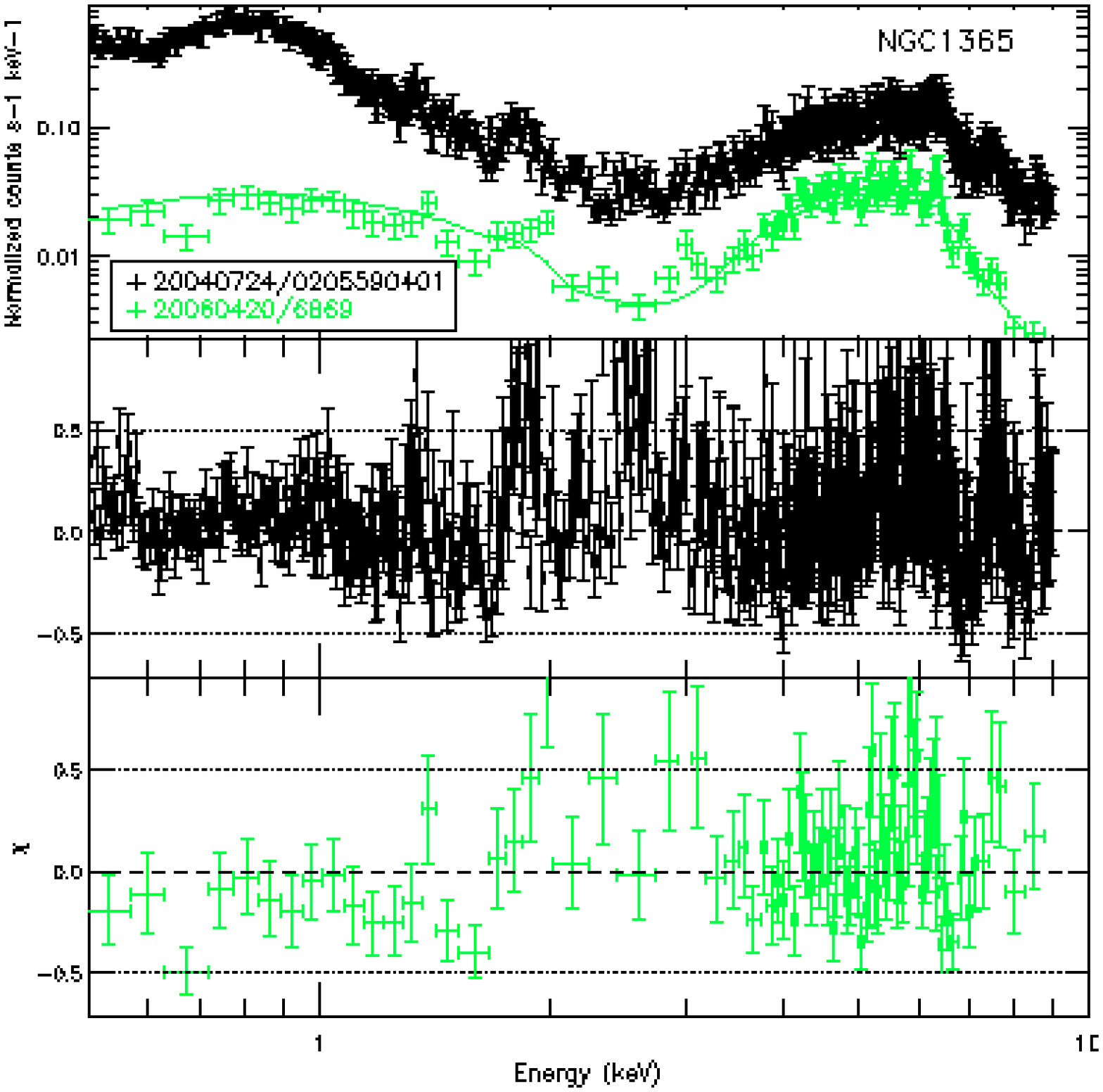}}
{\includegraphics[width=0.30\textwidth]{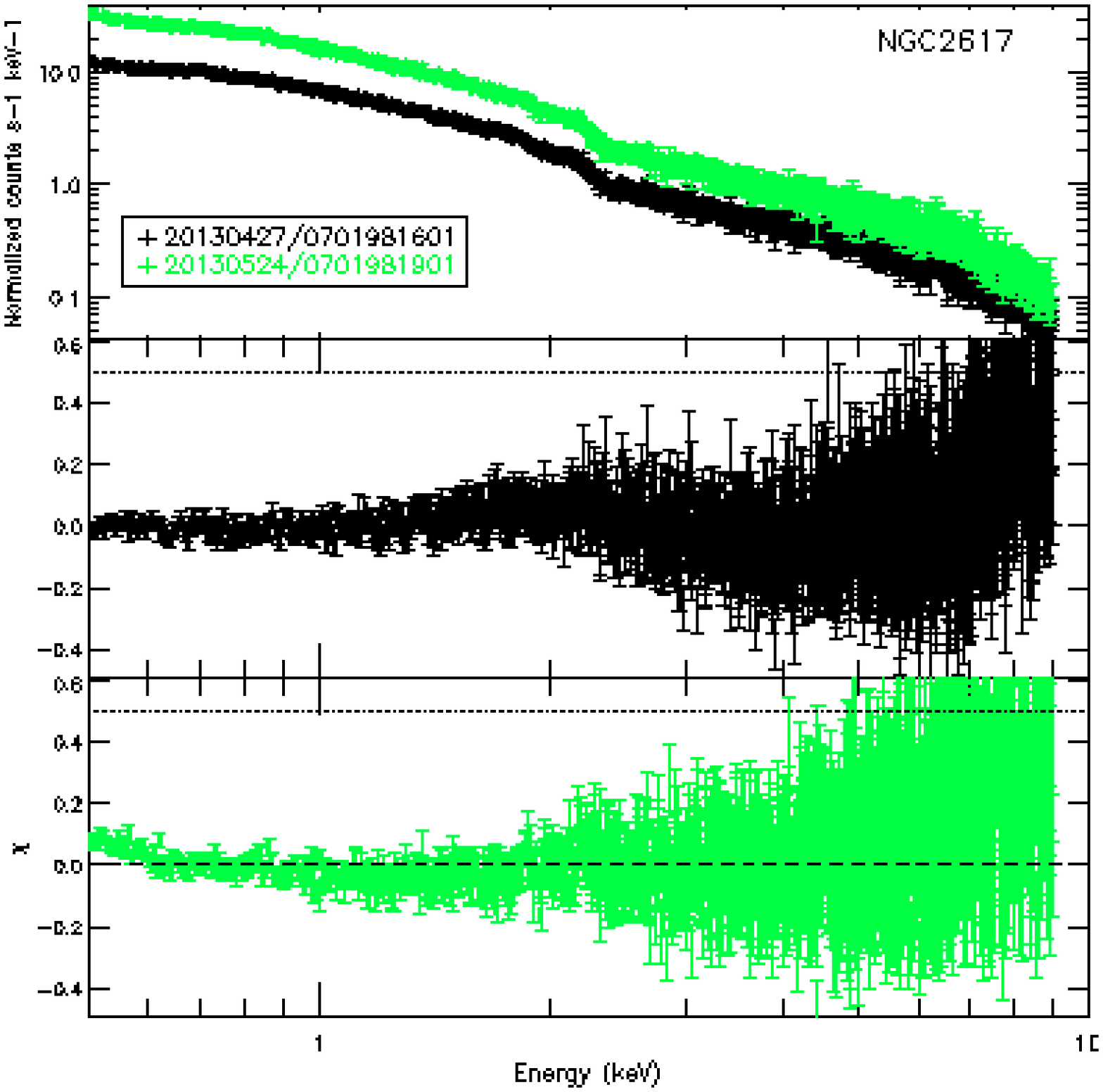}}
{\includegraphics[width=0.30\textwidth]{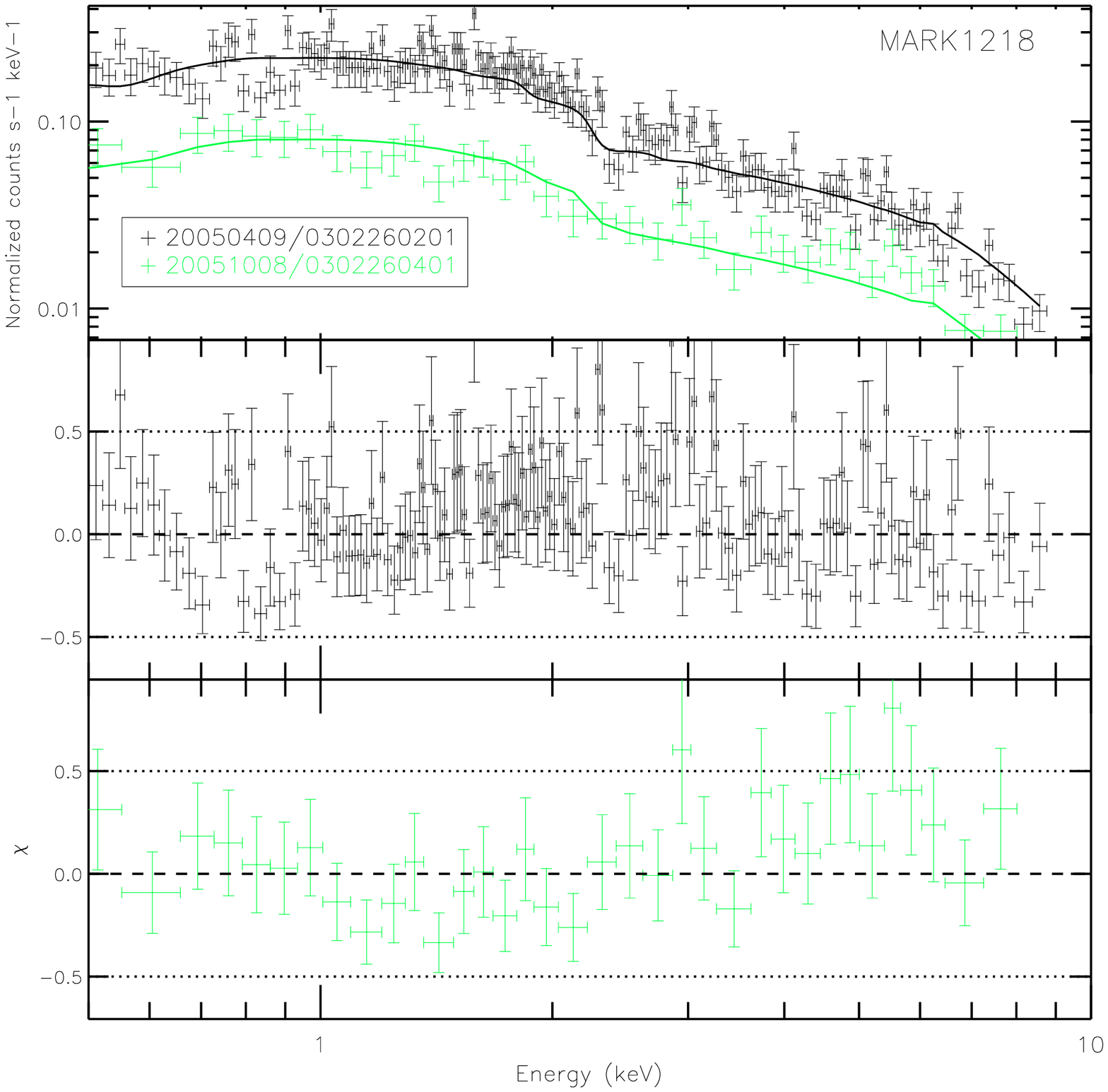}}

{\includegraphics[width=0.30\textwidth]{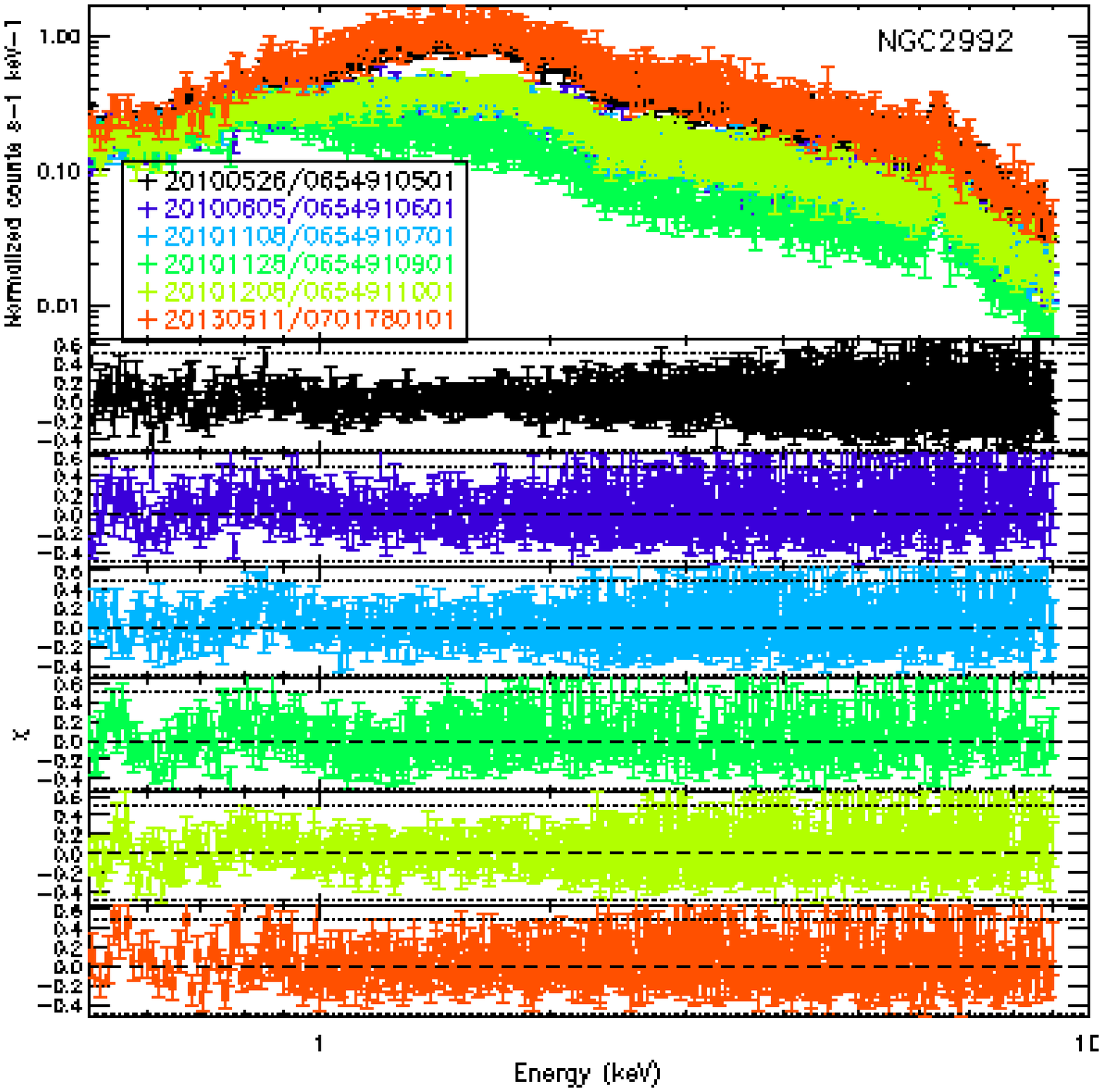}}
{\includegraphics[width=0.30\textwidth]{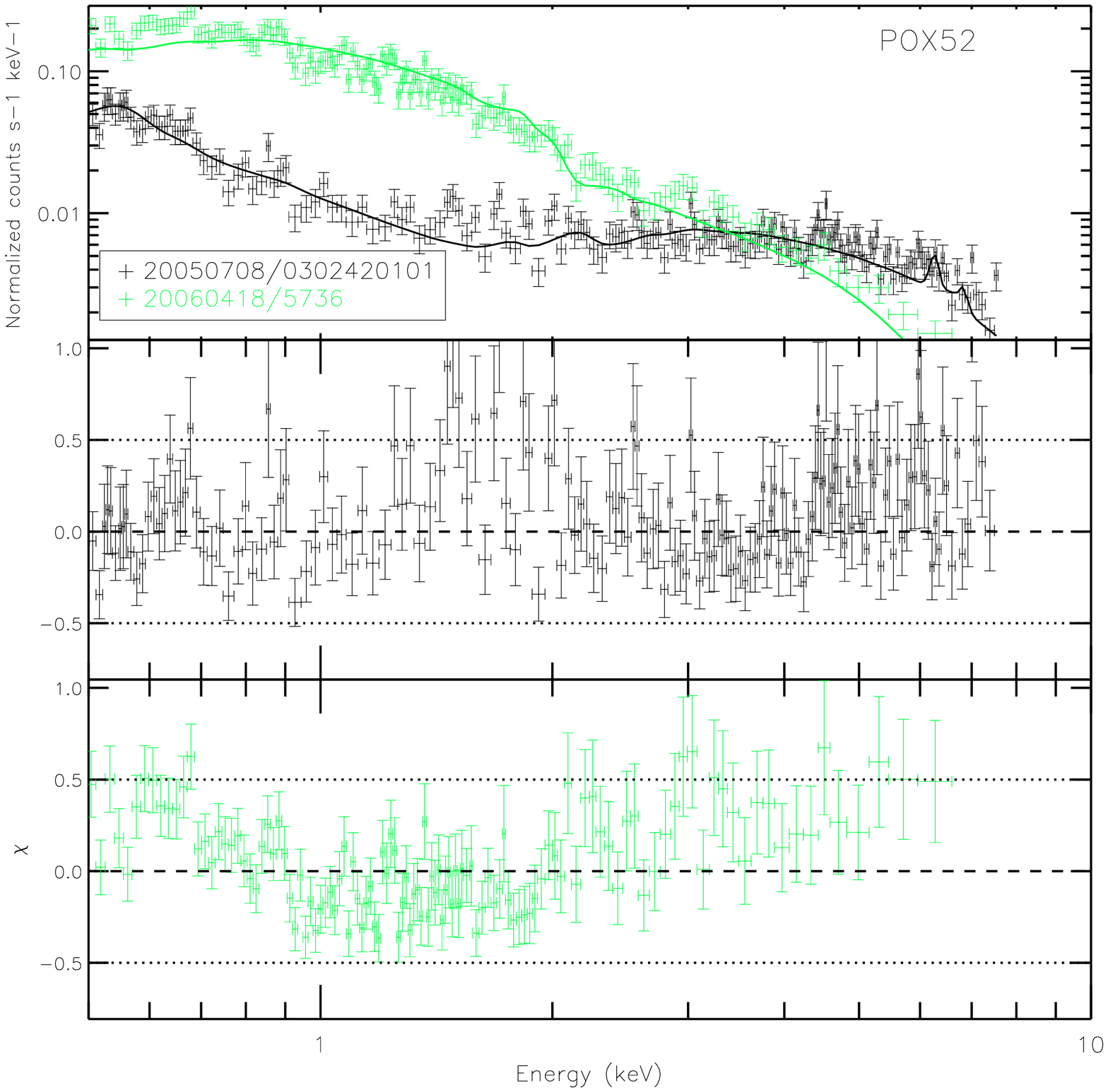}}
{\includegraphics[width=0.30\textwidth]{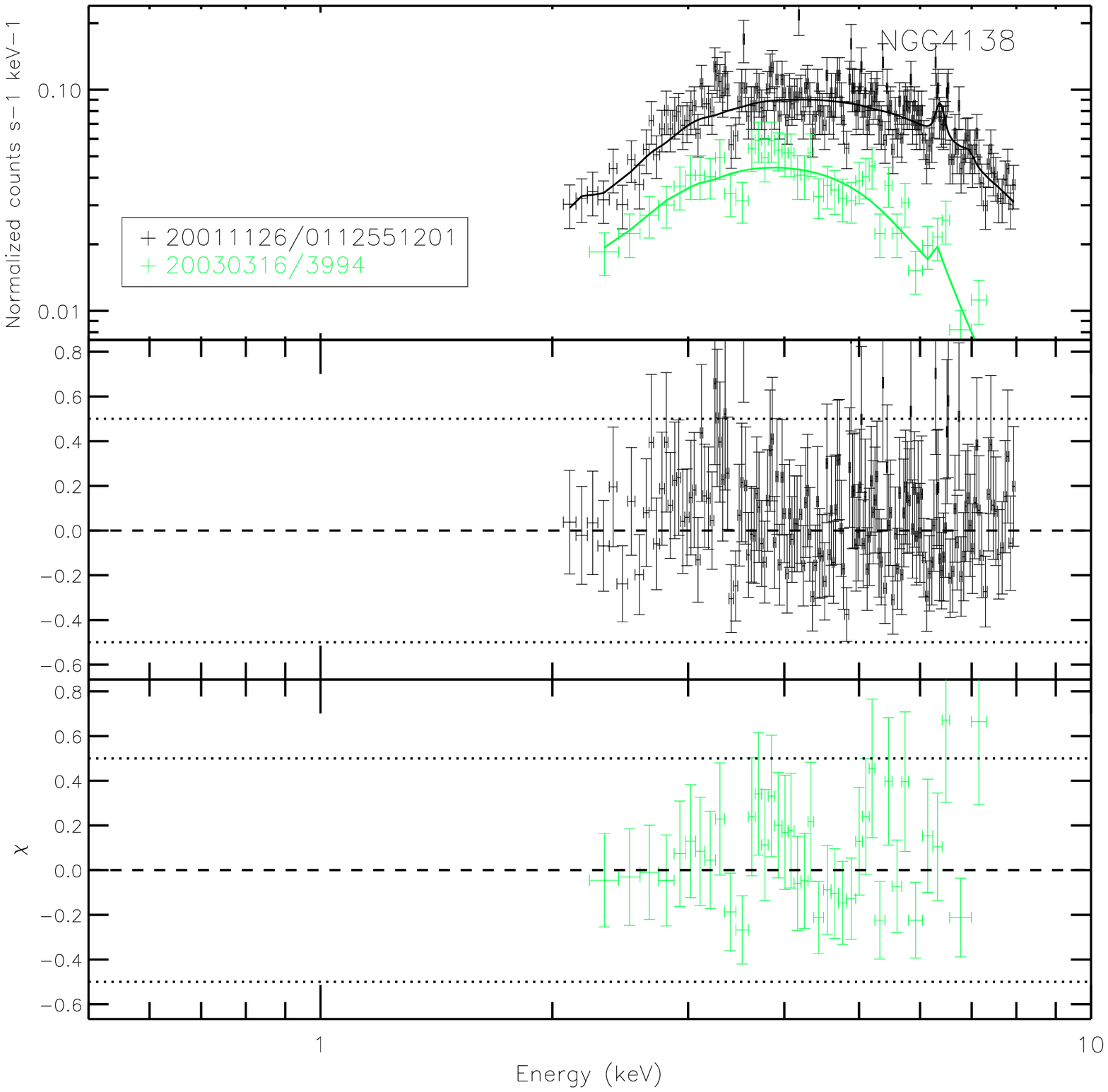}}
\caption{For each object, (top): simultaneous fit of X-ray spectra;
  (from second row on): residuals in units of $\sigma$. The legends
  contain the date (in the format yyyymmdd) and the obsID. Details are
  given in Table \ref{properties}.}
\label{bestfitSeyim}
\end{figure*}

\begin{figure*}
\setcounter{figure}{0}
\centering
{\includegraphics[width=0.30\textwidth]{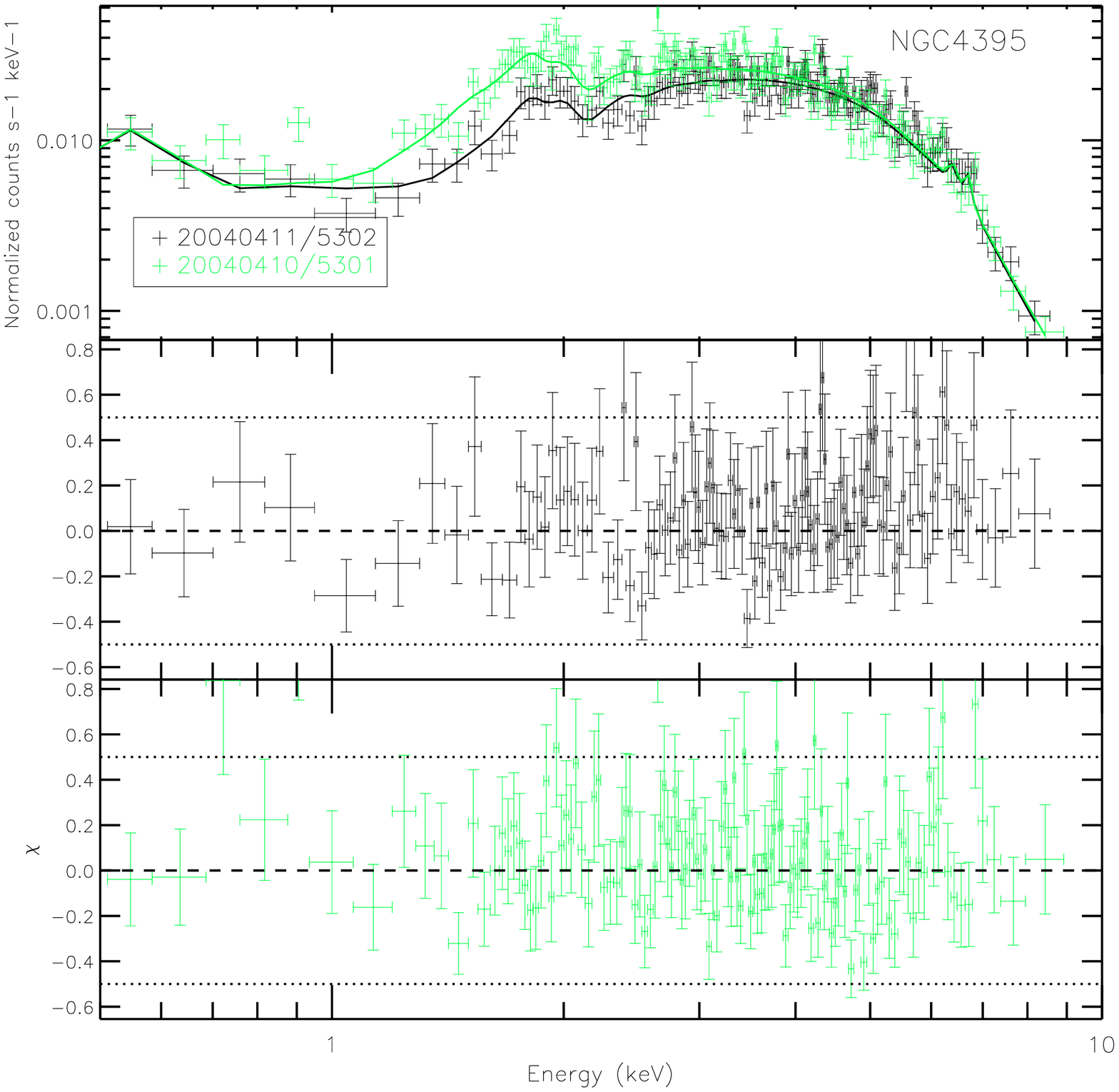}}
{\includegraphics[width=0.30\textwidth]{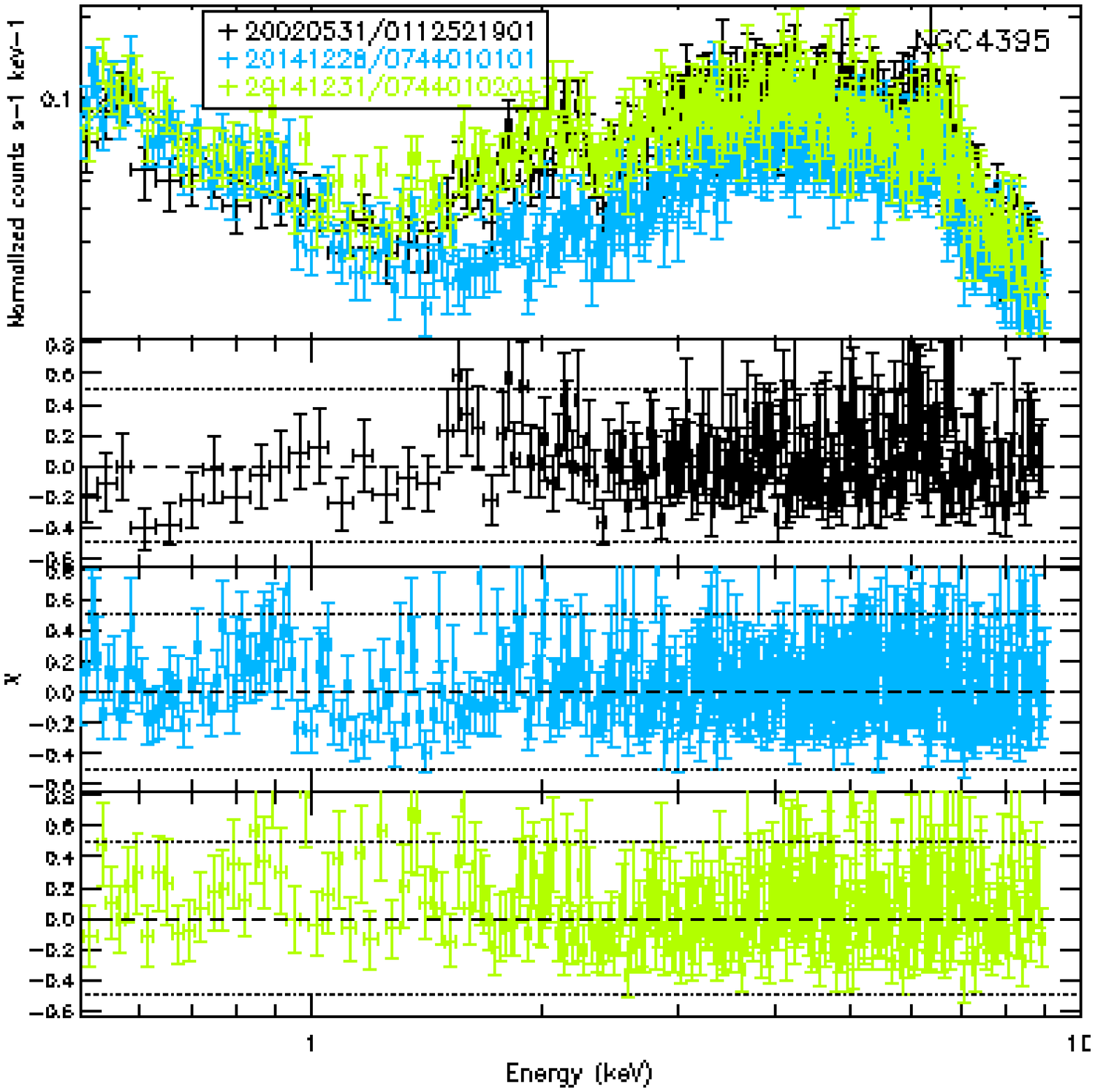}}
{\includegraphics[width=0.30\textwidth]{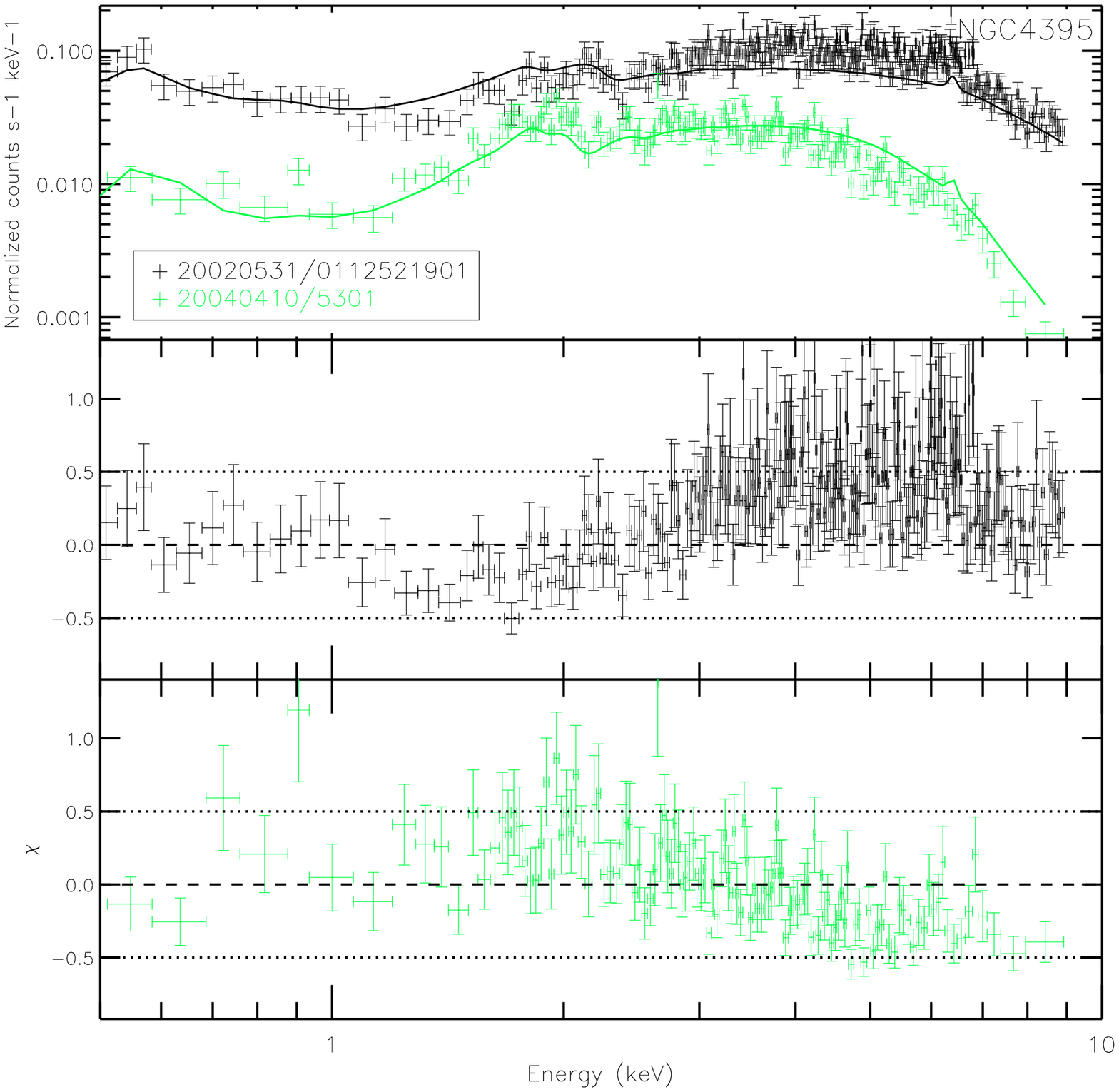}}

{\includegraphics[width=0.30\textwidth]{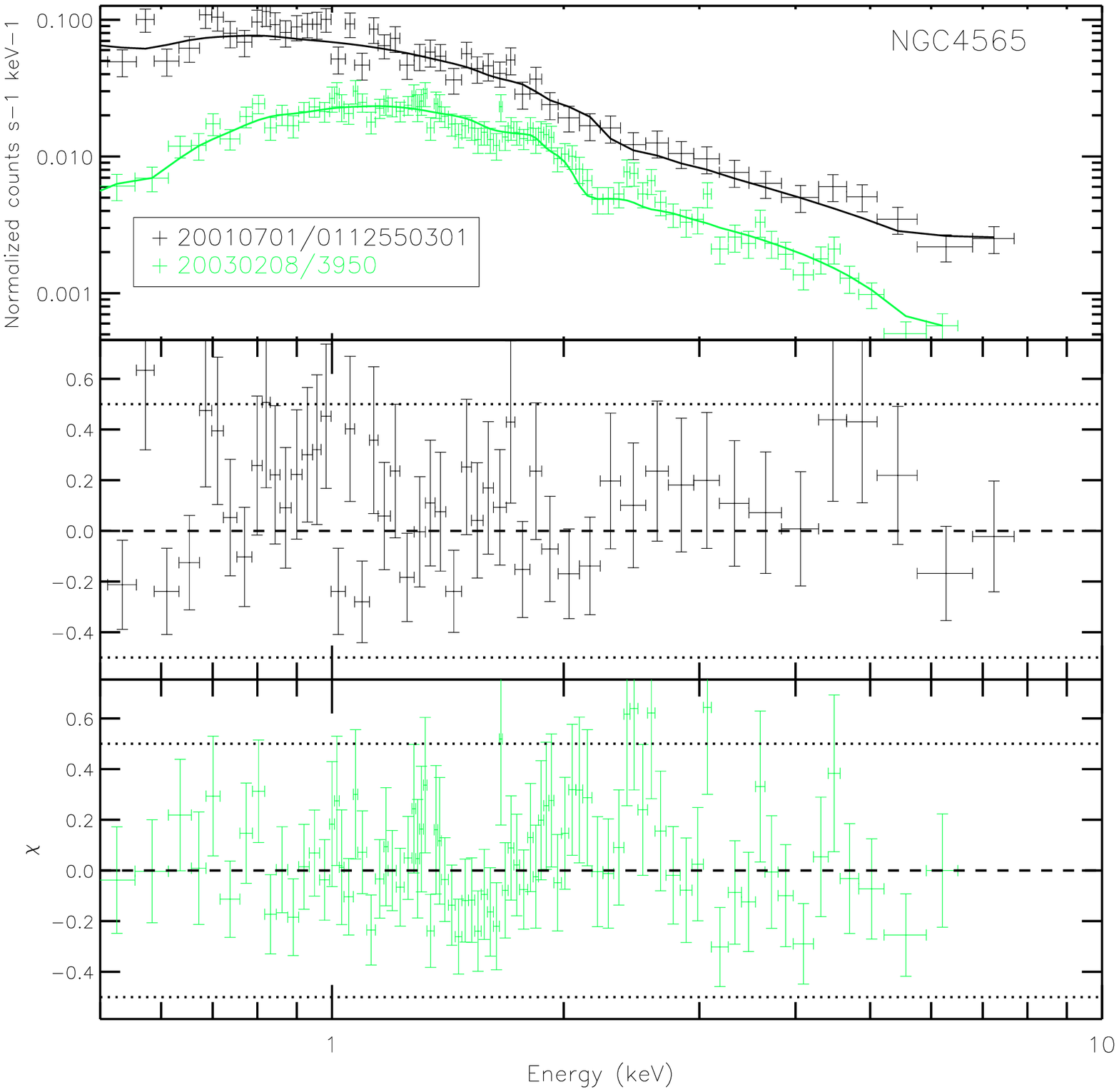}}
{\includegraphics[width=0.30\textwidth]{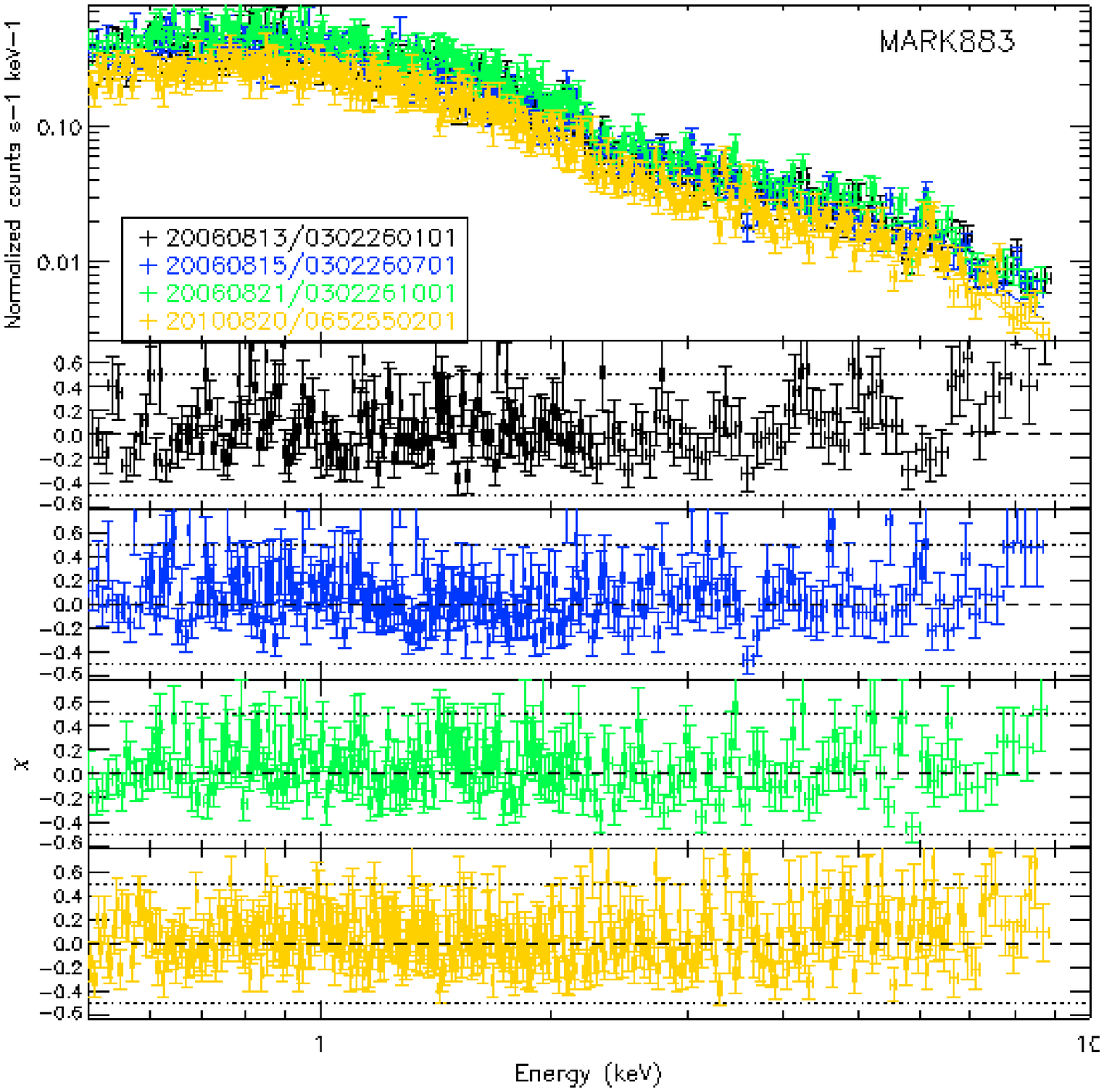}}
{\includegraphics[width=0.30\textwidth]{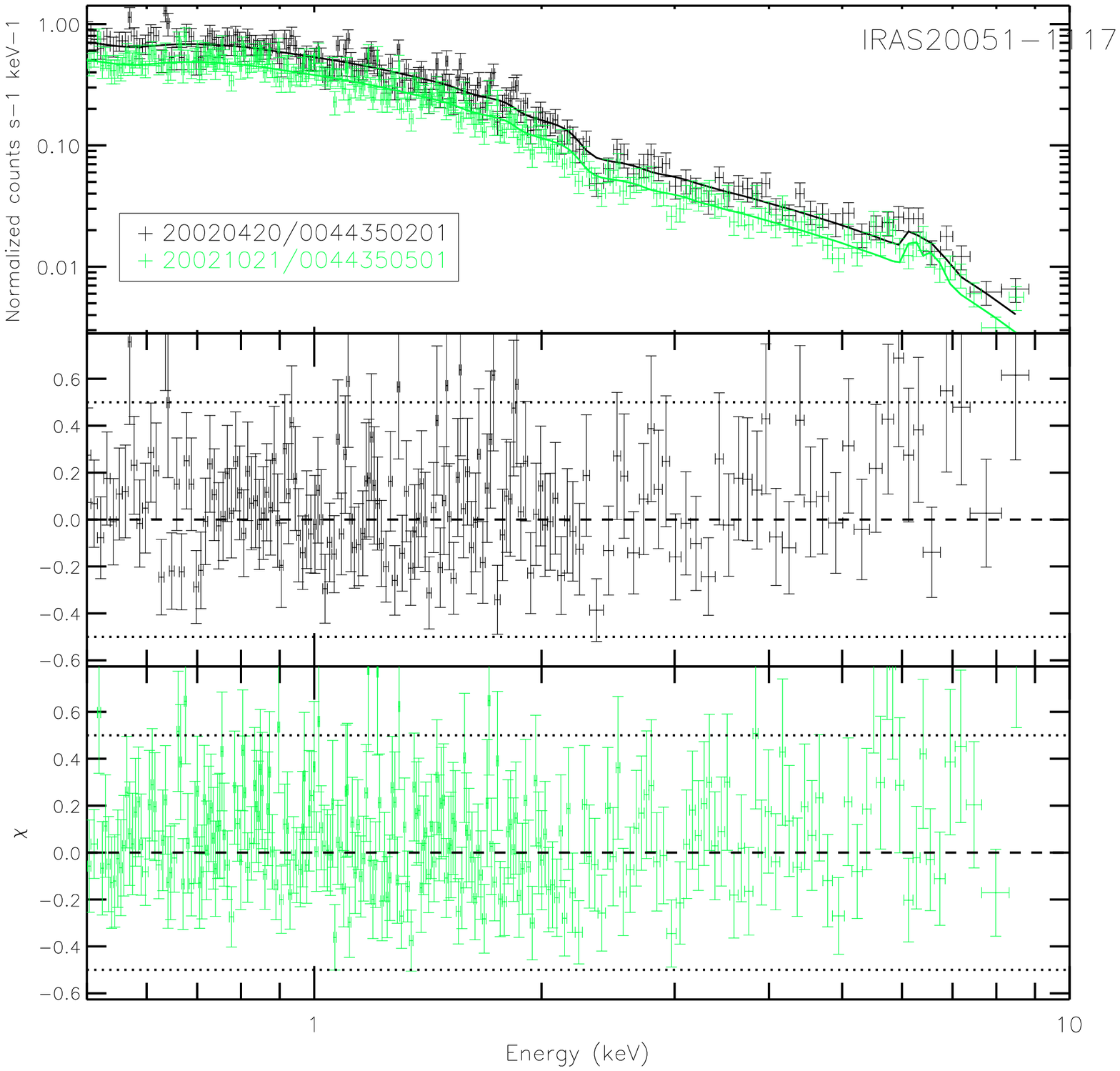}}

\caption{Cont.}
\end{figure*}

\section{\label{results}Results}

In this section we present the results of the spectral characteristics and variability patterns of the Seyfert 1.8/1.9 in the sample. For results on individual sources we refer the reader to Appendix \ref{indivnotes}, as well as for notes and comparison with previous studies.

\subsection{\label{shape}Spectral characteristics}

We used five different models to fit each spectrum individually. The best model for each source resulted to be the same in all the individual observations from the same satellite. When comparing data from different instruments, different best fit models were selected for two sources (NGC\,1365, and NGC\,4138), most probably because of the low count-rate in the \emph{Chandra} data, which required the simplest model. 
It is worth noticing that the \emph{XMM--Newton} spectrum of NGC\,4138 is best fitted with the ME2PL model, but the lack of counts in the \emph{Chandra} spectrum below 2 keV forced us to perform the analysis only above 2 KeV and thus using the PL model (see Appendix \ref{n4138}).
The ME model was not the best-fit for any of the spectra.

The median [25\% and 75\% percentiles] values of the spectral parameters are presented in Table \ref{meanscomp}.  Absorption at soft energies is usually compatible with the Galactic one (see Table \ref{bestfitSey}). Absorption at high energies is common in these sources, being obscured in the range of $10^{21}-10^{23}cm^{-2}$, with median of $N_{H2}=3.00[0.06-8.34] \times 10^{22} cm^{-2}$. The median value of the spectral indices is $\Gamma = 1.7[1.4-1.9]$, completely compatible with other AGN \citep[see e.g.,][]{brightman2011}. 
The thermal component has a median of kT = 0.19[0.09-0.62] keV.

The X-ray luminosity medians in our sample are logL(0.5--2.0 keV)=41.9[41.2-42.6] and logL(2--10 keV)=42.5[41.7-42.7].

\subsection{Compton-thickness}

We recall that a source was classified as a Compton-thick candidate within an observation when at least two out of the three criteria explained in Sect. \ref{thick} were met. None of the sources are classified as Compton-thick. Two of the sources are classified as changing-look candidates (NGC\,1365, and NGC\,2992), as already reported in the literature \citep{gilli2000,risaliti2009}. Another two sources have been classified as changing-look candidates in the literature (MARK\,609, \citealt{trippe2010} and NGC\,2617, \citealt{shappee2014}), but the present work does not detect these changes.
It is worth noticing that we did not find the flux of the $[O{\sc III}]$ in the literature for four sources (ESO\,540-G01, ESO\,195-IG21, ESO\,113-G10, and NGC\,2617), but the two other criteria were compatible with them being Compton-thin. 

\subsection{\label{spectral}Long-term X-ray spectral variability}

From the 15 nuclei in our sample, we compare spectra obtained from the same instrument in 10 cases, all of them observed by \emph{XMM--Newton}, and in one case (NGC\,4395) \emph{Chandra} data are also available. In the remaining five sources only one observation per instrument was available. 

\emph{Chandra} and \emph{XMM--Newton} data are available for the same source in eight cases (note that this analysis is independent of the one mentioned above, see Table \ref{obsSey}), thus the simultaneous analysis was carried out by using the methodology explained in Sect. \ref{simult}.

Long-term X-ray spectral variability is detected in all the 15 nuclei. Variations are detected in four parameters ($Norm_1$, $Norm_2$, $N_{H1}$, and $N_{H2}$). 
In nine objects the observed variability can be explained by varying only one parameter; in five nuclei varying two parameters is required (ESO\,195-IG21, NGC\,2617, NGC\,2992, POX\,52, and NGC\,4395), and in NGC\,1365 varying three parameters is required.
The most frequent variations are found in $Norm_2$, which are observed in ten nuclei (ESO\,540-G01, ESO\,195-IG21, NGC\,526A, NGC\,1365, MARK\,1218, NGC\,2992, NGC\,4138, NGC\,4395, MARK\,883, and IRAS\,20051-1117). Changes in $N_{H2}$ are also frequent, as they are observed in six nuclei (NGC\,1365, NGC\,2617, NGC\,2992, POX\,52, NGC\,4395, and NGC\,4565). Variations at soft energies are detected in six sources (ESO\,195-IG21, ESO\,113-G01, MARK\,609, NGC\,1365, NGC\,2617, and POX\,52). Among them, only in two objects (ESO\,195-IG21 and POX\,52) these variations are reported for a simultaneous fit using \emph{Chandra} and \emph{XMM--Newton} together, thus these variations cannot be ascribed to the comparison of data obtained from different instruments.

\subsection{\label{lightcurve}Short-term X-ray variability}

Short-term X-ray variations are analyzed in eight nuclei (Table \ref{estcurvasSey}). We recall that only light curves longer than 30 ksec were analyzed (see Sect. \ref{short}). Two sources do not show variations (POX\,52 and NGC\,4565) according to the $\chi^2/d.o.f$ and $\sigma^2_{NXS}$, whereas the remaining six are variable in at least one energy band. Four sources show variations in the total, soft, and hard energy bands (ESO\,113-G10, NGC\,526A, NGC\,1365, and NGC\,4395). Among these, all but NGC\,1365 show variations above the 3$\sigma$ confidence level in the three energy bands. Among the eight observations analyzed for NGC\,1365, four do not show variations in any band, two show variations in the total and hard bands, and another two show variations in the three energy bands (both from 2012).  
NGC\,2992 shows variations in the total and hard energy bands. Five observations were analyzed for this source, two of them not showing variations in any of the bands. 
NGC\,2617 shows variations in the total and soft energy bands, in both cases above the 3$\sigma$ confidence level.

\subsection{\label{flux}Long-term UV flux variability}

UV data are available for nine sources. The remaining six sources do not have more than one \emph{XMM--Newton} observations or different filters were used for the observations and thus cannot be directly compared. 

Among the nine sources, two of them do not show UV variability (MARK\,1218 and IRAS\,20051-1117), whereas the remaining seven sources show variations in at least one filter (see Table \ref{obsSey} and Fig. \ref{luminUVfigSey}). We remind that all the reported variations are above the 3$\sigma$ confidence level.

%

\begin{table}
\footnotesize
\caption{\label{meanscomp} Median values and 25\% and 75\% percentiles of the spectral parameters of Seyfert 1.8/1.9 presented in this work (Col. 2) and the Seyfert 2 sample (Col. 3) presented in \cite{lore2015}.}
\begin{tabular}{lcccccc} \hline
\hline
 & Seyfert 1.8/1.9 & Seyfert 2 \\
(1) & (2) & (3)  \\ \hline \vspace*{0.1cm}
log(L(0.5-2 keV) [erg \hspace{0.1cm} $s^{-1}$]) & 41.9$_{41.2}^{42.6}$ & 42.1$_{41.3}^{42.6}$ \\ \vspace*{0.1cm}
log(L(2-10 keV) [erg \hspace{0.1cm} $s^{-1}$])   & 42.5$_{41.7}^{42.7}$ & 42.7$_{42.5}^{42.8}$ \\ \vspace*{0.1cm}
$N_{H2}$ ($\times 10^{22} [cm^{-2}]$) & 3.00$_{0.06}^{8.34}$ & 22.2$_{9.8}^{38.4}$  \\ \vspace*{0.1cm}
$\Gamma$ & 1.7$_{1.4}^{1.9}$ & 1.7$_{1.5}^{2.0}$  \\ \vspace*{0.1cm}
kT [keV] & 0.19$_{0.09}^{0.38}$ & 0.71$_{0.67}^{0.81}$/0.15$_{0.12}^{0.18}$  \\ 
S/N(0.5--2 KeV) & 5.3$_{4.9}^{7.7}$ & $5.2_{3.8}^{6.4}$ \\
S/N(2--10 KeV) & $5.4_{4.7}^{6.6}$ & $3.8_{2.3}^{5.2}$ \\
\hline
\end{tabular}
\caption*{ {\bf Notes.} (Col. 1) Spectral parameter, (Col. 2) average values for Seyfert 1.8/1.9, and (Col. 3) average values of Seyfert 2 from \cite{lore2015}. The two temperatures represent the two thermal components in the model.}
\end{table}

\begin{figure}
\centering
{\includegraphics[width=0.50\textwidth]{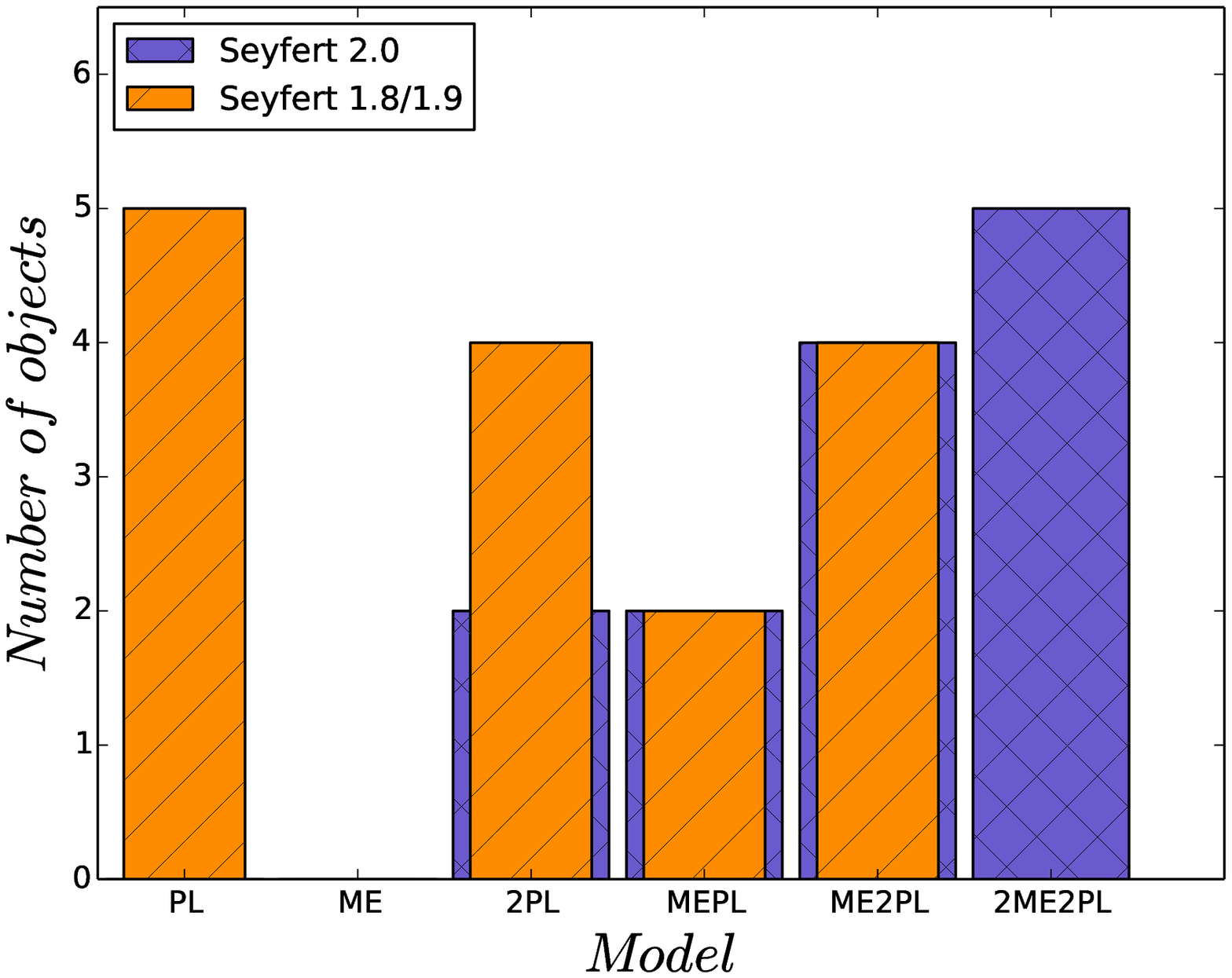}}
\caption{\label{histmodels}Histogram of the best fit models for Seyfert 1.8/1.9 and 2.}
\end{figure}

\section{\label{discusion}Discussion}

The optical and X-ray spectra of Seyfert 1.8/1.9 resembles those of Seyfert 2. For this reason, many studies aiming at analyzing the properties of obscured sources have included all these nuclei in the same samples \citep[e.g.,][]{risaliti1999, akylas2009, malizia2012}. 
However, a study of their variability properties compared to those of Seyfert 2 represents a powerful tool in revealing similarities and differences.
In this work we can compare these properties in an homogeneous way for the first time using the results for Seyfert 1.8/1.9 presented here and the sample of Seyfert 2 presented in \cite{lore2015}.

\subsection{Spectral properties}

\begin{figure}
\centering
{\includegraphics[width=0.5\textwidth]{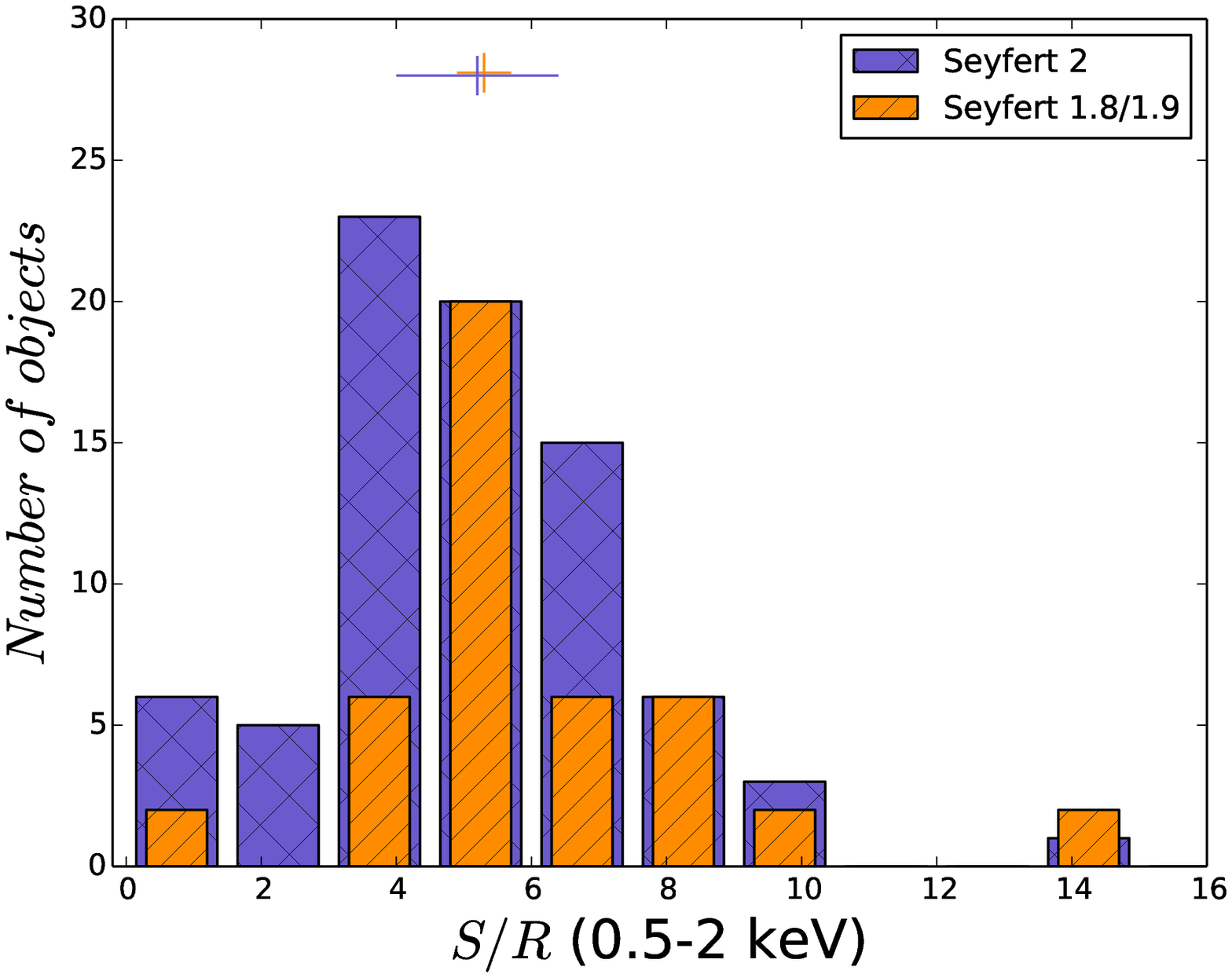}}
{\includegraphics[width=0.5\textwidth]{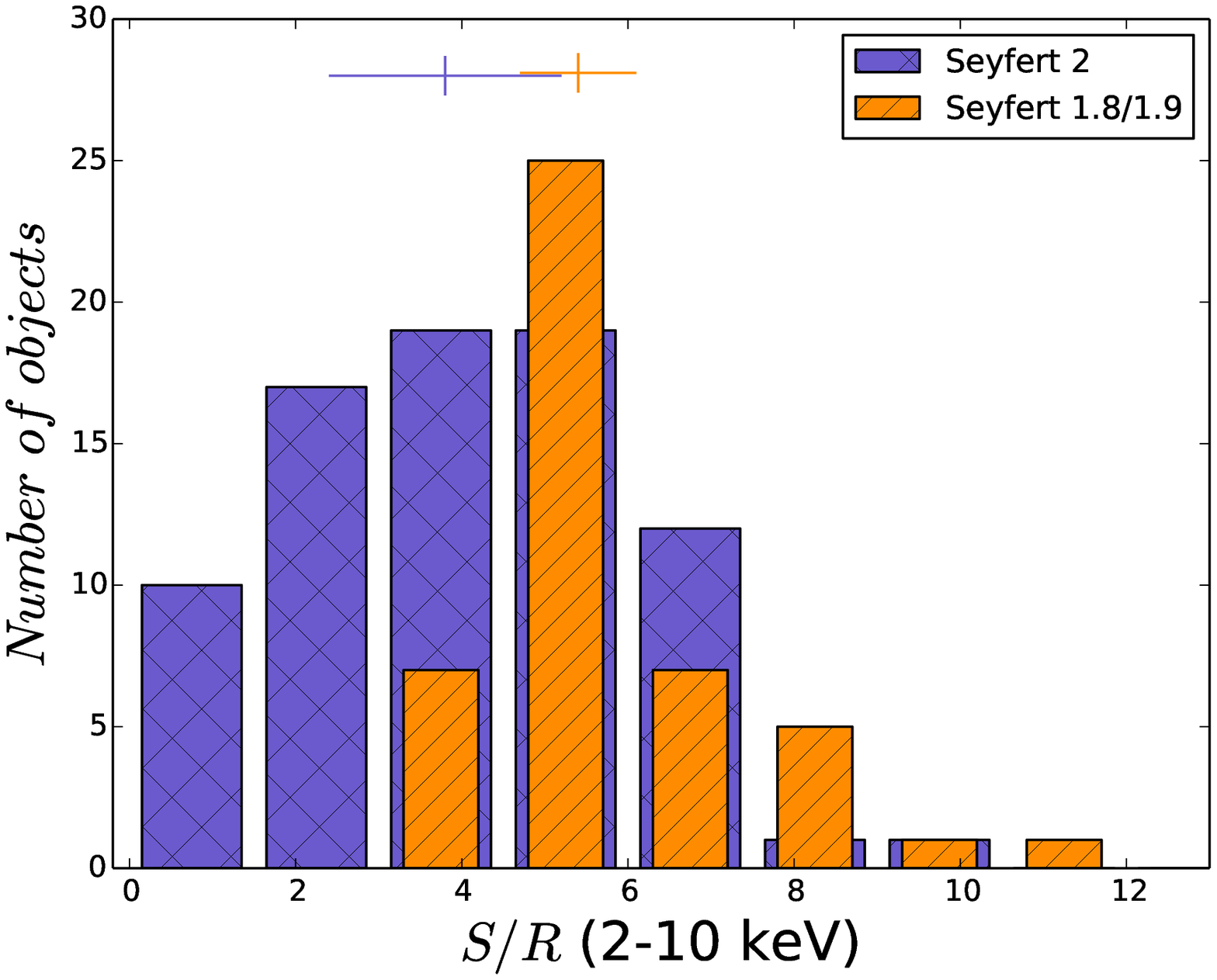}}
\caption{\label{snr}Histogram of the signal-to-noise ratio, $S/R$, for the individual spectra of Seyfert 1.8/1.9 and Seyfert 2 in the soft (0.5--2 keV, upper panel) and hard (2--10 keV, bottom panel) energy bands. The crosses represent the median values.}
\end{figure}

\begin{figure}
\centering
{\includegraphics[width=0.50\textwidth]{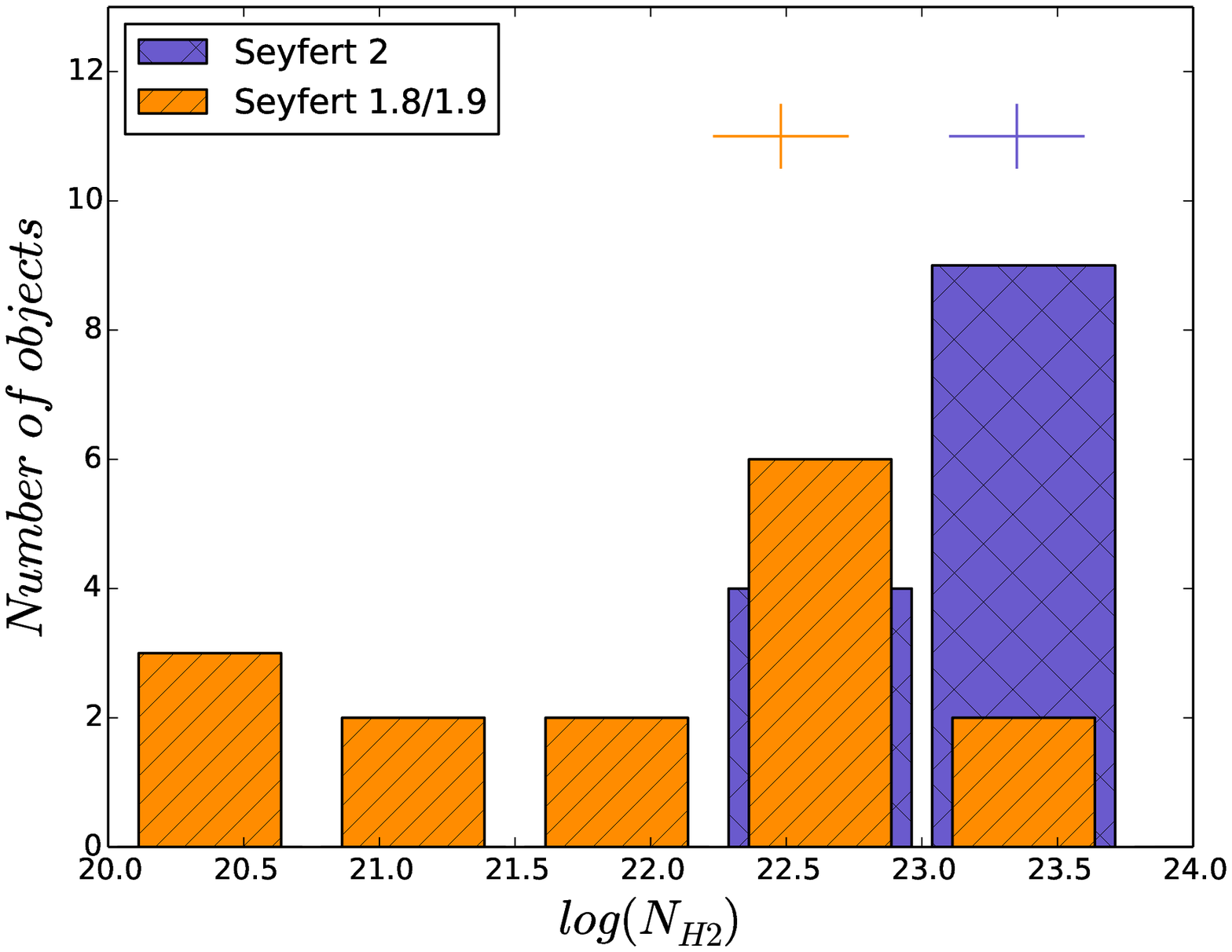}}
\caption{\label{nh2comp}Histogram of the column density, $N_{H2}$, for Seyfert 1.8/1.9 and Seyfert 2 in logarithmic scale. The crosses represent the median values.}
\end{figure}

Fig. \ref{histmodels} shows the models used for Seyfert 1.8/1.9 and 2.
The main conclusion is that Seyfert 2 require, in general, more complex 
models to fit the data.  The difference cannot be attributed to 
different spectral qualities since the average signal-to-noise ratio(S/N) -- which were estimated following \cite{stoehr2008} -- for each spectrum is similar for both type 1.8/1.9 and type 2 samples in the 0.5--2 keV and 2--10 keV energy bands (see Fig. \ref{snr}). In Table \ref{meanscomp}, we show 
the median values [25\% and 75\%  percentiles] of the spectral parameters for the Seyfert 1.8/1.9 and 
2 nuclei. It can be seen
that the main difference between these AGN families is in $N_{H2}$ (see
also Fig. \ref{nh2comp}), in agreement with previous studies
\citep{risaliti1999,guainazzi2001,akylas2009,brightman2011a}. This is in fact the main reason why Seyfert 1.8/1.9 have been explained as nuclei that are less obscured than Seyfert 2, their difference being ascribed only to the absorbing material.

However, another difference is observed in the
temperatures of the thermal emission, with Seyfert 2 having a bimodal distribution centred at
$KT_1 \sim 0.7$ keV and $KT_2 \sim 0.2$ keV whereas Seyfert 1.8/1.9
show only one temperature regime centred at $KT \sim 0.2$ keV. 
Moreover the comparison with the study by 
\cite{brightman2011,brightman2011a} on the 12 $\mu$m complete sample 
of Seyferts show that a thermal component is fitted in Seyfert 1-1.5 (Seyfert 2) 
in 15 (24) cases, with 12 (6) of them centred at 0.1 keV and three (18) centred
at 0.7 keV. This result might indicate that the thermal
component observed in Seyfert 1.8/1.9 is more similar to that observed
in Seyfert 1 than to that in type 2.

In this context, it should be noticed that a more realistic physical model for the absorbing material in Seyfert 1.8/1.9 might be represented by ionised absorption (i.e., winds) instead of neutral absorption (i.e., the torus). We changed the neutral absorption by an ionised absorption in our models but the spectral fits were statistically the same (the fits did not improve at the 99.9\% of confidence level using the F-test), except in the case of NGC\,1365, where the presence of ionised absorption is well established \citep{risaliti2005b, guainazzi2009}. Since we cannot differentiate between neutral or ionised absorption due to the resolution of the spectra presented in this work, this issue should be explored using high spectral resolution data.  

\subsection{Short-term variability}

\begin{figure}
\centering
{\includegraphics[width=0.54\textwidth]{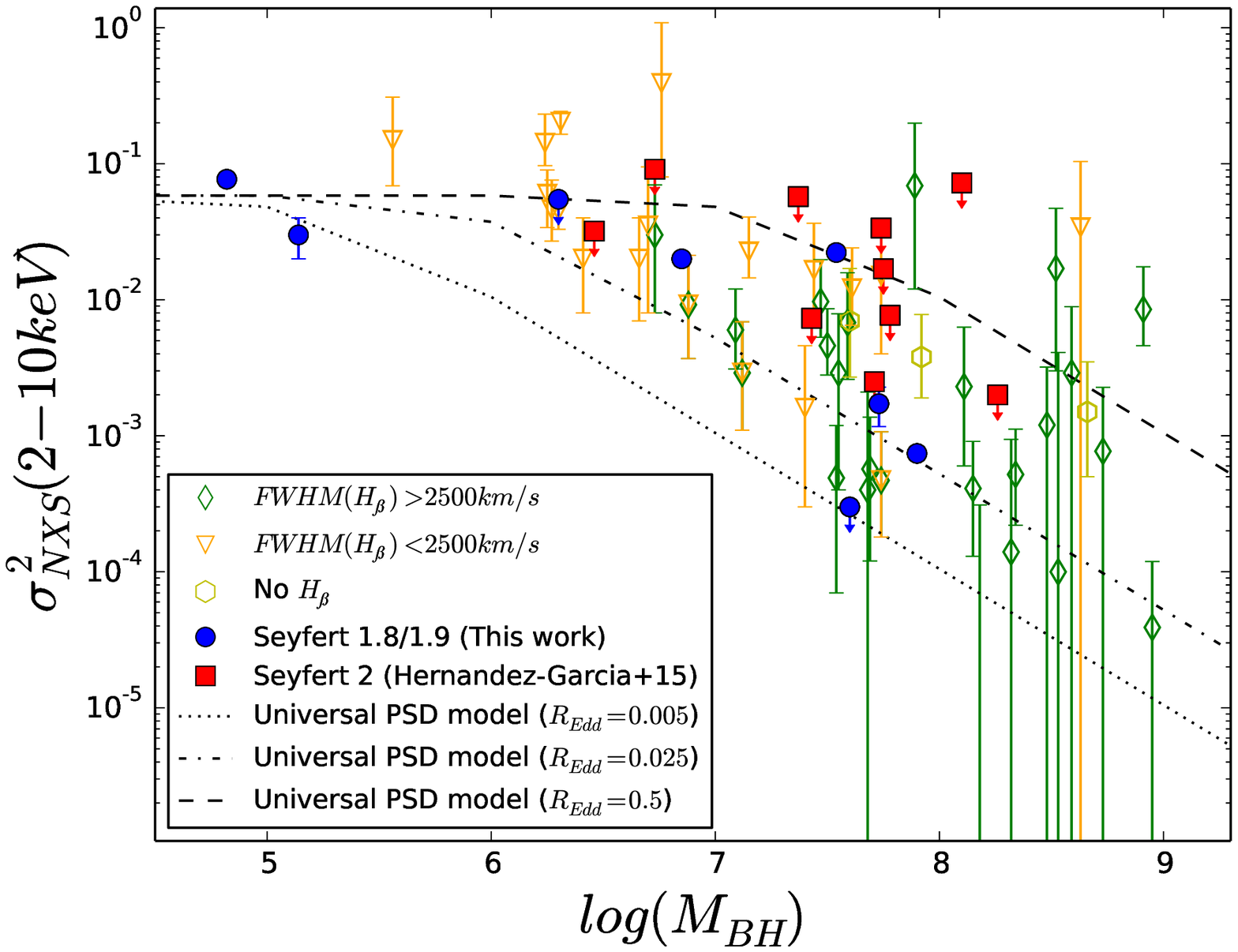}}
\caption{\label{sigmambh} Normalized excess variance, $\sigma^2_{NXS}$, against the black hole mass, $M_{BH}$ for the Seyfert 1.8/1.9 in the sample (blue circles) and Seyfert 2 in \cite[][red squares]{lore2015}. Results for the CAIXA sample using light curve segments of 40 ksec \citep{ponti2012} are presented, including sources with (FMWH($H_{\beta}>$2500 $km \hspace*{0.1cm} s^{-1}$ (green diamonds), FMWH($H_{\beta}$) $<$ 2500 $km \hspace*{0.1cm} s^{-1}$ (orange triangles), and those with no $H_{\beta}$ (yellow hexagons).
 The dotted, dash-dotted, and dashed lines are the universal PSD model, represented as a broken power law \citep{papadakis2004} using $R_{Edd}=$[0.005, 0.025, 0.5].}
\end{figure}

Further information can be obtained from the analysis of the variability.  
X-ray short-term (i.e., intra-day) variations in Seyfert 2 galaxies were not detected in \cite{lore2015}, and thus all the measurements of $\sigma^2_{NXS}$ are upper limits.
It is interesting to notice that most Seyfert 1 show changes at
these timescales \citep{nandra1997, turner1997}, but variations at these timescales in a few Seyfert 2 have also been reported in the literature \citep[e.g.,][]{omairavaughan2012}. Even though the
physical origin of these variations is not yet well settled, the involved
timescales imply that they occur very close to the SMBH, and it has
been suggested that accretion-disc/corona instabilities or variations
in the accretion rate may be involved \citep{uttley2005a, breedt2010,mchardy2010, soldi2014}. 
Fig. \ref{sigmambh} shows the 
 $\sigma^2_{NXS}$ in  the 2--10 keV energy band against $M_{BH}$ for 
the variable Seyfert 1.8/1.9 in our
sample and the upper limits for the Seyfert 2 together with the results reported by \cite{ponti2012} for the 
CAIXA's sample, which include unobscured sources. For a proper comparison, we used their $\sigma^2_{NXS}$ calculated for light curve segments of 40 ksec. The green squares represent those sources with a width at half maximum (FMWH) of the $H_{\beta}$ larger than 2500 $km \hspace*{0.1cm} s^{-1}$, the red triangles those with FMWH($H_{\beta}$) $<$ 2500 $km \hspace*{0.1cm} s^{-1}$, and the black hexagons the sources where they did not report the FMWH($H_{\beta}$).
It has been shown that the 2–10 keV power spectrum in AGN follows a power law of slope $\Gamma \sim$-2 at high frequencies, which then flattens to a slope of $\Gamma \sim$-1 below a break frequency \citep{papadakis2004,oneill2005,miniutti2009}.
This model is named the universal power spectrum density (PSD)
model because the integral of the PSD is equal to the $\sigma^2_{NXS}$ of a light curve \citep{vaughan2003}. The dotted line in Fig. \ref{sigmambh} represents this model as in \citet[][see also \citealt{omaira2011a}]{papadakis2004}, which was obtained for a sample of broad line Seyfert 1. For the relation, we used an Eddington ratio of $R_{Edd}=$[0.005,0.025,0.5], which is represented as a dotted, dash-dotted, and dashed lines in Fig. \ref{sigmambh}, respectively.
It can be observed that the Seyfert 1.8/1.9 in our sample fit well both with this model and with the results reported by \cite{ponti2012}, as well as Seyfert 2 do. Thereof the short-term variability in Seyfert 1, Seyfert 1.8/1.9 and Seyfert 2 is consistent within the statistical uncertainties and thus we cannot provide evidence for a difference among the classes in this respect until the variations in Seyfert 2 are rejected.

\subsection{Long-term variations}

We found differences in the spectral variations between Seyfert 1.8/1.9 and Seyfert 2 in timescales from weeks to years. Variations at soft energies are only detected in Seyfert 1.8/1.9. 
The origin of the soft X-ray emission in Seyferts is under debate, maybe related to e.g.,  
the possible dilution effect of scattering by photoionised gas \citep[e.g.,][]{omaira2010}
or to a warm
absorber (WA), consisting on ionized absorbing gas, that is observed
in about 65\% of Seyfert 1s \citep{laha2014}, and it is highly
variable \citep[e.g.,][and references therein]{winter2012,laha2016}.
However, the exposure times of the high resolution data (e.g., from the \emph{Reflection Grating Spectrometer}) of the sources discussed in this paper are too short to constrain the properties of the soft X-ray emission, and thus the nature of the soft X-ray variations \citep[e.g.,][]{laha2011,giustini2016}.

At long timescales we find that Seyfert 1.8/1.9, as well 
as Seyfert 2 and
LINERs, show hard X-ray variations in timescales between months
and years, these changes being related to intrinsic variations of the
hard nuclear continuum. This suggests that the emission
mechanism is the same in these AGN families. We have also computed the
variability dynamical timescale against the time between the
observations for the two families. We did not find differences in
the variability of these sources, neither any obvious trend against
the time difference. 
We notice that a caveat in our analysis might be that the observations were taken at random epochs, and thus the timescale between the observations do not refer to the variability timescales.

The AGN continuum at X-rays comes from the
Comptonization of photons from the inner parts of the accretion disk
\citep{shakura1973}. Thus, these long-term variations might be related to
fluctuations in the inner accretion disk \citep[e.g.,][]{lore2016}.
Moreover, variations due to absorbing clouds (i.e., $N_H$) are common in our sample, both in Seyfert 1.8/1.9 (six out of
the 15, i.e., 40\%) and 2s (four out of the 11 variable sources, i.e.,
36\%). However, these kind of eclipses are also observed in Seyfert 1s
\citep{risaliti2011,sanfrutos2013,markowitz2014,agis2014}. When
 it is possible to estimate the cloud velocity, and thus the
location of the absorbing material, the clouds
appear to be located very close to the BLR or within the borderline between the BLR and the torus \citep{risaliti2009, risaliti2013, nardinirisaliti2011, walton2014, connolly2014, markowitz2014, parker2015, giustini2016}.
The velocities of the clouds in Seyfert 1.8/1.9
(NGC\,1365, NGC\,2617, NGC\,2992, and NGC\,4395) and Seyfert 2 (Mark\,1210,
and NGC\,4507) in our sample are greater than $10^3
km/s$\footnote{Since the timescales between the observations were
  obtained at random, we can only estimate its velocities when the
  timescale is shorter than $\sim$ 1 month.}  \citep[following the procedure in][]{risaliti2010}. Therefore, these eclipses due to clouds passing through our line of sight seem to be happening at distances from the accretion disc consistent within or very close to the location of the BLR. 

Furthermore, while variations at UV frequencies are not detected in
Seyfert 2, these changes are observed in Seyfert 1.8/1.9 in the
present work. The same behaviour is also seen in Seyfert 1s
\citep[e.g.,][]{cardaci2010,netzer2013}. This result might indicate a less obstructed view of the accretion disk in Seyfert 1.8/1.9 compared to that of Seyfert 2, in agreement with the higher values of $N_{H2}$ in Seyfert 2 compared to those of Seyfert 1.8/1.9.

As a final remark, we would like to warn the reader for the simplistic association of Seyfert 1.8/1.9 to Seyfert 2, since different variability properties are observed in the two AGN groups both at X-ray and at UV frequencies. We leave open the possibility that Seyfert
1.8/1.9 behave more likely Seyfert 1, but a proper comparison between
the properties of these families need to be done in a systematic and
homogeneous way.

\section{\label{conclusion}Summary and conclusions}

We have performed a spectral and variability analysis at X-ray and UV frequencies of a sample of fifteen galaxies classified as Seyfert 1.8/1.9 based on optical spectroscopy. The main results of the study can be summarized as follows:

\begin{itemize}
\item X-ray long-term variability is found in all the 15 nuclei. None of the sources are classified as Compton-thick candidates, and two of them have been classified as changing-look candidates. 
\item The main variability pattern is related to intrinsic changes in the sources, which are observed in ten nuclei. Changes in the column density are also frequent, as they are observed in six nuclei. Variations at soft energies are detected in six sources.
\item X-ray short-term variations are detected in six out of the eight studied sources.
\item Variations at UV frequencies are detected in seven out of the nine sources where data were available.
\end{itemize} 

We have compared the properties of Seyfert 1.8/1.9 with the results of a sample of Seyfert 2 that were analyzed using the same methodology applied in this work \citep{lore2015}, allowing a homogeneous comparison. The main conclusions obtained from this work are the following:

\begin{itemize} 
\item The X-ray long term variations occur in a similar way in all the nuclei and are mainly related to intrinsic changes of the nuclear continuum. Variations in the absorbing column densities are also frequent in both AGN families.
\item X-ray short-term, soft X-ray, and UV long term variations are detected in Seyfert 1.8/1.9 but not in Seyfert 2, indicating that the view of the SMBH is unobstructed in Seyfert 1.8/1.9 and obstructed in Seyfert 2. 
\item We caution on the simplistic association of Seyfert 1.8/1.9 to Seyfert 2 to classify all of them as obscured sources, because they show different variability properties at X-ray and UV frequencies.
\end{itemize} 

A similar study of a sample of Seyfert 1 would be required in order to homogeneously compare their properties and to test whether Seyfert 1.8/1.9 are similar to these AGN or not.

\begin{acknowledgements}

We thank the anonymous referee for helpful comments. We acknowledge the AGN group at the
IAA for helpful comments during this work. This work was financed by
MINECO grant AYA 2010-15169, AYA 2013-42227-P, and Junta de Andaluc\'{i}a
TIC114. OGM acknowledges financial support from the UNAM PAPIIT IA100516 project. LHG and FP acknowledge the ASI/INAF agreement number 2013-023-R1. This research made use of data obtained from the
\emph{Chandra} Data Archive provided by the \emph{Chandra} X-ray
Center (CXC). This research made use of data obtained from the
\emph{XMM}-Newton Data Archive provided by the \emph{XMM}-Newton
Science Archive (XSA). This research made use of the NASA/IPAC
extragalactic database (NED), which is operated by the Jet Propulsion
Laboratory under contract with the National Aeronautics and Space
Administration. We acknowledge the usage of the HyperLeda data base
(http://leda.univ-lyon1.fr).

\end{acknowledgements}

\bibliographystyle{aa}
\bibliography{000referencias}

\newpage

\appendix

\onecolumn

\section{\label{tables}Tables}

\tiny
\renewcommand{\arraystretch}{1.4}
\begin{longtable}{lcccccccc}
\caption[Observational details]{\label{obsSey} Observational details.} \\   \hline \hline
 Name & Instrument & ObsID  & Date & R  & Net Exptime & Counts & OM Filter &   log($L_{UV}$)  \\
 &     &           &                    & ($\arcsec$)  & (ksec) &  & & (erg/s)   \\
 (1) & (2) & (3) & (4) & (5) & (6) & (7) & (8) & (9)  \\
\hline 
\endfirsthead
\caption{(Cont.)} \\
\hline \hline
 Name & Instrument & ObsID  & Date & R  & Net Exptime & Counts & OM Filter &   log($L_{UV}$)  \\
 &     &           &                    & ($\arcsec$)  & (ksec) &  & & (erg/s)   \\
 (1) & (2) & (3) & (4) & (5) & (6) & (7) & (8) & (9)  \\
 \hline  
\endhead
\endfoot
\\
\endlastfoot
 \hline
ESO\,540-G01 & \emph{Chandra}$^c$ & 2192 & 2001-08-31 & 3 &  20 & 1316 & -  \\
             & \emph{XMM-Newton}$^c$ & 0044350101 & 2002-05-29 & 25 &  11 & 2728 & UVW2 & 43.26$^+_-$0.04  \\
          &                   &            &             &   &    &         & UVW1 & 43.41$^+_-$0.01 \\
\hline
ESO\,195-IG21 & \emph{Chandra}$^c$ & 13898 & 2012-11-30 & 2 & 20 & 529 & -  \\
             & \emph{XMM-Newton}$^c$ & 0554500201 & 2008-06-21 & 20 &  25 & 6165 & & Not detected   \\
\hline
ESO\,113-G10 & \emph{XMM-Newton} & 0103861601 & 2001-05-03 & 25 &  4 & 10419 & UVW2 & 42.68$^+_-$0.06  \\
           & \emph{XMM-Newton} &  0301890101 & 2005-11-10 & 25 &  77 & 304379 & UVW2 & 42.87$^+_-$0.02  \\
          &                   &            &             &   &    &         & UVM2 & 42.94$^+_-$0.01 \\
          &                   &            &             &   &    &         & UVW1 & 43.02$^+_-$0.01 \\
\hline
NGC\,526A & \emph{XMM-Newton}$^c$ & 0109130201 & 2002-06-30 & 35 &  8 & 20949   & UVM2 & 42.13$^+_-$0.08  \\
          &                   &            &             &   &    &         & UVW1 & 42.27$^+_-$0.03 \\
          & \emph{XMM-Newton} & 0150940101 & 2003-06-21 & 35 &  35 & 118532 & UVW2 & 41.99$^+_-$0.07 \\
          &                   &            &             &   &    &         & UVM2 & 42.15$^+_-$0.04 \\
          &                   &            &             &   &    &         & UVW1 & 42.41$^+_-$0.01 \\
          & \emph{XMM-Newton} & 0721730301 & 2013-12-22 & 35 &  48 & 177241 & - \\
          & \emph{XMM-Newton} & 0721730401 & 2013-12-22 & 35 &  37 & 164811 & - \\
          & \emph{Chandra}$^c$ & 342 & 2000-02-07 & 4 &  9 & 1158 & -   \\
\hline
MARK\,609 & \emph{XMM-Newton} & 0103861001 & 2002-08-13 & 20 &  7 & 5054 & UVW2 & 43.32$^+_-$0.04 \\
          & \emph{XMM-Newton} & 0402110201 & 2007-01-27 & 20 &  16 & 10002 & UVM2 & 43.37$^+_-$0.01  \\
\hline
NGC\,1365 & \emph{Chandra}$^c$ & 6869 & 2006-04-20 & 2 & 15 & 2021 &  - \\
          & \emph{XMM-Newton} & 0151370101 & 2003-01-16 & 20 & 14 & 10599 & UVW2 & 41.71$^+_-$0.02  \\
          & \emph{XMM-Newton} & 0151370201 & 2003-02-09 & 20 & 3 & 1481 &  - \\
          & \emph{XMM-Newton} & 0151370701 & 2003-08-13 & 20  & 6 & 6337   & UVW2 & 41.71$^+_-$0.03  \\
          & \emph{XMM-Newton} & 0205590301 & 2004-01-17 & 20 & 50 & 83303 & UVW1 & 41.69$^+_-$0.01  \\
          &                   &            &             &   &    &         & UVM2 & 41.30$^+_-$0.02 \\
          & \emph{XMM-Newton}$^c$ & 0205590401 & 2004-07-24 & 20 & 29 & 30181 & UVW1 & 42.21$^+_-$0.01  \\
          &                   &            &             &   &    &         & UVM2 & 41.95$^+_-$0.01 \\
          & \emph{XMM-Newton} & 0505140201 & 2007-06-30 & 20 & 42 & 26349 & UVW1 & 42.07$^+_-$0.01  \\
          &                   &            &             &   &    &         & UVM2 & 41.35$^+_-$0.02 \\
          & \emph{XMM-Newton} & 0505140401 & 2007-07-02 & 20 & 78 & 49947 & UVW1 & 42.08$^+_-$0.01  \\
          & \emph{XMM-Newton} & 0505140501 & 2007-07-04 & 20 & 46 & 29528 & UVW1 & 42.24$^+_-$0.01  \\
          &                   &            &             &   &    &         & UVM2 & 42.04$^+_-$0.01 \\
          & \emph{XMM-Newton} & 0692840201 & 2012-07-25 & 20 & 106 & 115167 & UVW1 & 42.01$^+_-$0.01  \\
          & \emph{XMM-Newton} & 0692840301 & 2012-12-24 & 20 & 95 & 276846 & UVW1 & 41.94$^+_-$0.01  \\
          & \emph{XMM-Newton} & 0692840401 & 2013-01-23 & 20 & 89 & 366687 & UVW1 & 41.78$^+_-$0.01  \\
          & \emph{XMM-Newton} & 0692840501 & 2013-02-12 & 20 & 102 & 172790 & UVW1 & 41.77$^+_-$0.01  \\
\hline             
NGC\,2617 & \emph{XMM-Newton} & 0701981601 & 2013-04-27 & 25 &  39 & 574119 & UVW2 & 42.95$^+_-$0.02  \\
          &                   &            &             &   &    &         & UVM2 & 43.02$^+_-$0.01 \\
          &                   &            &             &   &    &         & UVW1 & 43.05$^+_-$0.01 \\
          & \emph{XMM-Newton} & 0701981901 & 2013-05-24 & 25 &  14 & 496297 & UVW1 & 43.24$^+_-$0.01  \\
\hline
MARK\,1218 & \emph{XMM-Newton} & 0302260201 & 20050409 & 20 & 6 & 3652 & UVW2 & 41.98$^+_-$0.25  \\
          &                   &            &             &   &    &         & UVM2 & 42.12$^+_-$0.09 \\
          &                   &            &             &   &    &         & UVW1 & 42.43$^+_-$0.03 \\
          & \emph{XMM-Newton} & 0302260401 & 20051008 & 20 &  3 & 935 & UVM2 &  42.06$^+_-$0.10 \\
\hline
NGC\,2992 & \emph{XMM-Newton} & 0654910501 & 2010-05-26 & 25  & 35 & 85124 & UVM2 & 41.14$^+_-$0.08 \\
          & \emph{XMM-Newton} & 0654910601 & 2010-06-05 & 25  & 33 & 35540 & UVM2 & 41.28$^+_-$0.05  \\
          & \emph{XMM-Newton} & 0654910701 & 2010-11-08 & 25  & 37 & 40296 & & Not detected   \\
          & \emph{XMM-Newton} & 0654910901 & 2010-11-28 & 25  & 33 & 20874 & & Not detected   \\
          & \emph{XMM-Newton} & 0654911001 & 2010-12-08 & 25  & 38 & 44100 & UVW2 & 40.70$^+_-$0.33 \\
          &                   &            &             &   &    &         & UVM2 & 41.24$^+_-$0.03 \\
          &                   &            &             &   &    &         & UVW1 & 41.87$^+_-$0.01 \\
          & \emph{XMM-Newton} & 0701780101 & 2013-05-11 & 25  &  9 & 26676 & UVM2 & 41.09$^+_-$0.06 \\
\hline
POX\,52 & \emph{Chandra}$^c$ & 5736 & 2006-04-18 & 2 &  22 & 4979 & -  \\
          & \emph{XMM-Newton}$^c$ & 0302420101 & 2005-07-08 & 20  & 69 & 4869 & UVM2 & 42.21$^+_-$0.03  \\
\hline
NGC\,4138 & \emph{Chandra}$^c$ & 3994 & 2003-03-16 & 2 &  6 & 889 & - \\
          & \emph{XMM-Newton}$^c$ & 0112551201 & 2001-11-26 & 25 &  8 & 4271 & - \\
\hline
NGC\,4395 & \emph{Chandra} & 5302 & 2004-04-11 & 2 &  28 & 2779 & -  \\
          & \emph{Chandra}$^c$ & 5301 & 2004-04-10 & 2 &  27 & 3105 & -  \\
          & \emph{XMM-Newton}$^c$ & 0112521901 & 2002-05-31 & 25 &  8 & 5209 & UVW1 & 40.09$^+_-$0.02  \\
          & \emph{XMM-Newton} & 0744010101 & 2014-12-28 & 25 & 36 & 16941 & UVW1 & 39.99$^+_-$0.01 \\
          & \emph{XMM-Newton} & 0744010201 & 2014-12-31 & 25 & 14 & 8819 &  UVW1 & 40.07$^+_-$0.01 \\
\hline        
NGC\,4565 & \emph{XMM-Newton}$^c$ & 0112550301 & 2001-07-01 & 20 &  9 & 1209 & - \\
          & \emph{Chandra}$^c$ & 3950 & 2003-02-08 & 2 &  59 & 2158 & - \\
\hline
MARK\,883 & \emph{XMM-Newton} & 0302260101 & 2006-08-13 & 25  & 8 & 5311 & UVW2 & 42.99$^+_-$0.08 \\
          &                   &            &             &   &    &         & UVM2 & 42.92$^+_-$0.04 \\
          &                   &            &             &   &    &         & UVW1 & 43.01$^+_-$0.02 \\
          & \emph{XMM-Newton} & 0302260701 & 2006-08-15 & 25  & 10 & 7307 & UVW2 & 42.85$^+_-$0.06 \\
          &                   &            &             &   &    &         & UVM2 & 42.91$^+_-$0.04 \\
          &                   &            &             &   &    &         & UVW1 & 43.02$^+_-$0.02 \\
          & \emph{XMM-Newton} & 0302261001 & 2006-08-21 & 25  & 8 & 7000 & UVW2 & 42.75$^+_-$0.06 \\
          &                   &            &             &   &    &         & UVM2 & 42.94$^+_-$0.04 \\
          &                   &            &             &   &    &         & UVW1 & 43.03$^+_-$0.02 \\
          & \emph{XMM-Newton} & 0652550201 & 2010-08-20 & 25  & 26 & 12409 & UVW1 & 43.13$^+_-$0.01 \\
\hline
IRAS\,2005-1117 & \emph{XMM-Newton} & 0044350201 & 2002-04-20 & 20 & 4 & 4535 & UVW2 & 42.65$^+_-$0.09 \\
          &                   &            &             &   &    &         & UVW1 & 43.03$^+_-$0.01 \\
             & \emph{XMM-Newton} & 0044350501 & 2002-10-21 & 20 & 10 & 7020 & UVW1 & 43.01$^+_-$0.02 \\
\hline     
\caption*{{\bf Notes.} (Col. 1) name, (Col. 2) instrument, (Col. 3)
  obsID, (Col. 4) date, (Col. 5) aperture radius for the nuclear
  extraction, (Col. 6) net exposure time, (Col. 7) number of counts in
  the 0.5-10 keV band, (Cols. 8 and 9) UV luminosity from the optical
  monitor and filter. The $c$ represents data from different
  instruments that were compared as explained in Sect. \ref{simult}.}
\end{longtable} 

\begin{figure*}
\centering
{\includegraphics[width=0.30\textwidth]{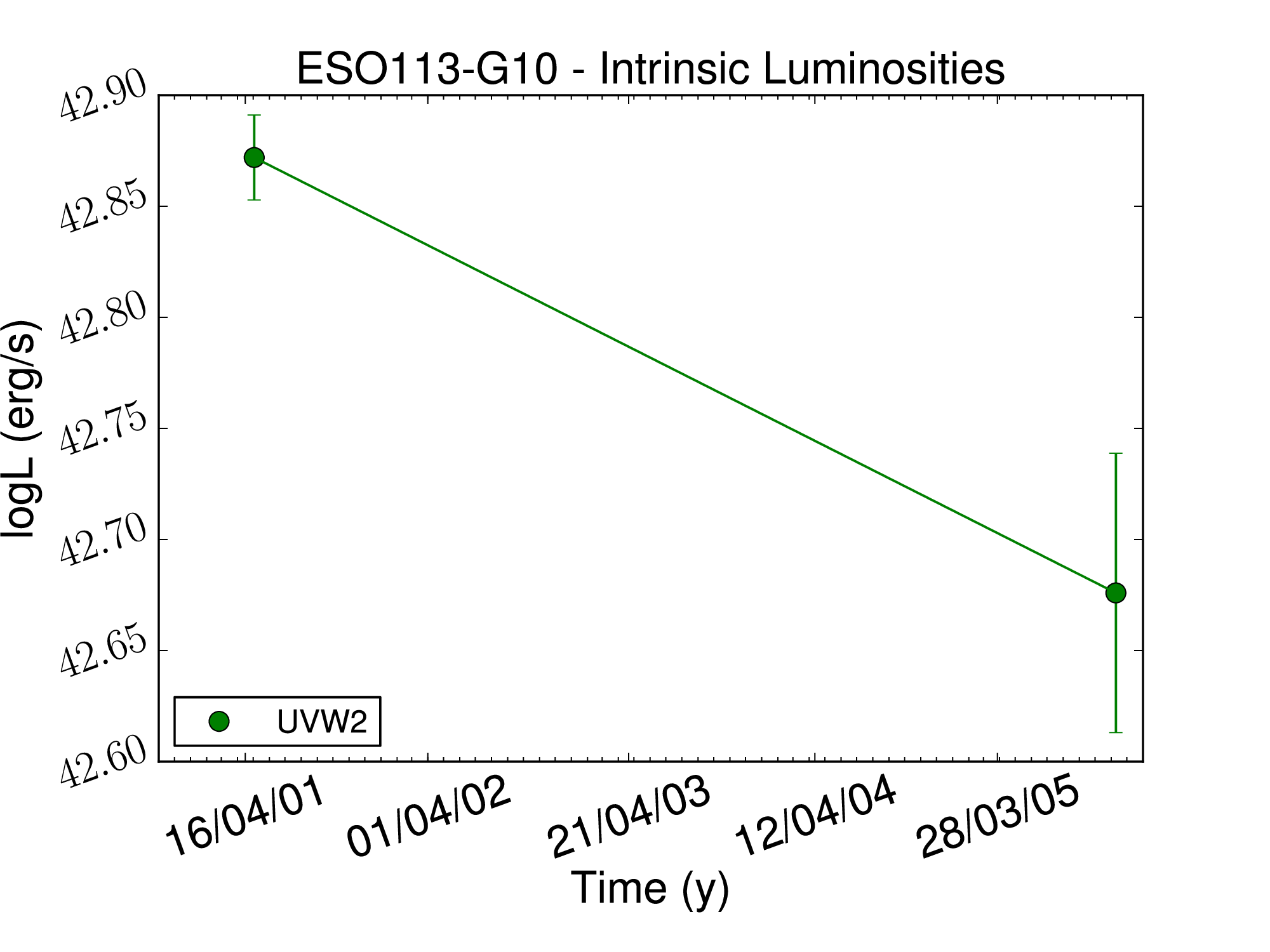}}
{\includegraphics[width=0.30\textwidth]{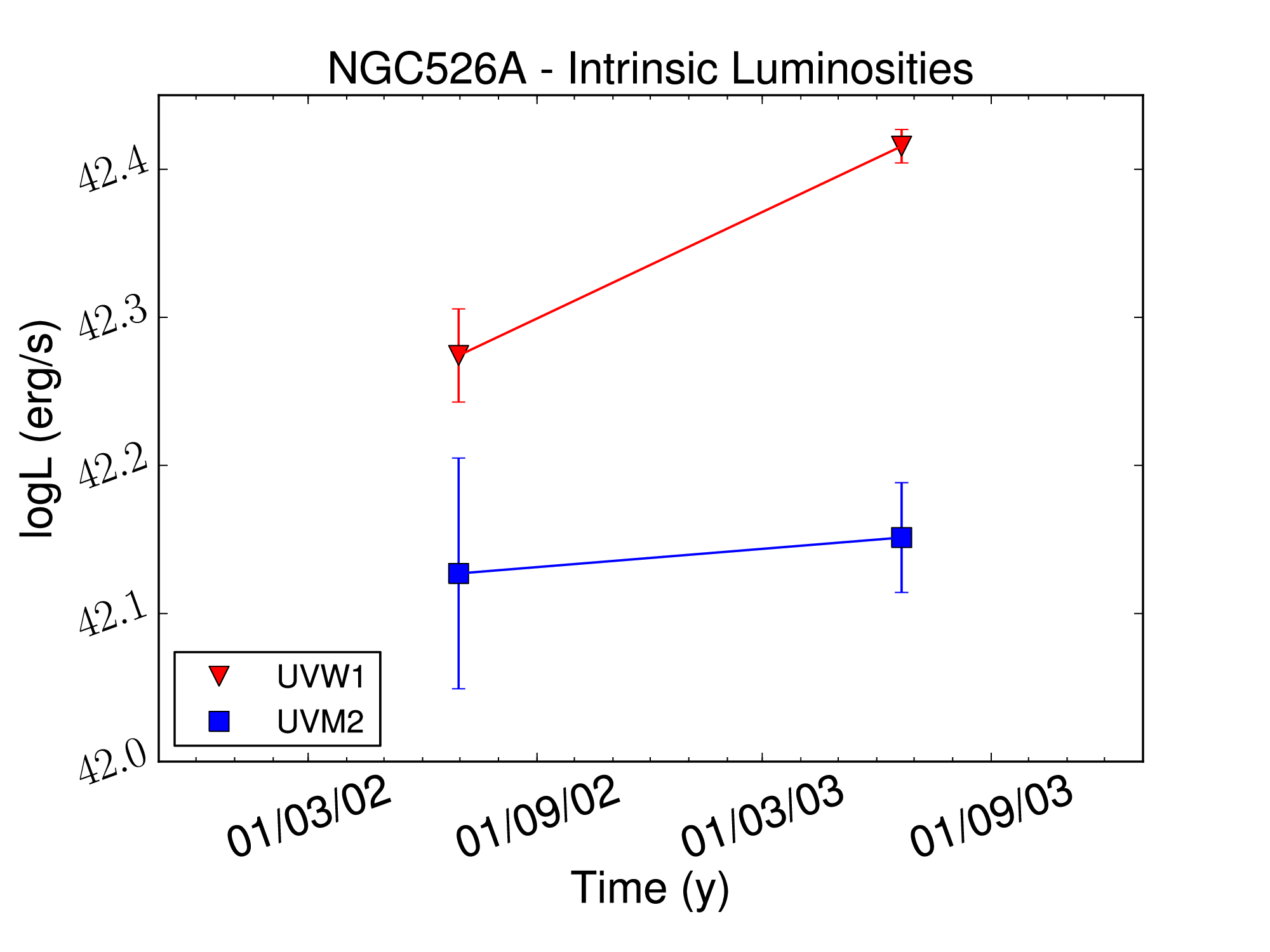}}
{\includegraphics[width=0.30\textwidth]{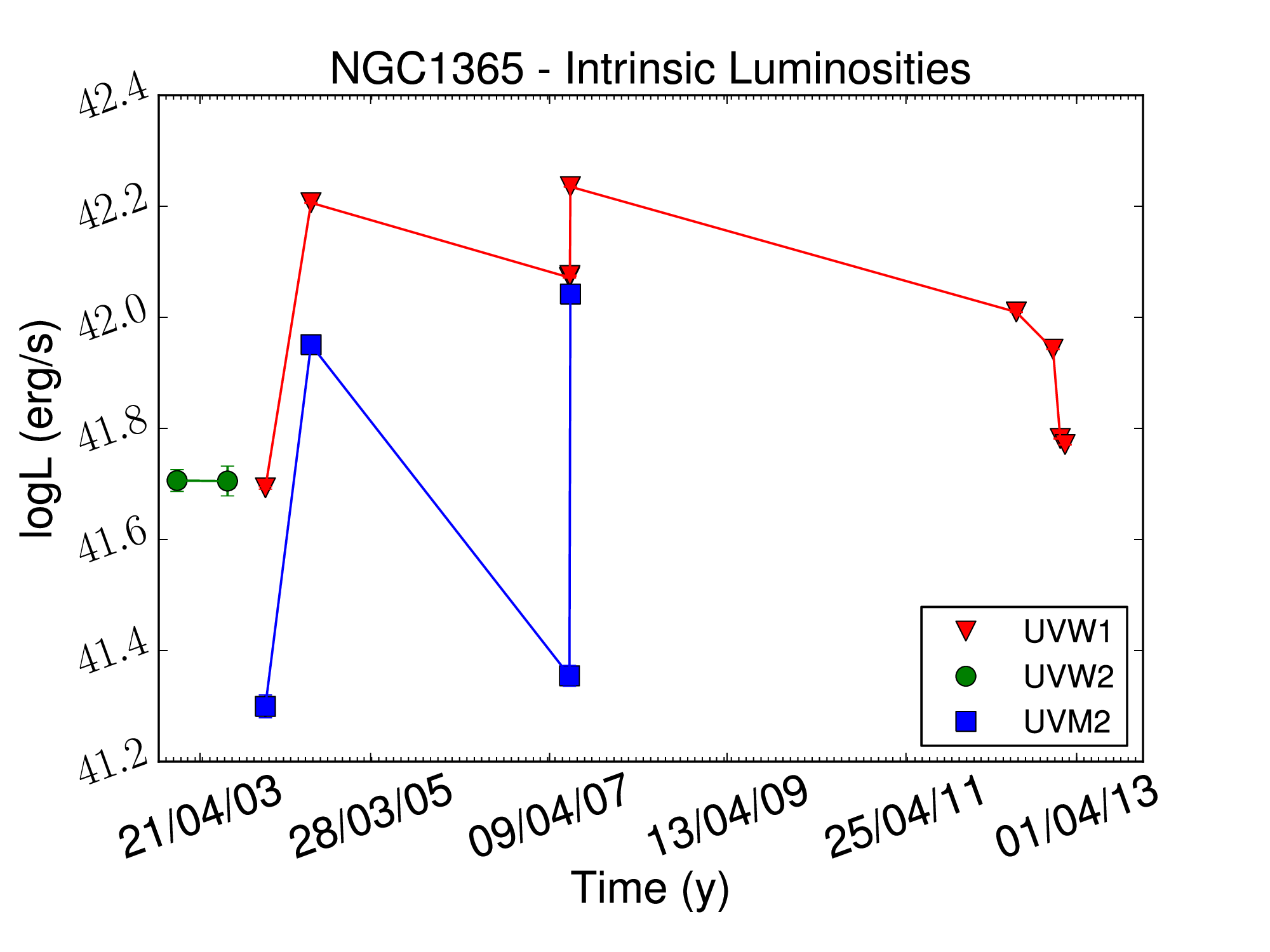}}

{\includegraphics[width=0.30\textwidth]{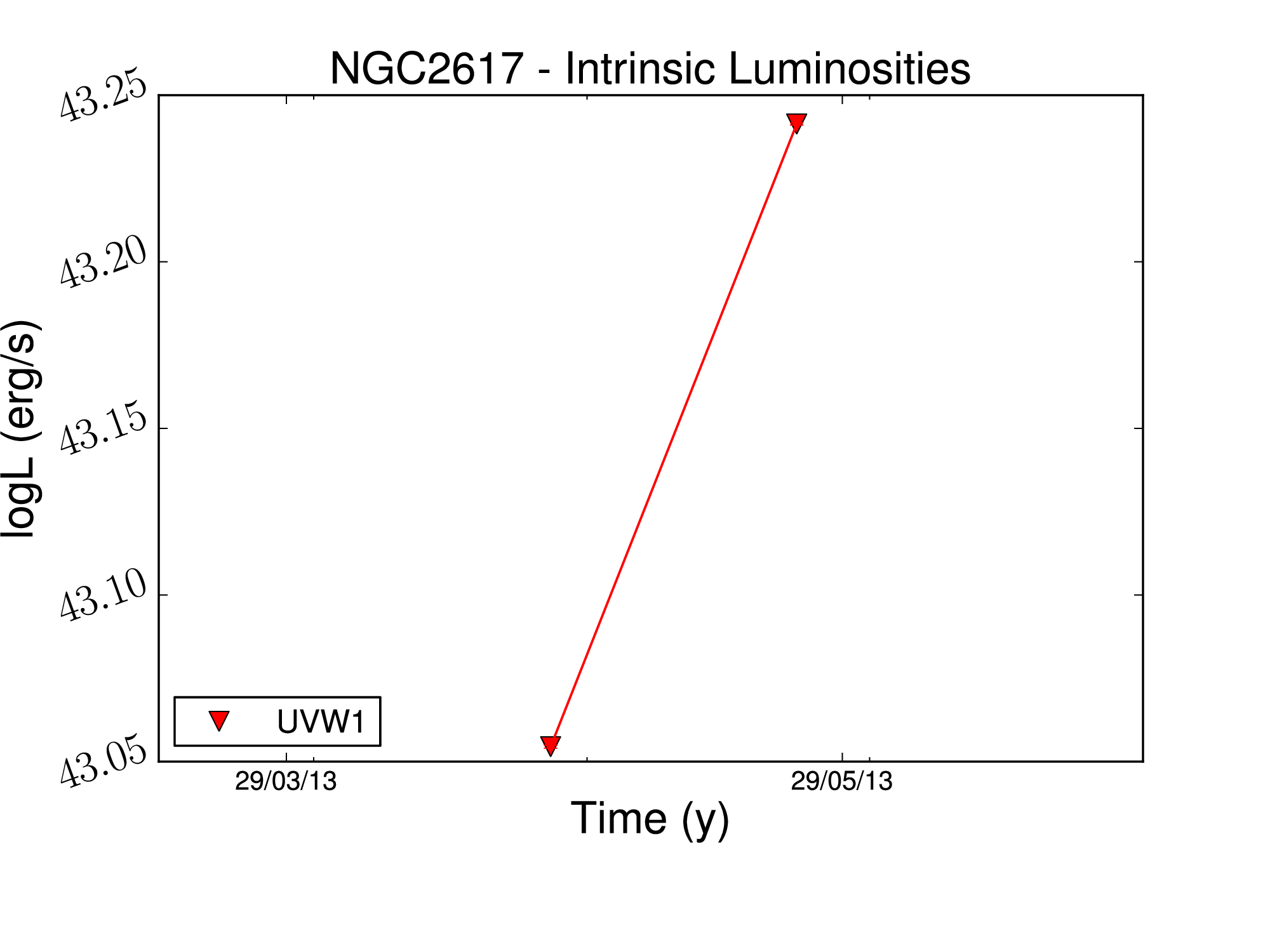}}
{\includegraphics[width=0.30\textwidth]{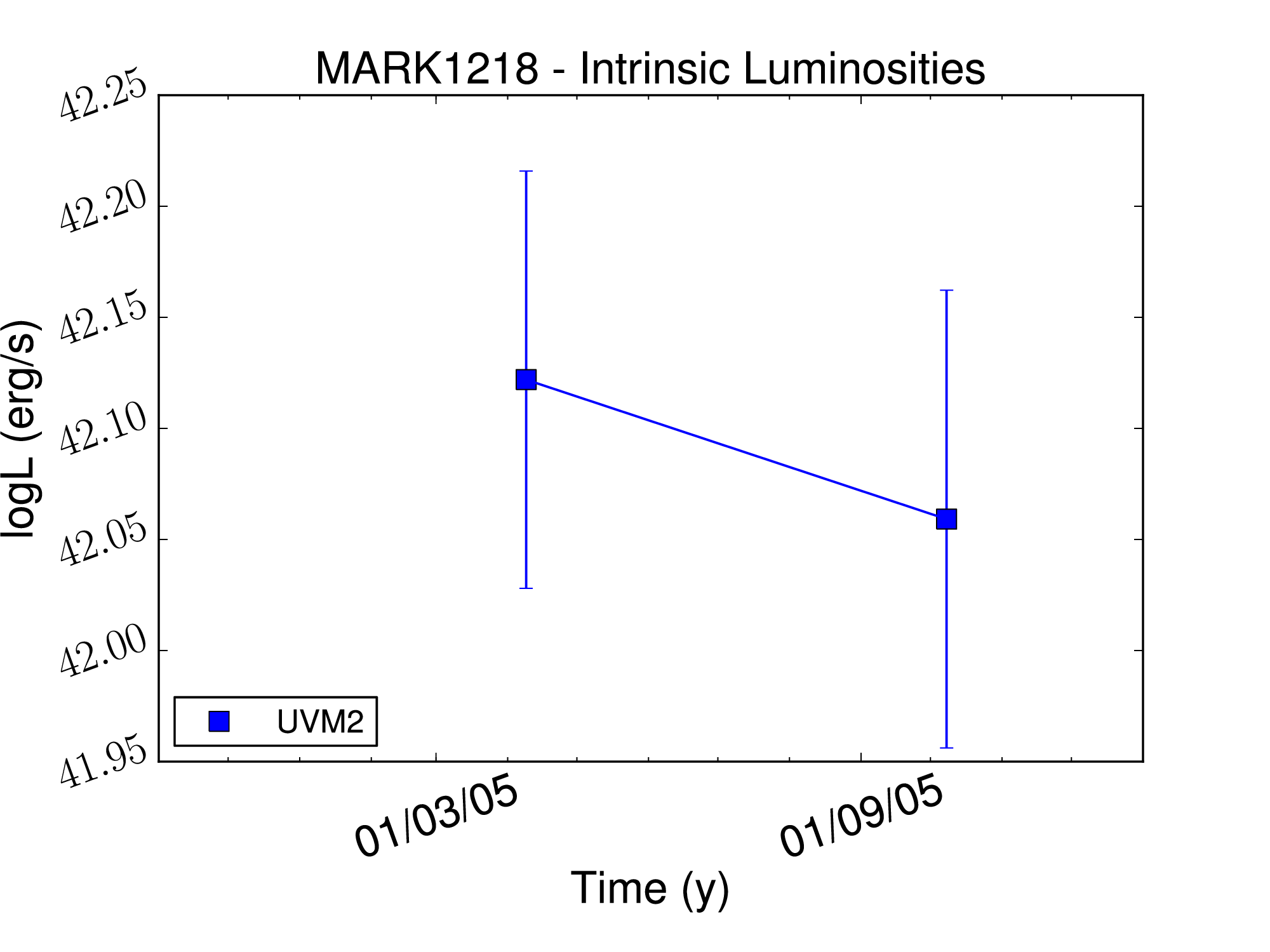}}
{\includegraphics[width=0.30\textwidth]{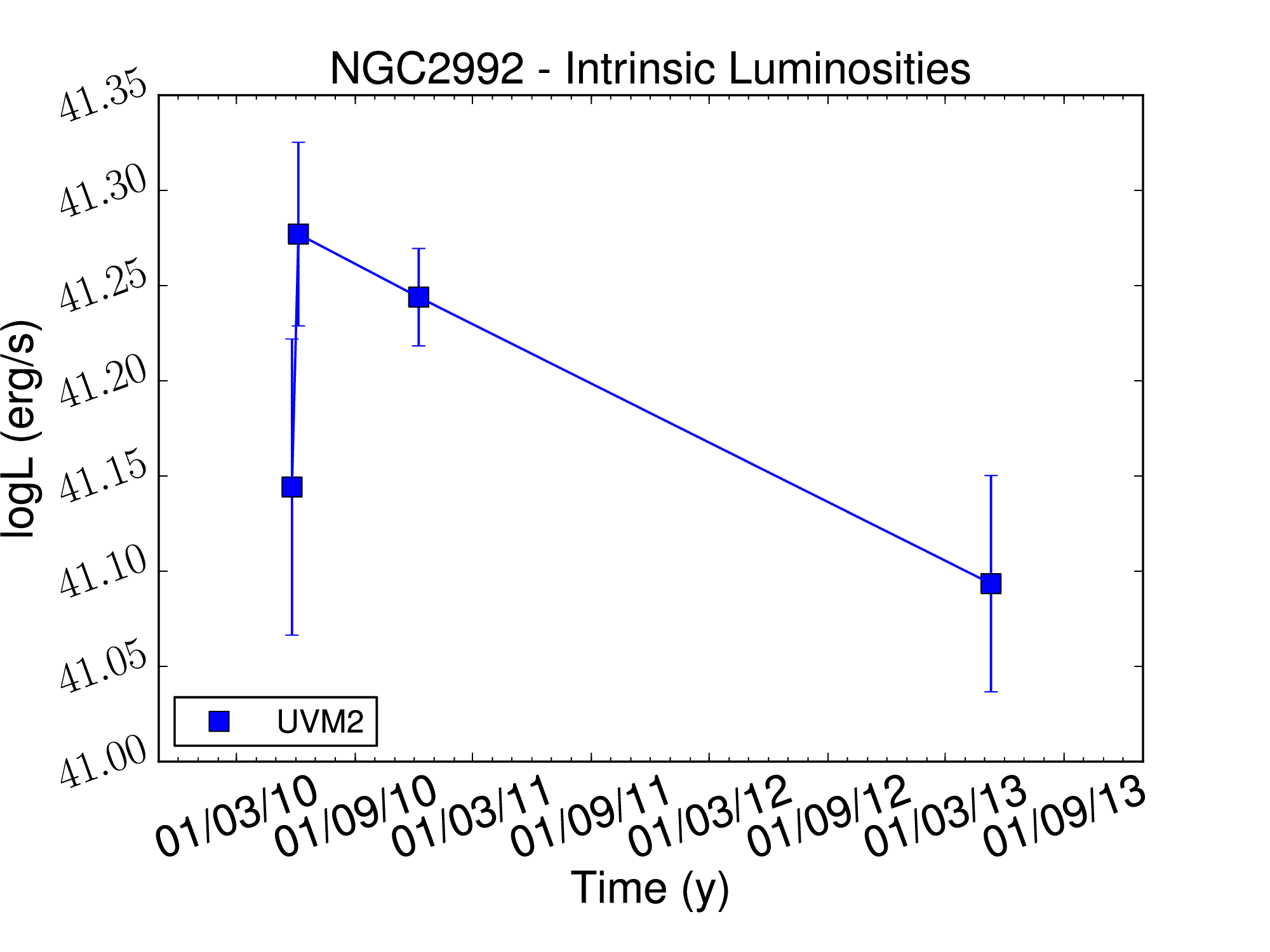}}

{\includegraphics[width=0.30\textwidth]{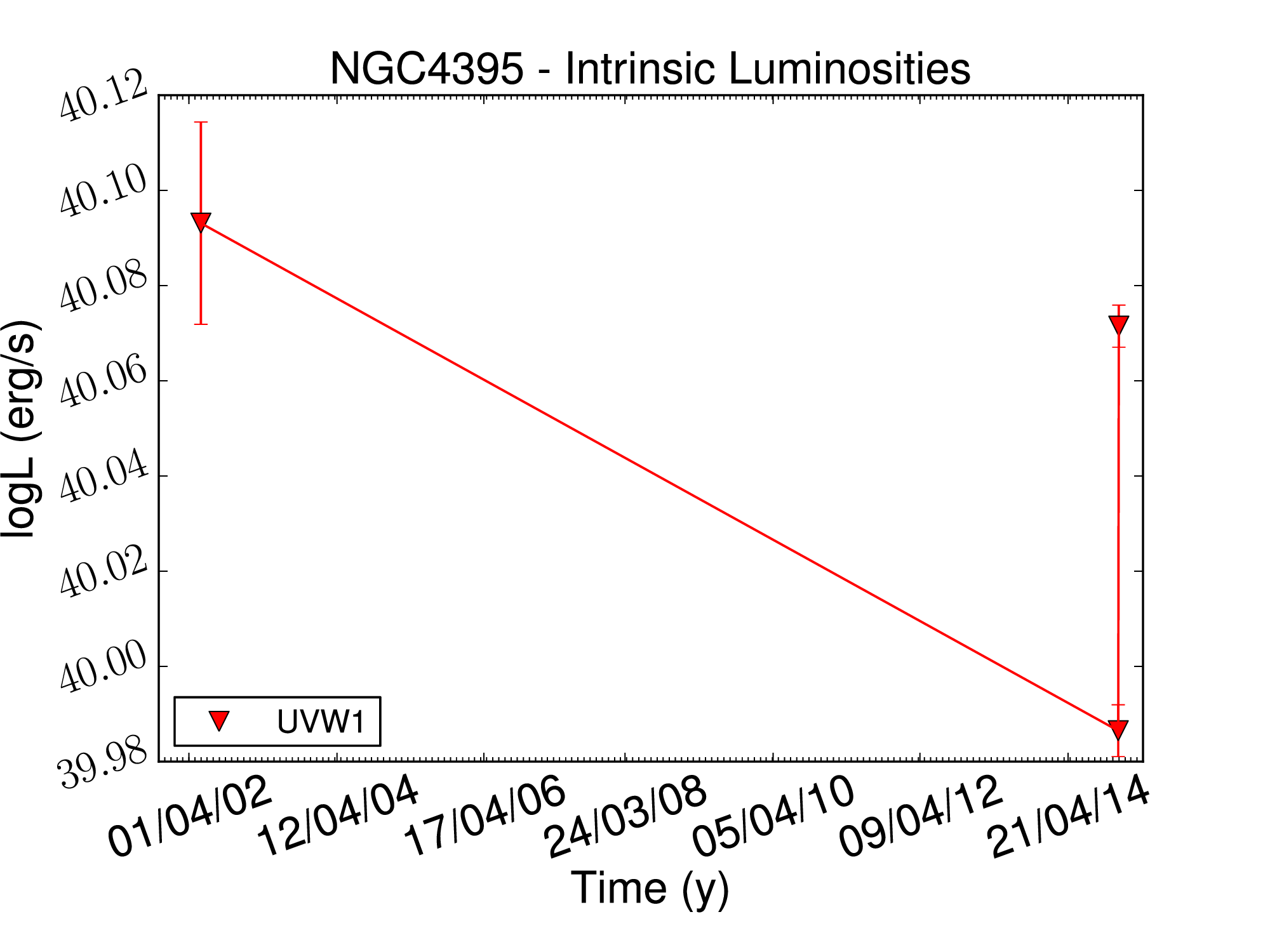}}
{\includegraphics[width=0.30\textwidth]{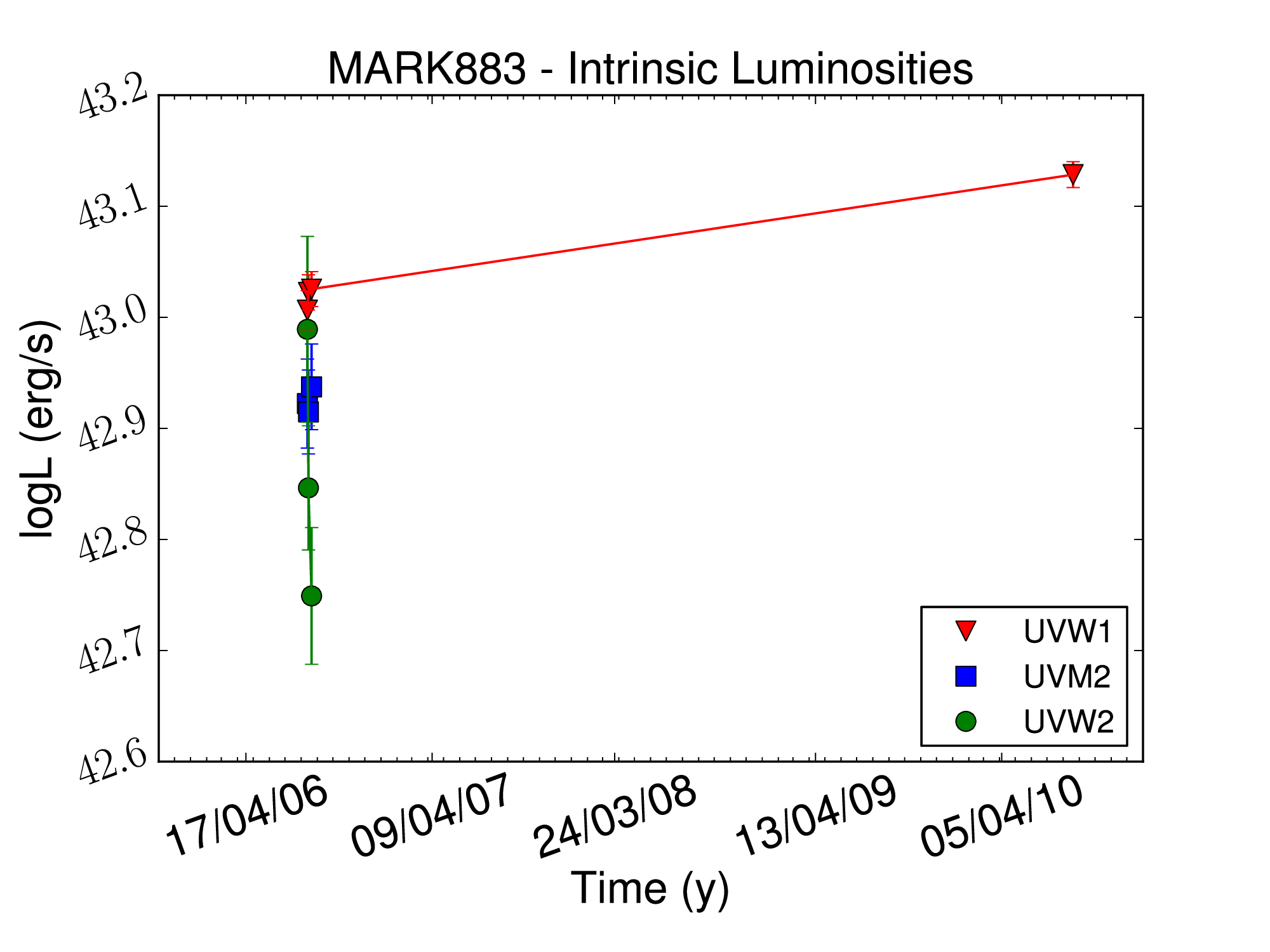}}
{\includegraphics[width=0.30\textwidth]{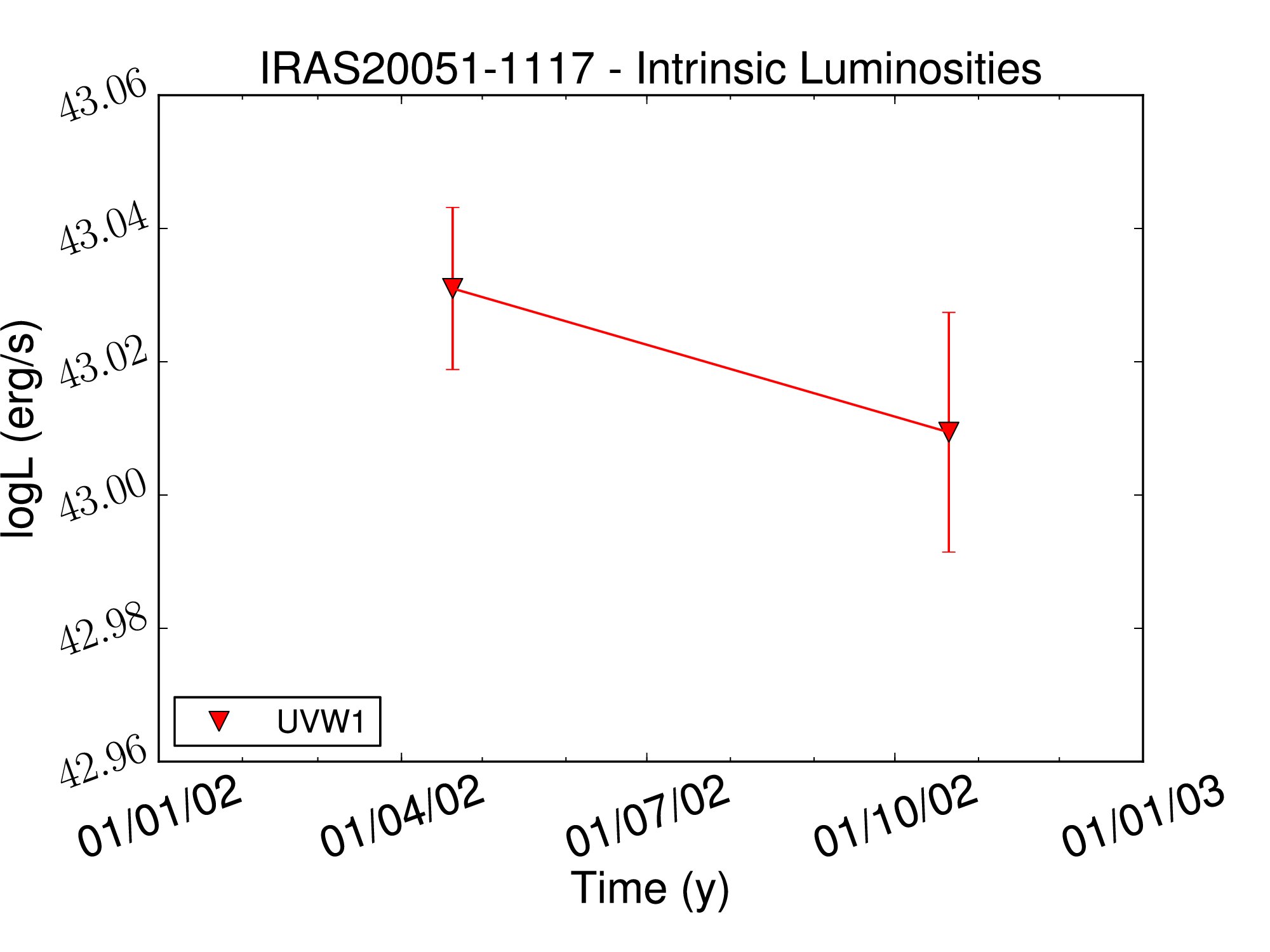}}
\caption{UV luminosities obtained from the data with the OM camera onboard \emph{XMM}--Newton, when available. Different filters have been used; UVW1 (red triangles), UVW2 (green circles), and UVM2 (blue squares).}
\label{luminUVfigSey}
\end{figure*}

\tiny
\renewcommand{\arraystretch}{1.4}
\begin{longtable}{lccccccccc}
\caption[Spectral fittings]{\label{bestfitSey} Final compilation of
  the best-fit models for the sample, including the individual
  best-fit model for each observation, and the simultaneous best-fit
  model (i.e., SMF0, SMF1, SMF2, or SMF3) with the varying parameters (except for SMF0). Note that when the column density is compatible with the Galactic absorption, a `-' is placed. }  \\ \hline \hline Analysis &
ObsID & Model & $N_{H1}$ & $N_{H2}$ & kT & $\Gamma$ & $Norm_1$ &
$Norm_2$ & $\chi^2/d.o.f$ \\ & & & & & keV & & ($10^{-4}$) &
($10^{-4}$) & F-test \\ (1) & (2) & (3) & (4) & (5) & (6) & (7) & (8)
& (9) & (10) \\ \hline \endfirsthead
\caption[]{(Cont.)} \\
\hline \hline
Analysis & ObsID & Model & $N_{H1}$ & $N_{H2}$ & kT & $\Gamma$ & $Norm_1$ & $Norm_2$ & $\chi^2/d.o.f$  \\
  & & &  &  & keV & & ($10^{-4}$) & ($10^{-4}$) & F-test \\
(1) & (2) & (3) & (4) & (5) & (6) & (7) & (8) & (9) & (10)  \\
\hline  
\endhead
\endfoot
\\
\endlastfoot
\multicolumn{10}{c}{ESO\,540-G01} \\ \hline 
Ind & 2192 & MEPL & 0$ _{ 0 }^{ 0.45 }$ & 0.04$ _{ 0 }^{ 0.14 }$ & 0.36$ _{ 0.20 }^{ 0.52 }$ & 1.89$ _{ 1.69 }^{ 2.15 }$ & 0.17$ _{ 0.08 }^{ 2.73 }$ & 0.69$ _{ 0.55 }^{ 0.84 }$ & 46.20/42 \\
Ind & 0044350101 & MEPL & - & 0.75$ _{ 0.63 }^{ 1.15 }$ & 0.65$ _{ 0.33 }^{ 0.80 }$ & 1.82$ _{ 1.74 }^{ 1.90 }$  & 0.90$ _{ 0.81 }^{ 0.99 }$ & 0.08$ _{ 0.01 }^{ 6.03 }$ & 87.51/92 \\ \hline
SMF (+ring) & 0044350101 & MEPL & - & - & 0.38$_{0.31}^{0.55}$ & 1.68$_{1.61}^{1.76}$ & 0.15$_{0.10}^{0.20}$ & 0.87$_{0.82}^{0.93}$ & 137.15/142  \\
& 2129 & & & & & & 0.57$_{0.53}^{0.62}$ & & \\
\hline
\multicolumn{10}{c}{ESO\,195-IG21} \\ \hline 
Ind & 13898 & MEPL & 5.34$ _{ 3.07 }^{ 8.25 }$ & 10.43$ _{ 5.98 }^{ 19.62 }$ & 0.27$ _{ 0.14 }^{ 1.11 }$ & 2.42$_{ 1.17 }^{ 0 }$ & 252.77$ _{ 2.03 }^{ 330.35 }$ & 9.58$ _{ 5.05 }^{ 135.83 }$ & 15.09/16 \\
Ind & 0554500201 & MEPL & 0.79$ _{ 0.68 }^{ 0.93 }$ & 3.39$ _{ 3.11 }^{ 3.69 }$ & 0.14$ _{ 0.11 }^{ 0.17 }$ & 1.32$ _{ 1.22 }^{ 1.43 }$ & 46.69$ _{ 11.98 }^{ 308.08 }$ & 4.87$ _{ 4.10 }^{ 5.82 }$ & 274.21/262 \\ \hline
SMF (+ring) & 0554500201 & MEPL & 0.27$_{0.00}^{2.11}$ & 3.81$_{3.15}^{4.65}$ & 2.67$_{0.38}^{4.85}$ & 1.35$_{1.24}^{1.48}$ & 0.47$_{0.13}^{2.23}$ & 5.15$_{4.21}^{6.43}$ & 270.70/255  \\
& 2012 & & & & & & 0.04$_{0.01}^{0.12}$ & 0.08$_{0.03}^{0.14}$ &   \\ 
\hline
\multicolumn{10}{c}{ESO\,113-G10} \\ \hline
Ind & 0103861601 & ME2PL & - & 1.57$ _{ 0.99 }^{ 2.82 }$ & 0.07 $_{ 0.07 }^{ 0.09 }$ & 3.27$ _{ 3.18 }^{ 3.40 }$ & 6.07$ _{ 6.07 }^{ 7.37 }$ & 6.12$ _{ 2.72 }^{ 12.09 }$ & 277.61/236 \\
Ind & 0301890101 & ME2PL & - & 4.02$ _{ 3.73 }^{ 4.35 }$ & 0.07$ _{ 0.07 }^{ 0.08 }$ & 2.63$ _{ 2.60 }^{ 2.66 }$ & 14.76$ _{ 14.76 }^{ 14.96 }$ & 19.33$ _{ 17.97 }^{ 20.76 }$ & 2080.8/1088 \\
\hline
SMF1 & 0103861601 & ME2PL & - & 3.62$ _{ 3.41 }^{ 3.87 }$ & 0.08$_{0.07}^{0.09}$ & 2.67$_{2.65}^{2.70}$ &  8.95$ _{ 8.73 }^{ 9.17 }$ & 20.48$_{19.13}^{21.91}$ & 2281.77/1302 \\
     & 0301890101 &       &  & & & & 14.81$_{14.70}^{14.91}$ & \\
\hline
\multicolumn{10}{c}{NGC\,526A} \\ \hline
Ind & 0109130201 & 2PL & 0.01$ _{ 0 }^{ 0.06 }$ & 1.14$ _{ 1.08 }^{ 1.24 }$ & - & 1.38$ _{ 1.34 }^{ 1.43 }$ & 1.08$ _{ 0.97 }^{ 1.58 }$ & 25.31$ _{ 23.72 }^{ 27.13 }$ & 706.22/732 \\
Ind & 0150940101 & 2PL & 0.03$ _{ 0 }^{ 0.05 }$ & 1.17$ _{ 1.14 }^{ 1.21 }$ & - & 1.48$ _{ 1.46 }^{ 1.50 }$ & 1.25$ _{ 1.04 }^{ 1.49 }$ & 39.74$_{ 38.67 }^{ 40.86 }$ & 1609.78/1538 \\
Ind & 0721730301 & 2PL & 0.03$ _{ 0.01 }^{ 0.05 }$ & 1.25$_{ 1.22 }^{ 1.28 }$ & - & 1.49$ _{ 1.47 }^{ 1.50 }$ & 1.37$ _{ 1.19 }^{ 1.58 }$ & 46.44$ _{ 45.37 }^{ 47.53 }$ & 1718.2/1638 \\
Ind & 0721730401 & 2PL & 0.06$ _{ 0.03 }^{ 0.09 }$ & 1.23$ _{ 1.20 }^{ 1.27 }$ & - & 1.50$ _{ 1.48 }^{ 1.51 }$ & 1.71$ _{ 1.44 }^{ 2.02 }$ & 55.94$ _{ 54.64 }^{ 57.30 }$ & 1747.04/1614 \\
Ind & 342 &  2PL & 0.73$ _{ 0.69 }^{ 1.04 }$ & 120.02$_{ 64.85 }^{ 185.04 }$ & - & 0.97$_{ 0.84 }^{ 1.25}$ & 2.40$ _{ 1.99 }^{ 2.78 }$ & 16.32$ _{ 4.02 }^{ 25.17 }$ & 61.32/45 \\ \hline
SMF1 & 0109130201 & 2PL & 0.03$_{0.02}^{0.04}$ & 1.21$_{1.20}^{1.23}$ & - & 1.48$_{1.47}^{1.49}$ & 1.36$_{1.24}^{1.48}$ & 28.49$ _{ 27.97 }^{ 29.02 }$ & 5863.99/5543 \\
     & 0150940101 & & & & & & & 40.16$ _{ 39.59 }^{ 40.74 }$ &   \\
     & 0721730301  & & & & & & & 45.72$ _{ 45.07 }^{ 46.38 }$   \\
     & 0721730401  & & & & & & & 55.12$ _{ 54.33 }^{ 55.93 }$   \\
\hline
\multicolumn{10}{c}{MARK\,609} \\ \hline
Ind & 0103861001 & 2PL & - & 10.43$ _{ 5.28 }^{ 149.71 }$ & - & 1.96$ _{ 1.89 }^{ 2.03 }$ & 3.78$ _{ 3.67 }^{ 3.89 }$ & 2.21$ _{ 1.18 }^{ 46.21 }$ & 183.79/173 \\
Ind & 0402110201 & 2PL & - & 12.62$ _{ 7.69 }^{ 24.95 }$ & - & 1.81$ _{ 1.78 }^{ 1.89 }$ & 2.90$ _{ 2.84 }^{ 2.96 }$ & 2.45$ _{ 1.80 }^{ 3.31 }$ & 359.17/319 \\ \hline
SMF1 & 0103861001 & 2PL & - & 11.41$_{7.16}^{17.41}$ & - & 1.89$_{1.85}^{1.93}$ & 3.71$ _{ 3.61 }^{ 3.81 }$ & 2.58$_{1.95}^{3.16}$ & 566.07/499 \\
     & 0402110201 &    &   &                        &   &                      & 2.91$ _{ 2.85 }^{ 2.97 }$ &                       &  \\
\hline 
\multicolumn{10}{c}{NGC\,1365} \\ \hline 
Ind & 6869 & 2PL & 0$ _{ 0 }^{ 0.05 }$  & 35.98$ _{ 33.47 }^{ 38.60 }$& - & 1.53$ _{ 1.35 }^{ 1.76 }$  & 0.46$ _{ 0.43 }^{ 0.54 }$ & 38.56$ _{ 27.19 }^{ 61.78 }$ & 126.46/89 \\
Ind & 0151370101 & ME2PL & - & 40.07$ _{ 37.58 }^{ 42.76 }$ & 0.64$ _{ 0.61 }^{ 0.66 }$ & 1.90$ _{ 1.83 }^{ 1.96 }$ & 1.60$ _{ 1.59 }^{ 1.76 }$ & 56.68$ _{ 48.52 }^{ 66.66 }$ & 421.94/359 \\
Ind & 0151370201 & ME2PL & - & 13.30$ _{ 8.30 }^{ 22.30 }$ & 0.73$ _{ 0.65 }^{ 0.82 }$ & 2.02$ _{ 1.83 }^{ 2.22 }$ & 1.57$ _{ 1.57 }^{ 2.01 }$ & 9.21$ _{ 5.56 }^{ 14.60 }$ & 70.21/49 \\
Ind & 0151370701 & ME2PL  & 0$ _{ 0 }^{ 0.04 }$ & 32.58$ _{ 30.26 }^{ 34.62 }$ & 0.68$ _{ 0.65 }^{ 0.71 }$ & 2.12$ _{ 2.04 }^{ 2.38 }$ & 1.71$ _{ 1.59 }^{ 2.02 }$ & 97.42$ _{ 97.41 }^{ 188.31 }$ &  335.83/232 \\
Ind & 0205590301 & ME2PL & 0.06$ _{ 0.05 }^{ 0.06 }$ & 11.95$ _{ 11.81 }^{ 12.12 }$ & 0.62$ _{ 0.60 }^{ 0.63 }$ & 2.18$ _{ 2.17 }^{ 2.22 }$ & 2.65$ _{ 2.65 }^{ 2.80 }$ & 130.56$ _{ 125.17 }^{ 144.97 }$ & 2751.27/1504 \\
Ind & 0205590401 & ME2PL & - & 25.86$ _{ 24.72 }^{ 26.41 }$ & 0.62$ _{ 0.61 }^{ 0.64 }$ & 2.03$ _{ 1.98 }^{ 2.08 }$ & 1.65$ _{ 1.65 }^{ 1.77 }$ & 80.55$ _{ 72.16 }^{ 88.22 }$ & 1181.26/778 \\
Ind & 0505140201 & ME2PL & - & 52.73$ _{ 46.76 }^{ 59.11 }$ & 0.65$ _{ 0.64 }^{ 0.66 }$ & 2.20$ _{ 2.16 }^{ 2.25 }$ & 1.53$ _{ 1.52 }^{ 1.62 }$ & 25.72$ _{ 21.58 }^{ 30.56 }$ & 632.85/444 \\
Ind & 0505140401 & ME2PL & - & 64.17$ _{ 60.03 }^{ 65.92 }$ & 0.64$ _{ 0.63 }^{ 0.65 }$ & 2.15$ _{ 2.13 }^{ 2.20 }$ & 1.57$ _{ 1.57 }^{ 1.63 }$ & 45.38$ _{ 41.01 }^{ 50.70 }$ & 977.71/681 \\
Ind & 0505140501 & ME2PL & - & 60.11$ _{ 54.24 }^{ 66.62 }$ & 0.63$ _{ 0.62 }^{ 0.65 }$ & 2.14$ _{ 2.10 }^{ 2.17 }$ & 1.57$ _{ 1.57 }^{ 1.65 }$ & 35.33$ _{ 30.12 }^{ 41.61 }$ & 700.17/499 \\
Ind & 0692840201 & ME2PL & - & 26.07$ _{ 25.71 }^{ 26.43 }$  & 0.65$ _{ 0.64 }^{ 0.66 }$ & 2.12$ _{ 2.10 }^{ 2.15 }$ & 1.69$ _{ 1.66 }^{ 1.72 }$ & 104.65$ _{ 104.65 }^{ 115.25 }$ &  2186.8/1529 \\
Ind & 0692840301 & ME2PL & - & 3.66$ _{ 3.60 }^{ 3.72 }$ & 0.61$ _{ 0.58 }^{ 0.64 }$ & 1.40$ _{ 1.39 }^{ 1.42 }$ & 4.91$ _{ 4.91 }^{ 5.06 }$ & 33.35$ _{ 32.41 }^{ 34.32 }$ & 5588.28/1703 \\
Ind & 0692840401 & ME2PL & 0.09$ _{ 0.09 }^{ 0.09 }$ & 1.99$ _{ 1.97 }^{ 2.02 }$ & 67.07$ _{ 0 }^{ 0 }$ & 1.65$ _{ 1.64 }^{ 1.67 }$ & 10.95$ _{ 10.95 }^{ 11.28 }$ & 46.00$ _{ 45.13}^{ 47.33 }$ & 4283.19/1697 \\
Ind & 0692840501 & ME2PL & - & 12.30$ _{ 12.14 }^{ 12.43 }$ & 0.63$ _{ 0.62 }^{ 0.64 }$ & 1.74$ _{ 1.73 }^{ 1.76 }$ & 2.73$ _{ 2.73 }^{ 2.80 }$ & 61.81$ _{ 59.72 }^{ 63.43 }$ & 2817.83/1687 \\ \hline
SMF3 & 0151370101 & ME2PL & - & 39.40$_{37.04}^{41.93}$ & 0.62$_{0.61}^{0.62}$ & 1.63$_{1.62}^{1.64}$ &  1.63$_{1.58}^{1.69}$ & 58.68$_{54.30}^{63.43}$ & 10738.3/6122 \\
 &  0151370701 & & & 31.17$_{29.45}^{33.02}$ &  &  & 1.74$_{1.66}^{1.83}$ & 73.99$_{69.02}^{79.27}$ &  \\
 & 0205590301 & & & 10.44$_{10.28}^{10.59}$ & &  & 2.07$_{2.03}^{2.11}$ & 79.99$_{77.90}^{82.13}$ &  \\
 & 0205590401 & & & 25.36$_{24.67}^{26.07}$ &  &  & 1.79$_{1.75}^{1.80}$ & 65.26$_{62.82}^{67.77}$ &  \\
 &  0692840201 & & & 25.18$_{24.84}^{25.52}$ &  &  & 1.77$_{1.75}^{1.80}$ & 74.12$_{72.03}^{76.25}$ & \\
 &  0692840501 & & & 12.77$_{12.62}^{12.93}$ &  &  & 2.52$_{2.49}^{2.55}$ & 87.64$_{85.42}^{89.91}$ & \\
\hline
SMF2 (+ring) & 0205590401 & ME2PL & - & 27.74$_{25.62}^{28.95}$ & $<$0.69 & 2.26$_{2.18}^{2.51}$ & 0.31$_{0.24}^{0.37}$ & 140.24$_{99.61}^{185.74}$ & 1388.72/873  \\
 & 6869 & & & 44.27$_{40.82}^{45.58}$ & & & & 199.22$_{158.50}^{259.85}$ &   \\  
\hline
\multicolumn{10}{c}{NGC\,2617} \\ \hline
Ind & 0701981601 & 2PL & - & 11.93$ _{ 10.44 }^{ 13.65 }$ & - & 2.01$ _{ 2.00 }^{ 2.02 }$ & 71.01$ _{ 70.80 }^{ 71.21 }$ & 39.95$ _{37.52 }^{ 42.55 }$ & 1903.22/1564 \\
Ind & 0701981901& 2PL & - & 10.07$ _{ 8.61 }^{ 11.77 }$ & - & 2.15$ _{ 2.14 }^{ 2.16 }$ & 0.02$ _{ 0.02 }^{ 0.02 }$ & 0.01$ _{ 0.01 }^{ 0.01 }$ & 1861.39/1469 \\ \hline
SMF2 & 0701981601 & 2PL & - & 6.88$ _{ 6.25 }^{ 7.54 }$ & - & 2.08$_{2.07}^{2.09}$ & 70.05$ _{ 69.81 }^{ 70.28 }$ & 49.41$_{47.49}^{51.41}$ & 4431.52/3039 \\
     & 0701981901 &     &  & 9.80$ _{ 8.75 }^{ 10.98 }$ &   &                      & 170.94$ _{ 170.46 }^{ 171.43 }$ & &  \\
\hline     
\multicolumn{10}{c}{MARK\,1218} \\ \hline
Ind & 0302260201 & PL & 0.09$ _{ 0.06 }^{ 0.12 }$ & - & - & 1.12$ _{ 1.06 }^{ 1.20 }$ & - & - & 187.44/146 \\
Ind & 0302260401 & PL & - & - & - & 0.81$ _{ 0.74 }^{ 0.94 }$ & - & - & 27.40/30 \\ \hline
SMF1 & 0302260201 & PL & 0.06$_{0.03}^{0.09}$ & - & - & 1.05$_{1.00}^{1.12}$ & 2.49$ _{ 2.31 }^{ 2.72 }$ & - & 231.136/181 \\
     &            &     &                     &    &  &                      & 0.91$ _{ 0.84 }^{ 1.00 }$ & &  \\
\hline
\multicolumn{10}{c}{NGC\,2992} \\ \hline
Ind & 0654910501 & 2PL & 0.16$ _{ 0.12 }^{ 0.19 }$ & 0.93$ _{ 0.85 }^{ 1.02 }$ & - & 1.50$ _{ 1.48 }^{ 1.53 }$ & 4.78$ _{ 3.74 }^{ 5.88 }$ & 20.42$ _{ 19.45 }^{ 21.35 }$ & 1460.82/1412 \\
Ind & 0654910601  & 2PL & 0.12$ _{ 0.10 }^{ 0.14 }$ & 1.18$ _{ 0.92 }^{ 1.47 }$ & - & 1.38$ _{ 1.32}^{ 1.44 }$ & 3.93$ _{ 3.42 }^{ 4.35 }$ & 4.05$ _{ 3.44 }^{ 4.70 }$ & 1078.77/947 \\
Ind & 0654910701 &  2PL & 0.11$ _{ 0.08 }^{ 0.14 }$ & 0.99$ _{ 0.77 }^{ 1.24 }$ & - & 1.37$ _{ 1.32 }^{ 1.42 }$ & 3.72$ _{ 3.09 }^{ 4.21 }$ & 4.22$ _{ 3.66 }^{ 4.78 }$ & 1012.9/1016 \\
Ind &  0654910901     &  2PL & 0.05$ _{ 0.03 }^{ 0.07 }$ & 8.13$ _{ 2.36 }^{ 57.00 }$ & - & 1.22$ _{ 1.10 }^{ 1.34 }$ & 2.65$ _{ 2.48 }^{ 2.83 }$ & 0.54$ _{ 0.48 }^{ 1.21 }$ & 676.2/664 \\
Ind &   0654911001   & 2PL & 0.08$ _{ 0.05 }^{ 0.11 }$ & 0.85$ _{ 0.69 }^{ 1.03 }$ & - & 1.37$ _{ 1.33 }^{ 1.42 }$ & 2.99$ _{ 2.29 }^{ 3.59 }$ & 5.79$ _{ 5.24 }^{ 6.33 }$ & 1106.19/1078 \\
Ind & 0701780101 & 2PL & 0.09$ _{ 0 }^{ 0.17 }$ & 0.81$ _{ 0.70 }^{ 0.95 }$ & - & 1.50$ _{ 1.46 }^{ 1.55 }$ & 3.36$ _{ 1.68 }^{ 5.75 }$ & 27.24$ _{ 25.32 }^{ 28.99 }$ & 818.68/821 \\ \hline
SMF2 & 0654910501 & 2PL & 0.09$_{0.08}^{0.09}$ & 0.76$ _{ 0.74 }^{ 0.78 }$ & - & 1.43$_{1.41}^{1.44}$ & 2.99$_{2.90}^{3.07}$ & 19.52$ _{ 19.06 }^{ 20.00 }$ & 6291.22/5968   \\
& 0654910601 &  & & 0.95$ _{ 0.89 }^{ 1.01 }$ & & & & 5.51$ _{ 5.30 }^{ 5.73 }$ &  \\
& 0654910701 &  & & 0.88$ _{ 0.83 }^{ 0.93 }$ &  & & & 5.53$ _{ 5.33 }^{ 5.74 }$ &  \\
& 0654910901 &  & & 7.44$ _{ 5.36 }^{ 10.37 }$ &  & & & 1.76$ _{ 1.53 }^{ 2.02 }$ \\
& 0654911001 &  & & 0.93$ _{ 0.88 }^{ 0.98 }$ &  & & & 6.56$ _{ 6.34 }^{ 6.79 }$ \\
& 0701780101 &  & & 0.72$ _{ 0.69 }^{ 0.74 }$ &  & & & 24.66$_{ 24.02 }^{ 25.32 }$ \\
\hline
\multicolumn{10}{c}{POX\,52} \\ \hline
Ind & 5736 & ME2PL & - & 3.23$ _{ 2.36 }^{ 4.44 }$ & 0.22$ _{ 0.20 }^{ 0.25 }$ & 2.62$ _{ 2.44 }^{ 2.78 }$ & 2.32$ _{ 2.32 }^{ 2.69 }$ & 4.25$ _{ 2.65 }^{ 6.13 }$ & 122.38/117 \\
Ind & 0302420101 & ME2PL & - & 6.59$ _{ 5.79 }^{ 7.39 }$ & 0.16$ _{ 0.14 }^{ 0.18 }$ & 1.79$ _{ 1.60 }^{ 1.98 }$ & 0.11$ _{ 0.11 }^{ 0.13 }$ & 1.33$ _{ 0.92 }^{ 1.89 }$ & 244.71/174 \\ \hline
SMF (+ring)  & 0302420101 & ME2PL & 8.38$ _{ 7.36 }^{ 10.43 }$ & 2.43$ _{ 1.30 }^{ 4.28 }$ & 0.15$_{0.11}^{0.17}$ & 2.30$_{2.26}^{2.34}$ & 2.76$_{2.05}^{2.98}$ & 0.30$_{0.10}^{1.00}$ & 378.76/290  \\
& 5736 & & - & 4.32$ _{ 3.51 }^{ 5.41 }$ & & & & &  \\
\hline
\multicolumn{10}{c}{NGC\,4138} \\ \hline
Ind & 3994 & PL & 7.92$ _{ 6.62 }^{ 9.95 }$ & - & - & 1.36$ _{ 1.01 }^{ 1.93 }$ & - & - & 34.66/35 \\
Ind &     0112551201 & PL(2--10 keV) & 7.89$ _{ 7.00 }^{ 8.82 }$ & - & - & 1.44$ _{ 1.26 }^{ 1.62 }$ &  & 0.14$ _{ 0.10 }^{ 0.19 }$ & 153.80/162 \\
    \hline
SMF (+ring) & 0112551201 & PL(2--10 keV) & 6.76$_{4.22}^{9.55}$ & - & - & 1.20$_{0.61}^{1.83}$ & 0.19$_{0.01}^{0.86}$ & - & 186.04/190  \\
& 2003 & & & & & & 8.91$_{3.16}^{26.70}$ & &  \\
\hline
\multicolumn{10}{c}{NGC\,4395} \\ \hline
Ind & 5302 & ME2PL & - & 4.00$ _{ 3.52 }^{ 4.52 }$ & 0.09$ _{ 0.05 }^{ 0.17 }$ & 1.13$ _{ 0.94 }^{ 1.33 }$ & 0.07$ _{ 0.07 }^{ 0.12 }$ & 3.99$ _{ 2.90 }^{ 5.52 }$ & 99.62/108 \\
Ind &  5301 & ME2PL & - & 3.40$ _{ 3.04 }^{ 3.76 }$  & 0.08$ _{ 0.03 }^{ 0.22 }$ & 1.52$ _{ 1.30 }^{ 1.83 }$ & 0.14$ _{ 0.11 }^{ 0.16 }$ & 5.98$ _{ 5.98 }^{ 10.70 }$ & 183.22/121 \\ \hline
SMF1 & 5302 & ME2PL & - & 4.34$ _{ 3.99 }^{ 4.69 }$ & 0.09$_{0.05}^{0.16}$ & 1.37$_{1.24}^{1.50}$ & 0.11$_{0.09}^{0.13}$ & 5.50$_{4.46}^{6.81}$ & 265.8/238 \\
     & 5301 & & & 3.01$ _{ 2.74 }^{ 3.29 }$ & & & & &  \\
\hline
Ind & 0112521901 & ME2PL & 5.71$ _{ 5.11 }^{ 6.45 }$ & 0.04$ _{ 0 }^{ 0.23 }$ & 0.16$ _{ 0.09 }^{ 0.21 }$ & 1.29$ _{ 1.12 }^{ 1.52 }$ & 6.75$ _{ 6.75 }^{ 11.53 }$ & 0.35$ _{ 0.29 }^{ 0.49 }$ & 261.70/220 \\
Ind & 0744010101 & ME2PL & - & 8.14$ _{ 7.67}^{ 8.63 }$ & 0.19$ _{ 0.18 }^{ 0.20 }$ & 1.27$ _{ 1.18 }^{ 1.35 }$ & 0.30$ _{ 0.29 }^{ 0.34 }$ & 8.06$ _{ 6.84 }^{ 9.51 }$ & 601.99/590 \\
Ind & 0744010201 & ME2PL & - & 4.78$ _{ 4.56 }^{ 5.20 }$ & 0.18$ _{ 0.14 }^{ 0.21 }$ & 1.16$ _{ 1.05 }^{ 1.26 }$ & 0.36$ _{ 0.35 }^{ 0.47 }$ & 7.49$ _{ 6.14 }^{ 8.93 }$ & 369.14/344 \\
\hline
SMF2 & 0112521901 & ME2PL & - & 5.81$ _{ 5.43 }^{ 6.21 }$ & 0.19$_{0.18}^{0.20}$ & 1.20$_{1.14}^{1.26}$ & 0.33$_{0.31}^{0.35}$ & 9.40$_{8.37}^{10.55}$ & 1217.45/1170 \\
     & 0744010101 &  & & 7.94$ _{ 7.56 }^{ 8.34 }$ & & & & 70.83$_{63.06}^{79.57}$ &  \\
     & 0744010201 &  & & 4.66$ _{ 4.38 }^{ 4.96 }$ & & & & 80.75$_{72.23}^{90.29}$ &  \\
\hline
SMF1 (+ring) & 0112521901 & ME2PL & 2.87$_{2.55}^{3.20}$ & 13.58$_{11.38}^{16.19}$ & 0.25$_{0.24}^{0.27}$ & 1.42$_{1.26}^{1.64}$ & 5.31$_{4.23}^{6.82}$ & 10.17$_{6.54}^{17.25}$ & 460.32/350  \\
 & 5301  & & & & & & & 0.76$_{0.01}^{3.00}$ &  \\
\hline
\multicolumn{10}{c}{NGC\,4565} \\ \hline 
Ind & 0112550301 & PL & 0.11$ _{ 0.07 }^{ 0.16 }$ & - & - & 1.80$ _{ 1.67 }^{ 1.99 }$ & - & - & 46.52/46 \\
Ind & 3950 & PL & 0.25$ _{ 0.22 }^{ 0.28 }$ & - & - & 1.96$ _{ 1.85 }^{ 2.05 }$ & - & - & 71.19/77 \\ \hline
SMF (+ring) &  0112550301 & PL & 0.13$ _{ 0.10 }^{ 0.17 }$ & - & - & 1.94$_{1.85}^{2.05}$ & 0.78$_{0.71}^{0.86}$ & - & 121.88/128  \\
& 3950 & & 0.25$ _{ 0.22 }^{ 0.29 }$ & & & & & &   \\
\hline
\multicolumn{10}{c}{MARK\,883} \\ \hline 
Ind & 0302260101 & PL & 0.04$ _{ 0.02 }^{ 0.06 }$ & - & - & 1.61$ _{ 1.54 }^{ 1.68 }$ & - & - & 206.37/191 \\
Ind & 0302260701 & PL & 0.04$ _{ 0.02 }^{ 0.06 }$ & - & - & 1.64$ _{ 1.59 }^{ 1.71 }$ & - & - & 269.11/249 \\
Ind & 0302261001 & PL & 0.07$ _{ 0.05 }^{ 0.08 }$ & - & - & 1.75$ _{ 1.69 }^{ 1.81 }$ & - & - & 221.43/243 \\
Ind & 0652550201 & PL & 0.04$ _{ 0.03 }^{ 0.05 } $ & - & - & 1.66$ _{ 1.62 }^{ 1.71 }$ & - & - & 438.40/407 \\
\hline
SMF1 & 0302260101 & PL & 0.05$_{0.04}^{0.05}$ & - & - & 1.67$_{1.64}^{1.70}$ & & 3.67$ _{ 3.54 }^{ 3.80 }$ & 1151.81/1105  \\
 & 0302260701 &  & & & & & & 3.64$ _{ 3.52 }^{ 3.76 }$ &  \\
 & 0302261001 &  & & & & & & 4.62$ _{ 4.47 }^{ 4.77 }$ & \\
 & 0652550201 &  & & & & & & 2.66$ _{ 2.59 }^{ 2.74 }$ & \\
\hline
\multicolumn{10}{c}{IRAS\,20051-1117} \\ \hline 
Ind & 0044350201 & PL & - & - & - & 1.89$ _{ 1.85 }^{ 1.94 }$ & - & - & 171.36/160 \\
Ind & 0044350501 & PL & - & - & - & 1.93$ _{ 1.89 }^{ 1.96 }$ & - & - & 246.53/245 \\ \hline
SMF1 & 0044350201 & PL & - & - & - & 1.91$_{1.88}^{1.94}$ & - &  5.88$ _{ 5.73 }^{ 6.04 }$ & 425.12/410 \\
& 0044350501 & & & & & & & 4.17$ _{ 4.08 }^{ 4.26 }$ &  \\
\hline
\newline
\end{longtable}

\begin{longtable}{lcccccc}
\caption[]{\label{lumincorrSey} X-ray luminosities.}  \\  \hline \hline
\multicolumn{7}{c}{ \hspace*{5.7cm} Individual \hspace*{3.5cm} Simultaneous } \\  \cmidrule(r){4-5} \cmidrule{6-7}
Name & Satellite & ObsID & log(L(0.5-2 keV))  & log(L(2-10 keV)) & log(L(0.5-2 keV))  & log(L(2-10 keV)) \\ 
(1) & (2) & (3) & (4) & (5) & (6) & (7)  \\ 
\hline 
\endfirsthead
\caption[]{ (Cont.)} \\
\hline \hline
\multicolumn{7}{c}{ \hspace*{5.7cm} Individual \hspace*{3.5cm} Simultaneous } \\ \cmidrule(r){4-5} \cmidrule{6-7}
Name & Satellite & ObsID & log(L(0.5-2 keV))  & log(L(2-10 keV)) & log(L(0.5-2 keV))  & log(L(2-10 keV)) \\ 
(1) & (2) & (3) & (4) & (5) & (6) & (7)  \\ 
\hline 
\endhead
\endfoot
\\
\endlastfoot
ESO\,540-G01 & \emph{XMM--Newton} & 0044350101 & 43.38 $_{    43.36 }^{    43.40  }$ &    41.83  $_{   41.79 }^{    41.86}$ & 41.63$_{41.62 }^{    41.65 }$ &   41.81$_{41.77}^{ 41.84}$ \\
             & \emph{Chandra} (3$\arcsec$) & 2192 & 41.49$_{     41.42 }^{    41.54  }$ &    41.55 $_{      41.02  }^{   44.12}$ & 41.41$_{41.39}^{     41.43}$ &     41.60$_{ 41.57}^{     41.64}$ \\
             & \emph{Chandra} (25$\arcsec$) & 2192    &   41.80$_{     41.73 }^{    41.87 }$ &     41.65$_{       40.99 }^{    43.67 }$ &   \\   
\hline
ESO\,195-IG21 & \emph{XMM--Newton} & 0554500201 & 43.46$_{     43.41 }^{    43.50 }$ &     43.31$_{     43.30  }^{   43.32 }$ & 42.83$_{     42.70    }^{ 42.93 }$ &   43.32$_{     43.31 }^{    43.34}$ \\
             & \emph{Chandra} (2$\arcsec$) & 13898 &  44.41$_{     44.22 }^{    44.55}$ &      42.88 $_{      42.31 }^{    43.29}$ & 41.08$_{       40.99 }^{    41.43 }$ &    41.76 $_{    41.62 }^{    41.87}$ \\
             & \emph{Chandra} (20$\arcsec$) & 13898 & 42.24$_{     42.06 }^{    42.37}$ &      42.74 $_{      42.25  }^{   43.48}$ & \\
\hline
ESO\,113-G10 & \emph{XMM--Newton} & 0103861601 & 42.67 $_{    42.55 }^{    42.76  }$ &   42.60 $_{    42.21 }^{    42.80}$ & 43.03$_{     43.01}^{     43.05 }$ &    42.66 $_{    42.66    }^{ 42.67}$ \\
     & \emph{XMM--Newton} & 0301890101 & 43.11  $_{   43.11 }^{    43.11 }$ &    42.76 $_{    42.75 }^{    42.76}$ & 43.11$_{     43.10 }^{    43.12  }$ &   42.74 $_{    42.73  }^{   42.74}$ \\
\hline
NGC\,526A & \emph{XMM--Newton} & 0109130201 & 42.69$_{     42.69 }^{    42.70}$ &     43.17$_{     43.16 }^{    43.17}$ & 42.74$_{42.74 }^{    42.75 }$ &    43.16$_{     43.15 }^{    43.16}$ \\
          & \emph{XMM--Newton} & 0150940101 & 42.88$_{     42.88 }^{    42.88}$ &     43.29$_{     43.29 }^{    43.30}$ & 42.89$_{42.88 }^{    42.89  }$ &   43.30$_{     43.29 }^{    43.30}$ \\
          & \emph{XMM--Newton} & 0721730301 & 42.95$_{     42.94 }^{    42.95}$ &     43.35 $_{    43.35 }^{    43.36}$ & 42.94$_{42.94 }^{    42.94  }$ &   43.35$_{     43.35 }^{    43.35}$ \\
          & \emph{XMM--Newton} & 0721730401 & 43.03$_{     43.03 }^{    43.03}$ &     43.43 $_{    43.42 }^{    43.43}$ & 43.02$_{43.02 }^{    43.02  }$ &   43.43$_{     43.43 }^{    43.43}$ \\
          & \emph{Chandra} (4$\arcsec$) & 342 & 42.55$_{     42.50 }^{    42.59 }$ &    43.09 $_{    42.97 }^{    43.19}$ \\
          & \emph{Chandra} (35$\arcsec$) & 342 & 41.87$_{     41.84 }^{    41.90 }$ &    42.58 $_{    42.55 }^{    42.62}$ & \\
\hline
MARK\,609 &   \emph{XMM--Newton} & 0103861001 & 42.55$_{     42.54 }^{    42.56 }$ &    42.67$_{     42.64  }^{   42.69}$ & 42.58 $_{    42.57}^{     42.59 }$ &    42.72 $_{    42.71 }^{    42.74}$ \\
          &   \emph{XMM--Newton} & 0402110201 & 42.50$_{     42.49 }^{    42.51 }$ &    42.69$_{     42.68 }^{    42.71}$ &   42.52$_{     42.51 }^{    42.53 }$ &    42.66 $_{    42.65 }^{    42.68}$ \\
\hline 
NGC\,1365  & \emph{XMM--Newton} & 0151370101 & 41.72$_{     41.71 }^{    41.73}$ &     41.86$_{     41.84 }^{    41.87}$ & 41.48$_{     41.47 }^{    41.49 }$ &    41.78$_{     41.77    }^{ 41.79}$ \\
 & \emph{XMM--Newton} & 0151370201 & 40.97$_{     40.95}^{     40.99 }$ &    41.08 $_{    41.04 }^{    41.12}$ & \\
 & \emph{XMM--Newton} & 0151370701 & 41.94$_{     41.93}^{     41.95 }$ &    41.97 $_{    41.96 }^{    41.99}$ & 40.45$_{     40.44 }^{    40.47 }$ &    42.82$_{     42.80 }^{    42.83}$ \\
 & \emph{XMM--Newton} & 0205590301 & 42.11$_{     42.11}^{     42.12}$ &     42.02 $_{    42.02 }^{    42.03 }$ & 41.62$_{     41.62 }^{    41.63 }$ &    41.93$_{     41.92 }^{    41.93}$ \\
 & \emph{XMM--Newton} & 0205590401 & 41.83$_{     41.83}^{     41.84 }$ &    41.90 $_{    41.89 }^{    41.91}$ & 39.98 $_{    39.98 }^{    39.99  }$ &   39.95$_{     39.94  }^{   39.96}$ \\
 & \emph{XMM--Newton} & 0505140201 & 41.37$_{     41.37}^{     41.38 }$ &    41.33 $_{    41.32}^{     41.34}$ & \\
 & \emph{XMM--Newton} & 0505140401 & 41.60$_{     41.60}^{     41.60 }$ &    41.59 $_{    41.58}^{     41.59}$ & \\
 & \emph{XMM--Newton} & 0505140501 & 41.50$_{     41.49}^{     41.50}$ &     41.49 $_{    41.48}^{     41.50}$ & \\
 & \emph{XMM--Newton} & 0692840201 & 41.97$_{     41.97}^{     41.98}$ &     41.98 $_{    41.97}^{     41.98}$ & 40.38 $_{    40.38 }^{    40.38 }$ &    41.50$_{     41.50  }^{   41.51}$ \\
 & \emph{XMM--Newton} & 0692840301 & 41.54$_{     41.54}^{     41.54}$ &     42.00 $_{    42.00 }^{    42.00}$ & \\
 & \emph{XMM--Newton} & 0692840401 & 41.71$_{     41.67 }^{    41.74 }$ &    42.00 $_{    41.98 }^{    42.01}$ & \\
 & \emph{XMM--Newton} & 0692840501 & 41.75$_{     41.74}^{     41.75 }$ &    41.99 $_{    41.98 }^{    41.99}$ & 40.76$_{     40.76 }^{    40.76  }$ &   41.90 $_{    41.90 }^{    41.90}$ \\
 & \emph{XMM--Newton} (2$\arcsec$) & 0205590401 & & & 42.10$_{     42.02 }^{    42.17 }$ &    41.94    $_{ 41.92  }^{   41.97}$  \\
 & \emph{Chandra} (2$\arcsec$) &  6869 & 41.45$_{     41.41}^{     41.49 }$ &    41.86$_{     41.81 }^{    41.90}$ & 39.69$_{     38.94 }^{    39.95}$ &     41.31$_{     41.27 }^{    41.34}$  \\
 & \emph{Chandra} (20$\arcsec$) &  6869 & 41.93$_{     41.92}^{     41.95}$ &     42.03$_{     42.01}^{     42.05}$ & \\
\hline
NGC\,2617 & \emph{XMM--Newton} & 0701981601 & 43.06$_{     43.05 }^{    43.06}$ &     43.12$_{     43.11 }^{    43.12}$ & 43.08$_{     43.08 }^{    43.08}$ &     43.10$_{     43.10 }^{    43.10}$ \\
          & \emph{XMM--Newton} & 0701981901 & 43.46$_{     43.46 }^{    43.46 }$ &    43.42$_{     43.42 }^{    43.43}$ & 43.35$_{     43.34 }^{    43.35}$ &     43.36$_{     43.36 }^{    43.37}$ \\
\hline
MARK\,1218 & \emph{XMM--Newton} & 0302260201 & 42.09 $_{    42.08 }^{    42.11 }$ &     42.70$_{     42.68  }^{   42.72}$ &   42.04 $_{    42.02 }^{    42.05}$ &      42.73$_{     42.71 }^{    42.75}$ \\
           & \emph{XMM--Newton} &  0302260401 &  41.56$_{     41.53}^{     41.59 }$ &     42.35 $_{      42.21 }^{    43.12 }$ &     41.60$_{     41.57 }^{    41.63 }$ &     42.29$_{     42.26    }^{ 42.33}$ \\
\hline
NGC\,2992 & \emph{XMM--Newton} & 0654910501 & 41.83$_{     41.83 }^{    41.84}$ &    42.23$_{     42.23 }^{    42.24}$ & 41.79$_{     41.79  }^{   41.79}$  &   42.24$_{     42.24 }^{    42.24}$ \\
          & \emph{XMM--Newton} & 0654910601 & 41.35$_{     41.34 }^{    41.35 }$ &   41.84$_{     41.83 }^{    41.84}$ & 41.37$_{     41.36  }^{   41.37}$   &  41.83$_{     41.82 }^{    41.84}$ \\
          & \emph{XMM--Newton} & 0654910701 & 41.34$_{     41.34 }^{    41.35 }$ &   41.83$_{     41.83 }^{    41.84}$ & 41.37$_{     41.36  }^{   41.37}$    & 41.83$_{     41.83 }^{    41.84}$ \\
          & \emph{XMM--Newton} & 0654910901 & 41.07$_{     41.07 }^{    41.08 }$  &  41.60$_{     41.59 }^{    41.60}$ & 41.11$_{     41.11 }^{    41.12}$  &   41.59$_{     41.58 }^{    41.60}$ \\
          & \emph{XMM--Newton} & 0654911001 & 41.38$_{     41.38 }^{    41.39 }$   & 41.88$_{     41.87 }^{    41.88}$ & 41.42$_{     41.41 }^{    41.42}$   &  41.88$_{     41.87  }^{   41.88}$ \\
          & \emph{XMM--Newton} & 0701780101 & 41.92$_{     41.91 }^{    41.92 }$   & 42.32$_{     42.31  }^{   42.33}$ & 41.88$_{     41.87  }^{   41.89}$   &  42.33$_{     42.32  }^{   42.33 }$ \\                
\hline
POX\,52 &   \emph{XMM--Newton} & 0302420101 & 41.55$_{     41.53}^{     41.57 }$ &    41.74 $_{    41.72 }^{    41.76}$ &41.89    $ _{ 41.87}^{     41.92 }$ &    41.76$ _{     41.74}^{     41.78}$ \\
        & \emph{Chandra} (2$\arcsec$) & 5736 & 42.26$_{     42.25 }^{    42.27 }$ &    41.92  $_{     41.89 }^{    42.78}$ & 41.88$ _{     41.86}^{     41.89}$ &     41.73$ _{     41.71}^{     41.75}$ \\
        & \emph{Chandra} (20$\arcsec$) & 5736 &  42.29$_{     42.28 }^{    42.30 }$ &    41.94$_{     41.90  }^{   41.98 }$ & \\
\hline
NGC\,4138 &  \emph{XMM--Newton} &     0112551201 & - &    41.47 $_{    41.46}^{     41.49}$ & - &    41.48$_{41.46}^{41.49}$ \\
          & \emph{Chandra} (2$\arcsec$) & 3994 & - &    41.54 $_{    41.36 }^{    41.66}$ & - &     41.58$_{     41.27}^{     41.76}$ \\
          & \emph{Chandra} (25$\arcsec$) &  3994 &   - &    41.54$_{     41.42  }^{   41.64 }$ & \\
\hline   
NGC\,4395 & \emph{Chandra} & 5302 & 39.37$_{39.37  }^{39.37 }$ &    39.96$_{39.96   }^{39.96}$ & 39.50$_{     39.46  }^{   39.54 }$ &    39.94 $_{    39.92 }^{    39.96}$ \\
         & \emph{Chandra} & 5301 & 39.65$_{39.65  }^{39.65 }$ &    39.93$_{39.93  }^{39.93}$ & 39.50$_{     39.47 }^{    39.53 }$ &    39.94 $_{    39.92 }^{    39.96}$ \\
         & \emph{XMM--Newton} & 0112521901 & 39.76$_{39.73 }^{39.78 }$ &    40.33$_{40.32  }^{40.34}$ & 39.75 $_{    39.73  }^{   39.76}$ &     40.34  $_{   40.33 }^{    40.35}$ \\
         & \emph{XMM--Newton} & 0744010101 & 39.69$_{39.67  }^{39.70 }$ &    40.23$_{40.22 }^{ 40.23}$ & 39.63$_{     39.62 }^{    39.64 }$ &    40.22 $_{    40.21}^{     40.23}$ \\
         & \emph{XMM--Newton} & 0744010201 & 39.65$_{39.63 }^{39.67 }$ &    40.28$_{ 40.27 }^{40.29}$ & 39.68 $_{    39.67  }^{   39.70  }$ &   40.28  $_{   40.27 }^{    40.29}$ \\
         & \emph{XMM--Newton} & 0112521901 & & & 39.94$_{39.91}^{39.96}$ & 40.38$_{40.34}^{40.42}$ \\
         & \emph{Chandra} & 5301 & & & 39.53$_{39.51}^{39.56}$ & 39.93$_{39.89}^{39.97}$ \\         
\hline  
NGC\,4565 &    \emph{XMM--Newton} & 0112550301 &  39.55 $_{    39.53 }^{    39.57 }$ &     39.76 $_{      39.87  }^{   40.72  }$ & 39.53$_{     39.51 }^{    39.55}$ &     39.70$_{     39.67 }^{    39.74}$ \\
            & \emph{Chandra} (2$\arcsec$) & 3950 & 39.47 $_{    39.45 }^{    39.49  }$ &    40.21  $_{     40.02  }^{   40.72   }$ & 39.48$_{     39.46 }^{    39.50 }$ &    39.60$_{     39.56 }^{    39.64}$ \\
           & \emph{Chandra} (20$\arcsec$) & 3950 &  39.52$_{     39.50  }^{   39.54 }$ &     39.66  $_{     39.12 }^{    40.39   }$ &  \\
\hline
MARK\,883  & \emph{XMM--Newton} & 0302260101 & 42.42$_{     42.41}^{     42.43 }$ &    42.73$_{     42.71 }^{    42.75}$ &  42.42$_{     42.41}^{     42.43 }$ &    42.72$_{     42.70}^{     42.73}$ \\
           & \emph{XMM--Newton} & 0302260701 & 42.42$_{     42.41}^{     42.43 }$ &    42.71$_{     42.69}^{     42.73}$ &  42.42$_{     42.41}^{     42.43 }$ &    42.71$_{     42.70}^{     42.72}$ \\
           & \emph{XMM--Newton} & 0302261001 & 42.56$_{     42.55 }^{    42.57}$ &     42.80$_{     42.78 }^{    42.81}$ &  42.52$_{     42.51}^{     42.53 }$ &    42.81$_{     42.80}^{     42.83}$ \\
           & \emph{XMM--Newton} & 0652550201 & 42.27$_{     42.27 }^{    42.28 }$ &    42.59$_{     42.58 }^{    42.60}$ &  42.28$_{     42.28}^{     42.29 }$ &    42.58$_{     42.57}^{     42.59}$ \\
\hline           
IRAS\,20051-1117  & \emph{XMM--Newton} & 0044350201 & 42.45$_{     42.44}^{     42.46}$ &     42.62$_{     42.60 }^{    42.65}$ &  42.46$_{     42.45 }^{    42.47 }$ &    42.60$_{     42.58    }^{ 42.62}$ \\
           & \emph{XMM--Newton} &        0044350501 & 42.30$_{     42.29 }^{    42.31 }$ &    42.46$_{     42.44 }^{    42.48 }$ & 42.31$_{     42.30 }^{    42.32 }$ &    42.45$_{     42.44    }^{ 42.47}$ \\
\hline           
\caption*{{\bf Notes.} (Cols. 4 and 5) soft and hard intrinsic luminosities for individual fits; (Cols. 6 and 7) soft and hard intrinsic luminosities for simultaneous fitting. Blanks mean observations that are not used for the simultaneous fittings.}
\end{longtable} 

\begin{figure*}
\centering
{\includegraphics[width=0.30\textwidth]{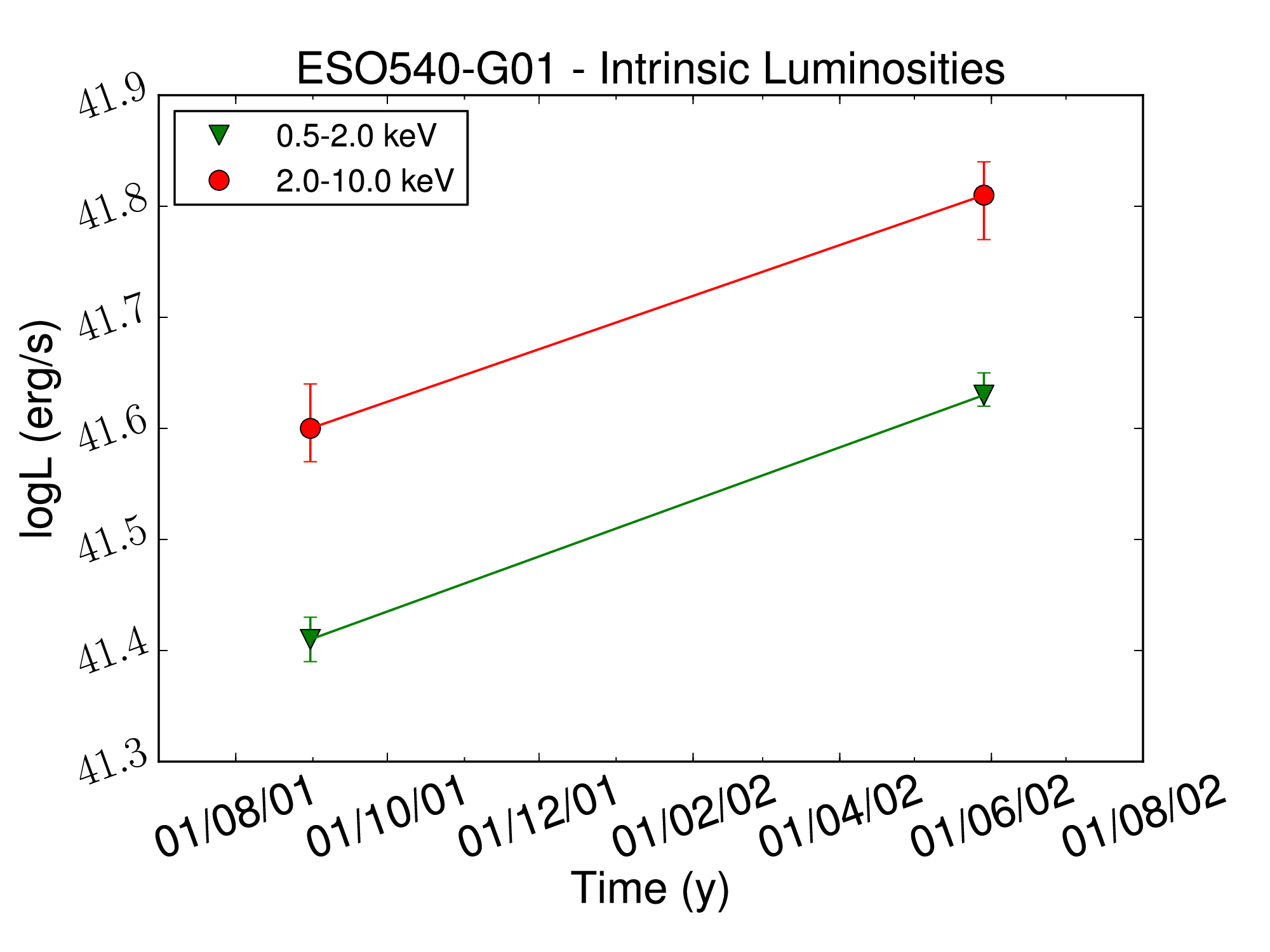}}
{\includegraphics[width=0.30\textwidth]{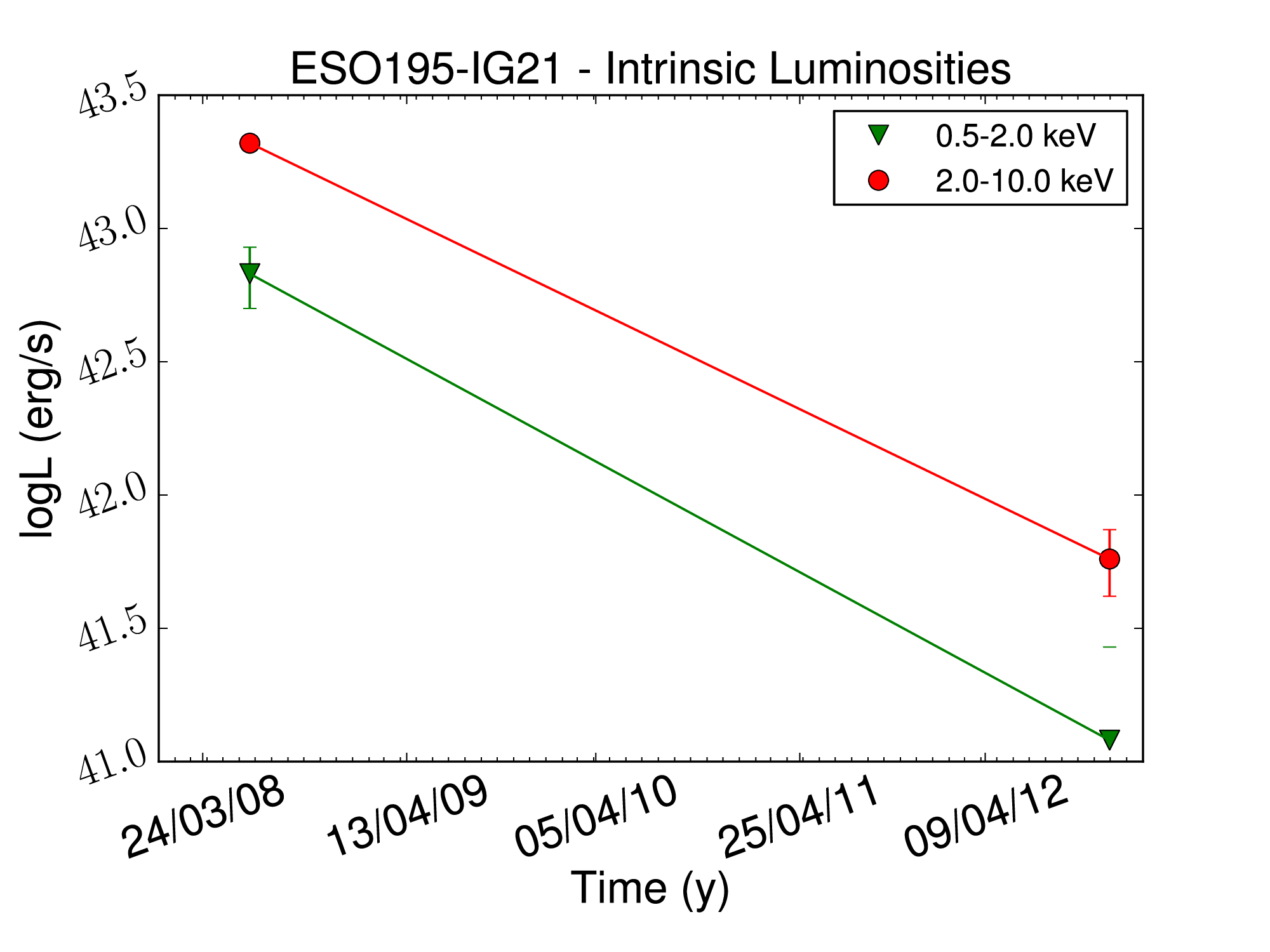}}
{\includegraphics[width=0.30\textwidth]{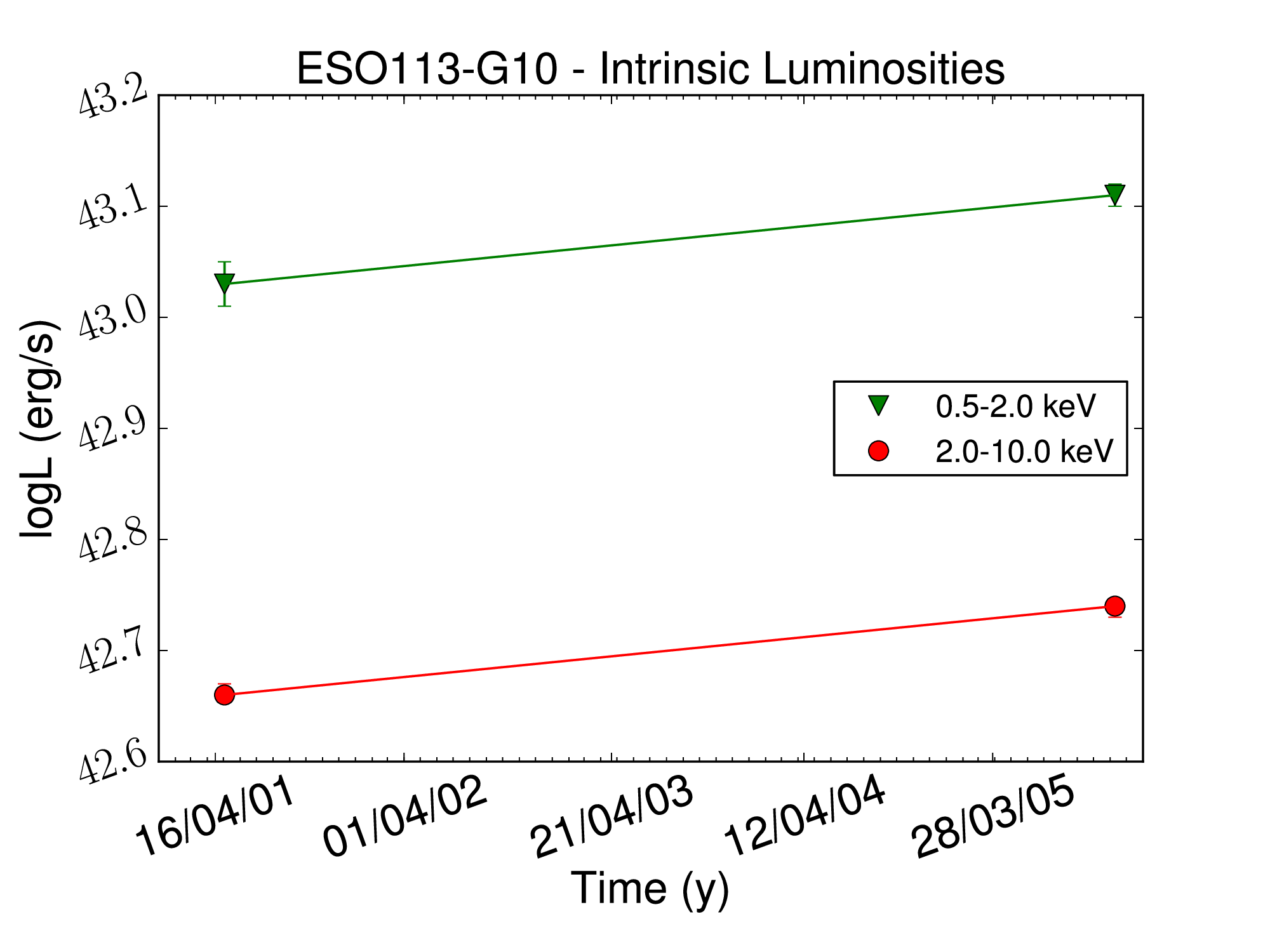}}

{\includegraphics[width=0.30\textwidth]{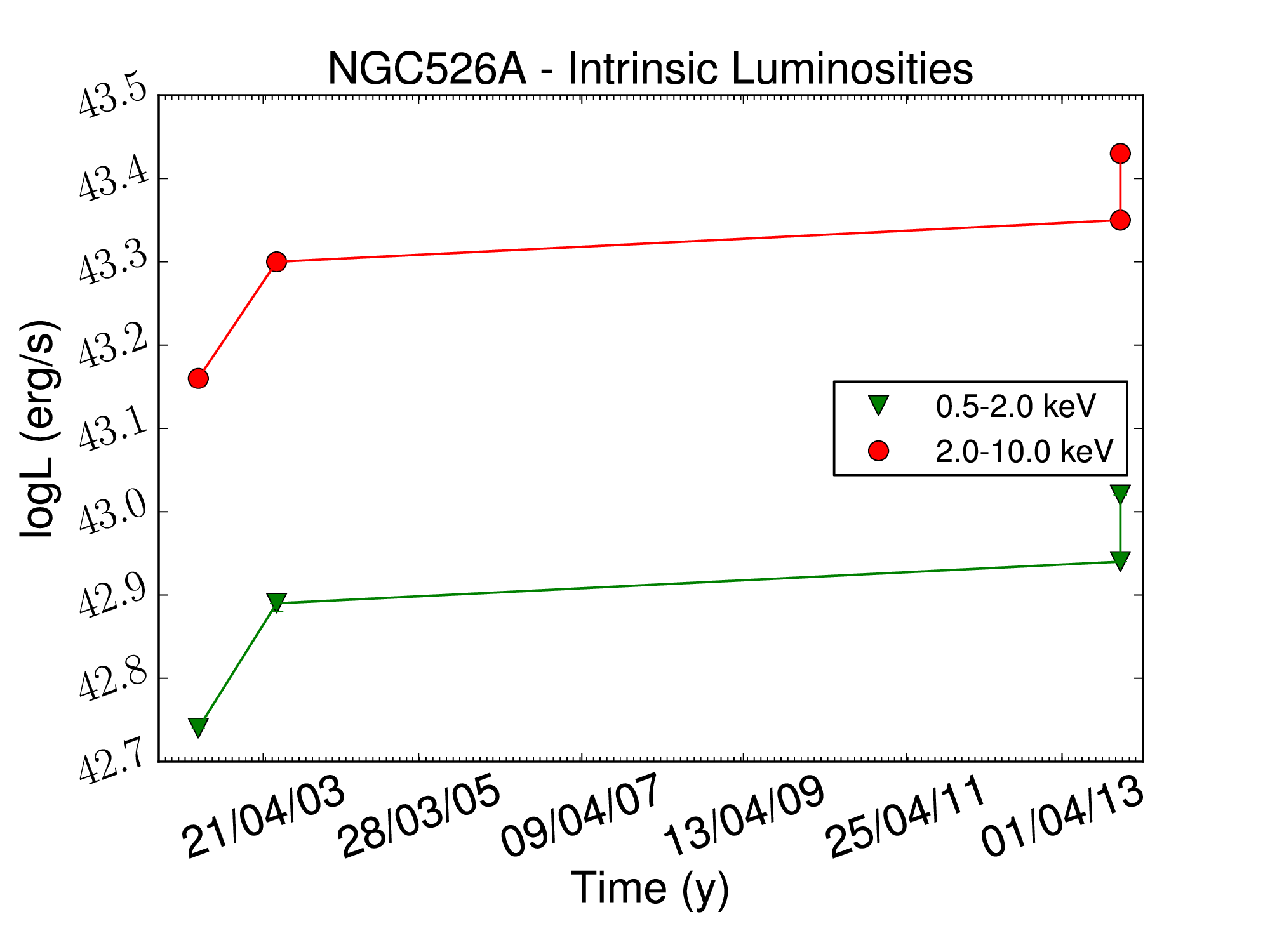}}
{\includegraphics[width=0.30\textwidth]{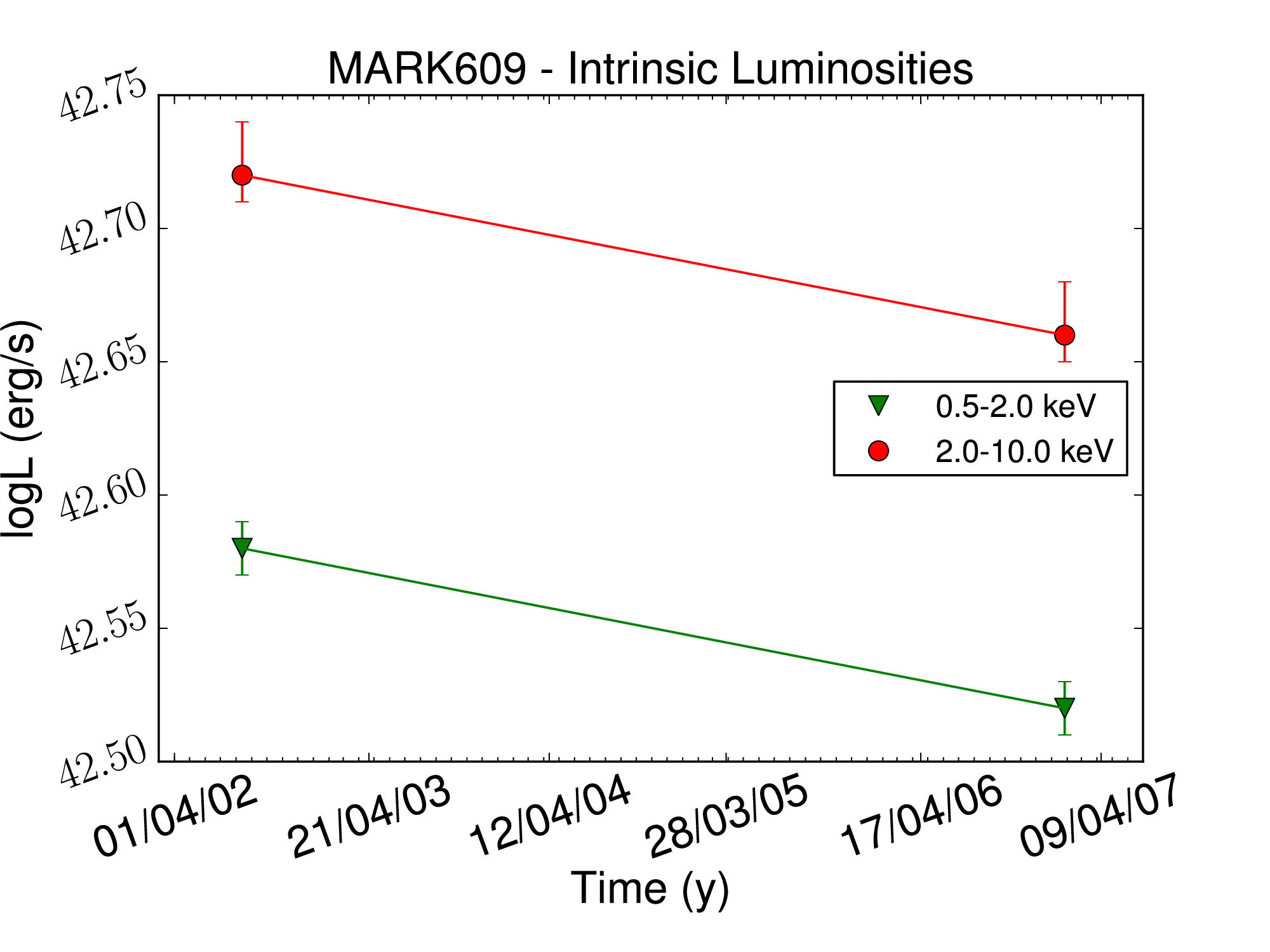}}
{\includegraphics[width=0.30\textwidth]{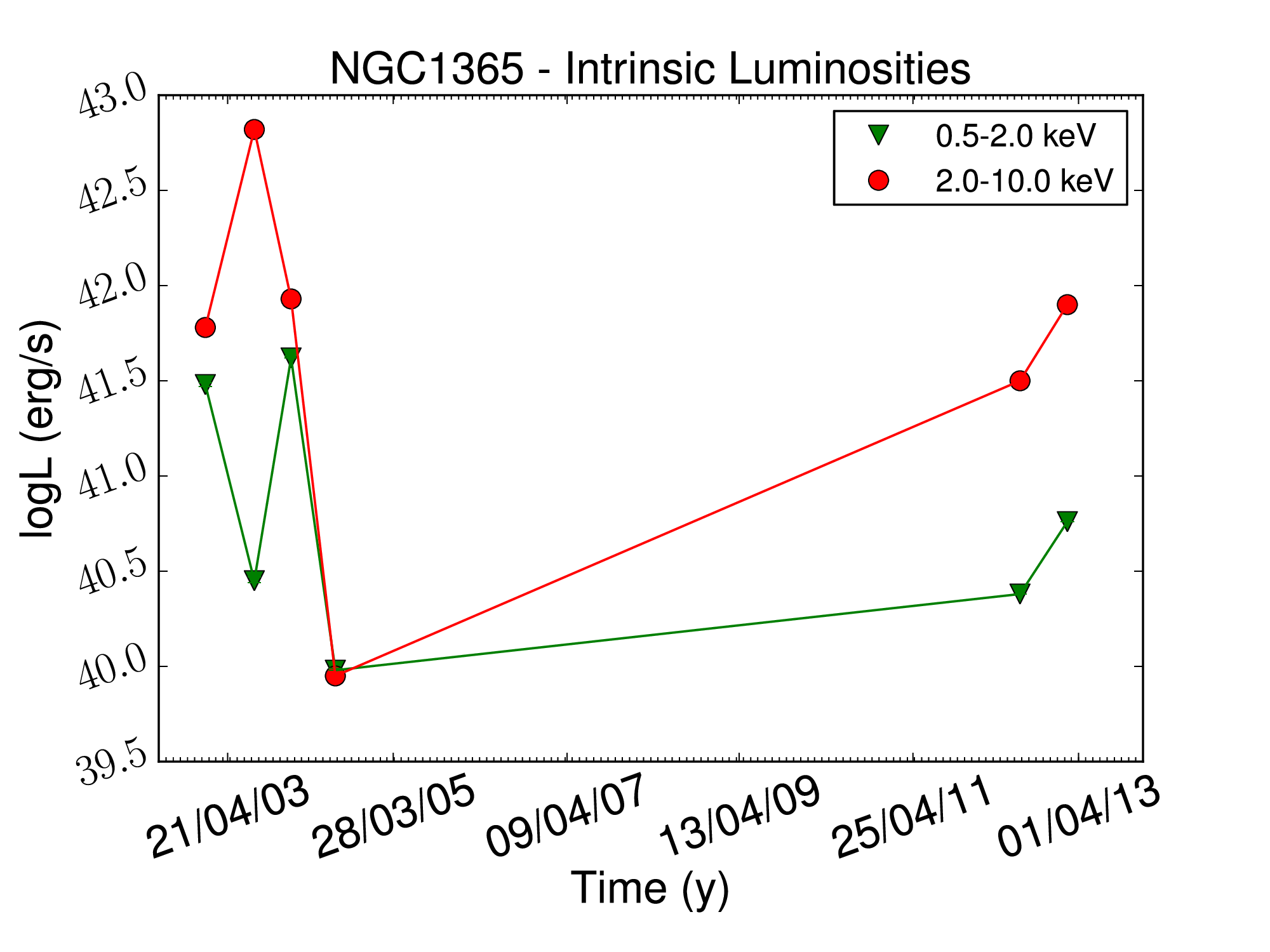}}

{\includegraphics[width=0.30\textwidth]{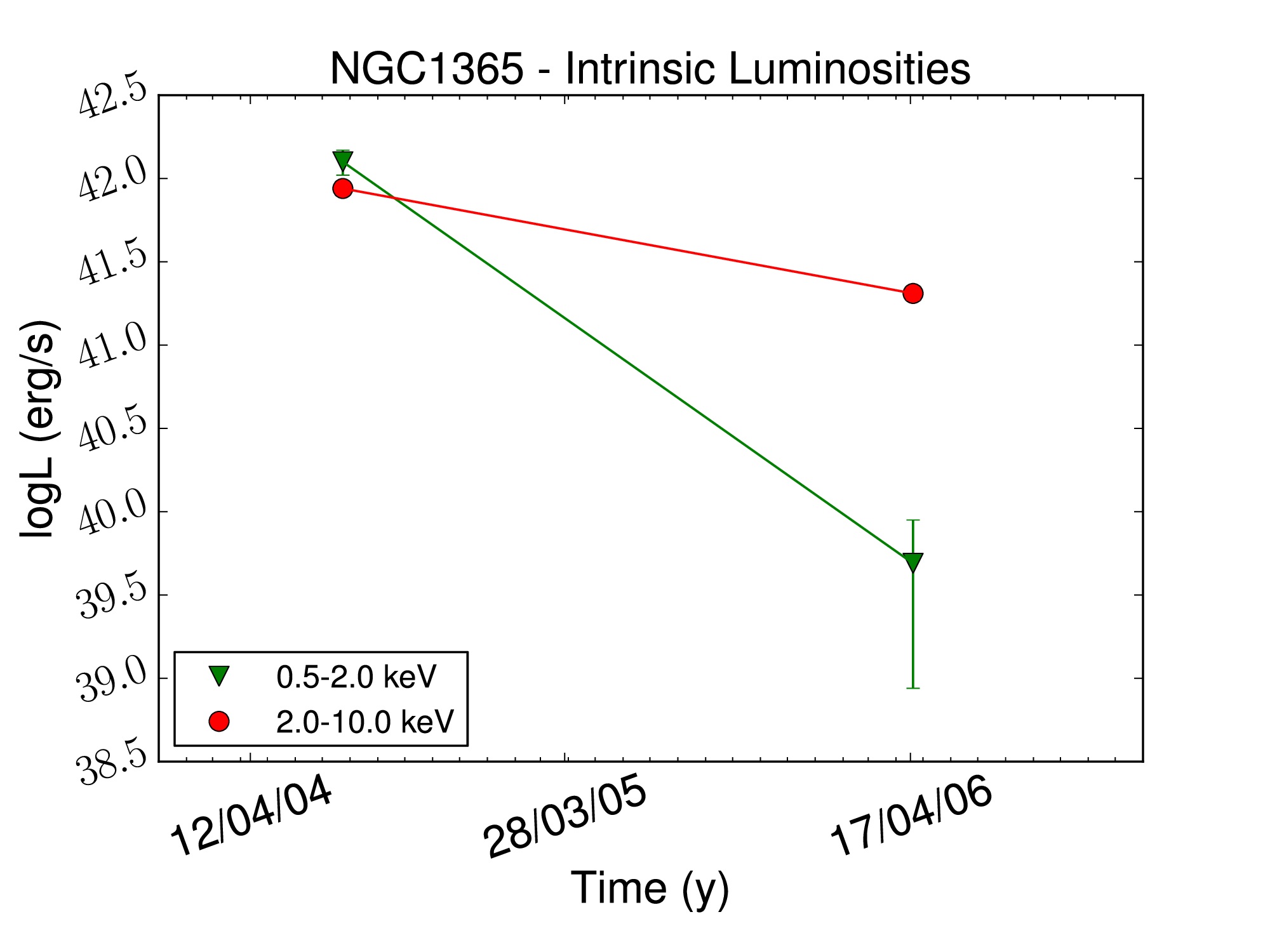}}
{\includegraphics[width=0.30\textwidth]{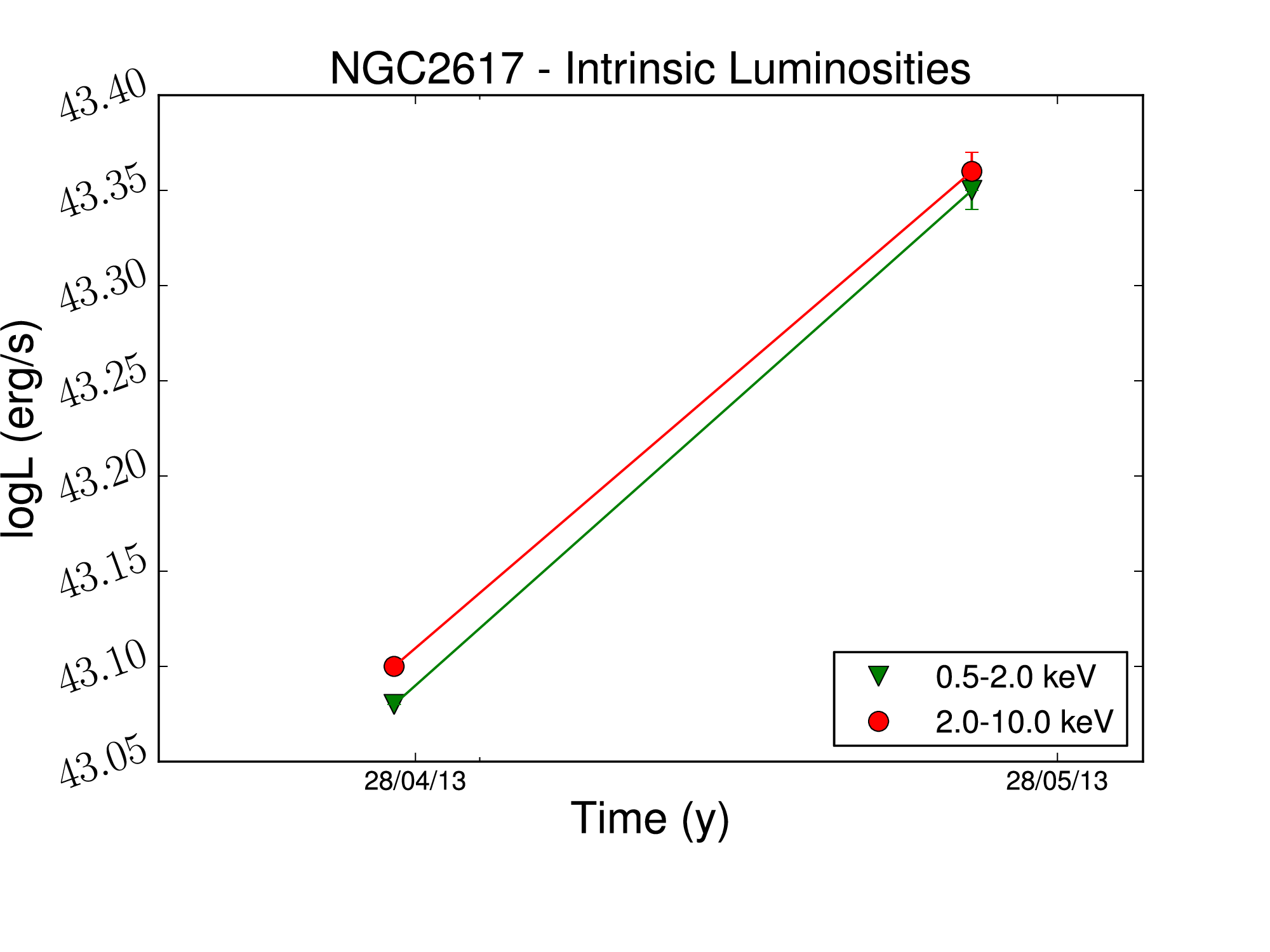}}
{\includegraphics[width=0.30\textwidth]{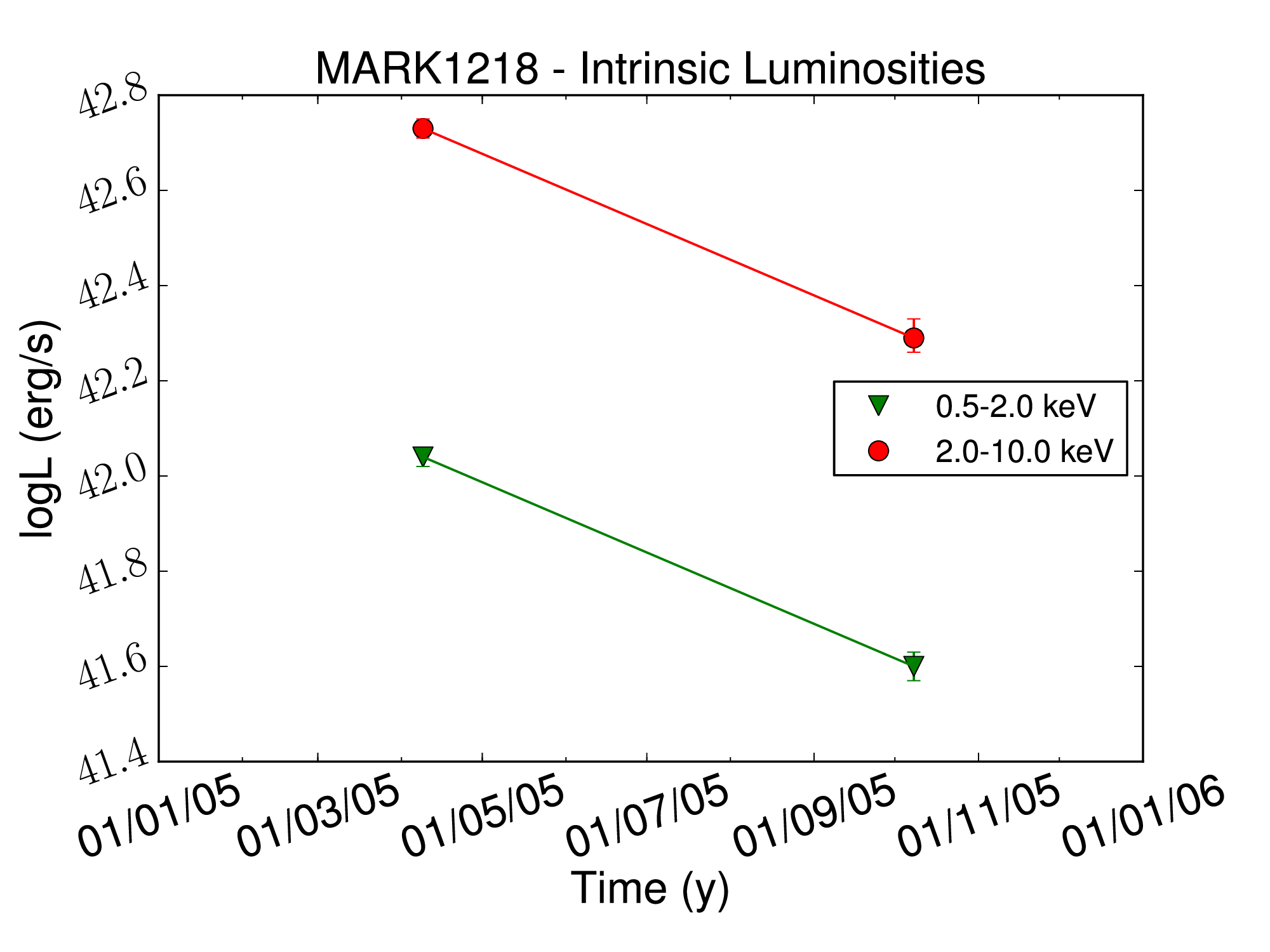}}

{\includegraphics[width=0.30\textwidth]{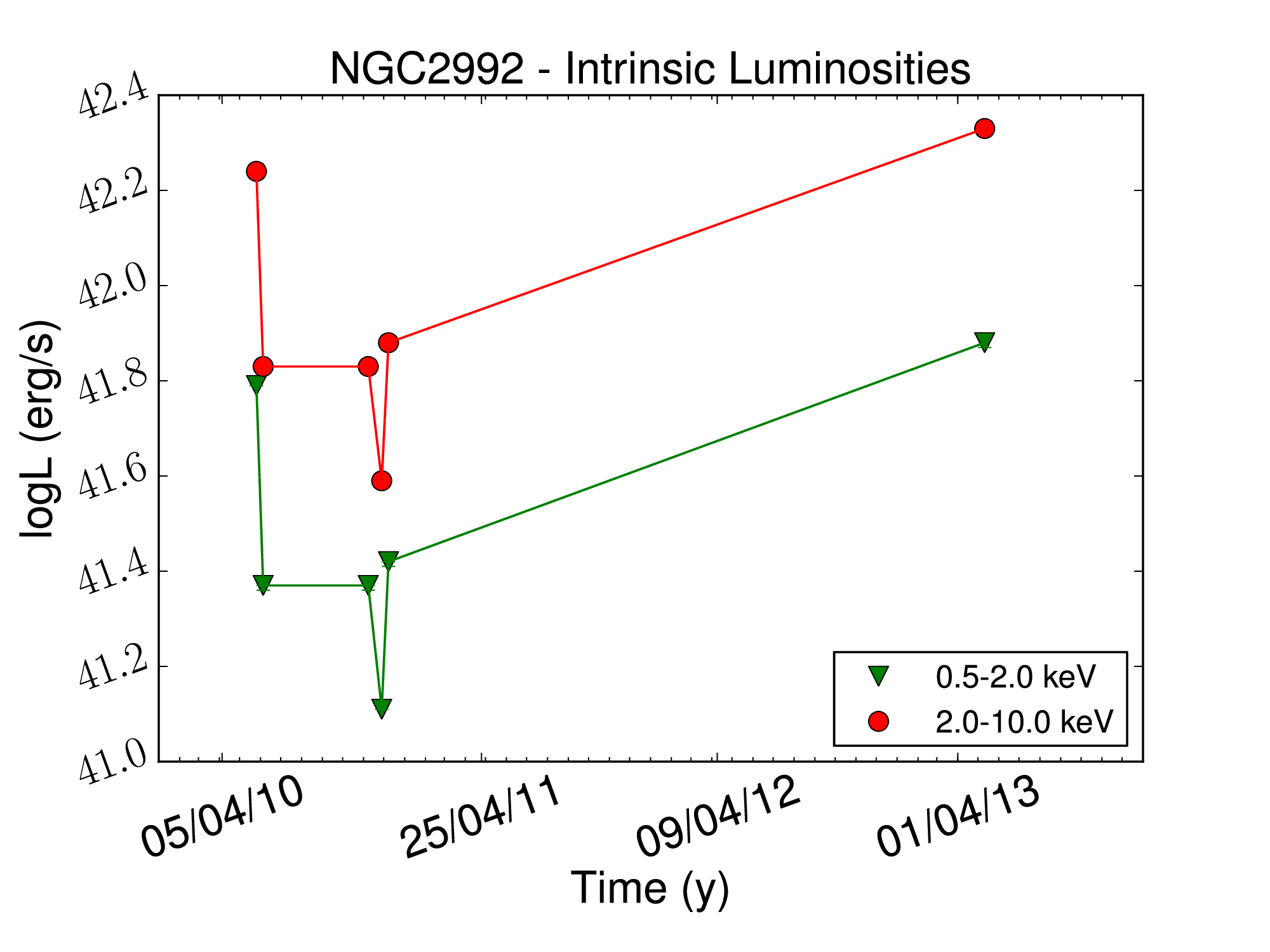}}
{\includegraphics[width=0.30\textwidth]{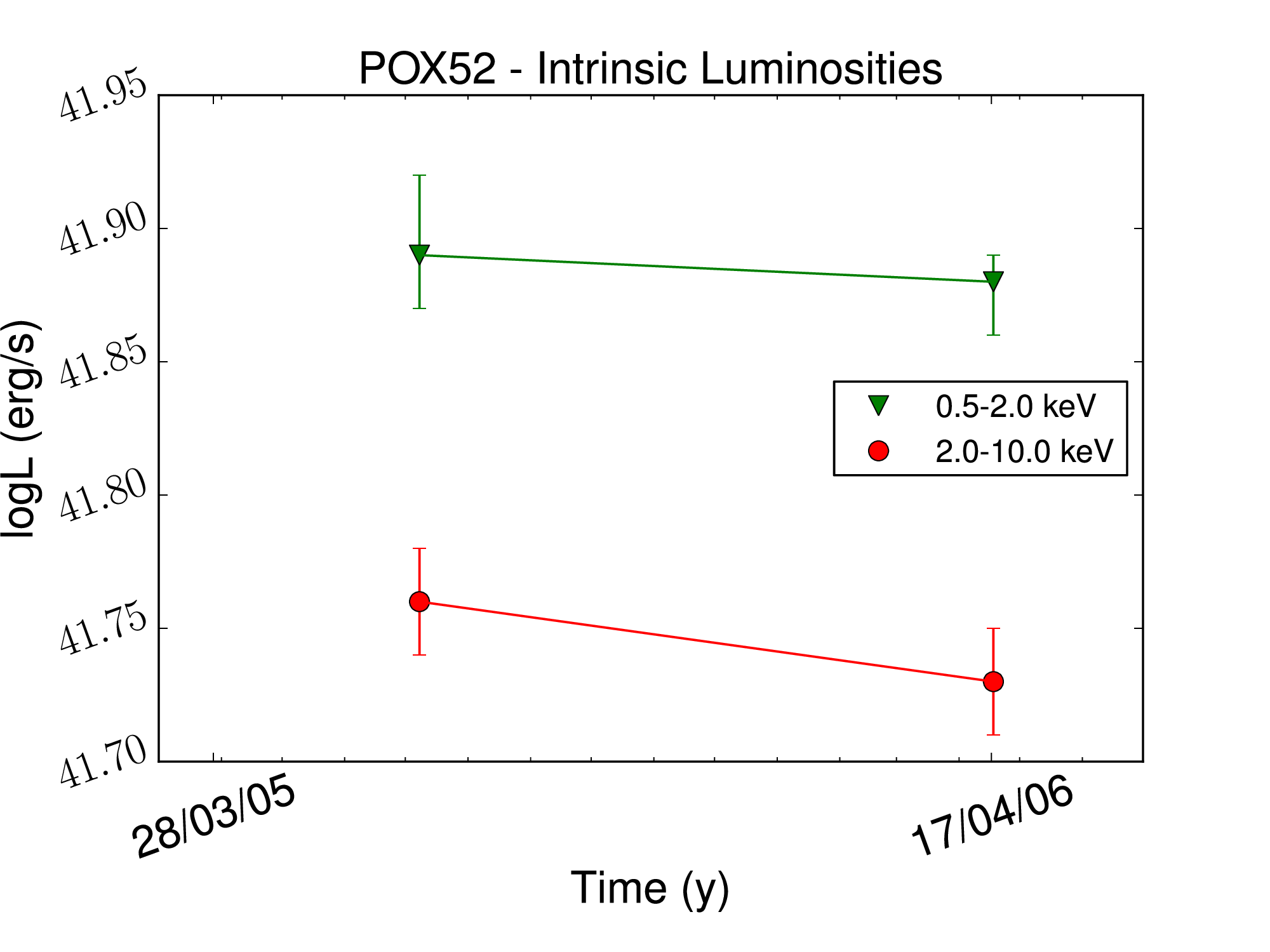}}
{\includegraphics[width=0.30\textwidth]{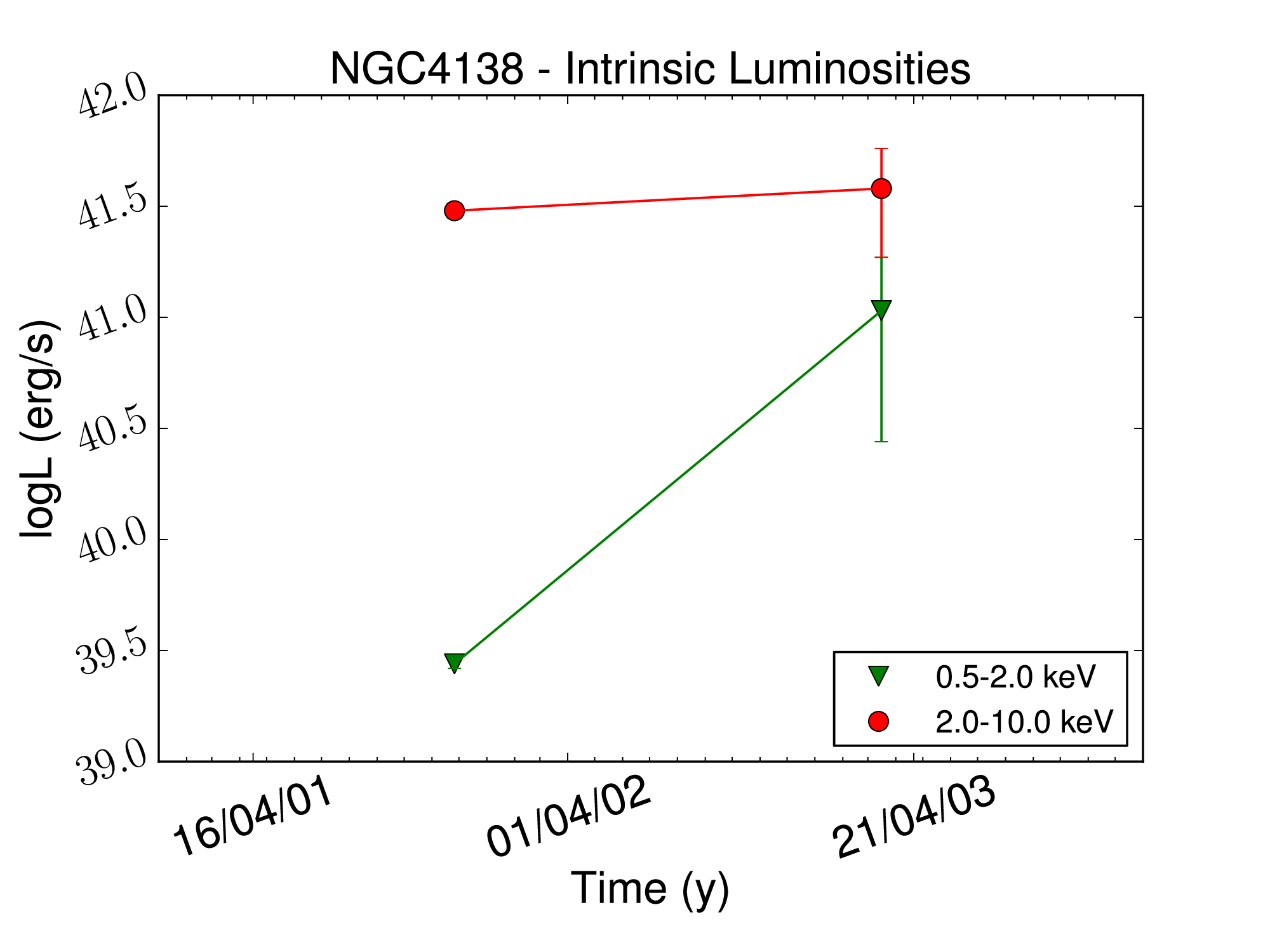}}

{\includegraphics[width=0.30\textwidth]{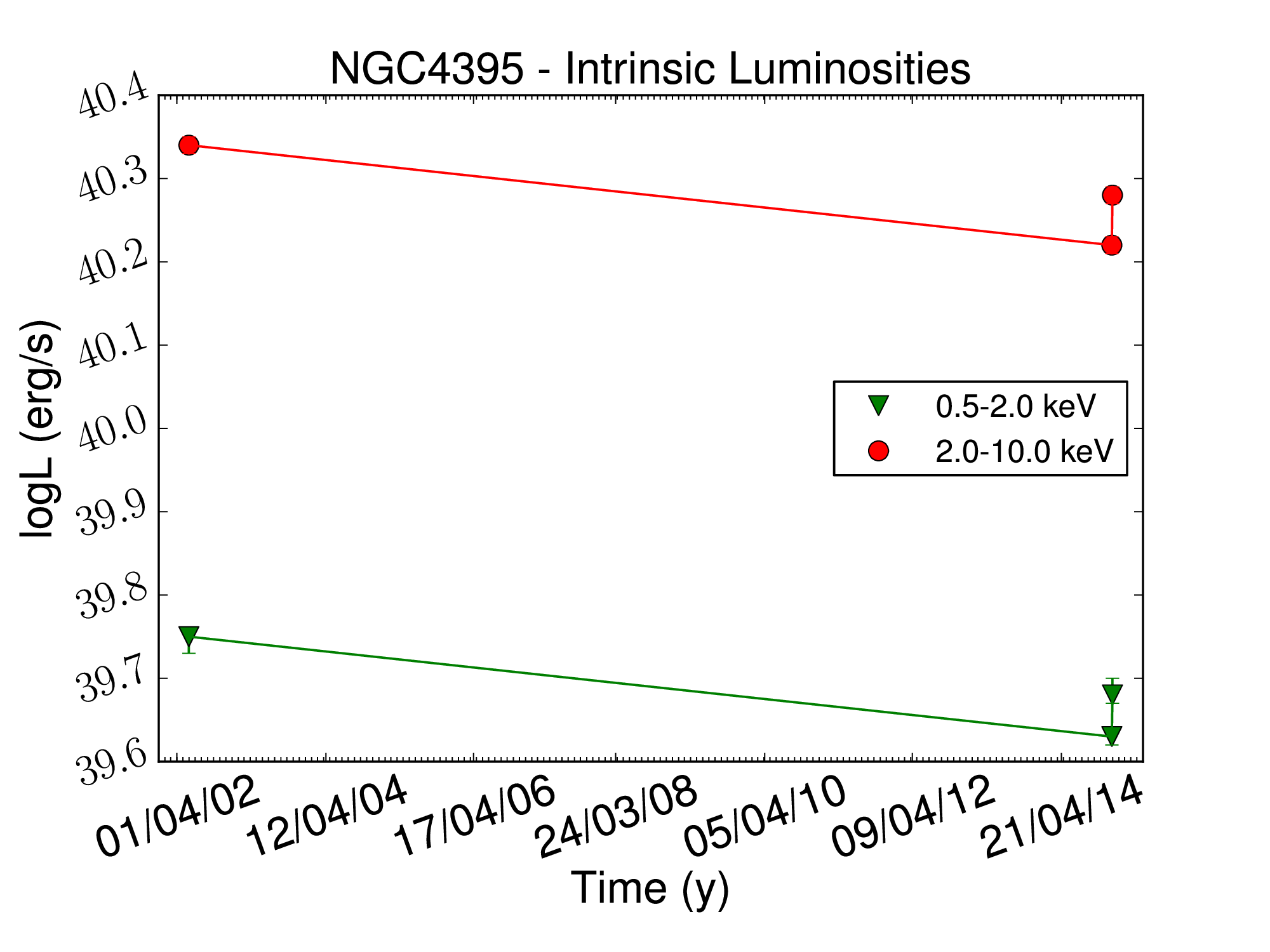}}
{\includegraphics[width=0.30\textwidth]{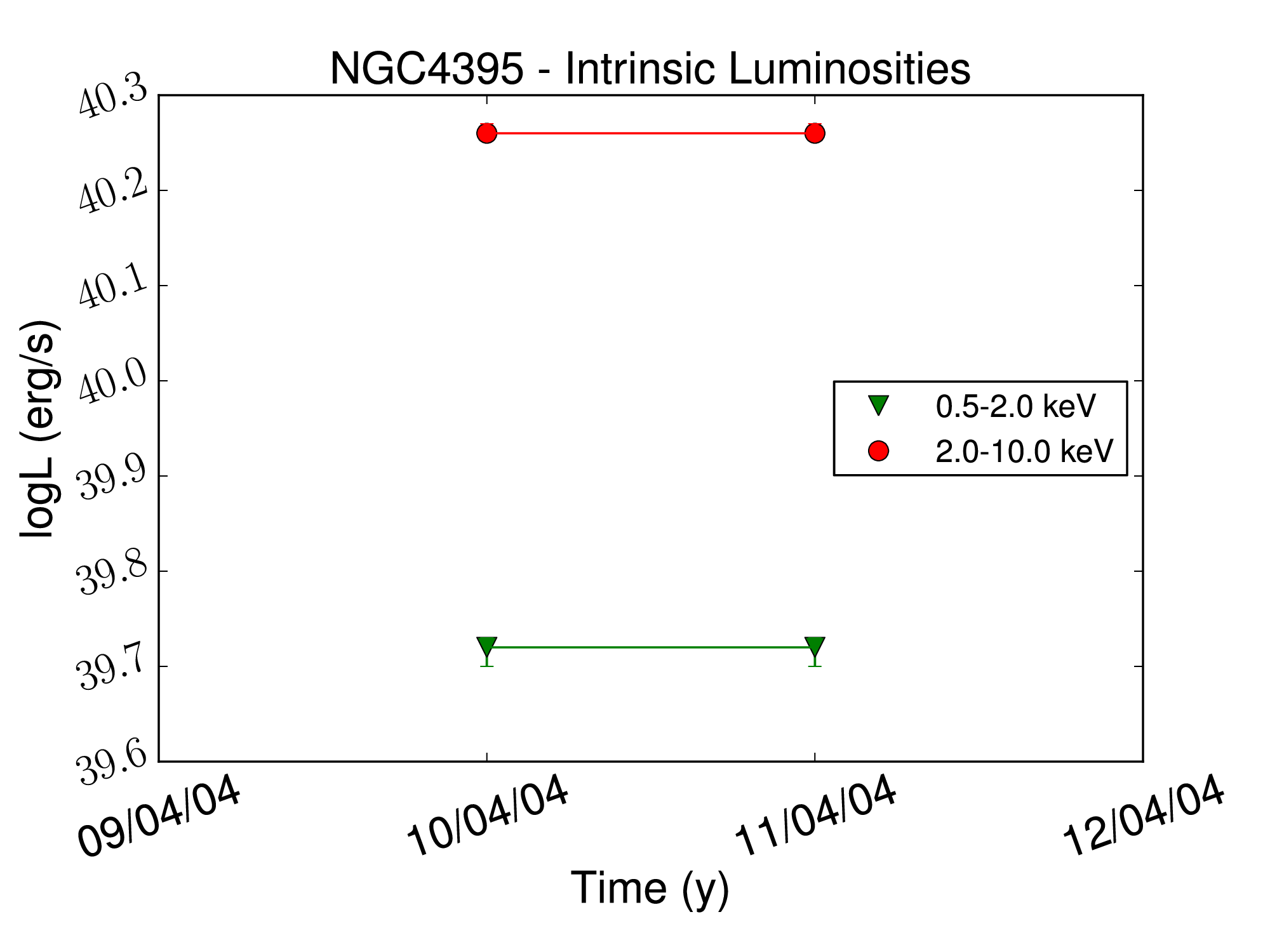}}
{\includegraphics[width=0.30\textwidth]{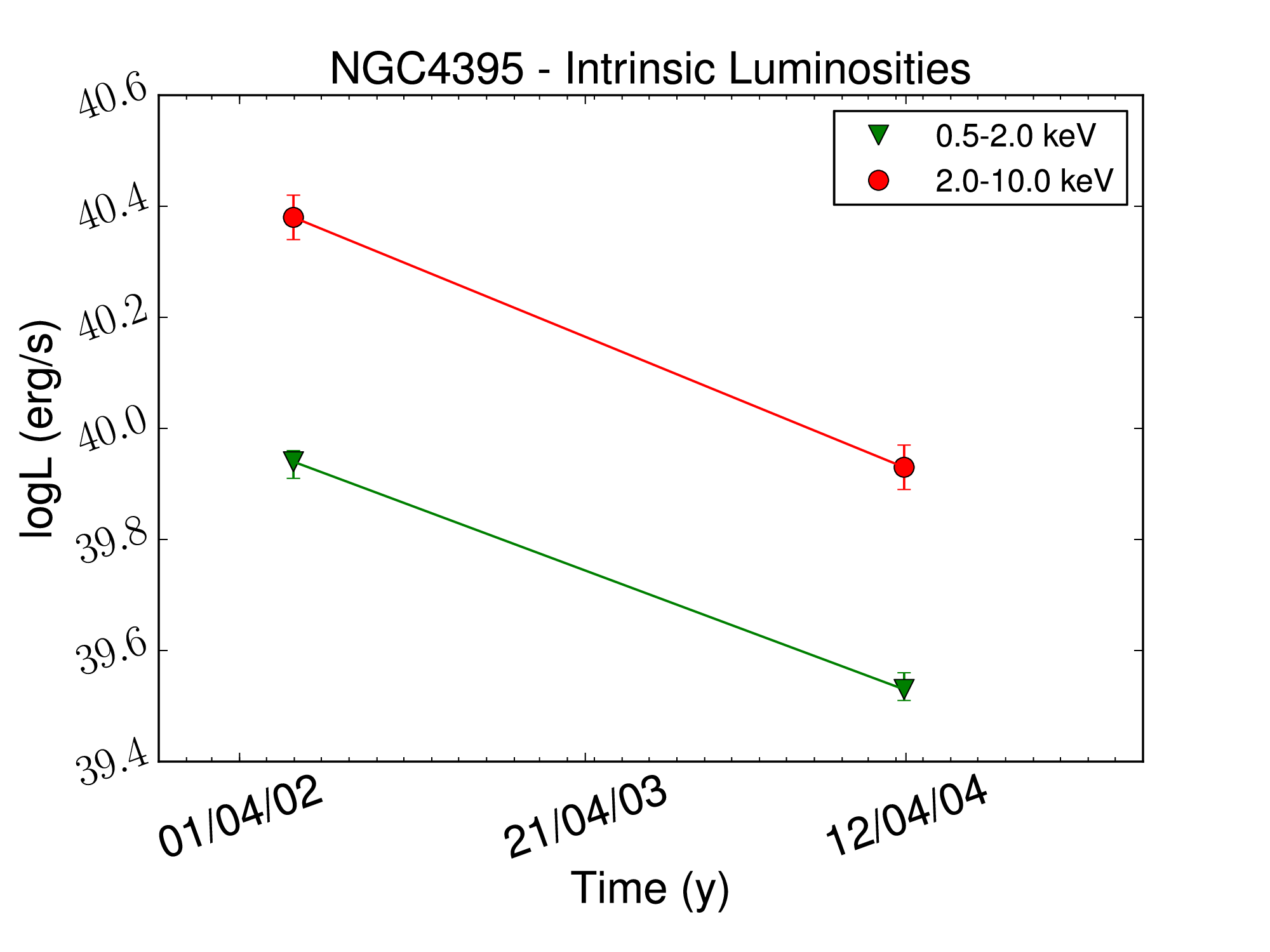}}

{\includegraphics[width=0.30\textwidth]{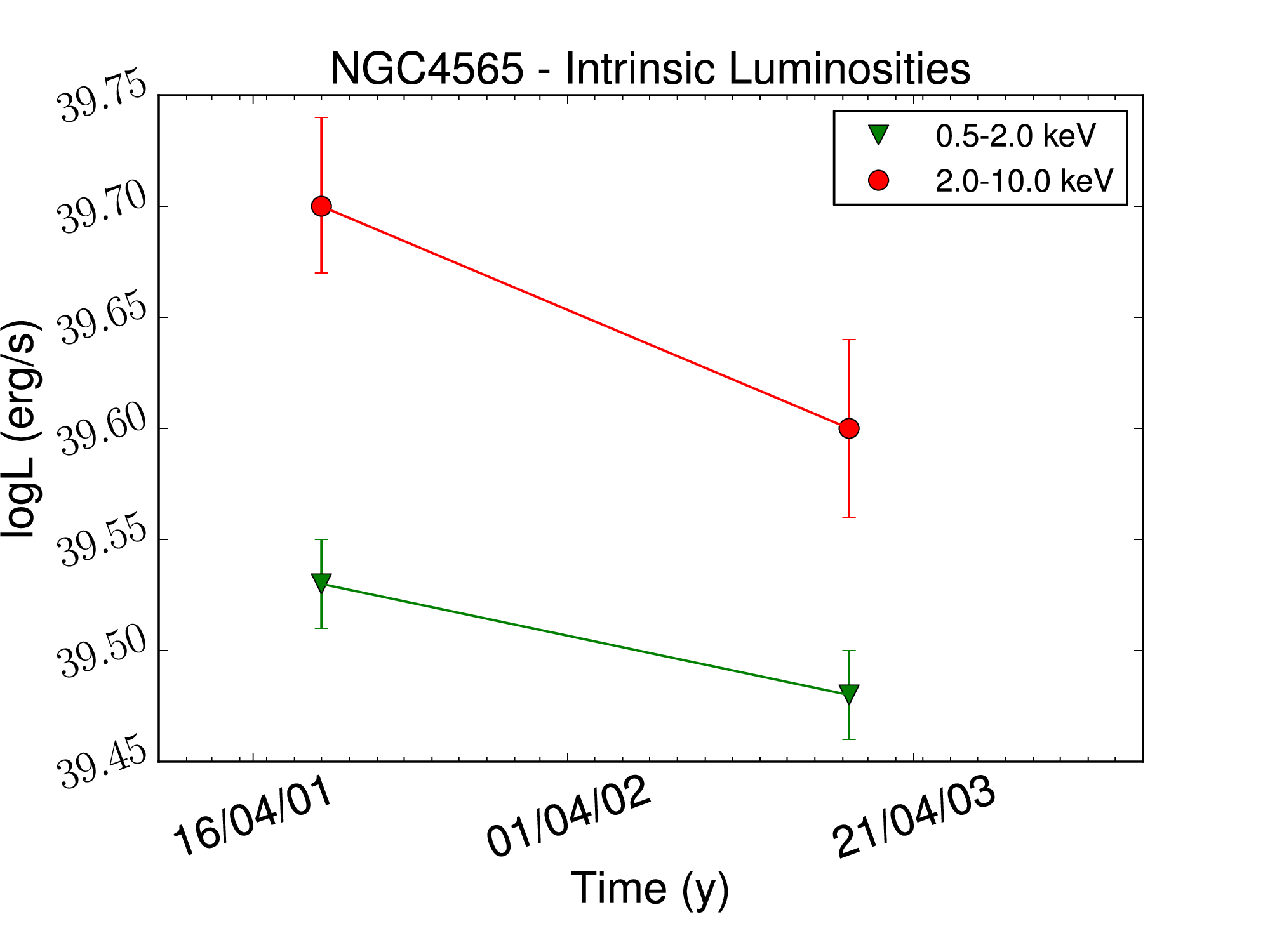}}
{\includegraphics[width=0.30\textwidth]{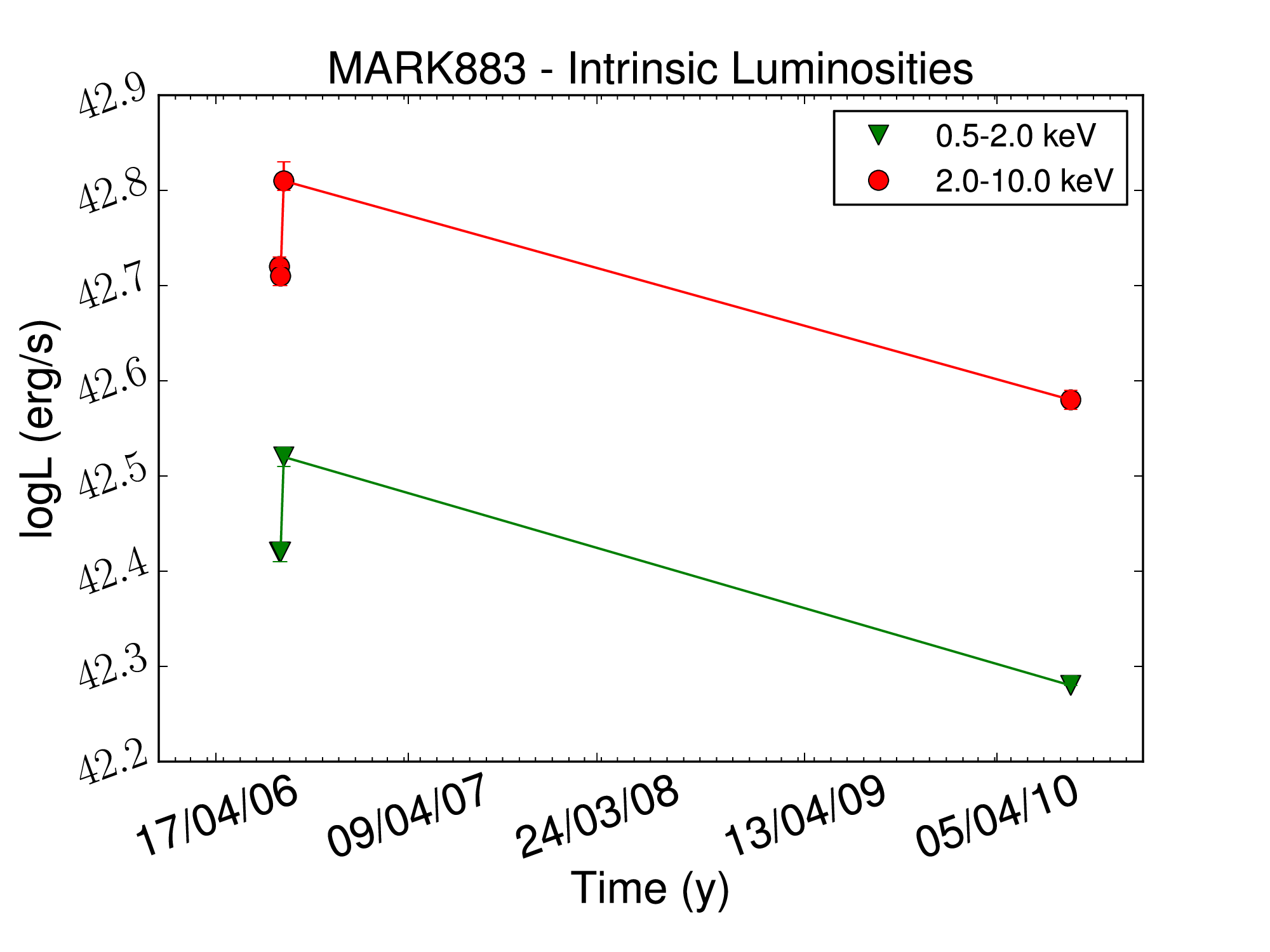}}
{\includegraphics[width=0.30\textwidth]{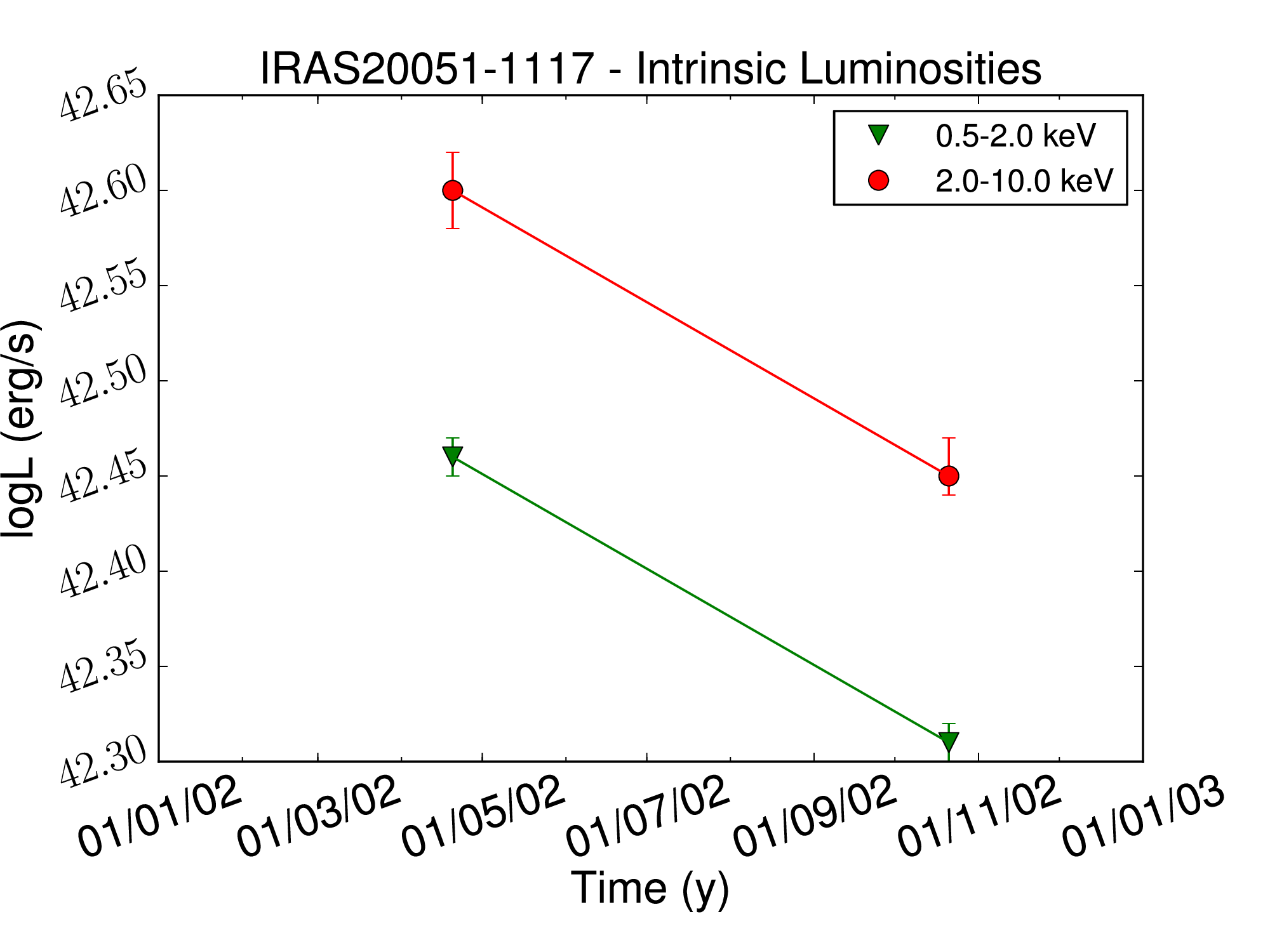}}
\caption{X-ray intrinsic luminosities calculated for the soft
  (0.5--2.0 keV, green triangles) and hard (2.0--10.0 keV, red
  circles) energies in the simultaneous fits, only for the variable
  objects.}
\label{luminXfigSey}
\end{figure*}

\begin{longtable}{lccccccc}
\caption[]{\label{estcurvasSey} Statistics of the light curves.}  \\  \hline \hline
Name &  ObsID & Energy & $\chi^2/d.o.f$  & Prob.(\%) & $\sigma_{NXS}^2$  &  $<\sigma^2_{NXS}>$ \\
(1) & (2) & (3) & (4) & (5) & (6) & (7)   \\ 
\hline 
\endfirsthead
\caption[]{(Cont.)} \\
\hline \hline
Name &  ObsID & Energy & $\chi^2/d.o.f$  & Prob.(\%) & $\sigma_{NXS}^2$  &  $<\sigma^2_{NXS}>$  \\
(1) & (2) & (3) & (4) & (5) & (6) & (7)  \\ 
\hline 
\endhead
\endfoot
\\
\endlastfoot
ESO\,113-G10 & 0301890101    & 0.5--10 (1) & 1194.3/32 & 100 & 0.0186$^+_-$0.0003 & 0.01856$^+_-$0.00005\\
          &            &   0.5--10 (2) & 1703.8/39 & 100  & 0.0185$^+_-$0.0003 & \\
          &            &   0.5--2 (1) & 1014.3/32 & 100 & 0.0185$^+_-$0.0003 & 0.0183$^+_-$0.0002 \\
          &            &   0.5--2 (2) & 1419.7/39 & 100  & 0.0181$^+_-$0.0003 & \\
           &             & 2--10 (1) &  222.5/32  & 100  & 0.019$^+_-$0.001 & 0.020$^+_-$0.001 \\
           &             & 2--10 (2) &  333.9/39  & 100  & 0.021$^+_-$0.001 & \\
NGC\,526A & 0150940101 & 0.5--10 &  228.9/28  & 100   &   0.0025$^+_-$0.0001 &  \\      
          &          &   0.5--2 & 130.6/28 & 100 & 0.0038$^+_-$0.0003 & \\
          &             &  2--10 & 128.4/28 & 100 & 0.0018$^+_-$0.0001 & \\
          & 0721730301 & 0.5--10 & 109.3/33 & 100 & 0.00068$^+_-$0.00007 & \\
          &            &  0.5--2 & 66.1/33 & 100 & 0.0009$^+_-$0.0002 & \\
          &            & 2--10 & 81.3/33 & 100 & 0.0006$^+_-$0.0001 & \\
          & 0721730401 & 0.5--10 & 376.2/38 & 100 & 0.00241$^+_-$0.00006 & \\
          &            & 0.5--2 & 75.4/38 &  100 & 0.0008$^+_-$0.0002 & \\
          &             &  2--10 & 341.1/38 & 100  & 0.00321$^+_-$0.00009 & \\
NGC\,1365 & 0205590301     &  0.5--10 & 1012.7/37 &  100 & 0.0206$^+_-$0.0004 & \\
          &           &   0.5--2 & 33.3/37 & 36  & $<$0.0014 \\
          &             & 2--10 &  1464.2/37 & 100    & 0.0412$^+_-$0.0009 & \\
          & 0505140201 &  0.5--10 & 41.3/27 & 99  & 0.0014$^+_-$0.0007 & \\
          &           &   0.5--2 & 37.2/27 & 99 & 0.0013$^+_-$0.0009 &  \\
          &             & 2--10 & 18.8/27 & 12 & 0.004$^+_-$0.004 & \\
          & 0505140401 &  0.5--10 (1) & 55.3/25 & 100  & 0.0030$^+_-$0.0007 & 0.0020$^+_-$0.0013 \\
          &            &  0.5--10 (2) & 18.7/36 & 1  & $<$0.0014 \\
          &           &   0.5--2 (1) & 25.6/25 & 57 & 0.0002$^+_-$0.0009 & 0.0001$^+_-$0.0001\\
          &           &   0.5--2 (2) & 31.6/36 & 32 & $<$0.0017 \\
          &             & 2--10 (1) & 67.6/25 & 100 & 0.018$^+_-$0.003 & 0.0012$^+_-$0.0008 \\
          &             & 2--10 (2) & 32.2/36 & 35 & $<$0.0068 \\
          & 0505140501 & 0.5--10 & 53.3/30 & 100 & 0.0020$^+_-$0.0007 &  \\
          &           &   0.5--2 & 45.0/30 & 99  & 0.0016$^+_-$0.0008 & \\
          &             & 2--10 & 24.1/30 & 33 & $<$0.0077 \\
          & 0692840201 & 0.5--10 (1) & 634.3/40 & 100  & 0.0183$^+_-$0.0005 & 0.0107$^+_-$0.0035 \\
          &            &  0.5--10 (2) & 384.1/40 & 100  &  0.0105$^+_-$0.0003 &  \\
          &            &  0.5--10 (3) & 228.6/38 & 100  & 0.0063$^+_-$ 0.0003 &  \\
          &           &   0.5--2 (1) & 52.6/40 & 99  & 0.0010$^+_-$0.0007 & 0.0006$^+_-$0.0003 \\
          &           &   0.5--2 (2) & 47.4/40 & 80  & 0.0006$^+_-$0.0007 &  \\
          &           &   0.5--2 (3) & 60.0/38 & 100 & 0.0019$^+_-$0.0007 &  \\
           &             & 2--10 (1) & 1291.7/40 & 100  & 0.058$^+_-$0.002 & 0.028$^+_-$0.009 \\
           &             & 2--10 (2) & 630.9/40 & 100 & 0.0281$^+_-$0.0009 &   \\
           &             & 2--10 (3) & 352.6/38 & 100  & 0.0178$^+_-$0.0007 &  \\
          & 0692840301  & 0.5--10 (1) & 3511.7/38 & 100  & 0.0426$^+_-$0.0006 & 0.0417$^+_-$0.0008 \\
          &            &  0.5--10 (2) & 3763.3/40 & 100  & 0.0409$^+_-$0.0005 &  \\
          &           &   0.5--2 (1) & 659.6/38 & 100  & 0.0257$^+_-$0.0007 & 0.0262$^+_-$0.0005 \\
          &           &   0.5--2 (2) & 791.5/40 & 100 & 0.0267$^+_-$0.0007 &  \\
          &             & 2--10 (1) & 3229.5/38 & 100  & 0.0538$^+_-$0.0009 & 0.052$^+_-$0.002 \\
          &             & 2--10 (2) & 3175.1/40 &  100 & 0.0495$^+_-$0.0008 &  \\
          & 0692840401 & 0.5--10 (1) & 5994.8/37 & 100  & 0.090$^+_-$0.001 & 0.06$^+_-$0.02 \\
          &            &  0.5--10 (2) & 5409.8/36 & 100  & 0.0453$^+_-$0.0005 &  \\
          &           &   0.5--2 (1) & 1458.4/37 &  100 & 0.067$^+_-$0.002 & 0.058$^+_-$0.006 \\
          &           &   0.5--2 (2) & 2447.0/36  & 100 & 0.054$^+_-$0.001 & \\
          &             & 2--10 (1) & 4977.2/37 & 100  & 0.106$^+_-$0.002 & 0.06$^+_-$0.02 \\
          &             & 2--10 (2) & 3164.0/36 & 100  & 0.0430$^+_-$0.0006 & \\
          & 0692840501 & 0.5--10 (1) & 6872.7/36 & 100 & 0.145$^+_-$0.002 & 0.03$^+_-$0.02 \\
          &            &  0.5--10 (2) & 1125.4/38 & 100  & 0.0276$^+_-$0.0006 &  \\
          &            &  0.5--10 (3) & 504.2/39 & 100  & 0.0115$^+_-$0.0003 &  \\
          &           &   0.5--2 (1) & 559.1/36 & 100 & 0.054$^+_-$0.002 & 0.01$^+_-$0.02 \\
          &           &   0.5--2 (2) & 21.9/38 & 2  & $<$0.0014 \\
          &           &   0.5--2 (3) & 24.9/39 & 5  & $<$0.0014 \\
          &             & 2--10 (1) & 7598.8/36 & 100  & 0.192$^+_-$0.004 & 0.05$^+_-$0.03 \\
          &             & 2--10 (2) & 1585.9/38 & 100  & 0.059$^+_-$0.001 &  \\
          &             & 2--10 (3) & 753.4/39 &  100 & 0.0271$^+_-$0.0007 & \\
NGC\,2617 & 0701981601    & 0.5--10 & 128.7/31 & 100 & 0.00046$^+_-$0.00004 &  \\
          &            &   0.5--2 & 138.3/31 & 100  & 0.00065$^+_-$0.00005 &  \\
           &             & 2--10 & 27.1/31 & 73  & $<$0.0003 \\
NGC\,2992 & 0654910501 & 0.5--10 & 41.2/40 & 58  & 0.00003$^+_-$0.00014 & \\
          &          &   0.5--2 & 44.9/40 & 73  & 0.0002$^+_-$0.0003 & \\
          &             &  2--10 & 27.6/40 & 7  & 0.00031$^+_-$0.0002 & \\
          & 0654910601 & 0.5--10 & 29.4/22 & 87  & 0.0006$^+_-$0.0005 & \\
          &          &   0.5--2 & 16.2/22 &  19 & $<$0.0024 & \\
          &             &  2--10 & 28.2/22 & 83  & 0.0011$^+_-$0.0009 & \\
          & 0654910701 & 0.5--10 & 70.4/36 & 100  & 0.0015$^+_-$0.0004 & \\
          &          &   0.5--2 & 36.2/36 & 54  & 0.00002$^+_-$0.00078 & \\
          &             &  2--10 & 61.8/36 &  100 & 0.0021$^+_-$0.0007 & \\
          & 0654910901 & 0.5--10 & 24.9/29 & 32 & $<$0.0017 & \\
          &          &   0.5--2 & 25.7/29 & 36 & $<$0.0033 \\
          &             & 2--10 & 25.6/29 & 35  & $<$0.0035 \\
          & 0654911001 & 0.5--10 & 61.1/32 & 100  & 0.0014$^+_-$0.0004 & \\
          &          &   0.5--2 & 45.2/32 & 99  & 0.0015$^+_-$0.0008 & \\
          &             & 2--10 & 50.9/32 & 99  & 0.0017$^+_-$0.0007 & \\
POX\,52 & 0302420101 & 0.5--10 (1) & 59.5/23 & 100  & 0.09$^+_-$0.02 & 0.05$^+_-$0.03 \\
          &            &   0.5--10 (2) & 50.9/33 & 99  & 0.03$^+_-$0.01 & \\
          &            &   0.5--2 (1) & 60.6/23 & 100  & 0.20$^+_-$0.06 & 0.08$^+_-$0.08\\
          &            &   0.5--2 (2) &    38.2/33      &  76  & 0.02$^+_-$0.03 & \\
           &             & 2--10 (1) & 34.6/23 & 95  & 0.05$^+_-$0.03 & 0.03$^+_-$0.01 \\
           &             & 2--10 (2) &       45.8/33      & 99  & 0.02$^+_-$0.02 & \\
NGC\,4395 & 0744010101 & 0.5--10 & 782.1/40 & 100  & 0.069$^+_-$0.003 & \\
          &            & 0.5--2 & 83.7/40 & 100 & 0.045$^+_-$0.01 & \\
          &            & 2--10 & 815.1/40 & 100 & 0.077$^+_-$0.003 & \\
NGC\,4565 & 3950 &  0.5--10 & 30.1/40 & 13  & $<$0.0161 \\
          &           &   0.5--2 & 43.5/40 & 68  & 0.004$^+_-$0.001 & \\
          &             & 2--10 & 40.0/40 & 53 & $<$0.0548 \\                 
\hline
\caption*{{\bf Notes.} (Col. 1) name, (Col. 2) obsID, (Col. 3) energy
  band in keV, (Cols. 4 and 5) $\chi^2/d.o.f$ and the probability of
  being variable in the 0.5-10.0 keV energy band of the total light
  curve, (Col. 6) normalized excess variance, $\sigma_{NXS}^2$, and
  (Col. 8) the mean value of the normalized excess variance,
  $<\sigma^2_{NXS}>$, for each light curve and energy band.}
\end{longtable}

\begin{longtable}{lcccccc|c}
\caption[]{\label{ew} Classification of the sample objects on the basis of the nuclear X-ray obscuration (and its variability).}  \\ \hline \hline Name & ObsID & $\Gamma$ & EW &
$F_x/F_{[OIII]}$ & Ref.$^1$ & CT? & Classification 
\\ & & & (keV) & & $[OIII]$ & &  \\ (1) & (2) & (3) & (4) & (5) & (6)
& (7) & (8)  \\ \hline \endfirsthead
\caption[]{(Cont.)} \\
\hline \hline
Name &  ObsID &  $\Gamma$  & EW  & $F_x/F_{[OIII]}$ & Ref.$^1$ & CT? & Classification   \\
     &        &  & (keV)  & & $[OIII]$ &  &   \\
(1) & (2) & (3) & (4) & (5) & (6) & (7) & (8)      \\ 
\hline 
\endhead
\endfoot
\\
\endlastfoot
ESO540-G01 & 0044350101 &  1.44$ _{ 1.07 }^{ 2.32 }$ & $<$0.48 & - & -  & \xmark  & Compton-thin \\
ESO195-IG21 & 0554500201 &  1.29$ _{ 1.06 }^{ 1.52 }$ & 0.11$ _{ 0.08 }^{ 0.15 }$ & - & -  & \xmark  & Compton-thin \\
 & 13898 & 2.06$ _{ 0.87 }^{ 2.57 }$ & 0.51$ _{ 0.23 }^{ 0.78 }$ & - &   & \xmark  \\
ESO113-G10 & 0103861601 & 1.62$ _{ 1.38 }^{ 1.89 }$ & $<$0.18 & - & -  & \xmark  & Compton-thin \\
 & 0301890101 & 1.80$ _{ 1.75 }^{ 1.87 }$ & $<$0.04 & - &   & \xmark  \\
NGC526A & 0109130201 & 1.30$ _{ 1.24 }^{ 1.41 } $ & 0.08$ _{ 0.05 }^{ 0.11 }$ & 75.4 & 1  & \xmark & Compton-thin \\
 & 0150940101 &  1.55$ _{ 1.48 }^{ 1.61 }$ & 0.06$ _{ 0.05 }^{ 0.07 }$ & 99.4 &   & \xmark   \\
 & 0721730301 &  1.48$ _{ 1.43 }^{ 1.53 }$ & 0.07$ _{ 0.06 }^{ 0.07 }$ & 114.2 &   & \xmark  \\
 & 0721730401 &  1.58$ _{ 1.52 }^{ 1.63}$ & 0.06$ _{ 0.05 }^{ 0.07 }$ & 137.3 &   & \xmark  \\
 & 342 &  $<$0.53 & $<$0.23 & 19.4 &   & \xmark  \\
MARK609 & 0103861001 &  1.59$ _{ 1.22 }^{ 1.86 }$ & $<$0.16 & 31.1 &  2 & \xmark  & Compton-thin \\
 & 0402110201 &  1.25$ _{ 1.08 }^{ 1.45 }$ & 0.11$ _{ 0.03 }^{ 0.19 }$ & 32.5 &   & \xmark  \\ 
NGC1365 & 0151370101 & 1.18$ _{ 0.91 }^{ 1.46 }$ & 0.20$ _{ 0.16 }^{ 0.25 }$ & 13.2 &  1 & \xmark  & Changing-look \\
 & 0151370201 & $<$1.36 & $<$0.31 & 2.2 &   & \xmark  \\
 & 0151370701  & 2.44$ _{ 2.20 }^{ 2.77 }$ & 0.14$ _{ 0.09 }^{ 0.18 }$ & 17.0 &   & \xmark  \\
 & 0205590301  & 2.88$ _{ 2.84 }^{ 2.94 }$ & 0.11$ _{ 0.10 }^{ 0.12 }$ & 19.1 &   & \xmark  \\
 & 0205590401  & 2.18$ _{ 2.08 }^{ 2.37 }$ & 0.17$ _{ 0.15 }^{ 0.19 }$ & 14.5 &   & \xmark  \\
 & 0505140201  & $<$0.39 & 0.46$ _{ 0.41 }^{ 0.52 }$ & 3.9 &   & \cmark  \\
 & 0505140401  & $<$0.09 & 0.42$ _{ 0.39 }^{ 0.54 }$ & 7.1 &   & \cmark  \\
 & 0505140501  & $<$0.17 & 0.43$ _{ 0.38 }^{ 0.58 }$ & 5.6 &   & \cmark  \\
 & 0692840201 & 2.08$ _{ 1.99 }^{ 2.12 }$ & 0.12$ _{ 0.11 }^{ 0.13 }$ & 17.4 &   & \xmark  \\
 & 0692840301  & 2.40$ _{ 2.38 }^{ 2.44 }$ & 0.09$ _{ 0.08 }^{ 0.10 }$ & 18.2 &   & \xmark  \\
 & 0692840401  & 2.32$ _{ 2.31 }^{ 2.38 }$ & 0.08$ _{ 0.07 }^{ 0.09 }$ & 18.2 &   & \xmark  \\
 & 0692840501  & 2.20$ _{ 2.13 }^{ 2.22 }$ & 0.09$ _{ 0.08 }^{ 0.10 }$ & 17.8 &   & \xmark  \\
 & 6869 &  1.34$ _{ 0.96}^{ 2.08 }$ & 0.13$ _{ 0.07 }^{ 0.19 }$ & 13.2 &   & \xmark  \\
NGC2617 & 0701981601 & 1.60$ _{ 1.58 }^{ 1.63 }$ & 0.06$_{ 0.05 }^{ 0.07 }$ & - & -  & \xmark  & Compton-thin \\
 & 0701981901 & 1.70$ _{ 1.67 }^{ 1.74 }$ & 0.05$ _{ 0.03 }^{ 0.06 }$ & - &   & \xmark  \\ 
MARK1218 & 0302260201 & 1.34$ _{ 1.13 }^{ 1.75 }$ & $<$0.09 & 18.1 & 2  & \xmark  & Compton-thin \\
 & 0302260401 &  1.670$ _{ 0.74 }^{ 2.99 }$ & $<$0.20 & 8.1  &   & \xmark \\ 
NGC2992 & 0654910501  & 1.54$ _{ 1.46 }^{ 1.63 }$ & 0.16$ _{ 0.14}^{ 0.18 }$ & 2.2 & 1  & \xmark & Changing-look \\
 & 0654910601  & 1.31$ _{ 1.20 }^{ 1.45 }$ & 0.33$ _{ 0.29 }^{ 0.36 }$ & 0.9 &   & \xmark  \\
 & 0654910701  & 1.26$ _{ 1.22 }^{ 1.36 }$ & 0.32$ _{ 0.29 }^{ 0.34 } $ & 0.9 &   & \xmark  \\
 & 0654910901 & 1.23$ _{ 1.05 }^{ 1.42 }$ & 0.50$ _{ 0.45 }^{ 0.55 }$ & 0.5 &   & \cmark  \\
 & 0654911001 & 1.30$ _{ 1.22 }^{ 1.42 }$ & 0.25$ _{ 0.22 }^{ 0.28 }$ & 1.0 &   & \xmark  \\
 & 0701780101  & 1.39$ _{ 1.33 }^{ 1.54 }$ & 0.17$ _{ 0.14 }^{ 0.21 }$ & 2.8 &   & \xmark  \\
POX52 & 0302420101 &  1.48$ _{ 1.14 }^{ 1.88 }$ & $<$0.11 & 8.2 &  3 & \xmark  & Compton-thin \\
 & 5736 & 1.79$_{ 1.44 }^{ 3.23 }$ & $<$0.45 & 12.5 &   & \xmark  \\
NGC4138 & 0112551201 & 1.23$ _{ 0.98 }^{ 1.49 }$ & 0.08$ _{ 0.04 }^{ 0.13 }$ & 478.6 &  4 & \xmark & Compton-thin  \\
 & 3994 &  1.29$ _{ 0.18 }^{ 1.94 }$ & 0.17$ _{ 0.04 }^{ 0.30 }$ & 478.6 &   & \xmark  \\
NGC4395 & 0112521901 & 1.28$ _{ 1.05 }^{ 1.53 }$ & 0.11$ _{ 0.07 }^{ 0.16 }$ & 64.8 & 4  & \xmark  & Compton-thin \\
 & 5301 & 1.24$ _{ 0.80 }^{ 1.65 }$ & $<$0.11 & 27.7 &   & \xmark  \\
 & 5302 & 1.26$ _{ 0.64 }^{ 1.67} $ & $<$0.10 & 25.8 &   & \xmark  \\

NGC4565 & 0112550301 &  1.55$ _{ 1.08 }^{ 2.49 }$ & $<$0.71 & 5.6 &  5 & \xmark  & Compton-thin \\
 & 3950 &  2.73$ _{ 1.51 }^{ 2.95 }$ & 0.03$ _{ 0.01 }^{ 0.05 }$ & 15.4 &   & \xmark  \\
MARK883 & 0302260101   & 1.22$ _{ 1.05 }^{ 1.49 }$ & $<$0.17 & 27.5 & 2  & \xmark  & Compton-thin \\
 & 0302260701  & 1.44$ _{ 1.26 }^{ 1.65 }$ & 0.12$ _{ 0.02 }^{ 0.21 }$ & 26.2 &   & \xmark  \\
 & 0302261001 & 1.58$ _{ 1.34 }^{ 2.13 }$ & 0.10$ _{ 0.02 }^{ 0.19 }$ & 32.3 &   & \xmark  \\
 & 0652550201  & 1.43$ _{ 1.33 }^{ 1.73 }$ & 0.18$ _{ 0.11 }^{ 0.24 }$ & 19.9 &   & \xmark  \\ 
IRAS20051-1117 & 0044350201 & 1.68$ _{ 1.36 }^{ 2.41 }$ & 0.21$ _{ 0.06 }^{ 0.35 }$ & 13.8 &   & \xmark  & Compton-thin \\
 & 0044350501 & 1.70$ _{ 1.48 }^{ 2.25 }$ & $<$0.16 & 9.5 & 5  & \xmark  \\ \hline
\caption*{{\bf Notes.} (Col. 1) name, (Col. 2) obsID, (Cols. 3 and 4)
  index of the power law and the equivalent width of the FeK$\alpha$
  line from the spectral fit (PL model) in the 3--10 keV energy band,
  (Col. 5) ratio between the individual hard X-ray luminosity (from
  Table \ref{lumincorrSey}) and the extinction-corrected [O III]
  fluxes, (Col. 6) references for the measure of $F_{[OIII]}$,
  (Col. 7) classification from the individual observation, (Col. 8)
  classification of the object, and (Col. 9) slope of the power law at
  hard energies for \emph{\emph{Compton}}-thick candidates from the
  simultaneous analysis (see Sect. \ref{thick}).  $^1$References: (1)
  \cite{bassani1999}; (2) \cite{dahari1988}; (3) \cite{whittle1992}; (4)
  \cite{panessa2006}; and (5) \cite{panessabassani2002}. }
\end{longtable}

\newpage

\twocolumn

\normalsize

\section{\label{indivnotes} Notes and comparisons with previous results for individual objects}

In this appendix we discuss the general characteristics of the
galaxies in our sample at different wavelengths, as well as
comparisons with previous variability studies. We recall that
long-term UV variability and short-term X-ray variations were studied
only for some sources (ten and seven sources, see Tables
\ref{properties} and \ref{estcurvasSey}, respectively), so comparisons
are only made in those cases. For the remaining objects, results from
other authors are mentioned, when available.

\subsection{ESO\,540-G01}

NGC\,540-G01 is the brightest member of the Hickson compact group 4 \citep{hickson1989}. It was classified as a Seyfert 1.8 using optical data \citep{coziol1993}, although also as a composite galaxy \citep{moran1996}. A radio counterpart was detected with \emph{VLA} data at 1.4 GHz \citep{condon1998}. It is another source classified as a type 1 AGN showing no X-ray absorption \citep{panessa2005}.

This galaxy was observed once by \emph{Chandra} in 2001 and once by \emph{XMM--Newton} in 2002. \cite{georgantopoulos2003a} studied the \emph{Chandra} data. They found long-term X-ray flux variations when comparing with previous \emph{ROSAT} data by a factor of 25 in ten years. When comparing \emph{Chandra} and \emph{XMM--Newton} data, the annular region contributed with 21\% in \emph{Chandra} data, and the best representation is obtained with SMF1 showing variations in $Norm_2$ (74\%) in one year period. This implies flux variability of 40\% (38\%) in the soft (hard) energy band.

\cite{georgantopoulos2003a} studied the \emph{Chandra} light curve and did not find short-term variability.

\subsection{ESO\,195-IG21}

ESO\,195-IG21 was classified as a Seyfert 1.8 using optical data, after its discovery using \emph{Swift} data \citep{baumgartner2008}. We did not find data at radio frequencies for this source in the literature.

This galaxy was observed once by \emph{XMM--Newton} in 2008 and once by \emph{Chandra} in 2012. Variability studies of this source were not found in the literature. In this case, the annular region contributed with 10\% to the \emph{Chandra} data. When comparing the data, SMF2 results in the best fit with $Norm_1$ (91\%) and $Norm_2$ (98\%) varying in a four years period. This results in flux variability of 98\% (97\%) in the soft (hard) energy band.

\subsection{ESO\,113-G10}

The nucleus was classified as a Seyfert 1.8 in the optical by \cite{pietsch1998}, who also reported X-ray variations of a factor of three between two \emph{ROSAT} observations obtained within six months. 


ESO\,113-G10 was observed twice with \emph{XMM--Newton} in 2001 and 2005.  Long-term X-ray variability studies of these source were not found in the literature.
We find that SMF1 with variations in $Norm_1$ (40\%) represents the data best in a four years period. This implies flux variations of 17\% in both the soft and hard energy bands. 

\cite{porquet2007} found short-term variations in the soft and hard light curve from 2005 in timescales lower than 500 s. 
These results agree well with those obtained by \cite{omairavaughan2012} and ours, because the analysis of the same light curve resulted in short-term variations of the total, soft, and hard energy bands.
Later, \cite{cackett2013} used the same observation to report a hard and a soft lag at low and high frequencies, respectively, between the 1.5--4.5 keV and 0.3--0.9 keV energy bands. 

At UV frequencies, we detected variations in the UVW2 (8$\sigma$) filter from the OM.

\subsection{\label{n526}NGC\,526A}

NGC\,526A is the west galaxy in the strongly interacting pair of galaxies in NGC\,526 \citep{mulchaey1996}. It was optically classified as a Seyfert 1.9 \citep{griffiths1979,winkler1992}. A radio counterpart was detected with \emph{VLA} data at 3.6 and 20 cm \citep{nagar1999}. 

X-ray flux and spectral variability was detected in this source as observed by \emph{HEAO 1}, \emph{Einstein}, \emph{EXOSAT} and \emph{GINGA} in timescales of years \citep{mushotzky1982, turner1989, polletta1996}. 

It has been observed with \emph{Chandra} five times between 1999 and 2003, and four times with \emph{XMM--Newton} between 2002 and 2013. Long-term variability analyses using these data are not reported in the current literature. SMF1 is the best representation of the \emph{XMM--Newton} data presented in this work, with variations in $Norm_2$ (48\%). This implies a flux variation of 48\% (46\%) in the soft (hard) energy band in 11 years period. When comparing with \emph{Chandra} data (the annular region contributed with 45\% in \emph{Chandra} data) out method cannot differentiate if variations in $N_{H2}$ ($\chi^2_r = 1.03$) or variations in $Norm_2$ ($\chi^2_r = 1.08$) are preferred in the two years period, both results being a good option. For this reason we do not report this analysis and we take into account the results from the \emph{XMM--Newton} data.

\cite{omairavaughan2012} studied the \emph{XMM--Newton} observation from 2003 and found that it showed short-term variations in the three analysed energy bands. We studied three \emph{XMM--Newton} observations (except the one from 2002) and detected short-term variations in all of them and the three energy bands, in good agreement with \cite{omairavaughan2012}. In short timescales, \cite{turner1997} also detected rapid variations using \emph{ASCA} data. Using \emph{RXTE} data between 2001 and 2003, \cite{zhang2011} estimated $\sigma^2_{NXS}$ = $8.16^+_-1.81 \times 10^{-2}$.

In the 14--195 keV energy band, \cite{soldi2014} estimated a variability amplitude of 35$^+_-$6\% using data from the \emph{Swift}/BAT 58-month survey.

UV data from the OM are available in two filters. Variations are detected in the UVW1 (11$\sigma$) but not in the UVM2 (1$\sigma$) filter.






\subsection{MARK\,609}

This galaxy was classified as a Seyfert 1.8 by \cite{osterbrock1981}, although it was later classified as a type 2 and 1.5, suggesting that this might be a changing-look candidate \citep[][and references therein]{trippe2010}. It has also been proposed to be a ``true'' Seyfert 2, i.e., an AGN that lack circumnuclear obscuration \citep{lamassa2011,lamassa2014}. A radio counterpart was detected with \emph{VLA} data at 6 cm \citep{ulvestadwilson1984}.

The source was observed twice with \emph{XMM--Newton} in 2002 and 2007. Variability studies using these data were not found in the literature. We find that SMF1 represents best the data, requiring changes in $Norm_1$ (22\%) within five years. This implies flux variations of 13\% in both the soft and hard energy bands.

\subsection{NGC\,1365}

NGC\,1365 is a barred spiral galaxy located in the Fornax I cluster. It was classified as a Seyfert 1.8 at optical wavelenghts \citep{maiolino1995}. This is the prototypical example of a changing-look source, having changed from reflection-dominated to transmission dominated states in different occasions, in good agreement with the results presented here \citep[e.g.,][]{risaliti2009}. A radio counterpart was observed with \emph{VLA} data \citep{jorsater1995}.

It was observed 13 times with \emph{XMM--Newton} between 2003 and 2013, and seven times with \emph{Chandra} between 2002 and 2006.

\cite{risaliti2005a} studied a \emph{Chandra} observation from 2002 and three \emph{XMM--Newton} observations from 2003. They reported variations from Compton-thin to Compton-thick in timescales of six weeks, which were attributed to changes in the absorber, whereas the soft thermal emission and the reflection component remained constant. The same authors also discovered four absorption lines between 6.7 and 8.3 keV (identified them as Fe XXV and Fe XXVI K$\alpha$ and K$\beta$ lines) related to absorption by a highly ionized gas \citep{risaliti2005b}, and detected an X-ray eclipse from clouds in the BLR from the \emph{XMM--Newton} observation from January 2004 \citep{risaliti2009}. \cite{guainazzi2009} analyzed the \emph{RGS} data and suggested that the relative weakness of the photoionisation might be related to the same material responsible for the X-ray abosrbing column density variations.

Simultaneous \emph{XMM--Newton} and \emph{Suzaku} data from 2012 and 2013 were studied by different authors. \cite{risaliti2013} and \cite{walton2014} found continuum variations related to absorption from reflection,  \cite{parker2014} reported variability in the absorber and intrinsic to the source using a principal component analysis (PCA), \cite{braito2014} found absorption variations in timescales of about 100 ksec in the 0.3--1.8 keV energy band related to a low ionization zone of a disk wind, and \cite{rivers2015} reported that changes in this source are mainly due to absorption.
Our method does not allow to perform the spectral fit to all the spectra simultaneously ($\chi^2_r>2.5$). Instead, we removed the spectra in the Compton-thick state \citep{risaliti2005b} and those in a `relativistic' state \citep{risaliti2013}. Therefore we used six \emph{XMM--Newton} observations. The best fit resulted from the use of SMF3 with $N_{H2}$ (68\%), $Norm_2$ (33\%), and $Norm_1$ (35\%) varying. These changes imply a change in flux of 81\% (24\%) in the soft (hard) energy band in a 10 years period. When comparing \emph{Chandra} and \emph{XMM--Newton} data, SMF2 was required with variatons in $N_{H2}$ (37\%) and $Norm_2$ (30\%) in a two years period. The annular region contributed with 53\% in \emph{Chandra} data.

\cite{connolly2014} studied 190 observations obtained with \emph{Swift} between 2006 and 2013. They reported variations in the normalizations of the soft and hard energy bands plus changes in the absorbing material, and interpreted these variations in terms of an AGN wind scenario. 

\cite{brenneman2013} studied three \emph{Suzaku} observations of NGC\,1365 obtained between 2008 and 2010. They reported variations both at short- and long-term, mainly due to absorption and continuum variations.

\cite{omairavaughan2012} studied three \emph{XMM--Newton} observations between 2004 and 2007. They found short-term variations in the hard and total energy bands of the three observations, whereas the soft band did not show variations. We analyzed two of these light curves and found a good agreement. In total, we were able to study short-term variations in eight observations. We found that variations in the hard and total energy bands are very common in all the observations, but variability in the soft band is observed only in observations from 2012 and 2013. The light curve presented by \cite{connolly2014} using \emph{Swift} data also showed short-term variations. They attributed these variations to clouds passing throught the line of sight of the observer.
Using the simultaneous \emph{XMM--Newton} and \emph{NuSTAR} data, \cite{kara2015} found a Fe K lag, plus another lag at low frequencies, probably due to absorption variations.

Data at UV frequencies were available from the OM in the three UV filters. Strong variations were detected with the UVW1 (62$\sigma$) and UVM2 (76$\sigma$) filters. Variations were not detected with the UVW2 (1$\sigma$) filter, but data were available only in two epochs separated by about seven months. 

\subsection{NGC\,2617}

NGC\,2617 is a spiral galaxy. It was optically classified as a Seyfert 1.8 \citep{moran1996}, although it was classified as a type 1 in 2014, when it was also classified as a changing-look AGN \citep{shappee2014}. A radio counterpart was detected with the \emph{VLA} \citep{condon1998}.

This galaxy was observed twice with \emph{XMM--Newton} in 2013. 
These data were studied by \cite{giustini2016}, who reported changes in flux, spectral shape ($\Delta \Gamma \sim 0.1$) and in $N_{H}$ within the month timescale. 
Our analysis of the same \emph{XMM--Newton} data showed that SMF2 is the best representation of the data set, with changes in $Norm_1$ (59\%) and $N_{H2}$ (30\%) within a one month period. This implies flux variations of 46\% (45\%) in the soft (hard) energy band.
A follow up of this source of $\sim$ 70 days using X-ray, UV, optical and NIR data was conducted by \cite{shappee2014} after a transient source alert on 2013 April. They observed an increasing X-ray flux of about one order of magnitude followed by an increase in its optical/UV continuum flux. By cross-correlating the light curves, they found that the UV (2--3 days) and the NIR (6--9 days) lagged behing the X-rays, and explained this variability behaviour due to X-ray radiation driven by the disk.

\cite{giustini2016} analyzed the \emph{XMM--Newton} light curves and reported modest variations in the 0.3-10 keV energy band. They also reported hints of a soft band delay on time scales larger than 5 ks between the soft and the hard energy bands.
From the analysis of one of the \emph{XMM--Newton} light curves, we found short-term variations in the soft and total energy bands.

In the UV, we detected variations in the UVW1 (30$\sigma$) filter with data from the OM.

\subsection{MARK\,1218}

MARK\,1218 (also named NGC\,2622) was optically classified as a Seyfert 1.8 by \cite{osterbrock1983a}, but it changed to a Seyfert 1 spectrum a few years later \citep{smith2004} and has also been classified as a type 1.9 \citep{trippe2010}. A radio counterpart was detected with \emph{VLA} data at 6 and 20 cm \citep{ulvestad1986}.

This source was observed with \emph{XMM--Newton} twice in 2005. Variability studies of this source were not found in the literature. \cite{singh2011} performed a spectral fit of the obsID. 0302260201 using an absorbed power law with warm absorption, and obtained an X-ray luminosity of logL(2--10 keV) = 42.68, in very good agreement with our estimation. We found that the best representation of the data is SMF1 with changes in $Norm$ (63\%). This implies flux changes of 64\% in both the soft and hard enery bands.

In the UV, variations were not detected with the UVM2 (1$\sigma$) filter from the OM.

\subsection{NGC\,2992}

NGC\,2992 is interacting with NGC\,2993, which is located at 2.9$\arcmin$. It was classified as a Seyfert 1.9 by \cite{ward1980} using optical data. A radio counterpart was detected with \emph{VLA} data at 5 GHz \citep{condon1982}. In good agreement with our results, it is also a changing-look candidate \citep{gilli2000, trippe2008}.

Historically, X-ray flux variability by a factor of 20 was found in this source between different satellites \citep[][and references therein]{shu2010}.

It was observed four times with \emph{Chandra} between 2003 and 2010, and ten times with \emph{XMM--Newton} between 2003 and 2013. \cite{shu2010} studied the \emph{XMM--Newton} data from 2003 and reported flux variability when comparing with \emph{Suzaku} data from 2005. They attributed the changes to intrinsic variability of the source.
\cite{parker2015} used the nine \emph{XMM--Newton} observations between 2003 and 2010 to study variability through principal component analysis. They reported variations in two components, one corresponding to intrinsic source variability, and a more ambigous one, corresponding either to changes in the power law, the column density, or the soft excess.
From the analysis of the \emph{XMM--Newton} data, the best representation is obtained with SMF2 varying $N_{H2}$ (5\%) and $Norm_2$ (21\%) in a three years period. This results in a flux variation of 19\% in the soft and hard energy bands.

The covering of this source with \emph{RXTE} during one year (24 observations) was presented by \cite{murphy2007}, who reported flux variations by a factor of 10 in timescales of days to weeks. They related these variations with changes in the inner accretion disk.

\cite{omairavaughan2012} studied the nine \emph{XMM--Newton} observations between 2003 and 2010 and reported short-term variations in three energy bands but not in all the observations, being more frequent in the total and hard energy bands than in the soft band. We analysed five \emph{XMM--Newton} observations and detected variations in the total and hard energy bands in a few observations, but none in the soft band.

At higher energies, a combined study with \emph{INTEGRAL}, \emph{Swift}, and \emph{BeppoSAX} data published by \cite{beckmann2007a} showed that variations in the normalization of the power law were needed when using an absorbed broken power law model to fit the data simultaneously. They found a constant $\Gamma$ and flux variations by a factor of 11 in timescales of months/years.

Variations in the near infrared were reported by \cite{glass1997} due to an outburst between March 1988 and April 1992 observed with the 1.9m telescope at Sutherland.

At UV frequencies, variations in the UVM2 filter of 6$\sigma$ were detected with the OM.

\subsection{POX\,52}

POX\,52 is a dwarf elliptical galaxy. It was optically classified as a Seyfert 1.8 by \cite{barth2004}. A radio counterpart was observed with \emph{VLA} data at 5 GHz \citep{thornton2008}.

This source was observed once by \emph{XMM--Newton} in 2005 and once by \emph{Chandra} in 2006. \cite{thornton2008} studied these observations. They found variations in the column density due to partial covering during this period. This result agrees well with our analysis, where variations in $N_{H2}$ (44\%) and $N_{H1}$ (it passes from a value of $\sim 8 \times 10^{22} cm^{-2}$ to the Galactic value) are needed to explain the observed variations in the same dataset. We note however that our values of $N_{H2}$ are larger than theirs, most probably due to the different models used (we used two absorbers instead of one). The annular region contributed with 4\% to the \emph{Chandra} data.

\cite{omaira2011a} studied a sample of ultraluminous X-ray sources (ULX) using \emph{XMM--Newton} data and included this source because the AGN has a low black hole mass. They reported short-term variations through the estimation of $\sigma_{NXS}^2$ in the 2--10 keV energy band.
\cite{dewangan2008} also estimated the $\sigma_{NXS}^2$ and found it to be compatible with short-term variations for the same \emph{XMM--Newton} observation.
The \emph{Chandra} and \emph{XMM--Newton} light curves were analysed by \cite{thornton2008}, who found variations in timescales of 500 s and $10^{4}$ s, respectively, in the 0.5--10 keV energy band. 
These results agree well with our analysis of the \emph{XMM--Newton} light curve, where we also detected these variations.

At UV frequencies, \cite{thornton2008} studied the OM data and also \emph{GALEX} data, but variations were not detected.

\subsection{\label{n4138}NGC\,4138}

NGC\,4138 is a spiral galaxy that was classified as a Seyfert 1.9 by \cite{ho1997}. A radio counterpart was detected using \emph{VLA} data at 2cm \citep{nagar2002}.

This galaxy was observed once with \emph{XMM--Newton} in 2001 and once with \emph{Chandra} in 2003. Variability studies of this source were not found in the literature. 
The \emph{XMM--Newton} spectrum is best fitted with the ME2PL model, but the \emph{Chandra} spectrum did not have enough counts below $\sim$2 KeV, and thus the PL model was used to fit both spectra individually and also for the simultaneous fit in the 2--10 keV energy band. The annular region contributed with 83\% in \emph{Chandra} data. SMF1 with variations in $Norm$ (98\%) represents best the data in a two years period. This implies flux variations of 97\% (21\%) in the soft (hard) energy band.

\subsection{NGC\,4395}

NGC\,4395 is a late-type spiral galaxy that holds an intermediate mass black hole. Its nucleus was classified as a Seyfert 1.8 by \cite{ho1997}. A nuclear radio source was detected with VLBA data \citep{wrobel2006}.

NGC\,4395 was observed four times with \emph{Chandra} between 2000 and 2004 and six times with \emph{XMM--Newton} between 2002 and 2014.
\cite{oneill2006} studied three \emph{Chandra} spectra plus the \emph{XMM--Newton} spectrum from November 2003. They reported flux variations of a factor 2 between the \emph{Chandra} and \emph{XMM--Newton} observations.
\cite{nardinirisaliti2011} studied the \emph{XMM--Newton} data from November 2003 and \emph{Suzaku} data obtained in June 2007. They reported variations related to the covering fraction of the neutral absorber and discussed that this absorber is located in the BLR. These studies agree well with our results. The best representation of the \emph{Chandra} data shows variations in $N_{H2}$ (31\%) within one day when fitting SMF1. \emph{XMM--Newton} data require SMF2 with $N_{H2}$ (20\%) and $Norm_2$ (88\%) varying in 12 years period. This implies flux variability of 15\% (13\%) in the soft (hard) energy band. When comparing \emph{Chandra} and \emph{XMM--Newton} data, the annular region contributes with 14\% in \emph{Chandra} data, and SMF1 is used with variations in $N_{H2}$ (93\%) in a two years period. 


\cite{dewangan2008} studied two \emph{XMM--Newton} light curves (2002 May and 2003 November) and reported short-term variations only in the 2003 data. \cite{omairavaughan2012} analyzed the \emph{XMM--Newton} observation from November 2003 and reported variations in the total, soft, and hard energy bands. 
\cite{vaughan2005} studied the 2003 \emph{XMM--Newton} light curves together and found a high variability amplitude, with the fractional rms exceeding the unity. They found that this source follows the rms-flux relation usually observed in accreting black holes \citep[e.g.,][]{lore2015b}.
\cite{moran2005} studied the \emph{Chandra} data from June 2000 and reported short-term variations from the analysis of the light curve, with changes of a factor of 2--3. They suggested that these changes are related to variations of the absorbing medium.
\cite{oneill2006} found short-term variations during one of the \emph{Chandra} light curves.
One \emph{XMM--Newton} light curve is analyzed in the present work, which shows variations in the soft, hard, and total energy bands. 

Time-lags between optical/UV frequencies and X-rays have been studied by \cite{mchardy2016} (using \emph{XMM--Newton}, OM (UVW1 filter) and optical data in the g-band), \cite{cameron2012} (using \emph{Swift} data), and \cite{oneill2006} (using \emph{Chandra} and \emph{HST}/STIS data) and found lags of 473 and 788 s, 400 s, and a zero-lag correlation, respectively, relative to the X-rays.


Variations of this source have also been detected at near-infrared frequencies in timescales shorter than a day by \cite{minezaki2006}. They observed the source in 2004 with the 2 m telescope at the Haleaka Observatories and found variations in he J and H bands correlated with the V band, whereas variations in the K band were not detected.

UV variations are detected in the UVW1 (9$\sigma$) filter using the available OM data.

\subsection{NGC\,4565}

NGC\,4565 is an edgewise spiral galaxy. It was classified as a Seyfert 1.9 in the optical \citep{ho1997}. A compact radio core was detected using \emph{VLA} data at 2cm \citep{nagar2002}. \cite{houlvestad2001} suggested that the nucleus may be variable at radio frequencies since flux measurements fluctuated by a factor of 2 between their \emph{VLA} data and previous \emph{FIRST} measurements.

This source was observed once by \emph{XMM--Newton} in 2001 and twice by \emph{Chandra} in 2000 and 2003. \cite{cappi2006} analyzed the \emph{XMM--Newton} data. They fitted the spectrum with an absorbed power law model with $N_H = 0.12 ^+_- 0.04 \times 10^{22} cm^{-2}$, $\Gamma = 1.8 ^+_- 0.2$, and logL(2--10 keV) = 39.4. They compared their spectral fit to the one reported by \cite{terashimawilson2003} of the \emph{Chandra} data from 2000, finding a good agreement between the spectral parameters and intrinsic luminosities. Our analysis agree well with their spectral fits. We find that SMF1 is the best representation of the data with changes in $N_H$ (48\%) in two years period. The annular region contributed with 17\% to \emph{Chandra} data.

We examined a \emph{Chandra} light curve and detected possible variations in the soft energy band.

\subsection{MARK\,883}

It was optically classified as Seyfert 1.9 by \cite{osterbrock1983a}, who stated that the broad component is `barely detected', and later classified as a Seyfert 2 \citep{trippe2010}. A radio counterpart was detected at 6 cm with \emph{VLA} data \citep{ulvestad1986}.

This nucleus was observed four times with \emph{XMM--Newton} between 2006 and 2010. Variability studies of this source were not found in the literature. From our analysis we find that SMF1 represents the data best, showing changes in $Norm$ (28\%) in a timescale of four years. This implies flux variations of 28\% in both the soft and hard energy bands.

Data at UV frequencies with the OM are available in the three filters. Variations are detected in the UVW1 (13$\sigma$) and UVW2 (5$\sigma$) filters but not in the UVM2 (1$\sigma$) filter.

\subsection{IRAS\,20051-1117}

IRAS\,20051--1117 is a luminous spiral galaxy. The line ratios of this galaxy located it in the boundarie between a composite and a Seyfert galaxy \citep{moran1996}. It shows a broad component, so it was classified as a Seyfert 1.9 using optical data \citep{georgantopoulos2004, shi2010}. A radio counterpart was detected with \emph{VLA} data at 1.4 GHz \citep{condon1998}.
This is one of the cases where a source is classified as a type 2 AGN at optical wavelenghts but shows no absorption at X-ray frequencies \citep{panessabassani2002}.

It was observed once with \emph{Chandra} and twice with \emph{XMM--Newton} in 2002. \cite{georgantopoulos2004} and \cite{shi2010} studied the \emph{Chandra} and one \emph{XMM--Newton} observations and reported no variations between the two epochs, separated by only three weeks. They obtained a luminosity of logL(2--10 keV) = 42.60, in perfect agreement with our estimated luminosity for the same \emph{XMM--Newton} spectrum. Using the two \emph{XMM--Newton} observations, we found that changes in $Norm_2$ (29\%) are required in the SMF1 within half a year. This results in flux variations of 29\% in both the soft and hard energy bands.

\cite{georgantopoulos2004} did not find short-term variations in the \emph{Chandra} nor the \emph{XMM--Newton} light curves. 

At UV frequencies variations were not detected with the UVW1 (2$\sigma$) filter.

\section{Images}

\subsection{\label{multiimages} Optical spectra, and X-ray, 2MASS and optical \emph{HST} images}

In this appendix we present images at different wavelengths for each
energy and the optical spectrum when available from NED.  In X-rays we
extracted \emph{Chandra} data in four energy bands: 0.6-0.9 keV (top
left), 1.6-2.0 keV (top middle), 4.5-8.0 keV (top right), and 0.5-10.0
keV (bottom left). The {\sc csmooth} task included in CIAO was used to
adaptatively smooth the three images in the
top panels (i.e., the images in the 0.5-10.0 keV energy band are not
smoothed), using a fast Fourier transform algorithm and a minimum and
maximum significance level of the signal-to-noise of 3 and 4,
respectively.  When data from \emph{Chandra} was not available,
\emph{XMM}--Newton images were extracted in the same energy bands, and
the {\sc asmooth} task was used to adaptatively
smooth the images.  At infrared frequencies, we retrieved an image
from 2MASS in the $K_s$
filter\footnote{http://irsa.ipac.caltech.edu/applications/2MASS/IM/interactive.html}.
At optical frequencies we used images from the \emph{Hubble} Space
telescope (\emph{HST})\footnote{http://hla.stsci.edu/}, preferably in
the F814W filter, but when it was not available we retrieved an image
in the F606W filter.  \emph{HST} data have been processed following
the sharp dividing method to show the internal structure of the
galaxies \citep{marquez1996}.  The red squares in the bottom images
represent the area covered by the \emph{HST} image (presented in the
bottom right panel when available).  In all images the gray levels
extend from twice the value of the background dispersion to the
maximum value at the center of each galaxy. We used
IRAF \footnote{http://iraf.noao.edu/} to estimate these values.

\begin{figure*}
\begin{center}
\includegraphics[width=\textwidth]{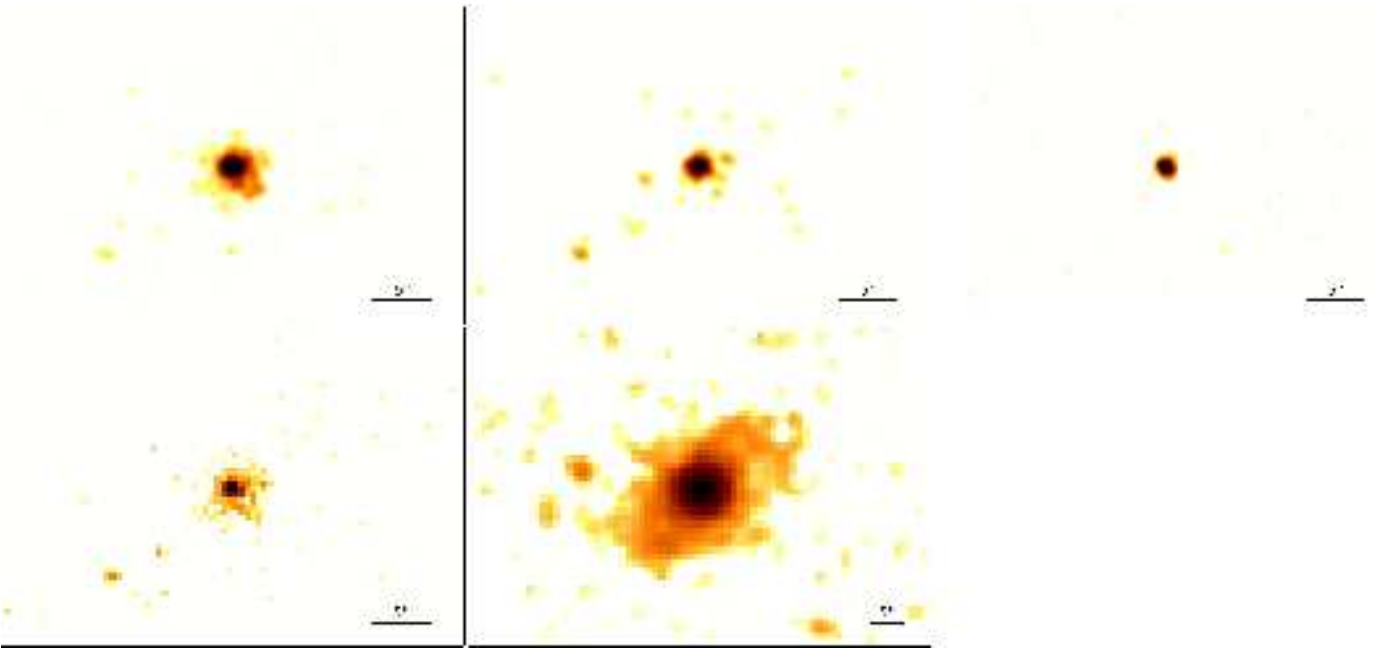} 
\end{center}
 \caption{Images of ESO\,540-G01. (Top left): Smoothed X-ray 0.6-0.9 keV
   energy band; (top center): smoothed X-ray 1.6-2.0 keV energy band;
   (top right): smoothed X-ray 4.5-8.0 keV energy band; (bottom left):
   X-ray 0.5-10.0 keV energy band without smoothing; (bottom center):
   2MASS image in the $K_s$ band.}
\end{figure*}

\begin{figure*}
\begin{center}
\includegraphics[width=\textwidth]{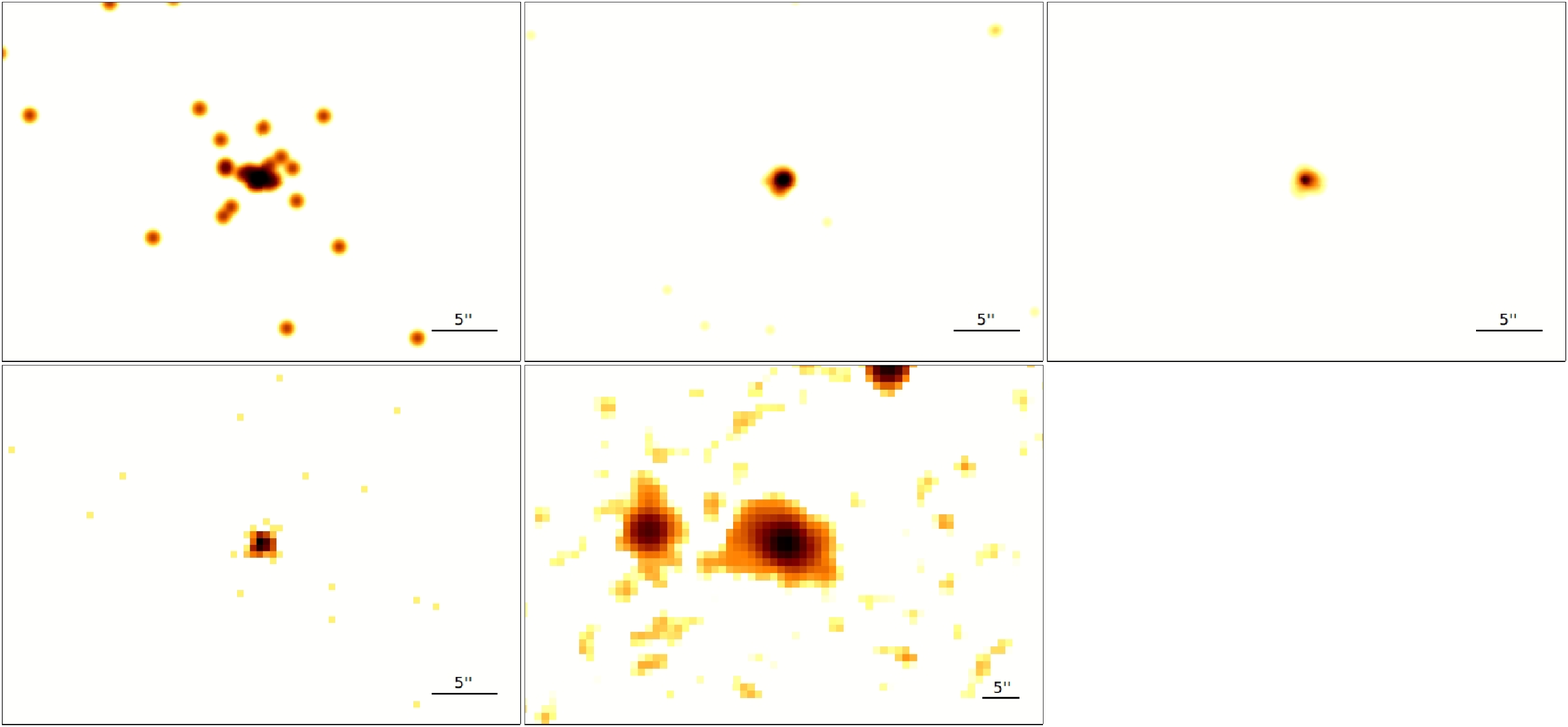} 
\end{center}
 \caption{Images of ESO\,195-IG21. (Top left): Smoothed X-ray 0.6-0.9 keV
   energy band; (top center): smoothed X-ray 1.6-2.0 keV energy band;
   (top right): smoothed X-ray 4.5-8.0 keV energy band; (bottom left):
   X-ray 0.5-10.0 keV energy band without smoothing; (bottom center):
   2MASS image in the $K_s$ band.}
\end{figure*}

\begin{figure*}
\begin{center}
\includegraphics[width=\textwidth]{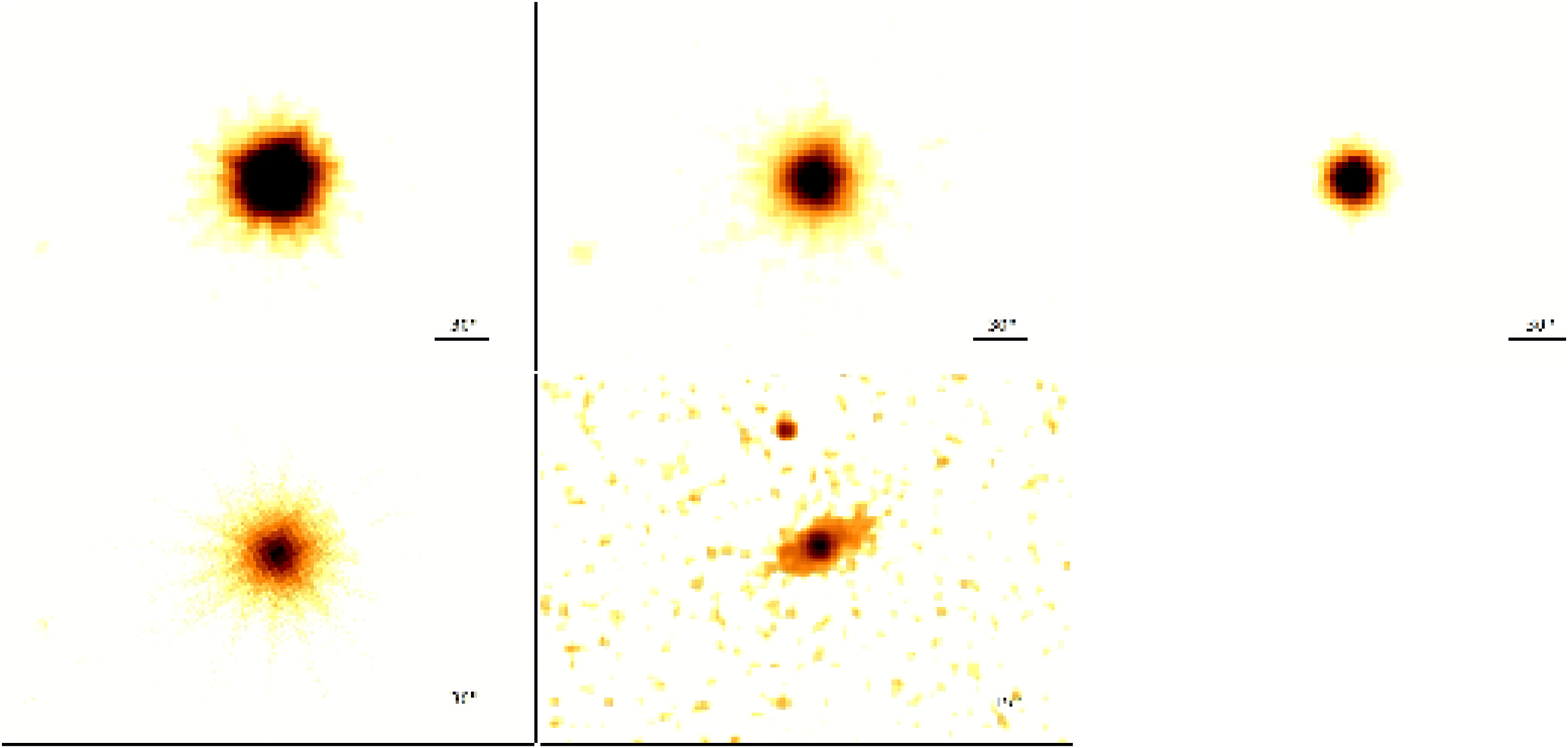} 
\end{center}
 \caption{Images of ESO\,113-G10. (Top left): Smoothed X-ray 0.6-0.9 keV
   energy band; (top center): smoothed X-ray 1.6-2.0 keV energy band;
   (top right): smoothed X-ray 4.5-8.0 keV energy band; (bottom left):
   X-ray 0.5-10.0 keV energy band without smoothing; (bottom center):
   2MASS image in the $K_s$ band.}
\end{figure*}

\onecolumn
\begin{figure*}
\begin{center}
\includegraphics[width=\textwidth]{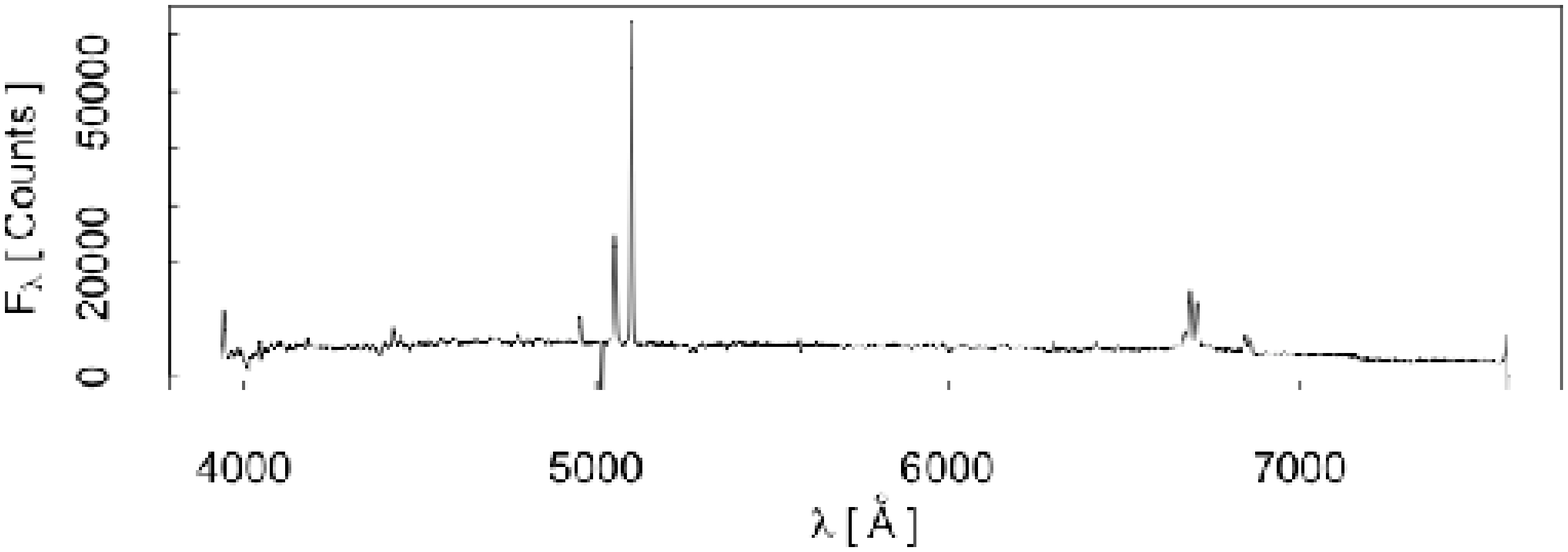}

\includegraphics[width=\textwidth]{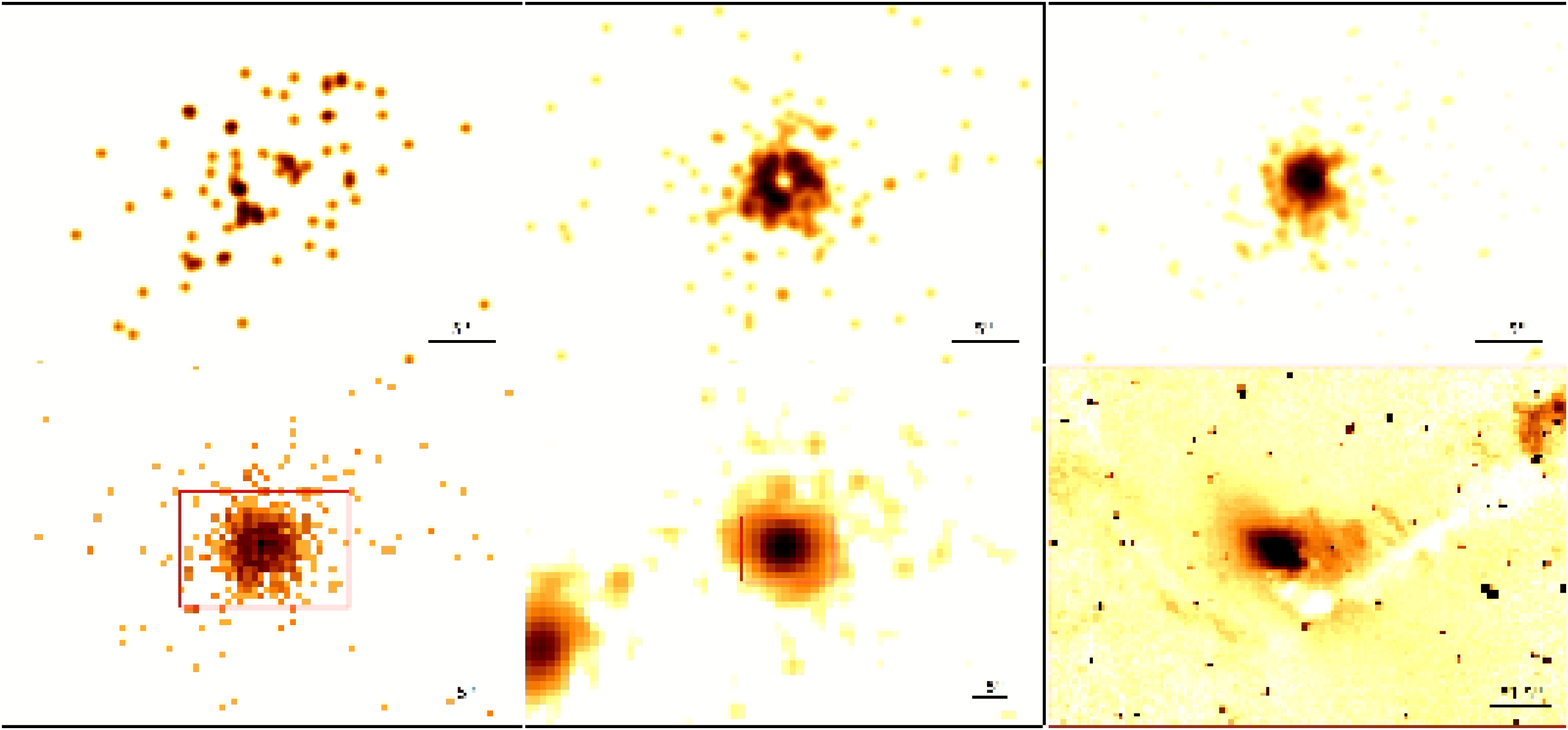} 
\end{center}
 \caption{Up: Optical spectrum (from NED); bottom: images of
   NGC\,526A. (Top left): Smoothed X-ray 0.6-0.9 keV energy band; (top
   center): smoothed X-ray 1.6-2.0 keV energy band; (top right):
   smoothed X-ray 4.5-8.0 keV energy band; (bottom left): X-ray
   0.5-10.0 keV energy band without smoothing; (bottom center): 2MASS
   image in the $K_s$ band; (bottom right): Sharp divided Hubble image in the F606W
   filter.}
\end{figure*}

\onecolumn
\begin{figure*}
\begin{center}
\includegraphics[width=\textwidth]{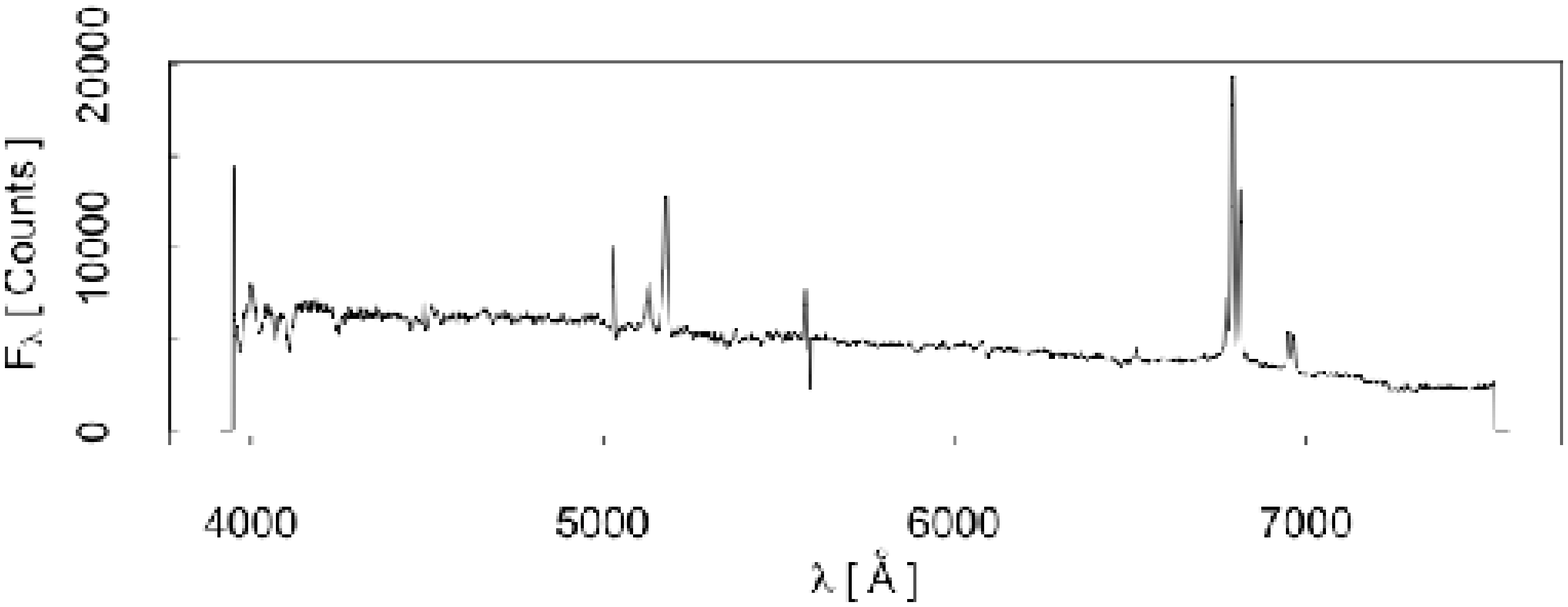}

\includegraphics[width=\textwidth]{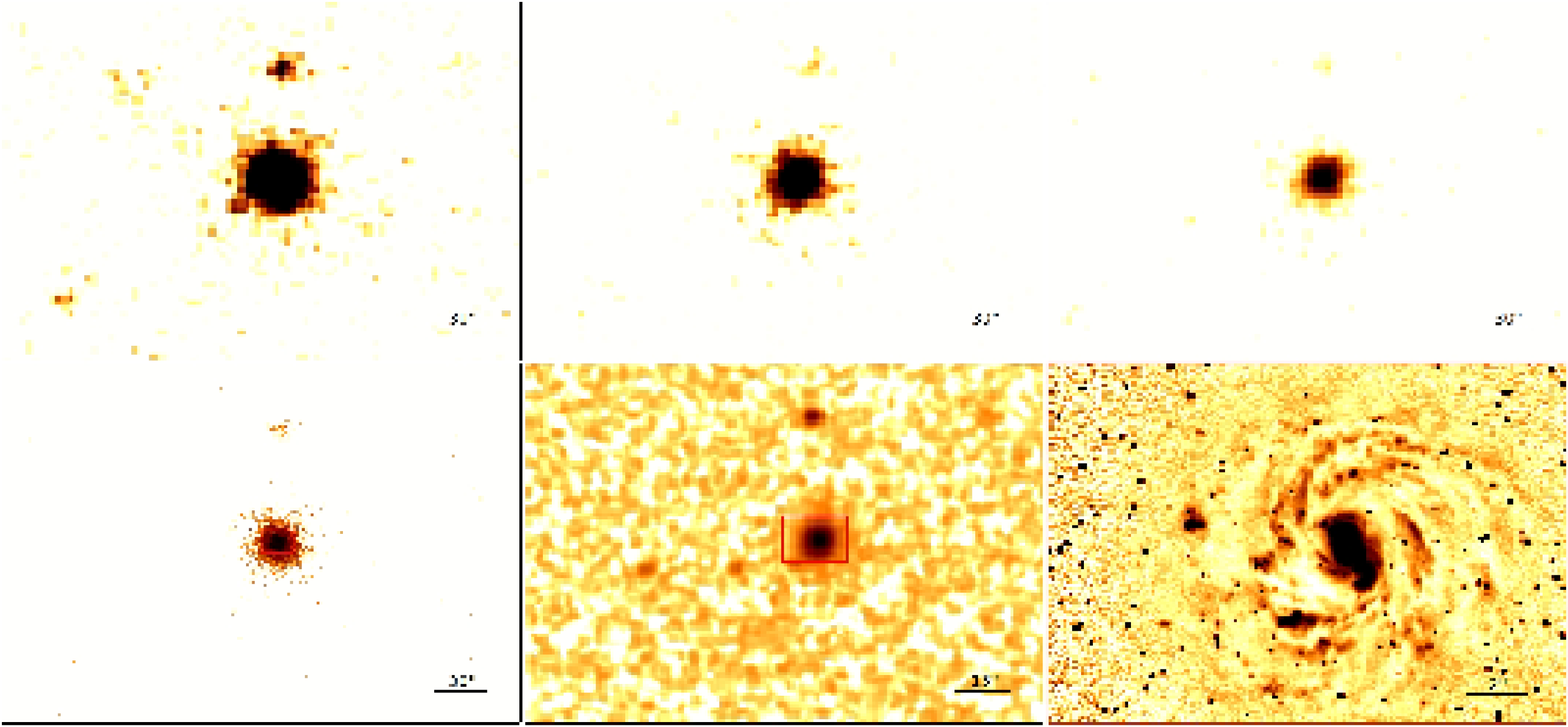} 
\end{center}
 \caption{Up: Optical spectrum (from NED); bottom: images of
   MARK\,609. (Top left): Smoothed X-ray 0.6-0.9 keV energy band; (top
   center): smoothed X-ray 1.6-2.0 keV energy band; (top right):
   smoothed X-ray 4.5-8.0 keV energy band; (bottom left): X-ray
   0.5-10.0 keV energy band without smoothing; (bottom center): 2MASS
   image in the $K_s$ band; (bottom right): Sharp divided Hubble image in the F606W
   filter.}
\end{figure*}

\onecolumn
\begin{figure*}
\begin{center}
\includegraphics[width=\textwidth]{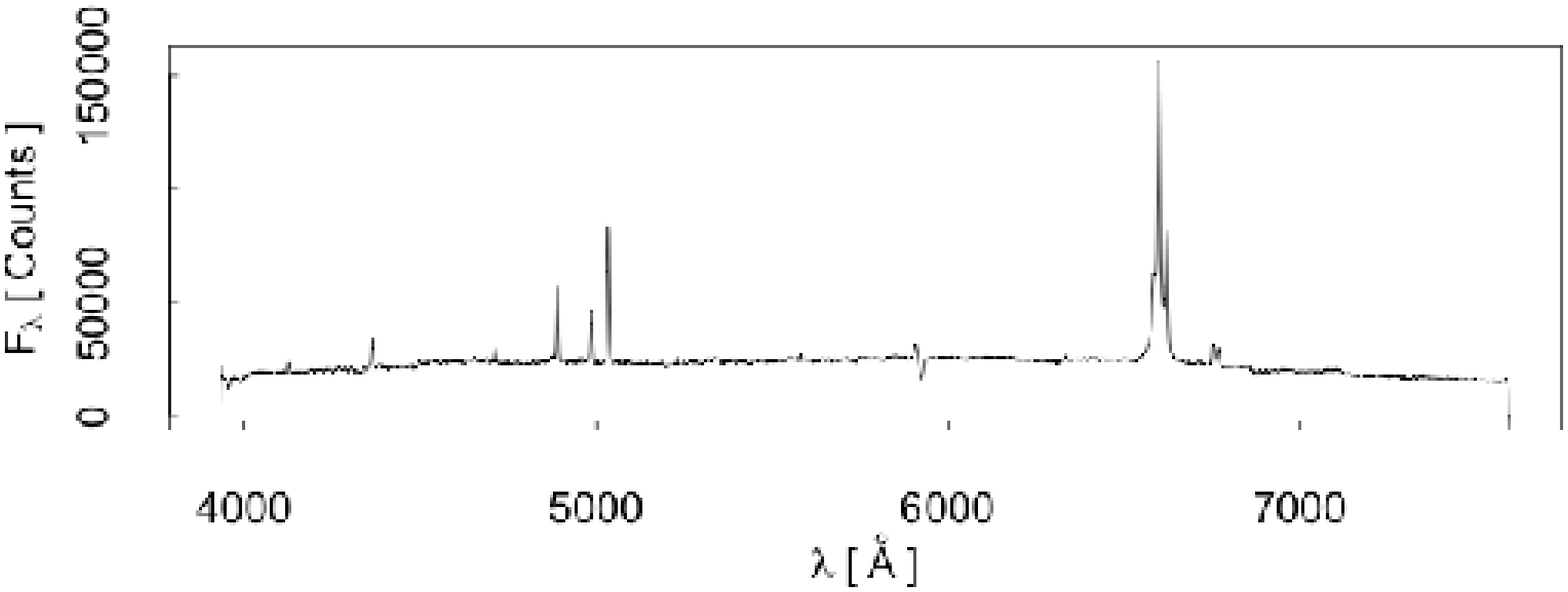}

\includegraphics[width=\textwidth]{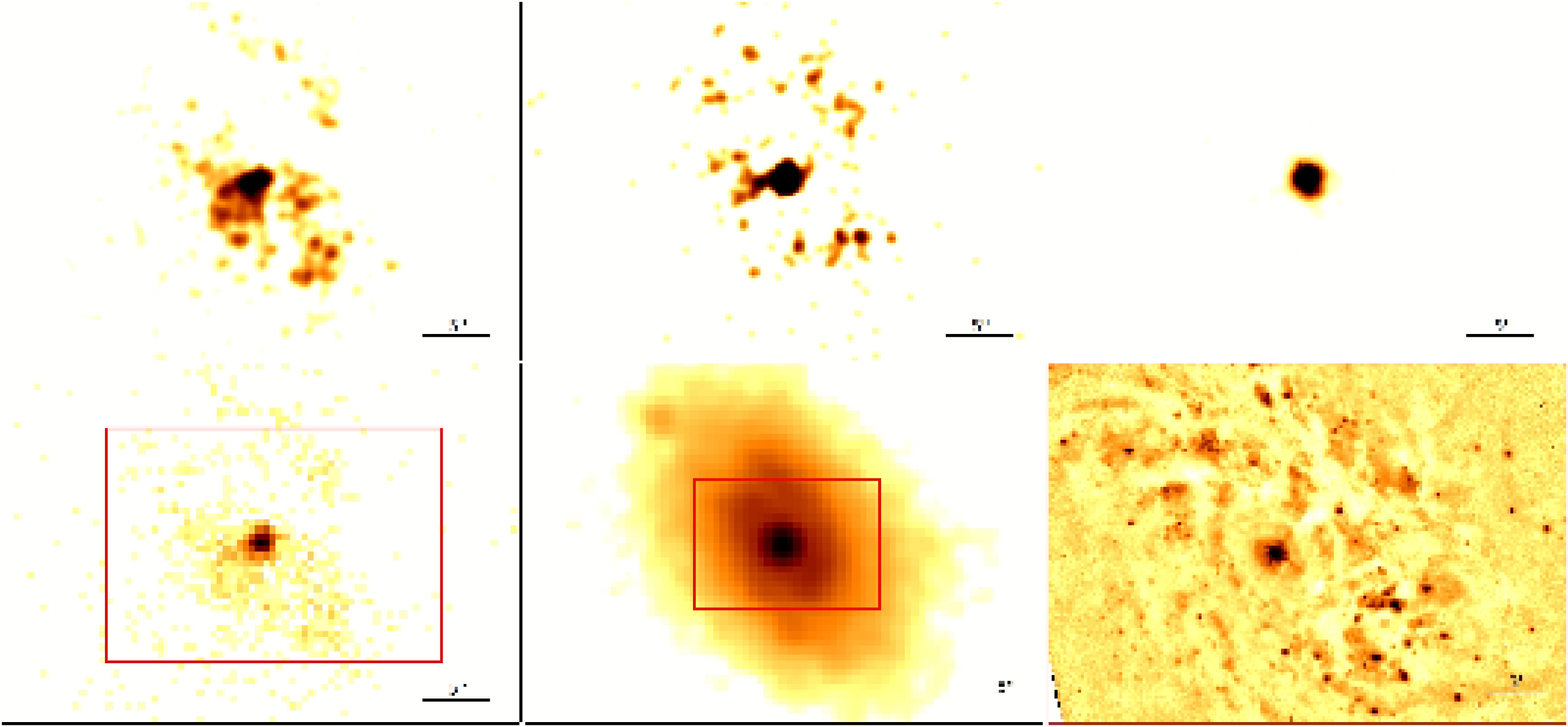} 
\end{center}
 \caption{Up: Optical spectrum (from NED); bottom: images of
   NGC\,1365. (Top left): Smoothed X-ray 0.6-0.9 keV energy band; (top
   center): smoothed X-ray 1.6-2.0 keV energy band; (top right):
   smoothed X-ray 4.5-8.0 keV energy band; (bottom left): X-ray
   0.5-10.0 keV energy band without smoothing; (bottom center): 2MASS
   image in the $K_s$ band; (bottom right): Sharp divided Hubble image in the F814W
   filter.}
\end{figure*}

\onecolumn
\begin{figure*}
\begin{center}
\includegraphics[width=\textwidth]{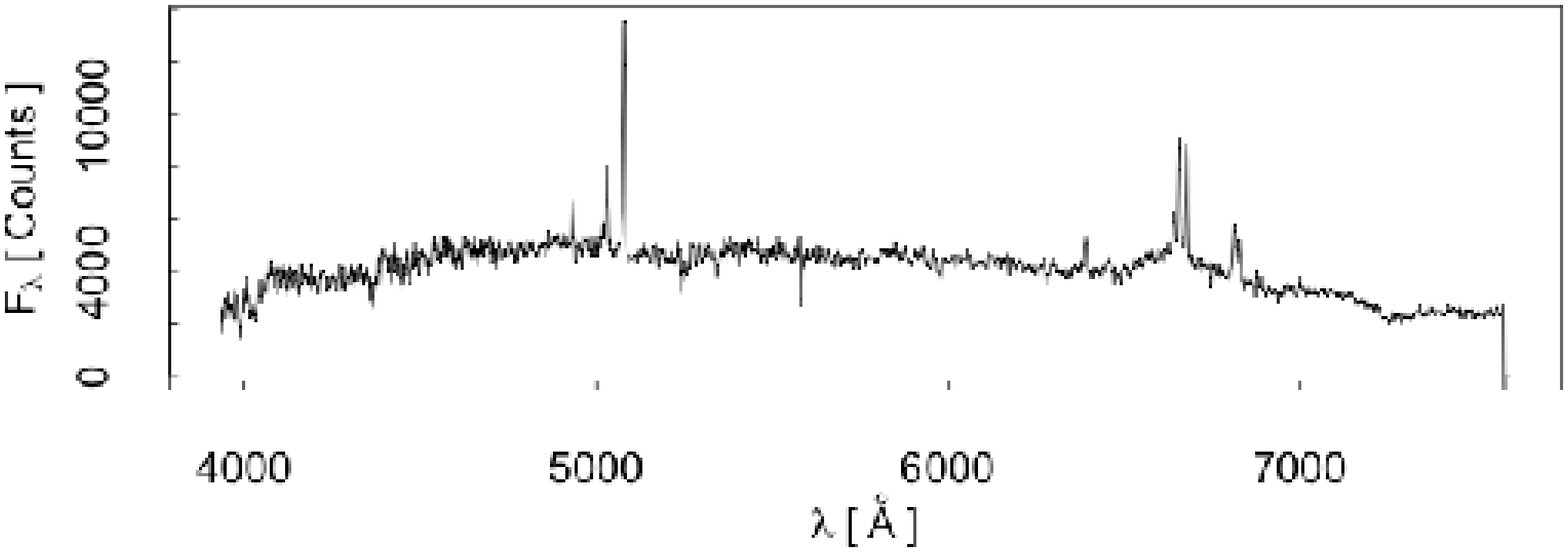}

\includegraphics[width=\textwidth]{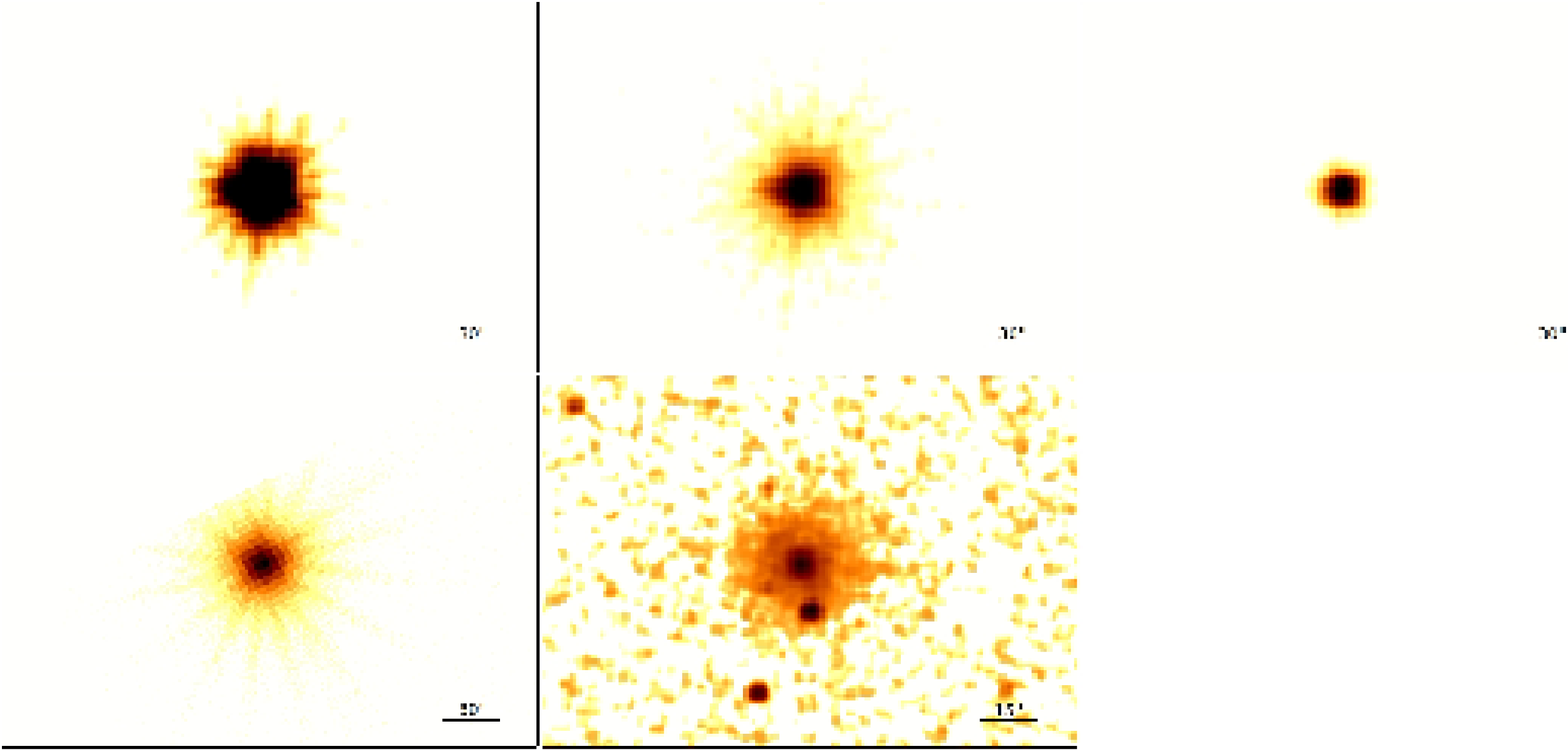} 
\end{center}
 \caption{Up: Optical spectrum (from NED); bottom: images of
   NGC\,2617. (Top left): Smoothed X-ray 0.6-0.9 keV energy band; (top
   center): smoothed X-ray 1.6-2.0 keV energy band; (top right):
   smoothed X-ray 4.5-8.0 keV energy band; (bottom left): X-ray
   0.5-10.0 keV energy band without smoothing; (bottom center): 2MASS
   image in the $K_s$ band.}
\end{figure*}

\onecolumn
\begin{figure*}
\begin{center}
\includegraphics[width=\textwidth]{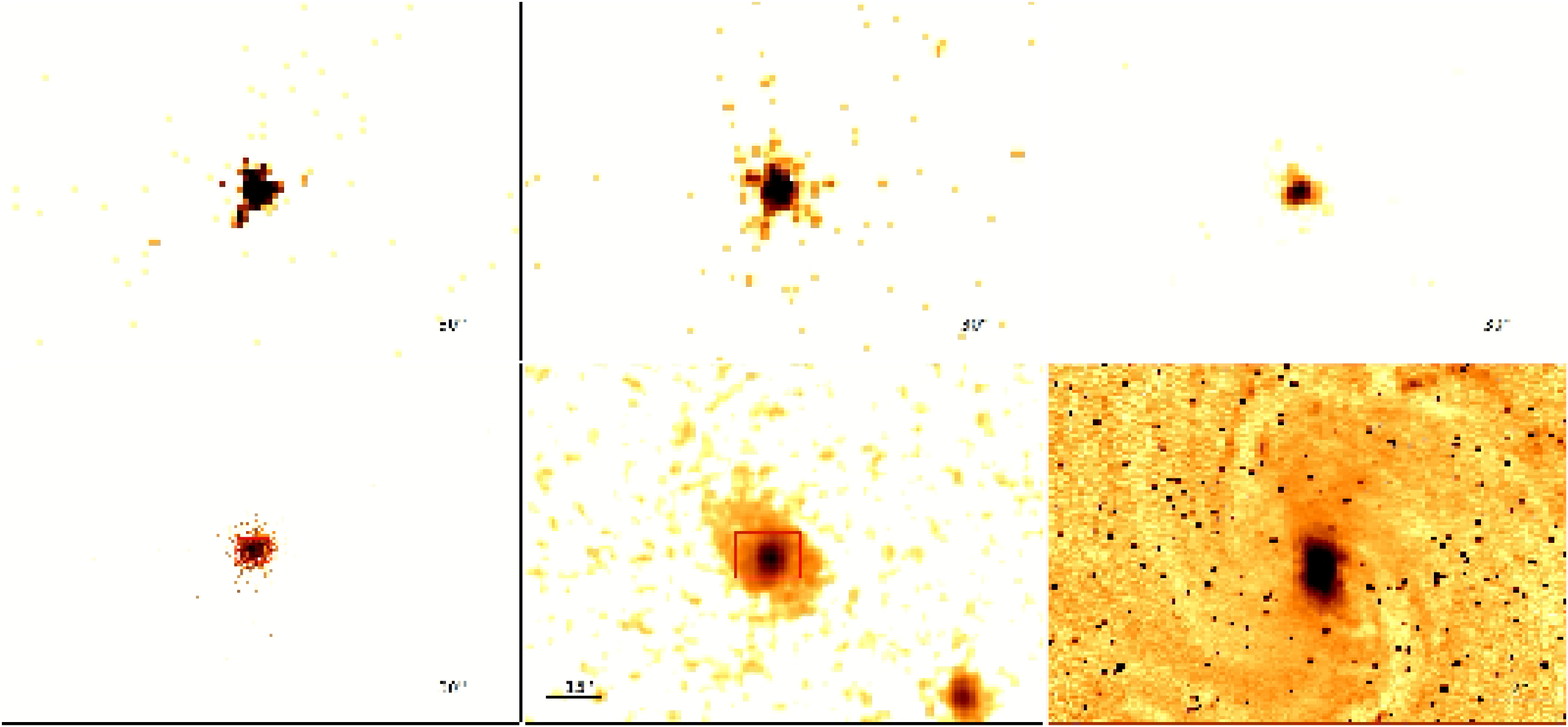} 
\end{center}
 \caption{Images of
   MARK\,1218. (Top left): Smoothed X-ray 0.6-0.9 keV energy band; (top
   center): smoothed X-ray 1.6-2.0 keV energy band; (top right):
   smoothed X-ray 4.5-8.0 keV energy band; (bottom left): X-ray
   0.5-10.0 keV energy band without smoothing; (bottom center): 2MASS
   image in the $K_s$ band; (bottom right): Sharp divided Hubble image in the F606W
   filter.}
\end{figure*}

\onecolumn
\begin{figure*}
\begin{center}
\includegraphics[width=\textwidth]{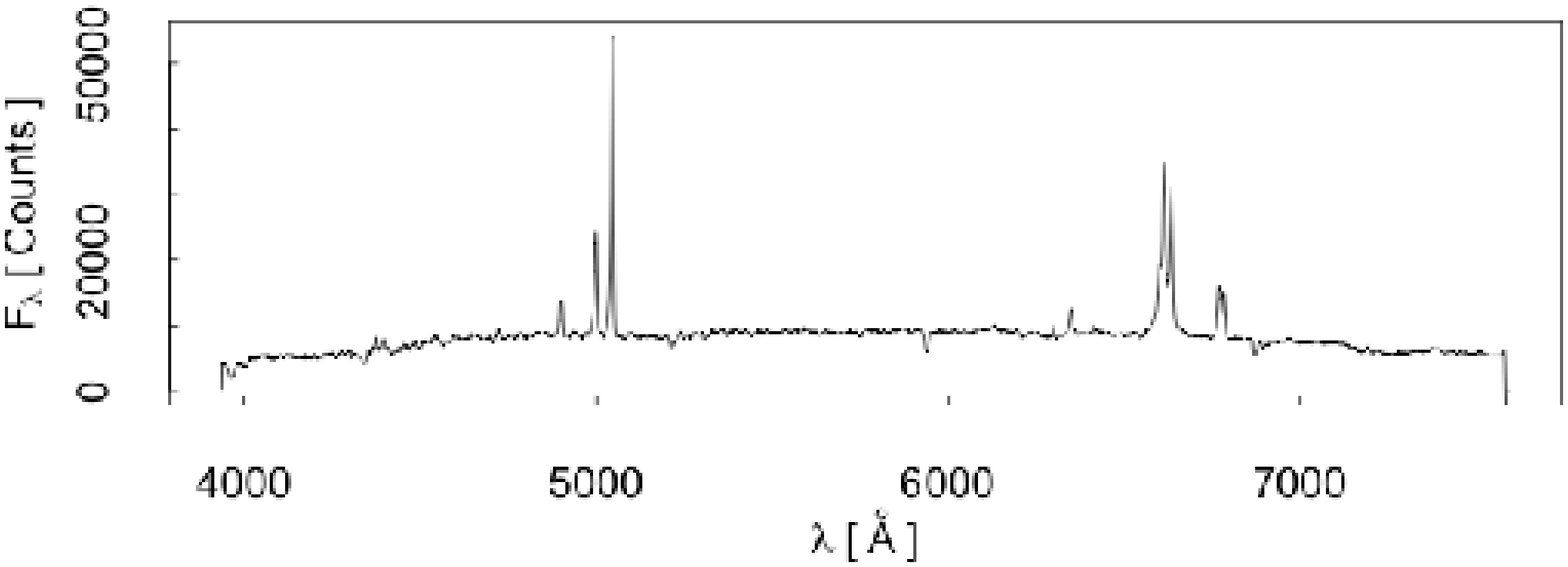}

\includegraphics[width=\textwidth]{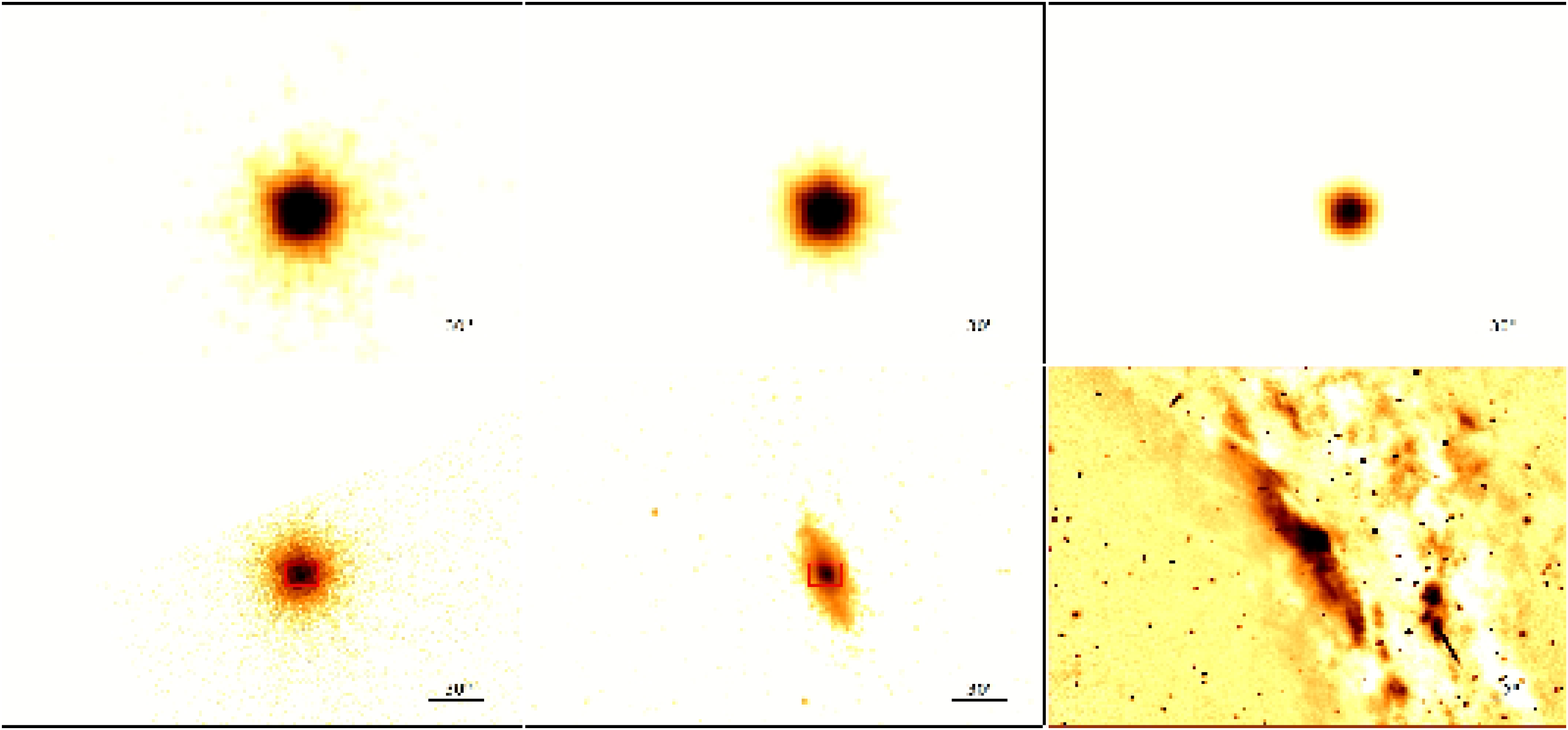} 
\end{center}
 \caption{Up: Optical spectrum (from NED); bottom: images of
   NGC\,2992. (Top left): Smoothed X-ray 0.6-0.9 keV energy band; (top
   center): smoothed X-ray 1.6-2.0 keV energy band; (top right):
   smoothed X-ray 4.5-8.0 keV energy band; (bottom left): X-ray
   0.5-10.0 keV energy band without smoothing; (bottom center): 2MASS
   image in the $K_s$ band; (bottom right): Sharp divided Hubble image in the F606W
   filter.}
\end{figure*}

\begin{figure*}
\begin{center}
\includegraphics[width=\textwidth]{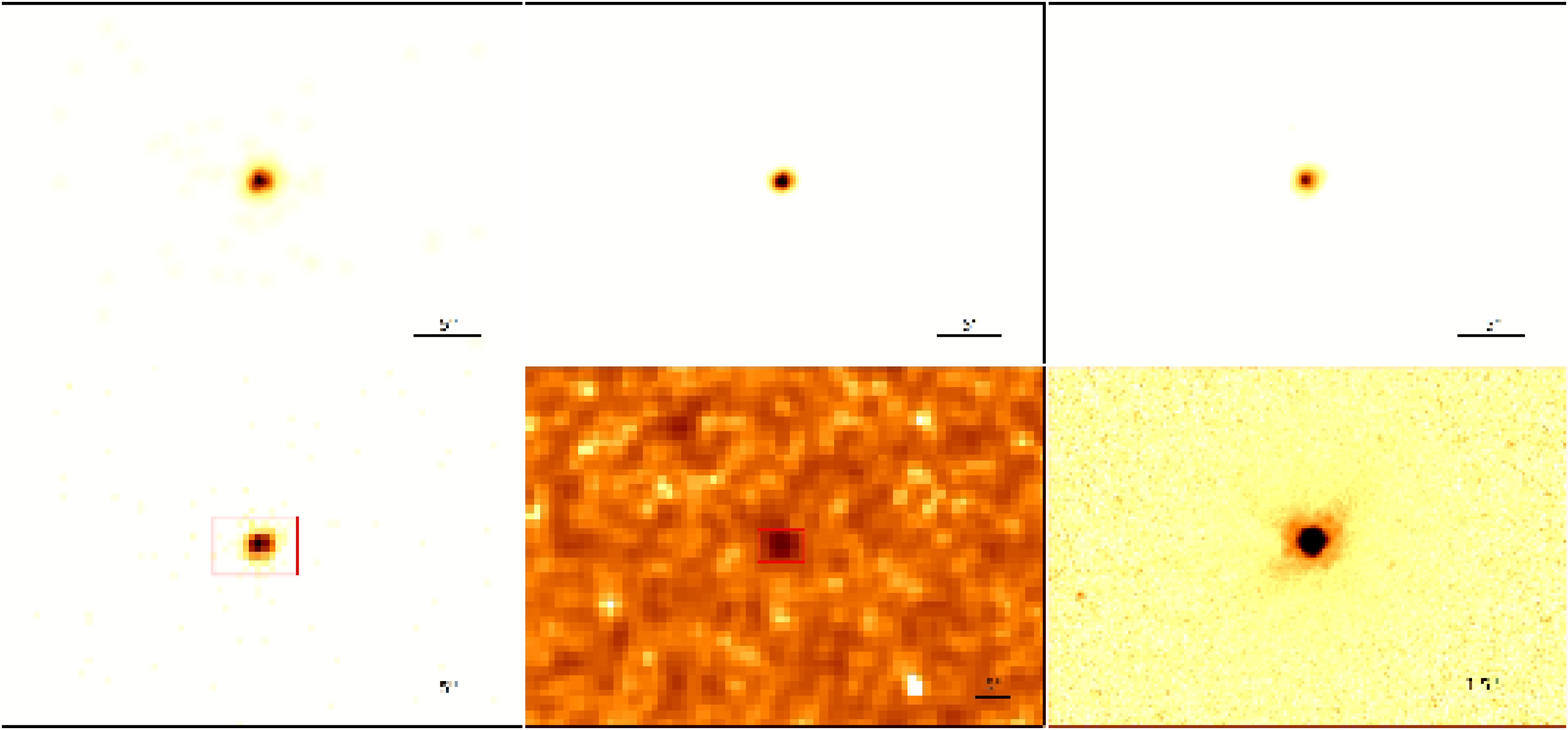} 
\end{center}
 \caption{images of
   POX\,52. (Top left): Smoothed X-ray 0.6-0.9 keV energy band; (top
   center): smoothed X-ray 1.6-2.0 keV energy band; (top right):
   smoothed X-ray 4.5-8.0 keV energy band; (bottom left): X-ray
   0.5-10.0 keV energy band without smoothing; (bottom center): 2MASS
   image in the $K_s$ band; (bottom right): Sharp divided Hubble image in the F814W
   filter.}
\end{figure*}

\onecolumn
\begin{figure*}
\begin{center}
\includegraphics[width=\textwidth]{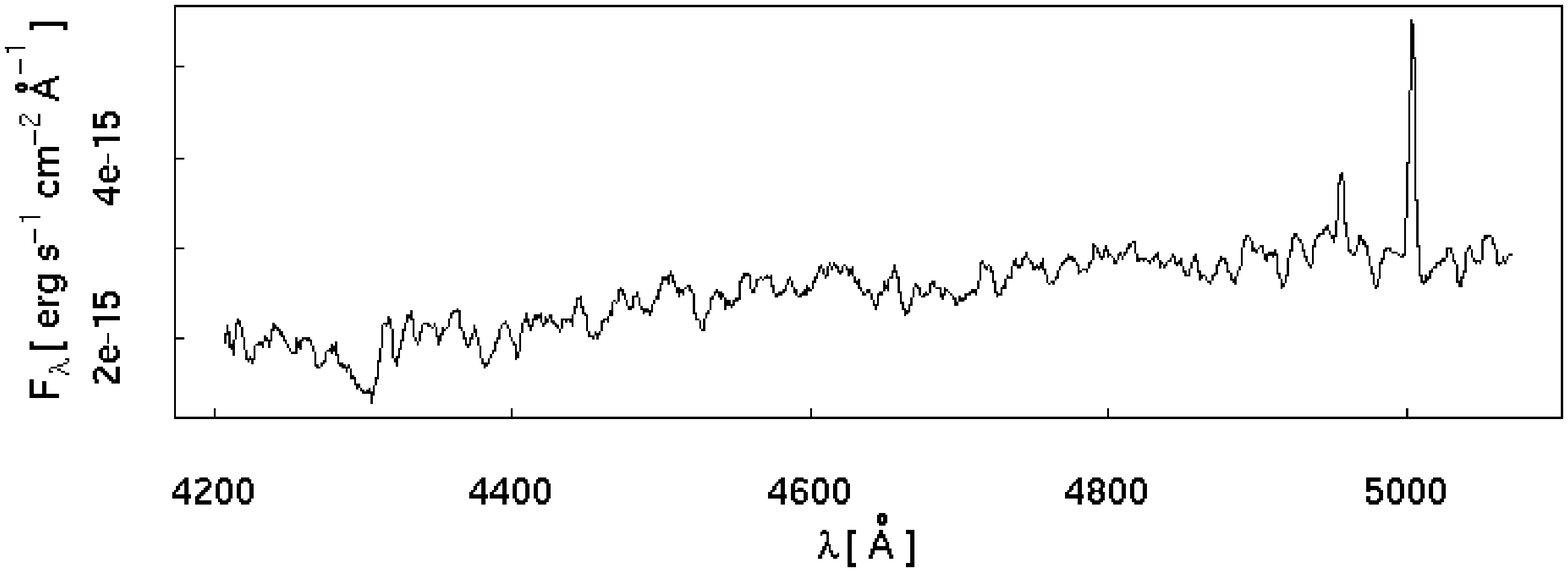} \vspace*{-2.1cm}

\includegraphics[width=\textwidth]{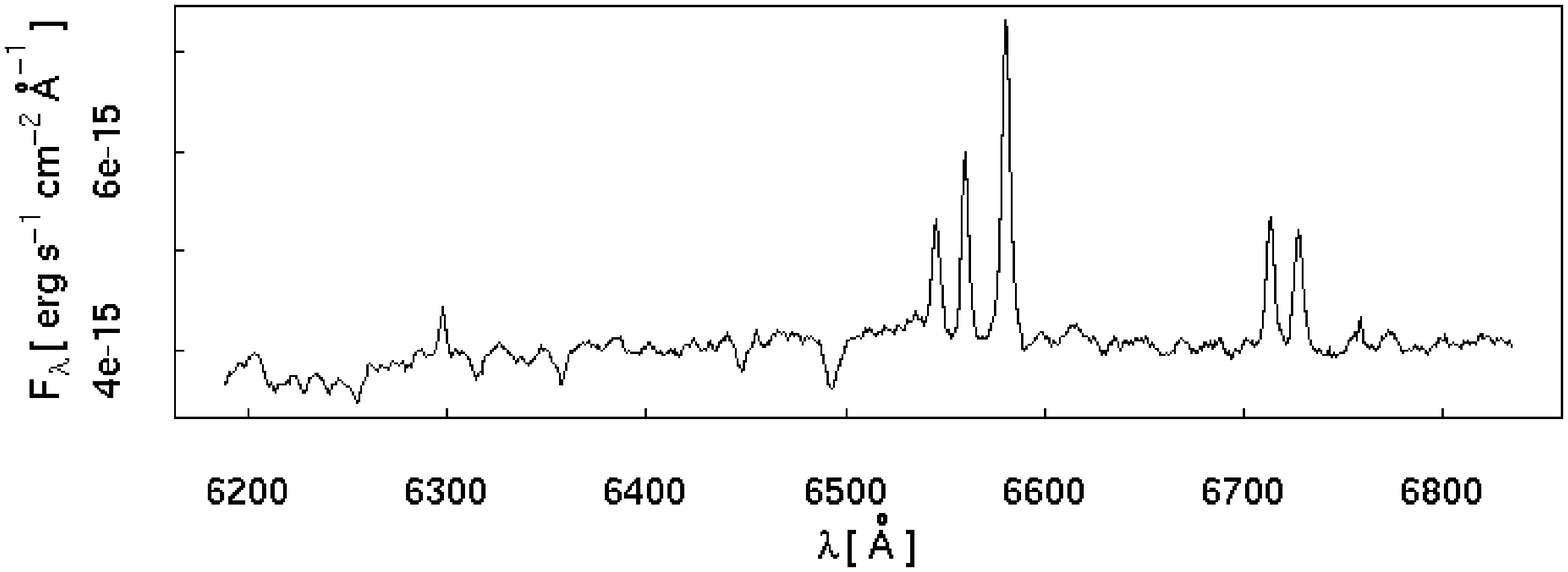}

\includegraphics[width=\textwidth]{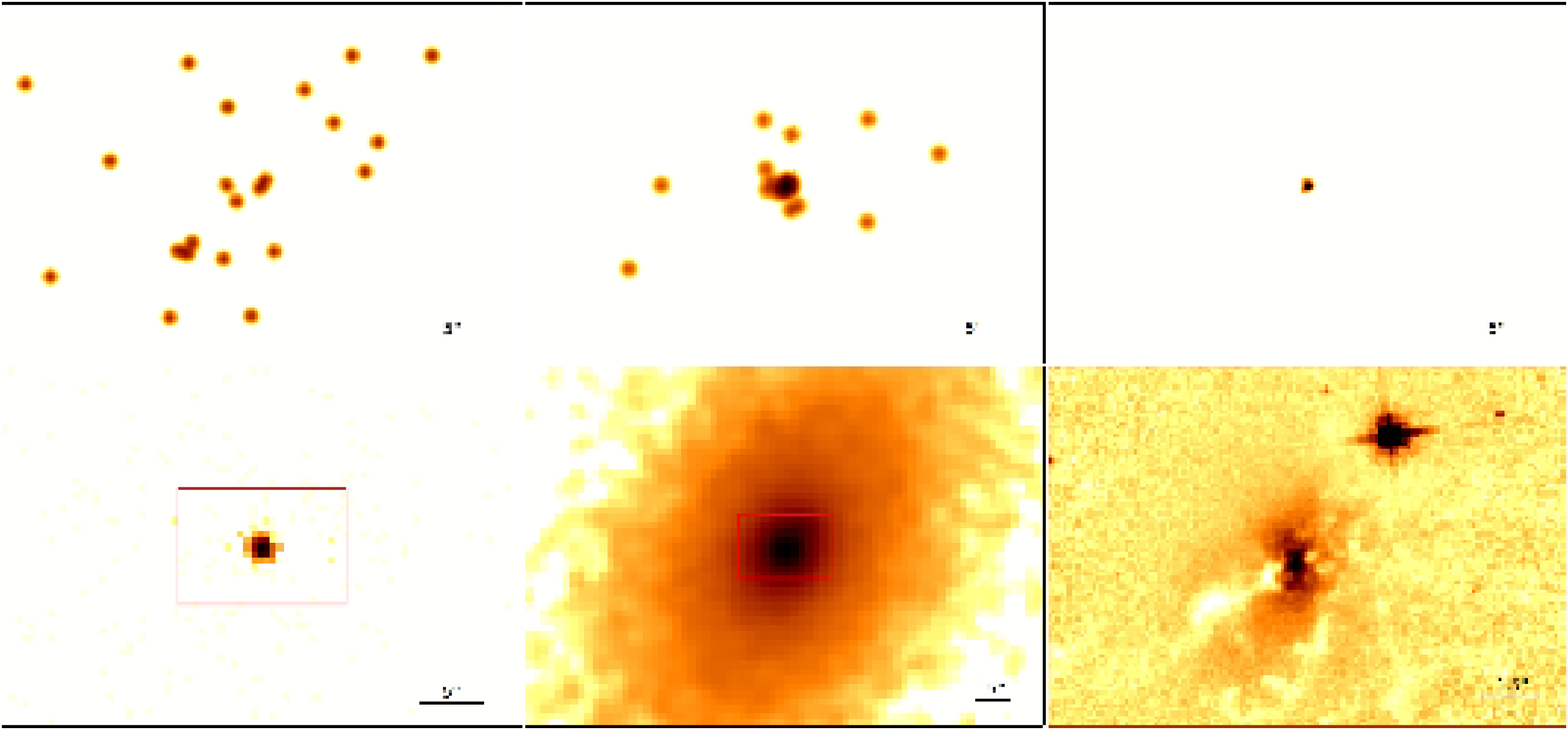} 
\end{center}
 \caption{Up: Optical spectra (from NED); bottom: images of
   NGC\,4138. (Top left): Smoothed X-ray 0.6-0.9 keV energy band; (top
   center): smoothed X-ray 1.6-2.0 keV energy band; (top right):
   smoothed X-ray 4.5-8.0 keV energy band; (bottom left): X-ray
   0.5-10.0 keV energy band without smoothing; (bottom center): 2MASS
   image in the $K_s$ band; (bottom right): Sharp divided Hubble image in the F547W
   filter.}
\end{figure*}

\onecolumn
\begin{figure*}
\begin{center}
\includegraphics[width=\textwidth]{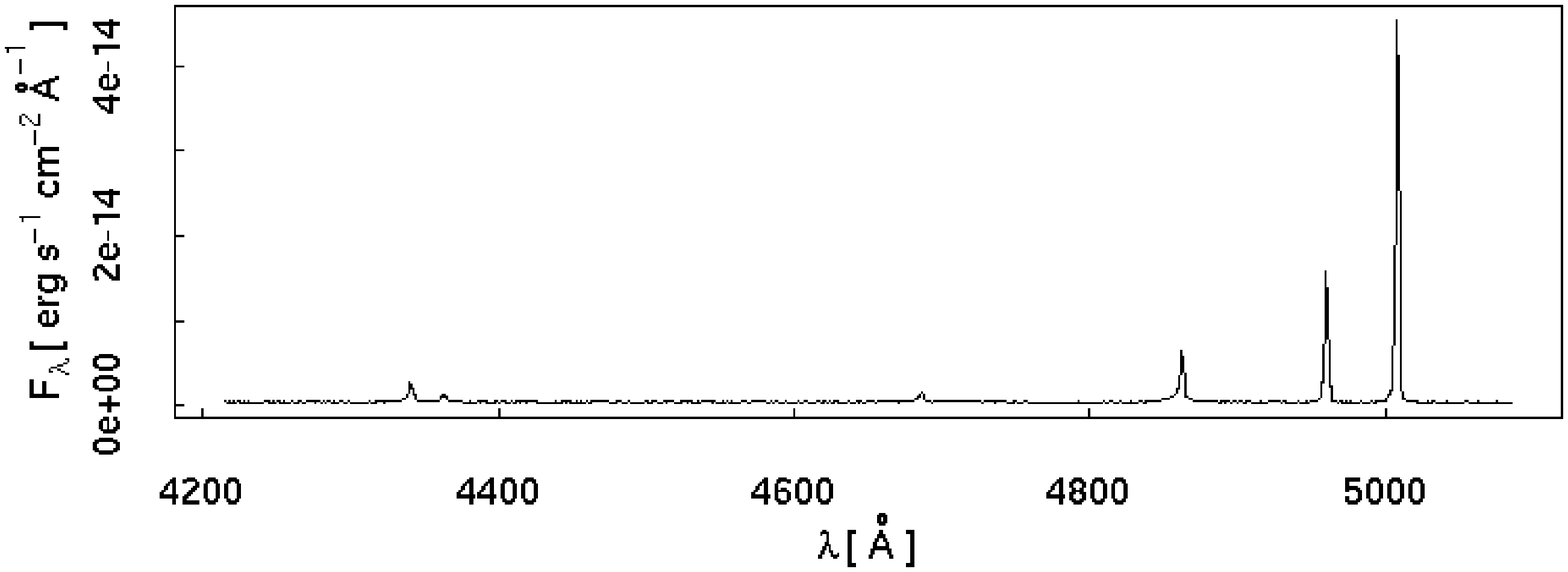} \vspace*{-2.cm}

\includegraphics[width=\textwidth]{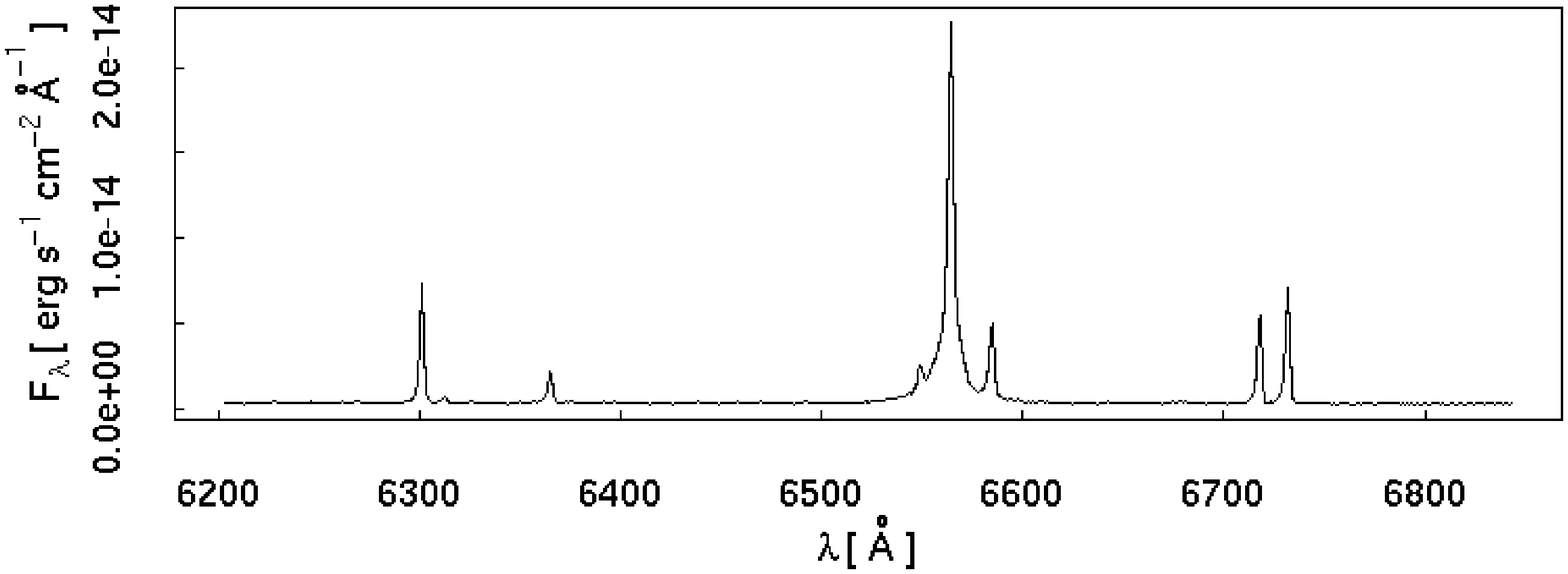}

\includegraphics[width=\textwidth]{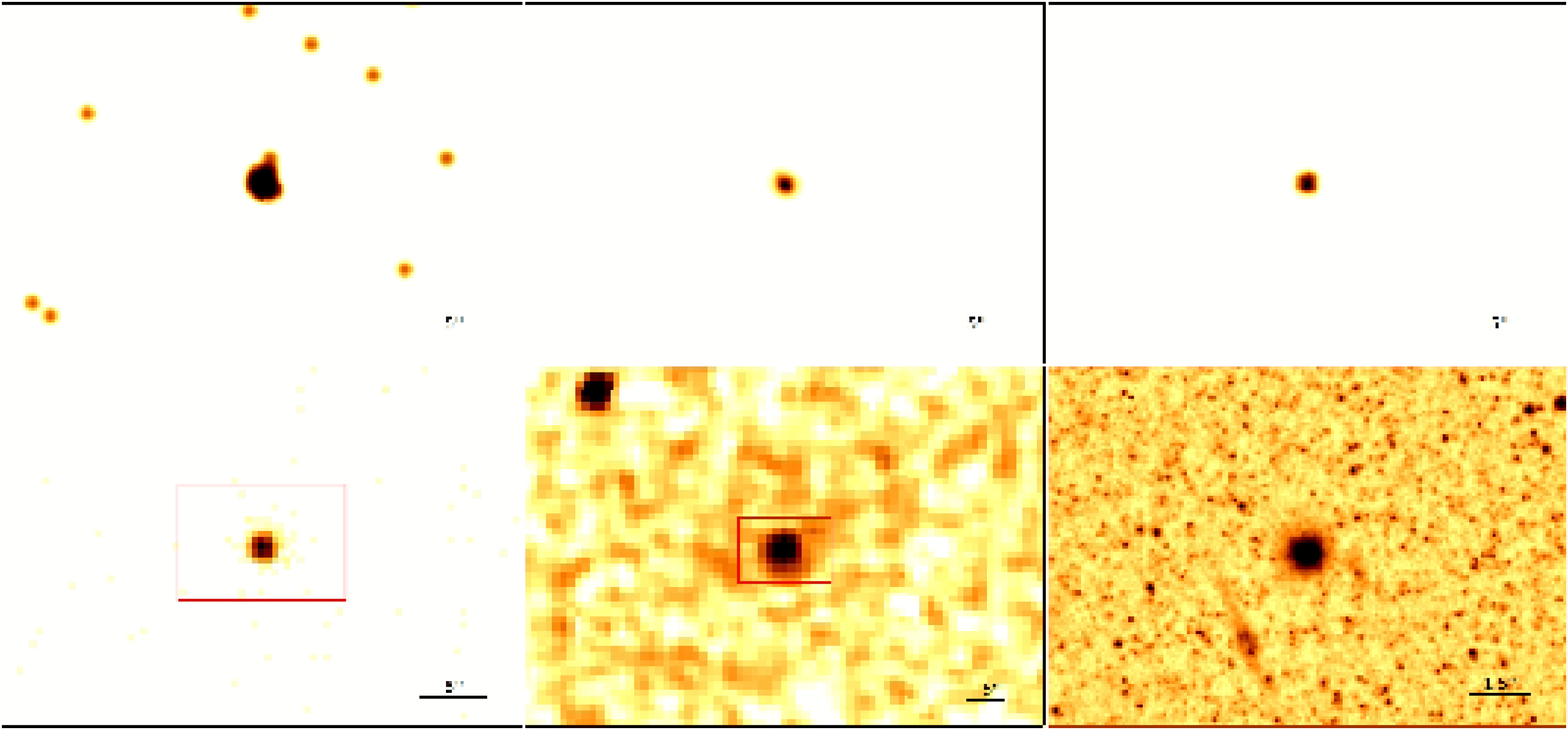} 
\end{center}
 \caption{Up: Optical spectra (from NED); bottom: images of
   NGC\,4395. (Top left): Smoothed X-ray 0.6-0.9 keV energy band; (top
   center): smoothed X-ray 1.6-2.0 keV energy band; (top right):
   smoothed X-ray 4.5-8.0 keV energy band; (bottom left): X-ray
   0.5-10.0 keV energy band without smoothing; (bottom center): 2MASS
   image in the $K_s$ band; (bottom right): Sharp divided Hubble image in the F814W
   filter.}
\end{figure*}

\onecolumn
\begin{figure*}
\begin{center}
\includegraphics[width=\textwidth]{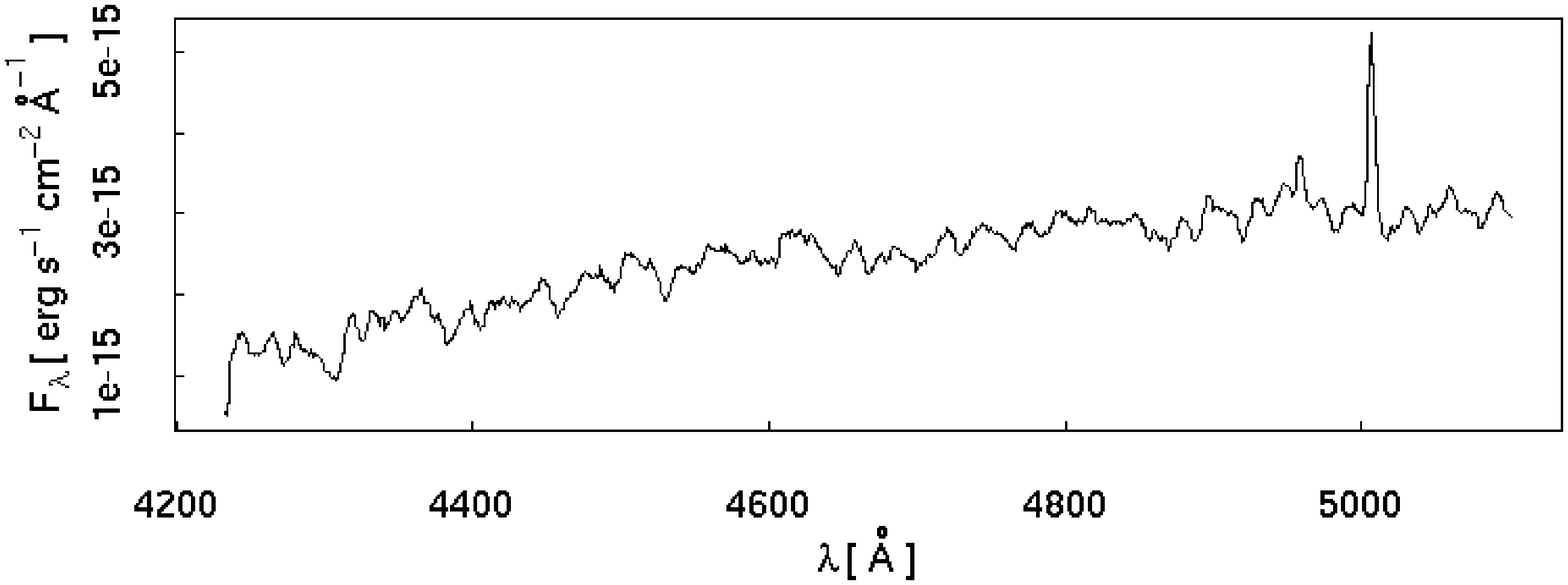} \vspace*{-2.cm}

\includegraphics[width=\textwidth]{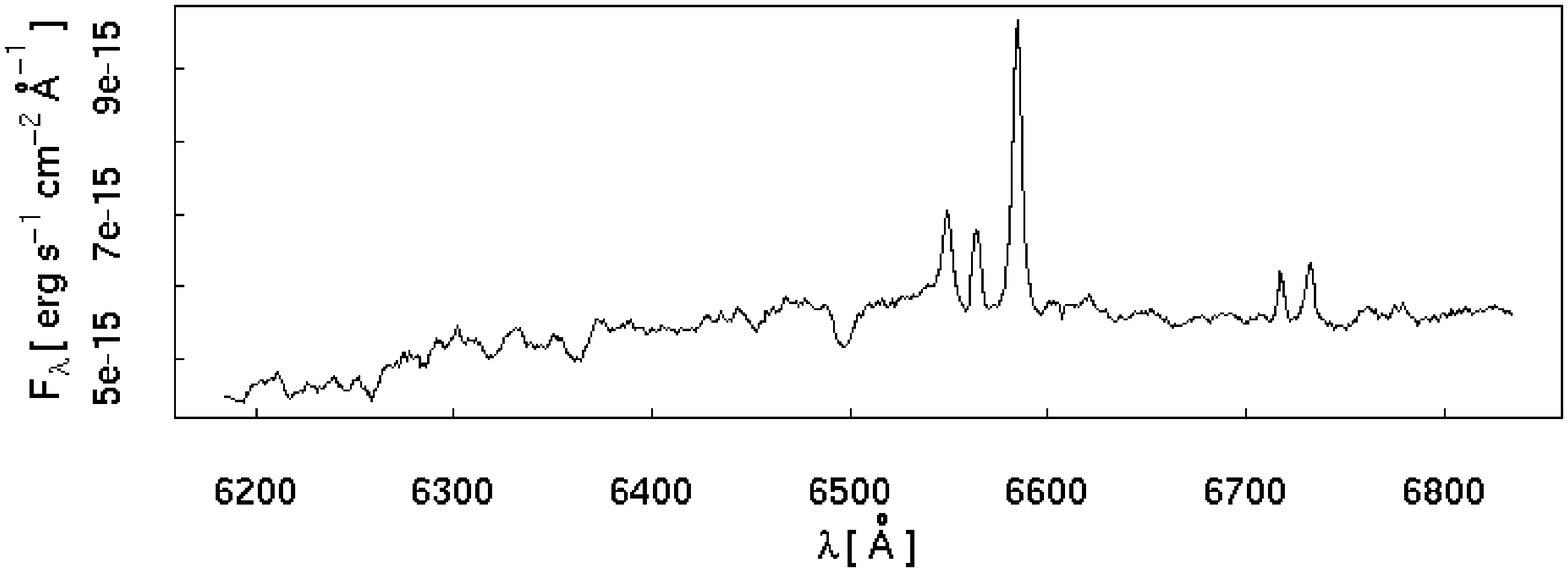}

\includegraphics[width=\textwidth]{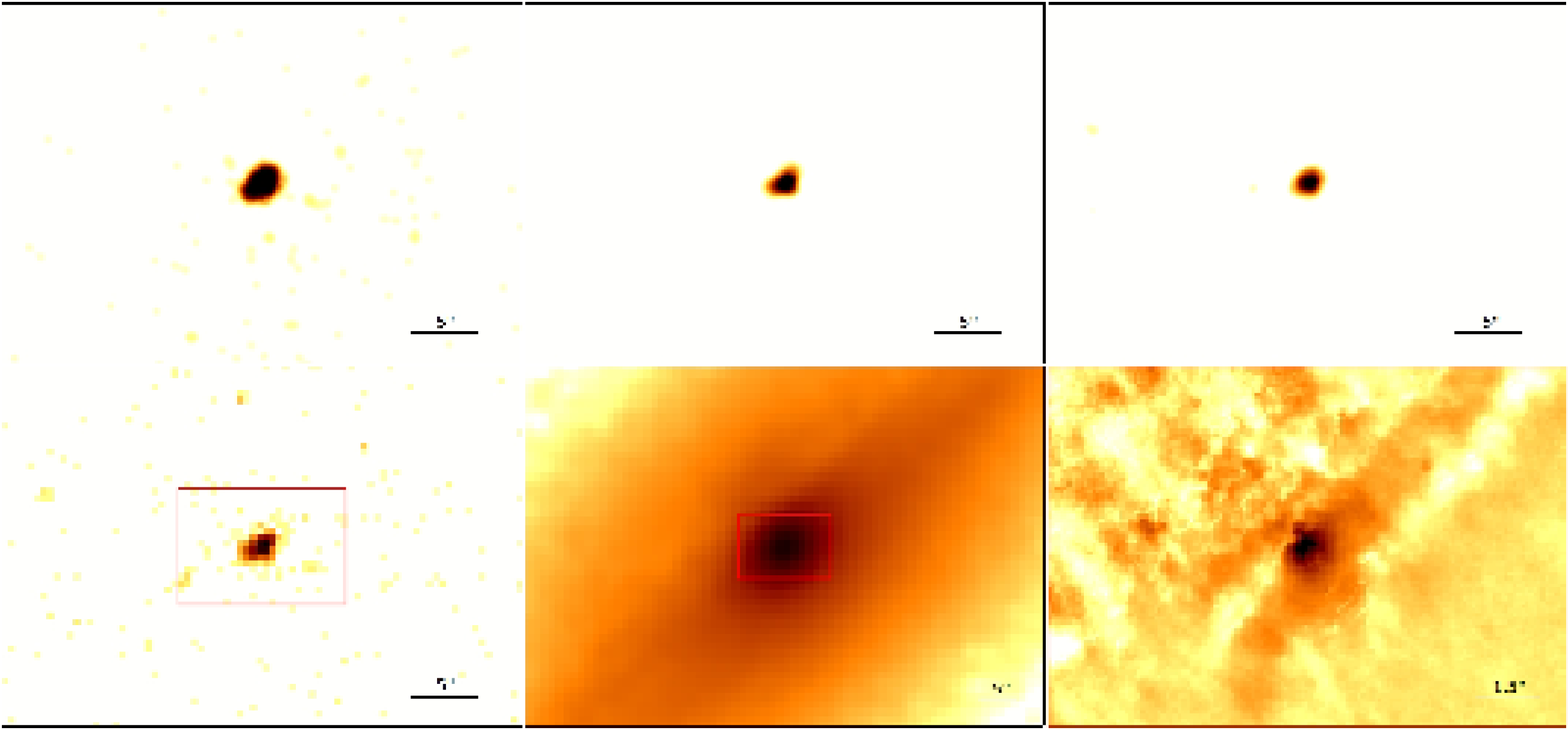} 
\end{center}
 \caption{Up: Optical spectra (from NED); bottom: images of
   NGC\,4565. (Top left): Smoothed X-ray 0.6-0.9 keV energy band; (top
   center): smoothed X-ray 1.6-2.0 keV energy band; (top right):
   smoothed X-ray 4.5-8.0 keV energy band; (bottom left): X-ray
   0.5-10.0 keV energy band without smoothing; (bottom center): 2MASS
   image in the $K_s$ band; (bottom right): Sharp divided Hubble image in the F814W
   filter.}
\end{figure*}

\begin{figure*}
\begin{center}
\includegraphics[width=\textwidth]{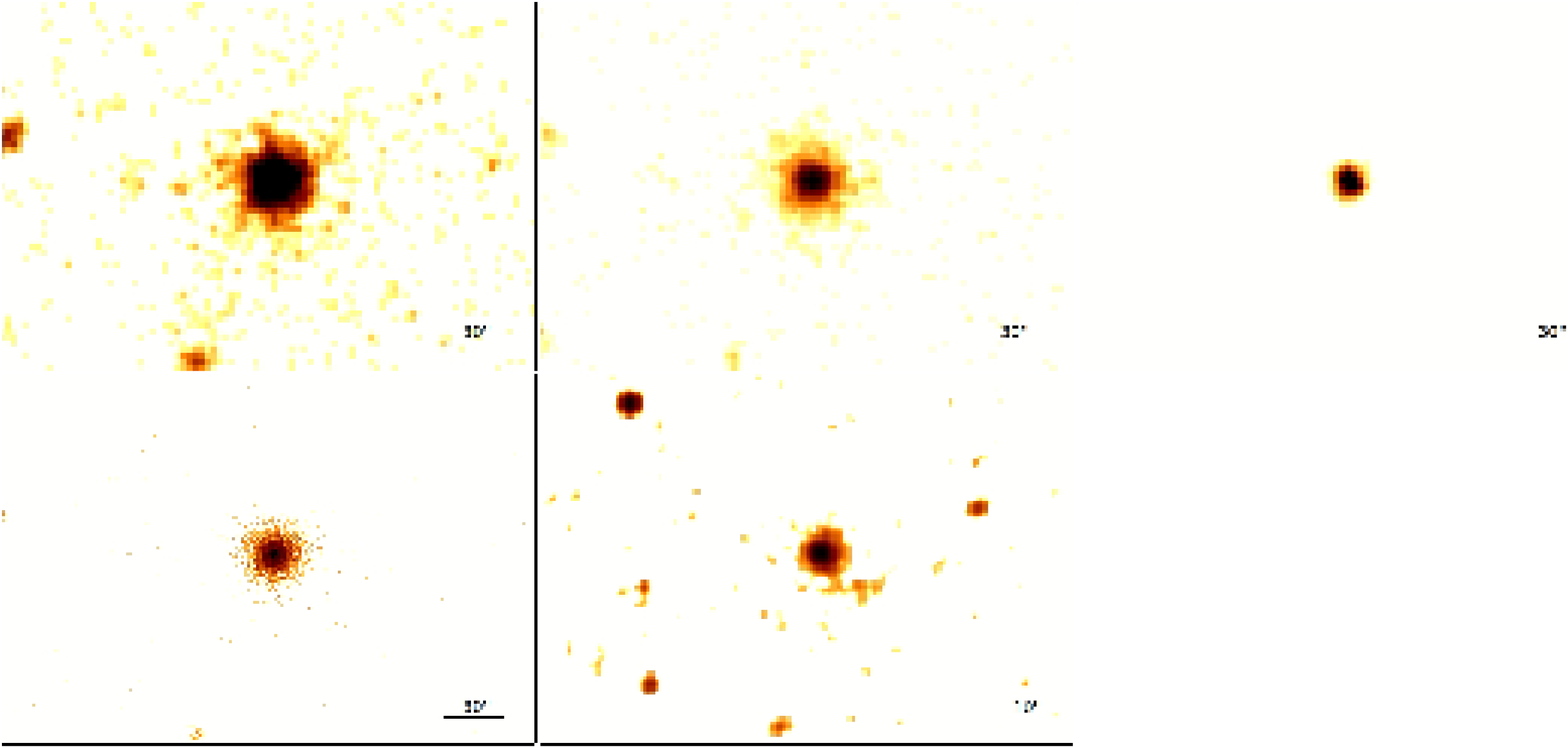} 
\end{center}
 \caption{Images of MARK\,883. (Top left): Smoothed X-ray 0.6-0.9 keV
   energy band; (top center): smoothed X-ray 1.6-2.0 keV energy band;
   (top right): smoothed X-ray 4.5-8.0 keV energy band; (bottom left):
   X-ray 0.5-10.0 keV energy band without smoothing; (bottom center):
   2MASS image in the $K_s$ band.}
\end{figure*}

\onecolumn
\begin{figure*}
\begin{center}
\includegraphics[width=\textwidth]{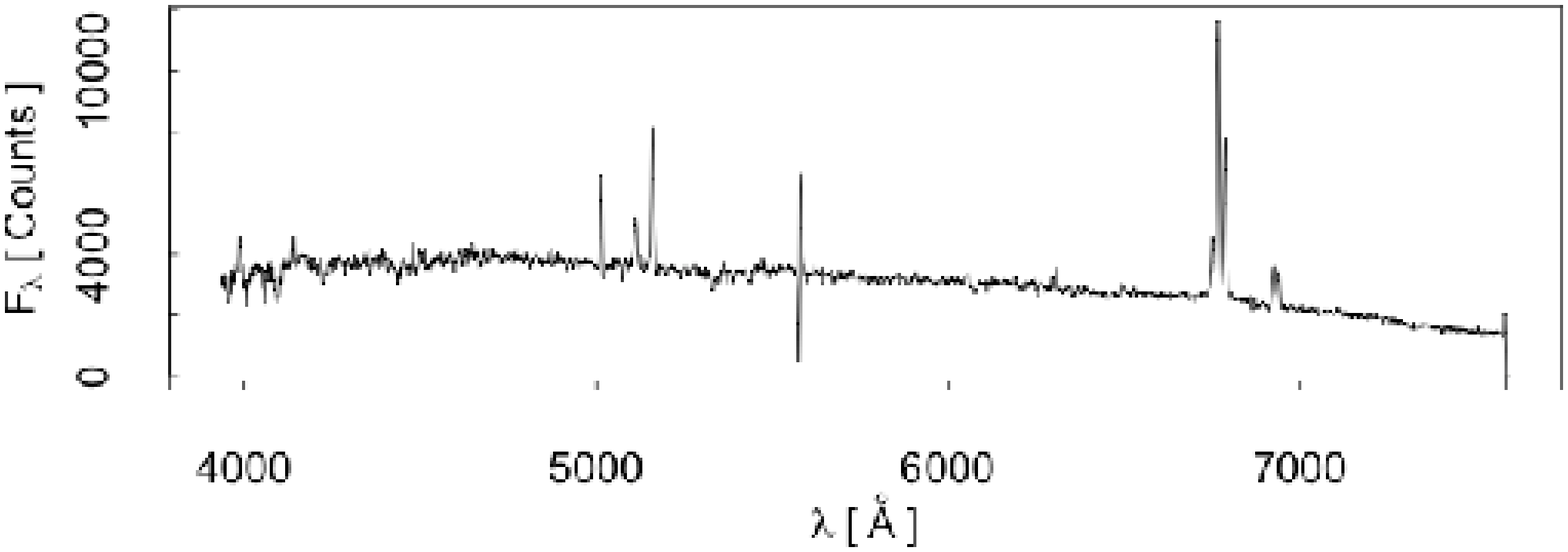}

\includegraphics[width=\textwidth]{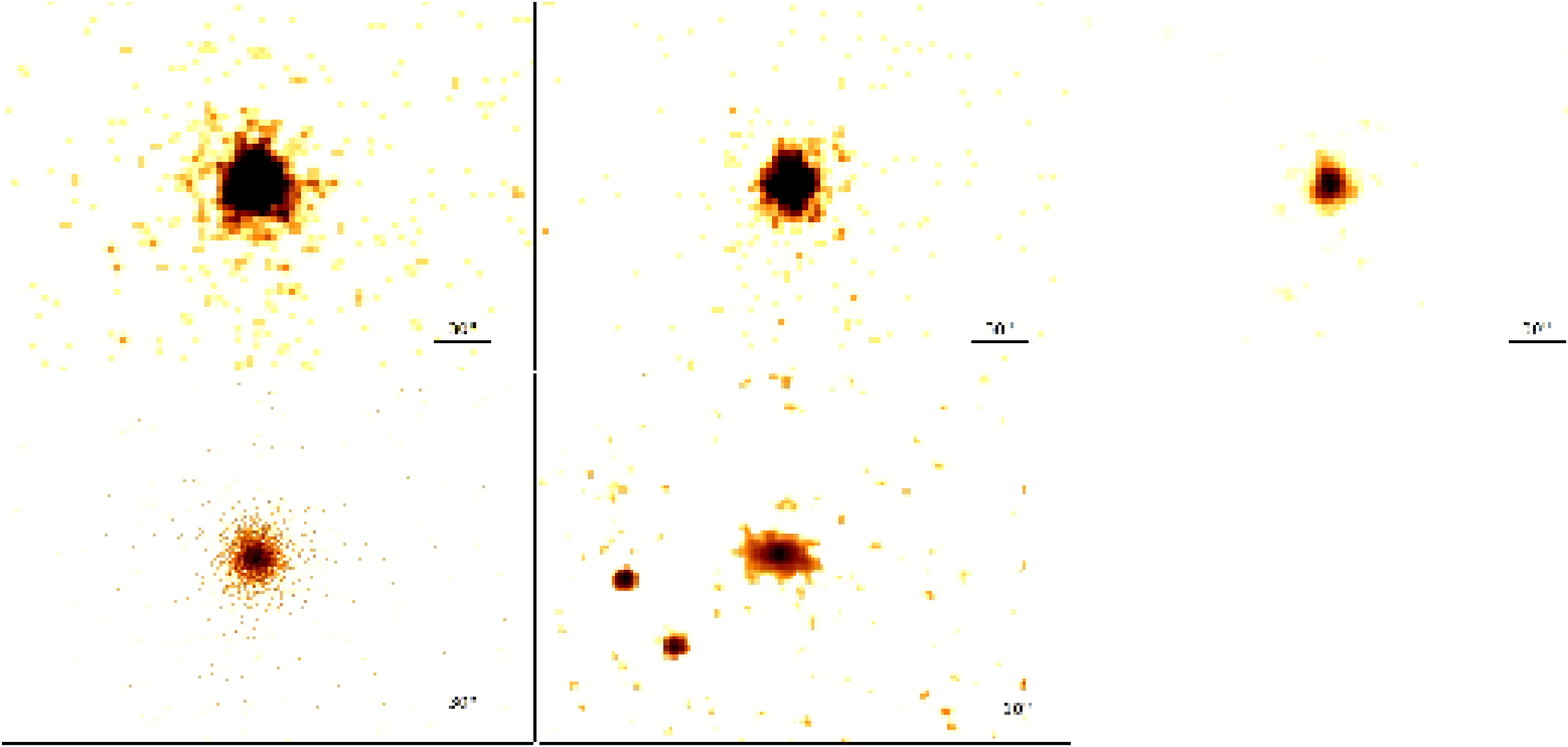} 
\end{center}
 \caption{Up: Optical spectrum (from NED); bottom: images of
   IRAS\,20051-1117. (Top left): Smoothed X-ray 0.6-0.9 keV energy band; (top
   center): smoothed X-ray 1.6-2.0 keV energy band; (top right):
   smoothed X-ray 4.5-8.0 keV energy band; (bottom left): X-ray
   0.5-10.0 keV energy band without smoothing; (bottom center): 2MASS
   image in the $K_s$ band.}
\end{figure*}

\subsection{\label{Ximages} \emph{Chandra} and  \emph{XMM}--Newton images }

In this appendix we present the images from \emph{Chandra} (left) and
\emph{XMM}-Newton (right) that were used to compare the spectra from
these two instruments in the 0.5-10 keV band. In all cases, the gray
scales extend from twice the value of the background dispersion to the
maximum value at the center of each galaxy.

\onecolumn

\begin{figure}
\caption{ \label{images} Images for \emph{Chandra} data (left) and
  \emph{XMM--Newton} data (right) for the sources in the 0.5-10 keV
  band. Big circles represent \emph{XMM-Newton} data apertures. Small
  circles in the figures to the left represent the nuclear extraction
  aperture used with \emph{Chandra} observations (see Table
  \ref{obsSey}). } \centering
\includegraphics[width=\textwidth]{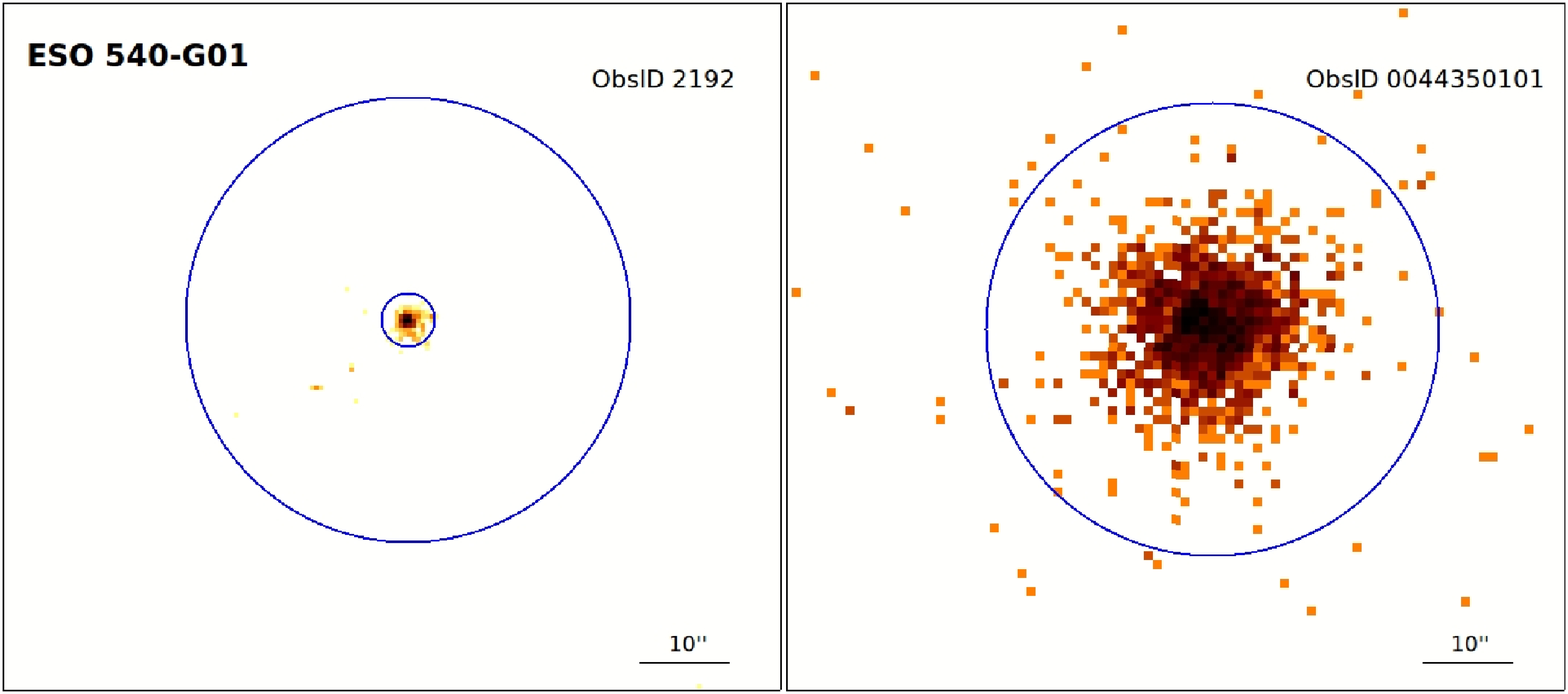}

\includegraphics[width=\textwidth]{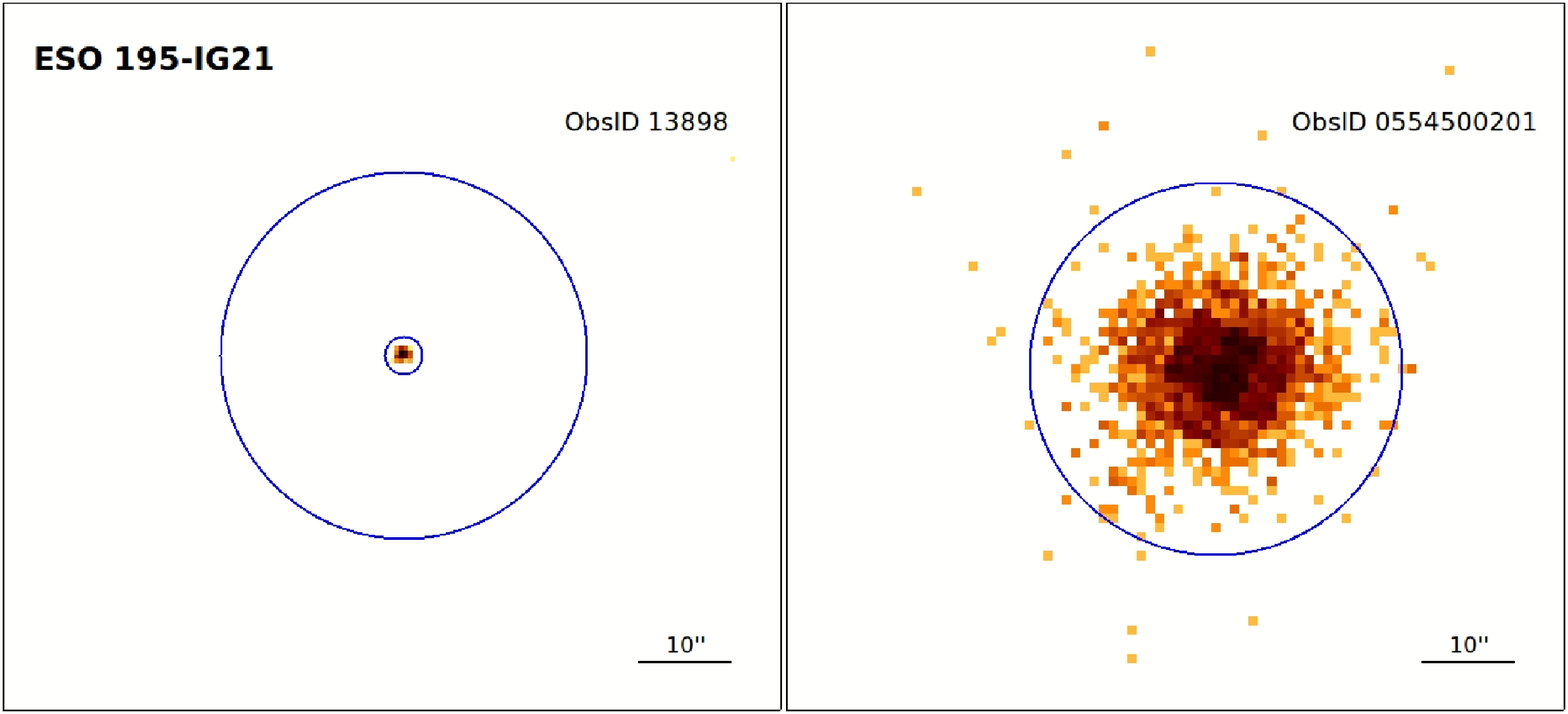}

\includegraphics[width=\textwidth]{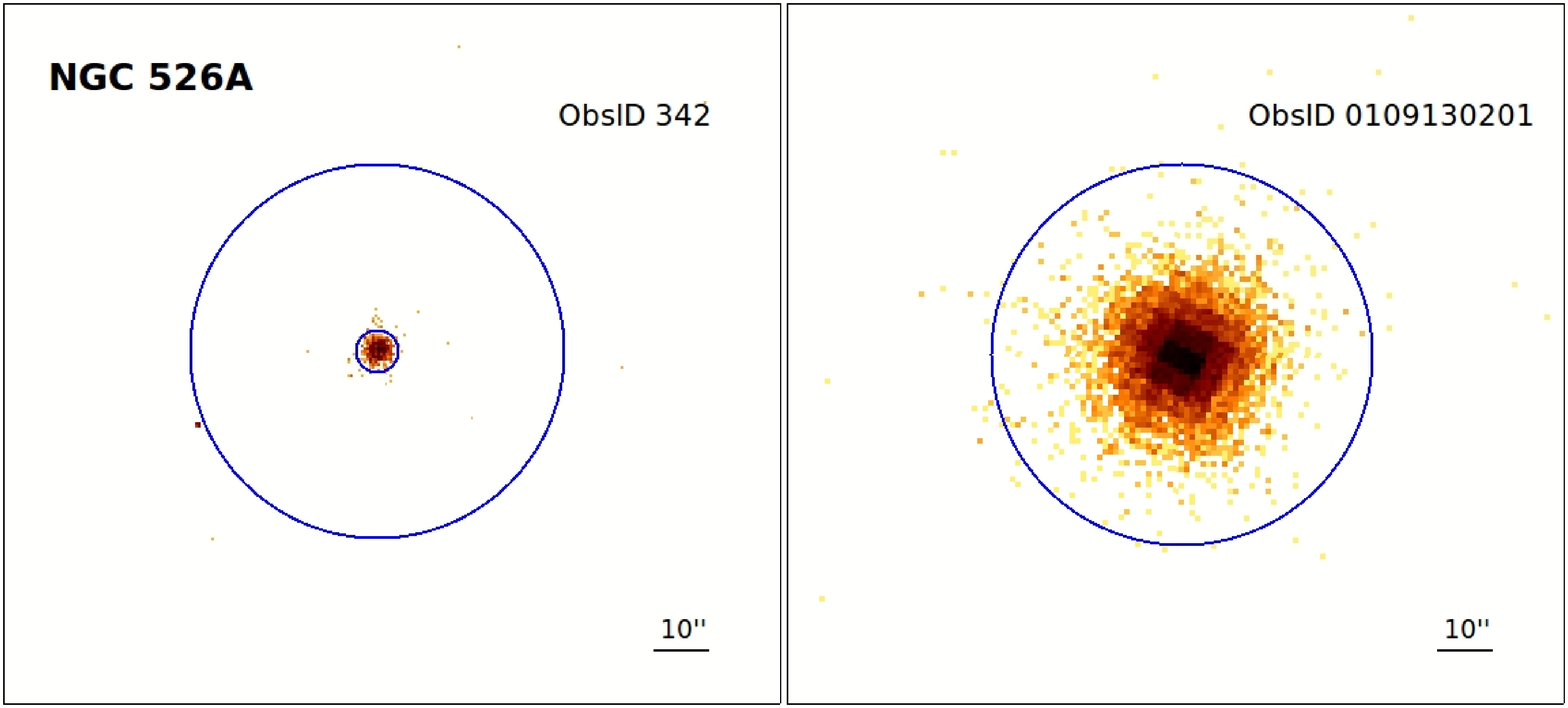}

\end{figure}

\begin{figure}

\centering
\includegraphics[width=\textwidth]{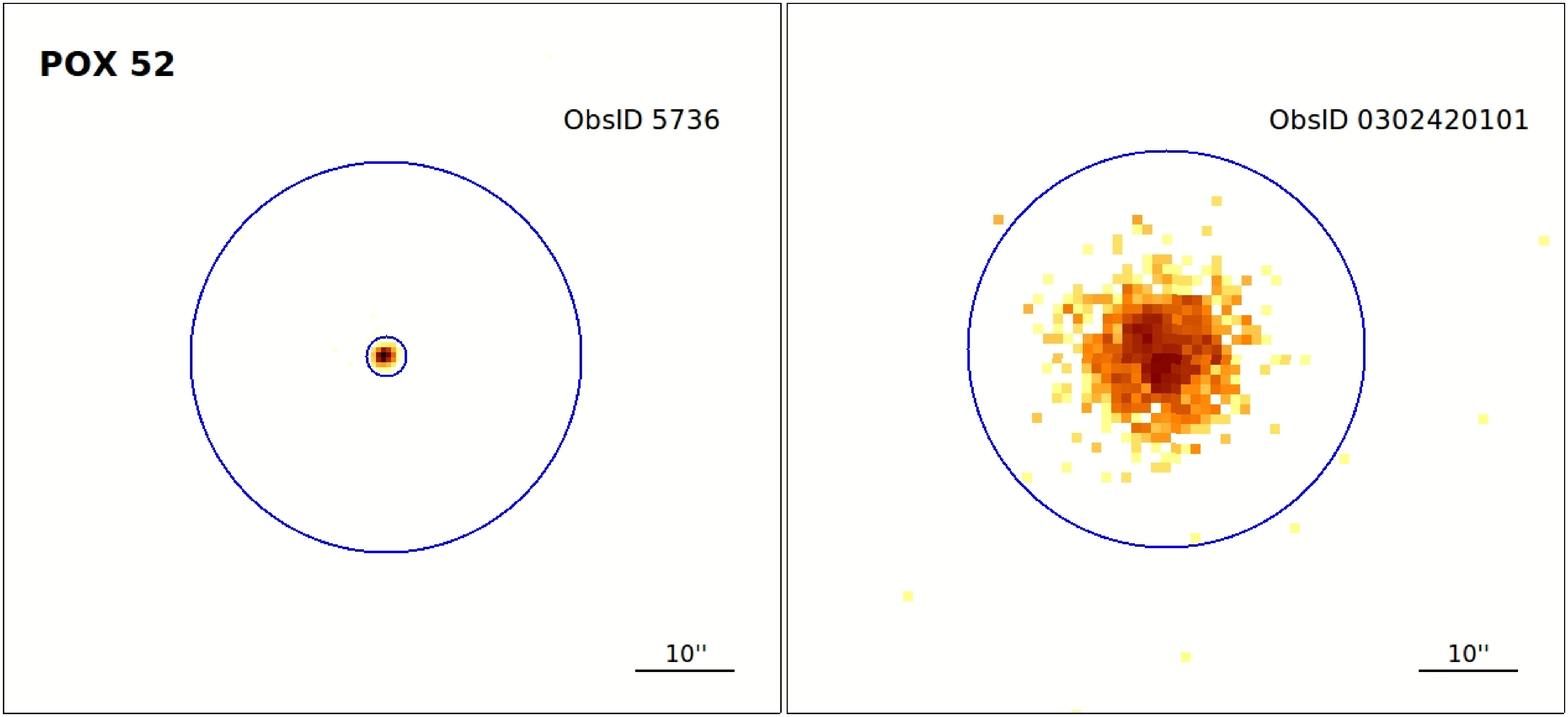}

\includegraphics[width=\textwidth]{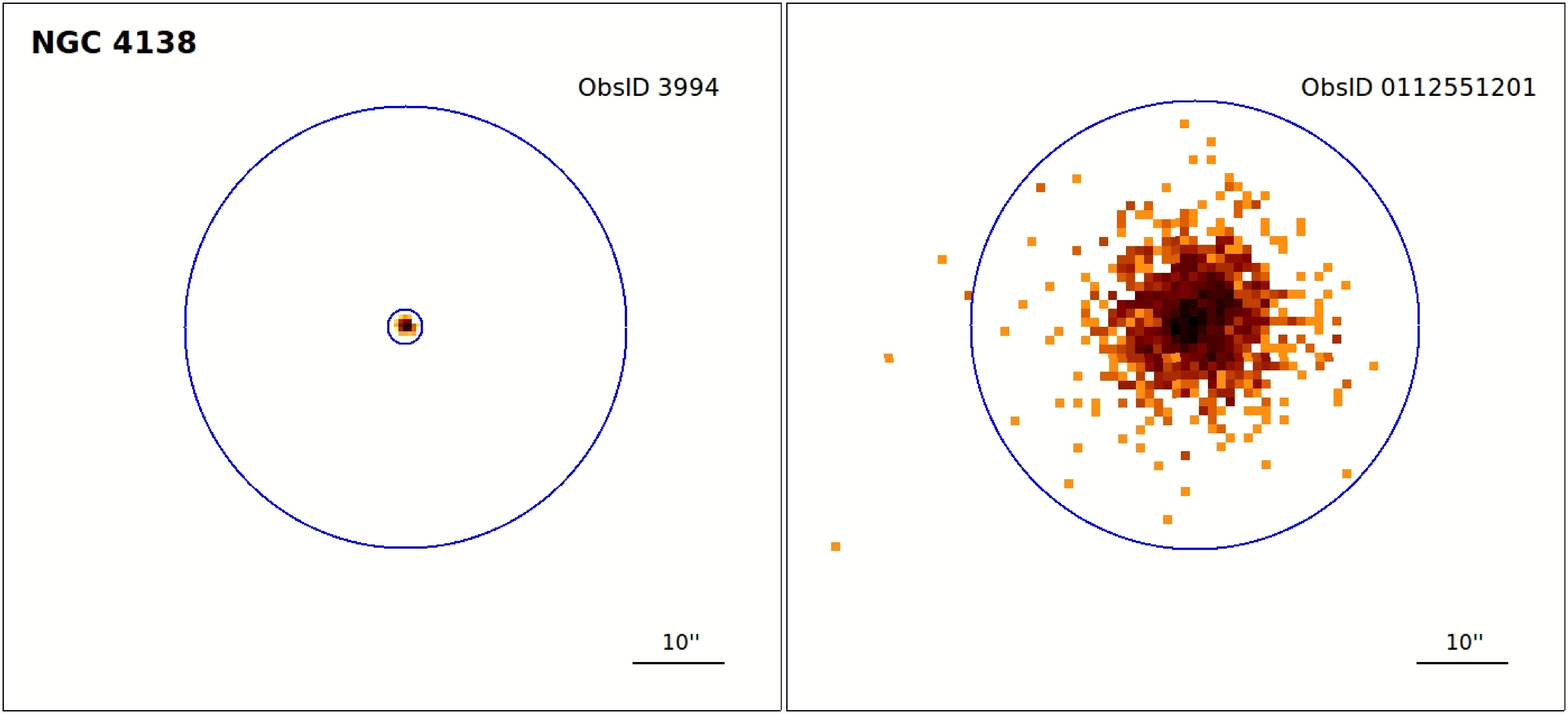}

\includegraphics[width=\textwidth]{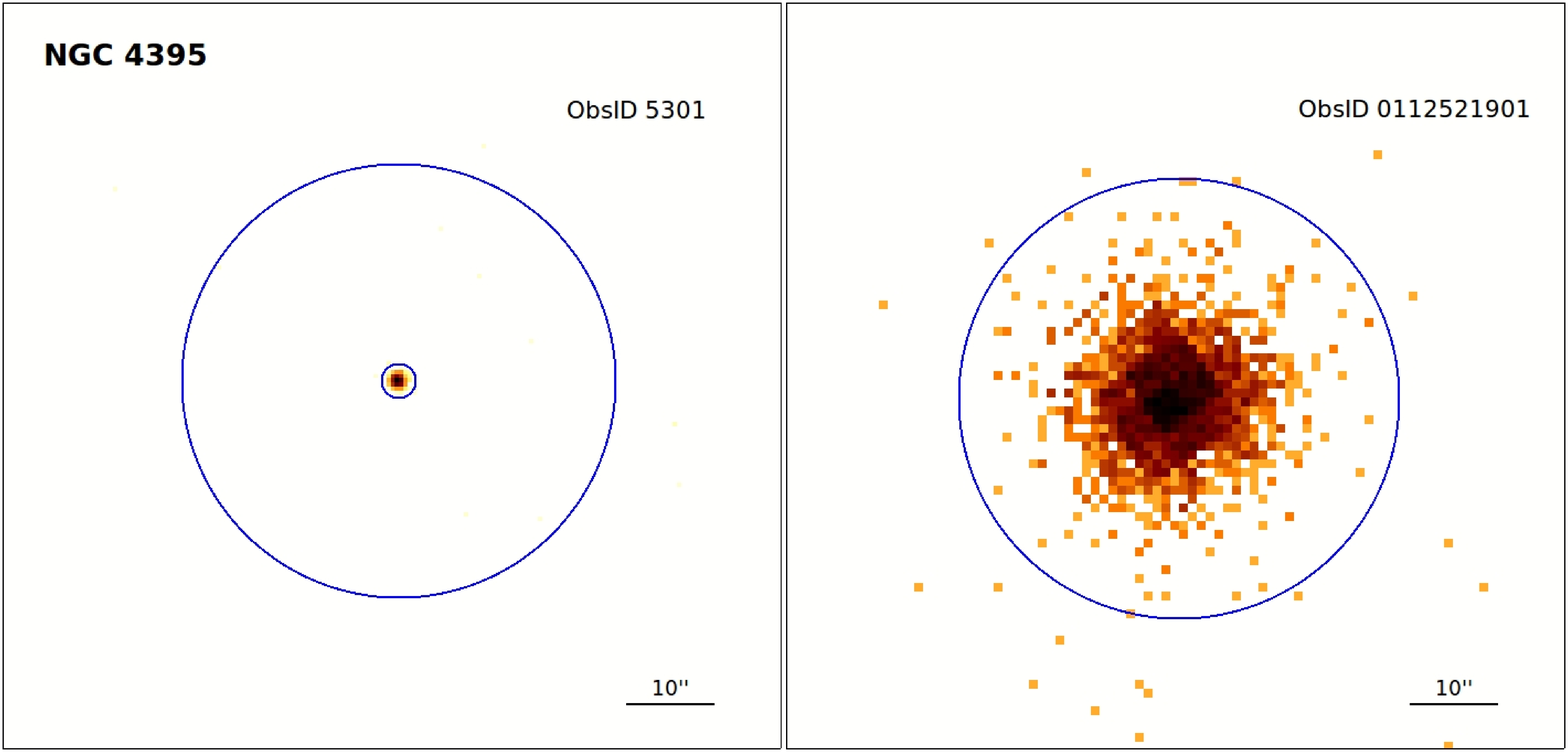}
\end{figure} 

\begin{figure}

\centering
\includegraphics[width=\textwidth]{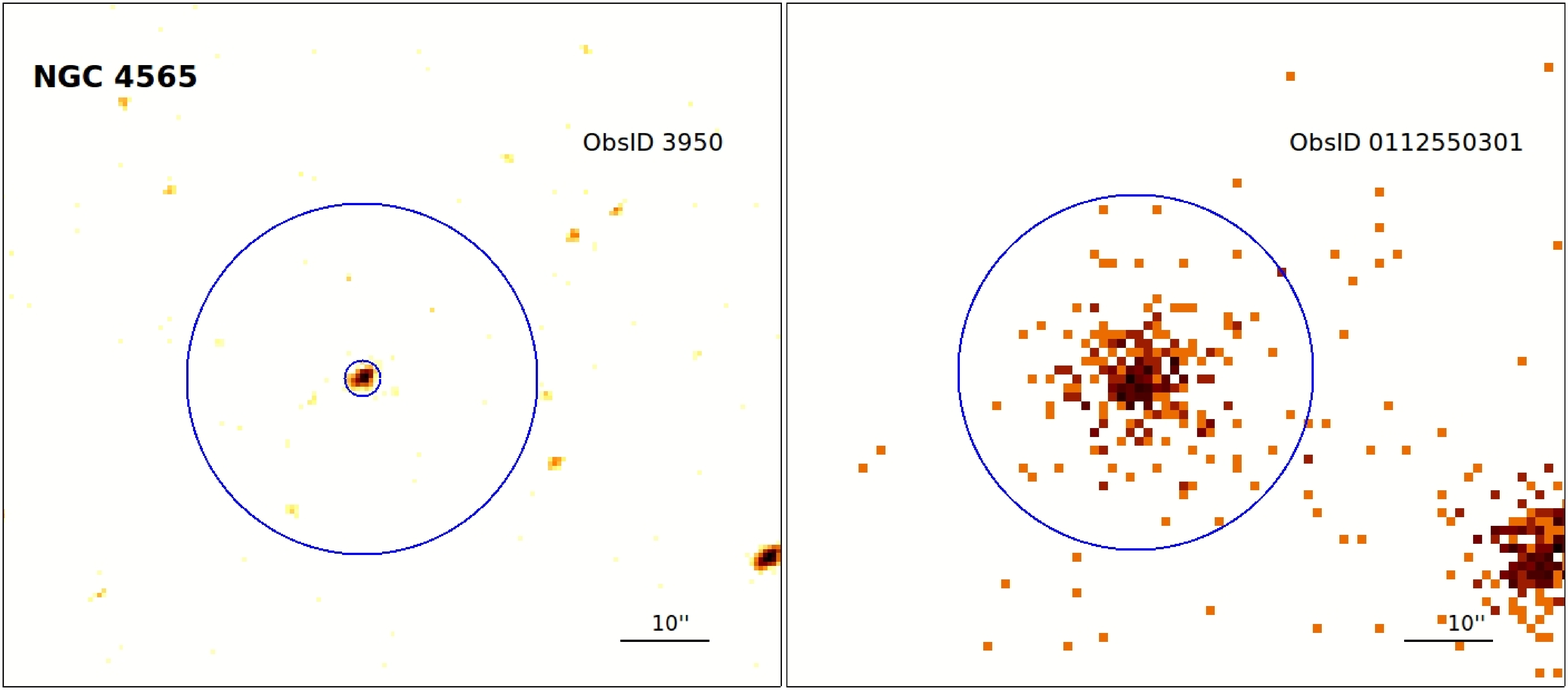}

\end{figure}

\onecolumn

\section{\label{lightcurves} Light curves}

This appendix provides the plots corresponding to the light
curves. Three plots per observation are presented, corresponding to
soft (left), hard (middle), and total (right) energy bands. Each light
curve has a minimum of 30 ksec (i.e., 8 hours) exposure time, while
long light curves are divided into segments of 40 ksec (i.e., 11
hours). Each segment is enumerated in the title of the light
curve. Count rates versus time continua are represented. The solid
line represents the mean value, dashed lines the $^+_-$1$\sigma$ from
the average.

\begin{figure}
\centering
{\includegraphics[width=0.30\textwidth]{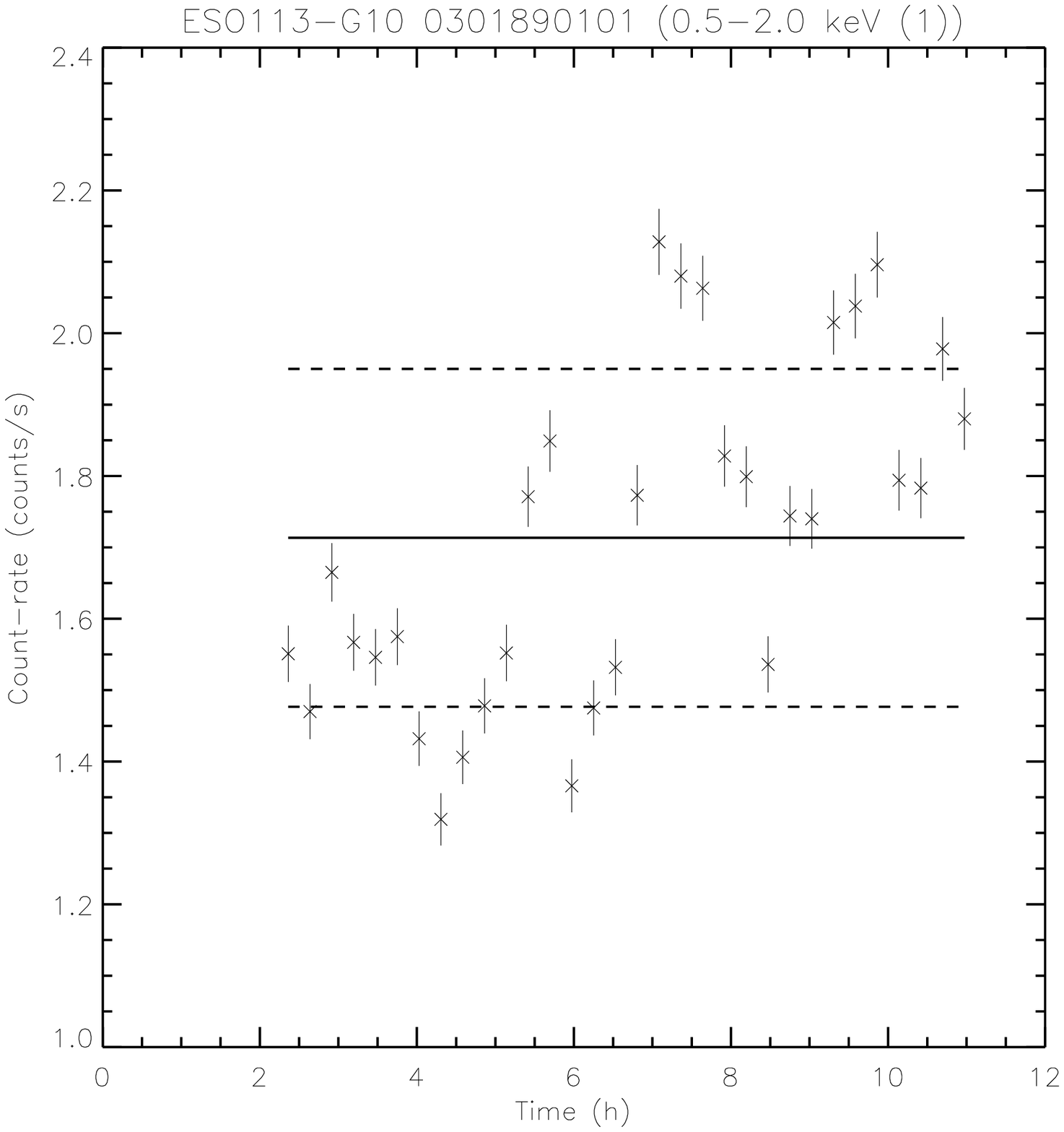}}
{\includegraphics[width=0.30\textwidth]{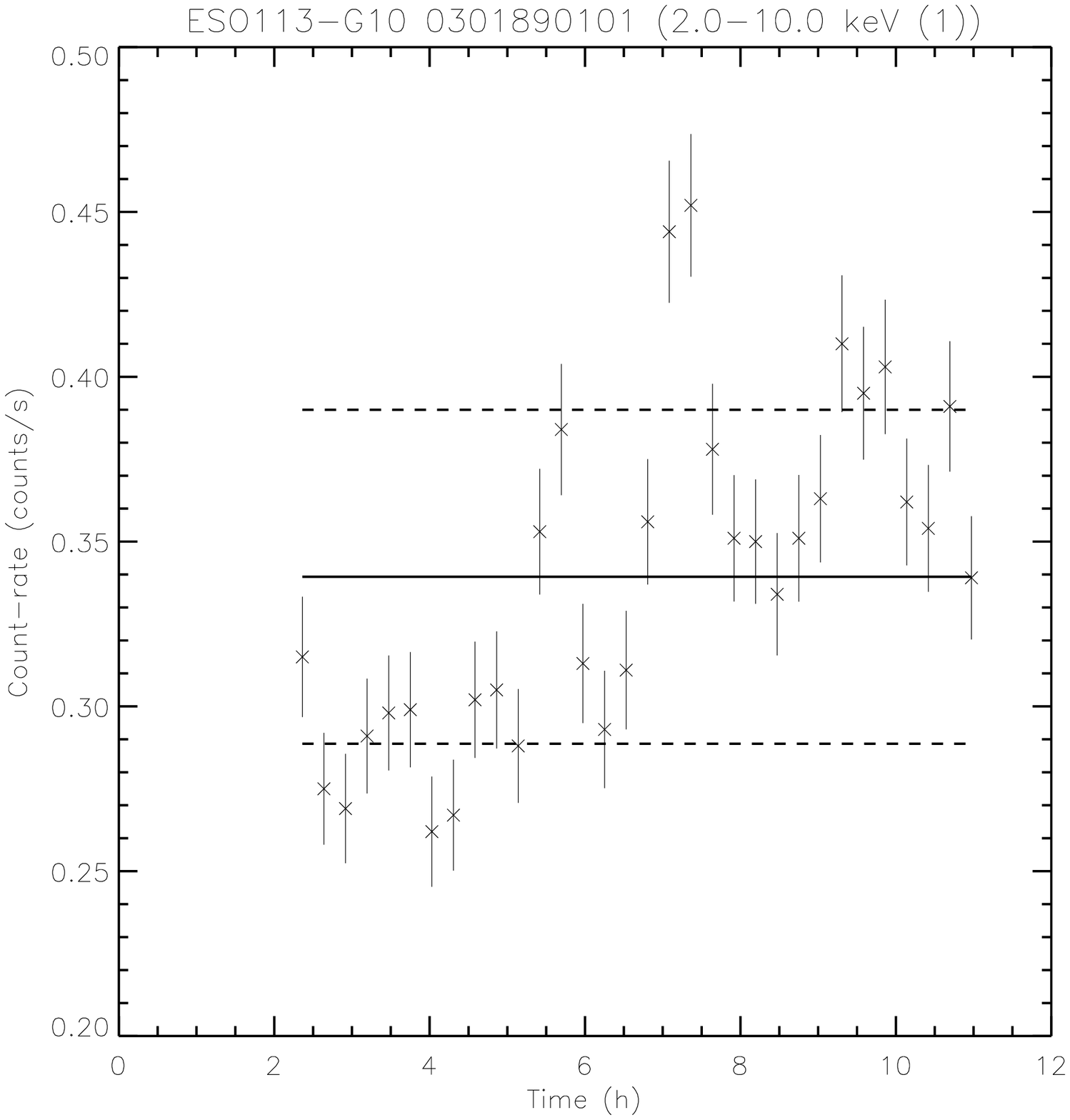}}
{\includegraphics[width=0.30\textwidth]{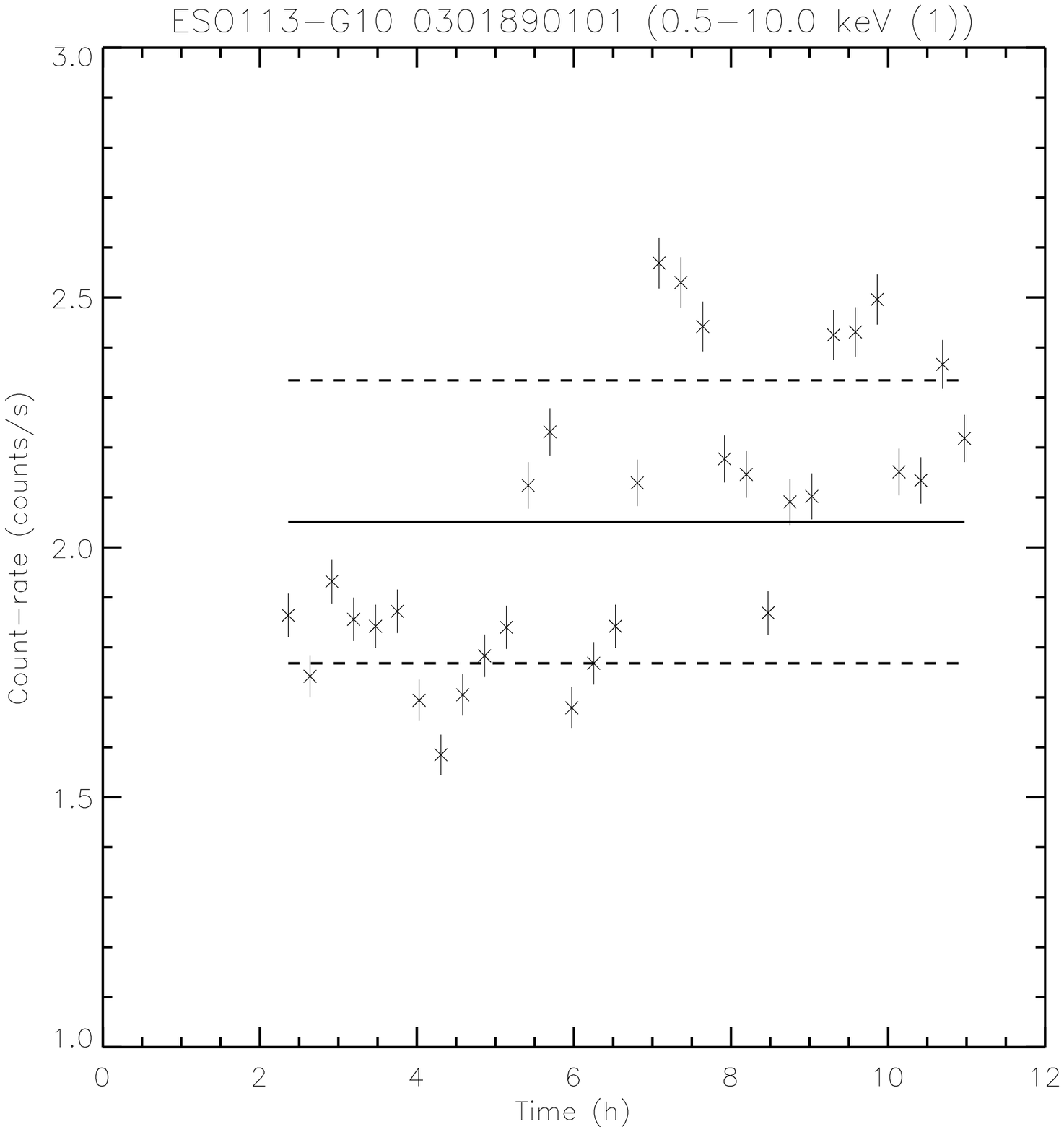}}

{\includegraphics[width=0.30\textwidth]{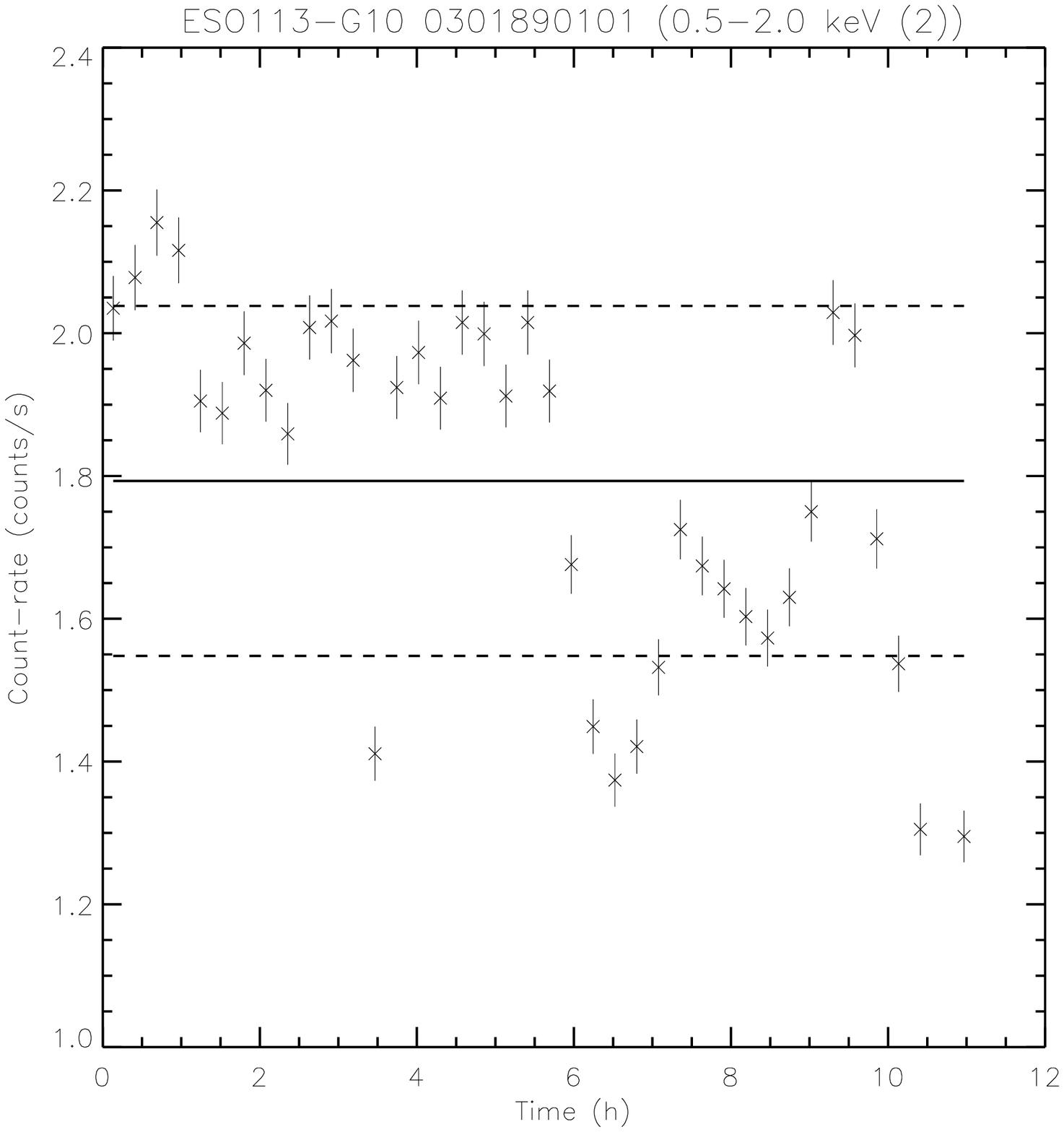}}
{\includegraphics[width=0.30\textwidth]{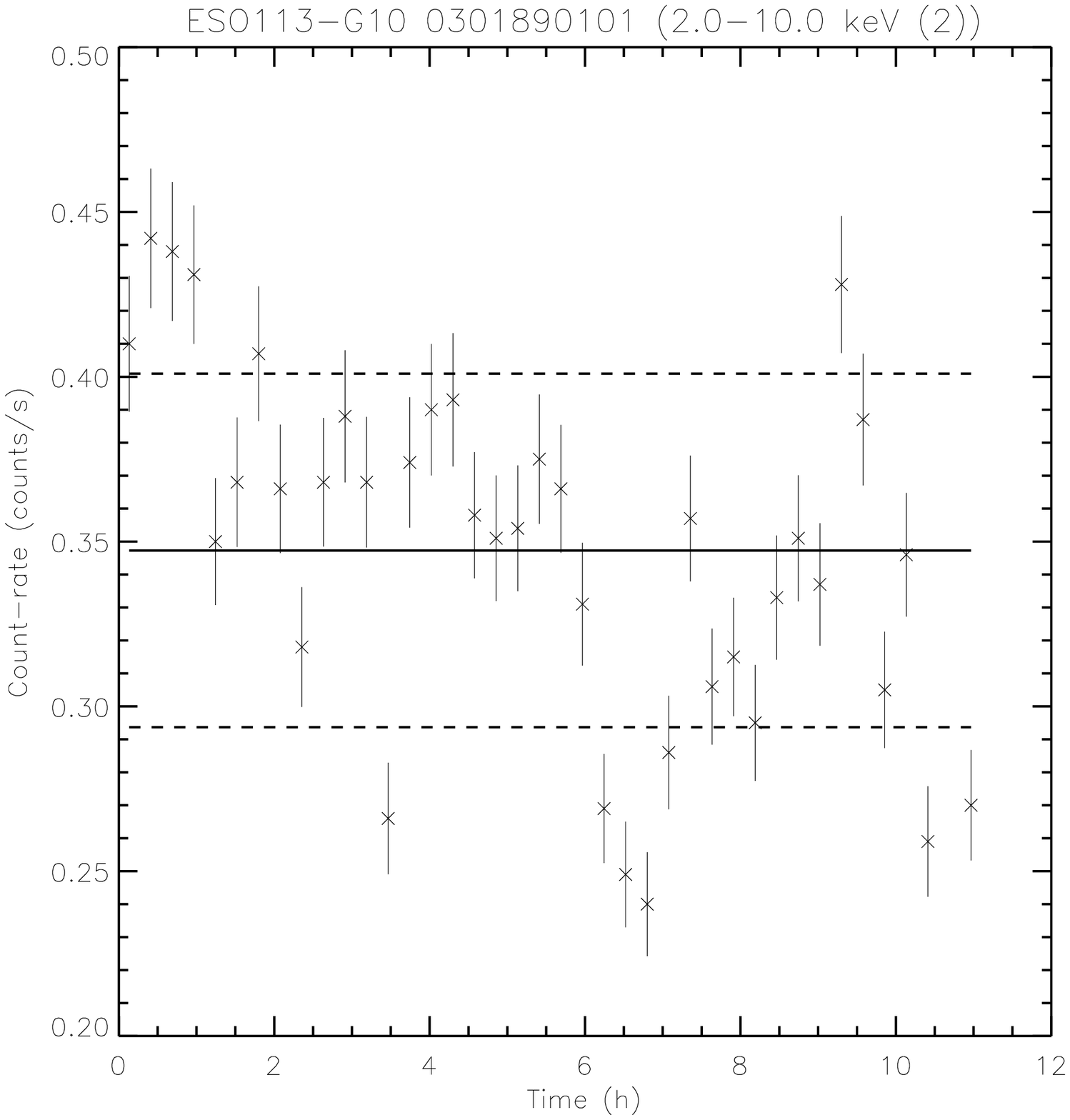}}
{\includegraphics[width=0.30\textwidth]{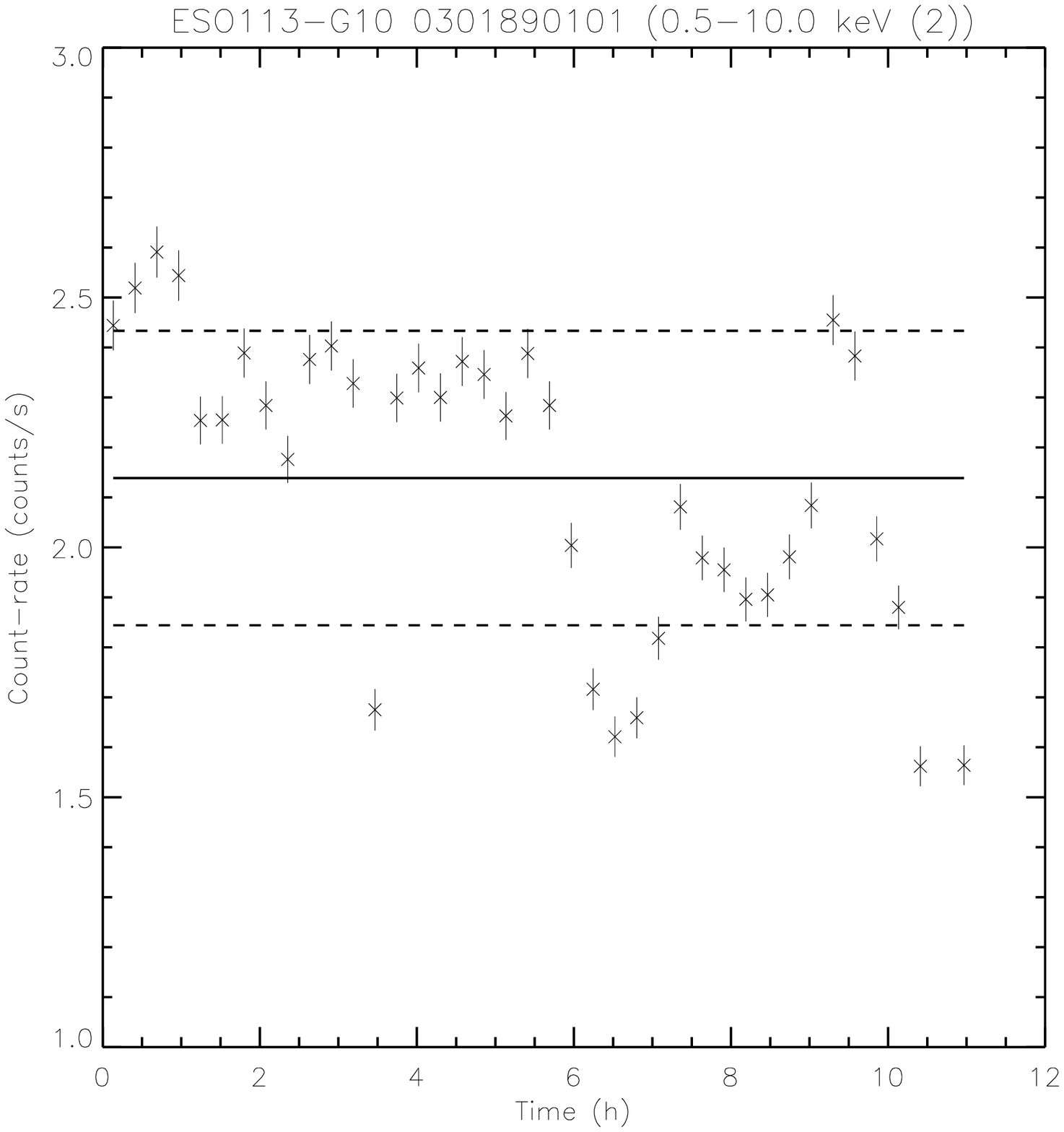}}
\caption{Light curves of ESO\,113-G10 from \emph{XMM--Newton} data.}
\label{leso113}
\end{figure}

\begin{figure}
\centering
{\includegraphics[width=0.30\textwidth]{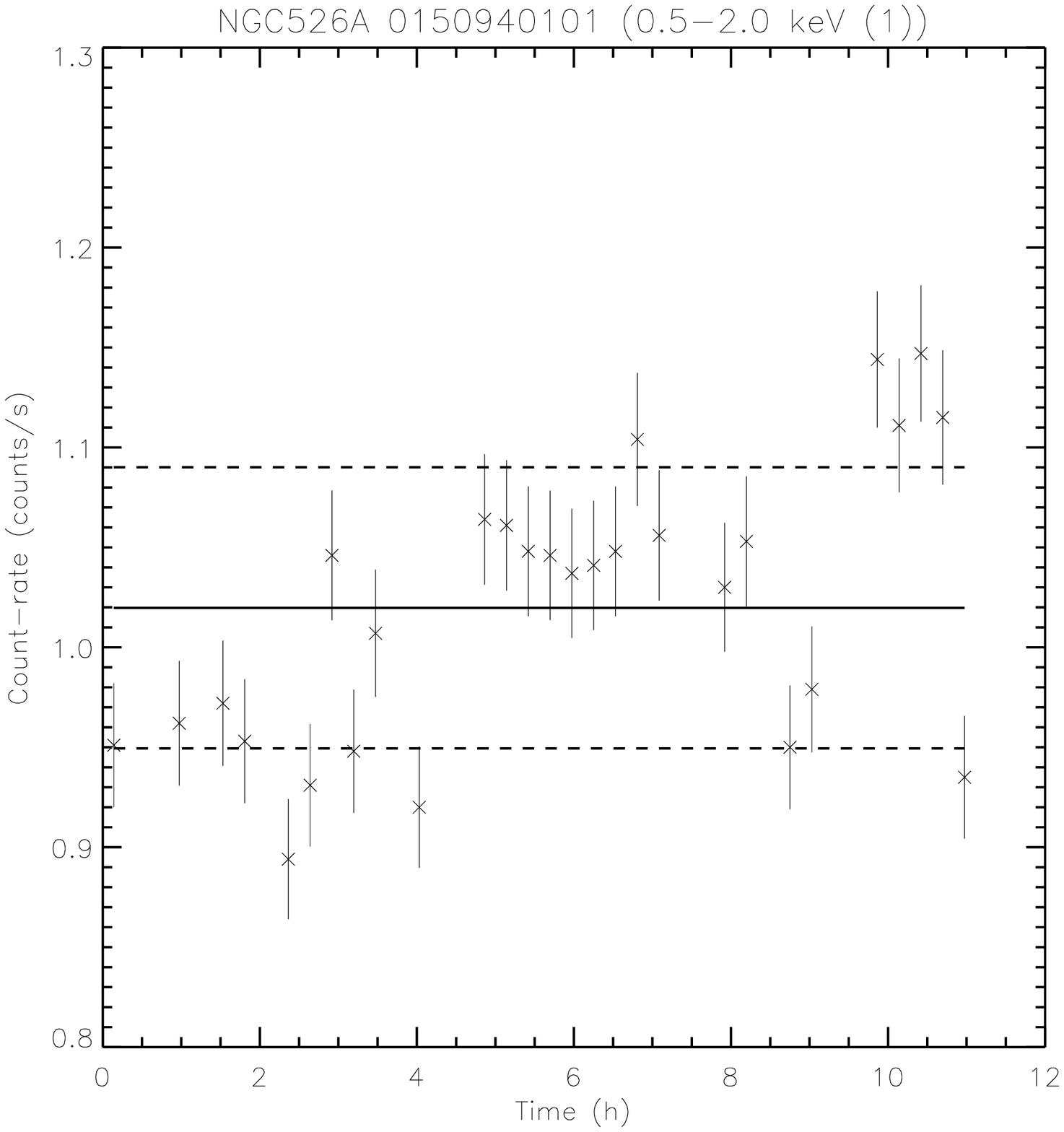}}
{\includegraphics[width=0.30\textwidth]{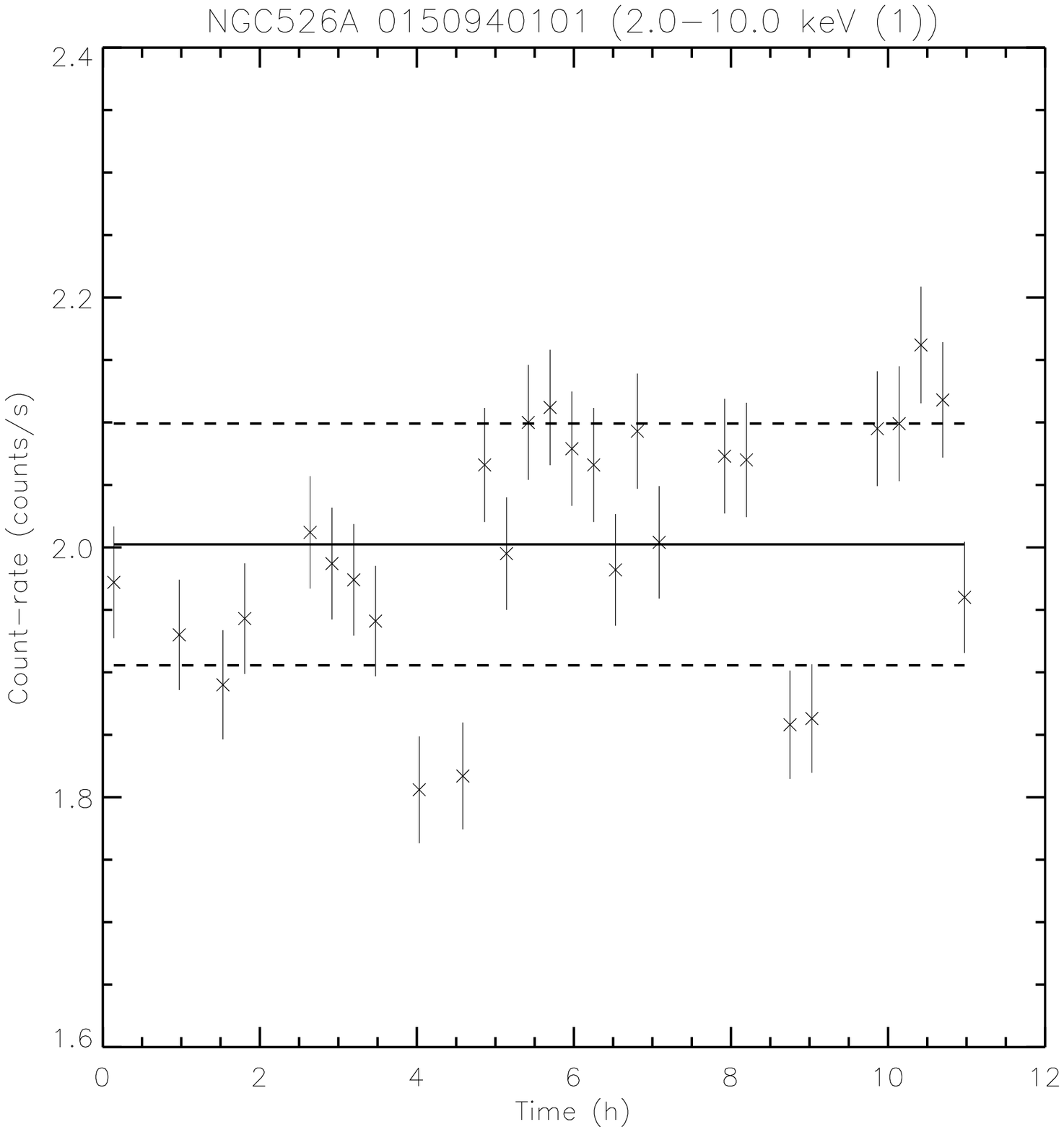}}
{\includegraphics[width=0.30\textwidth]{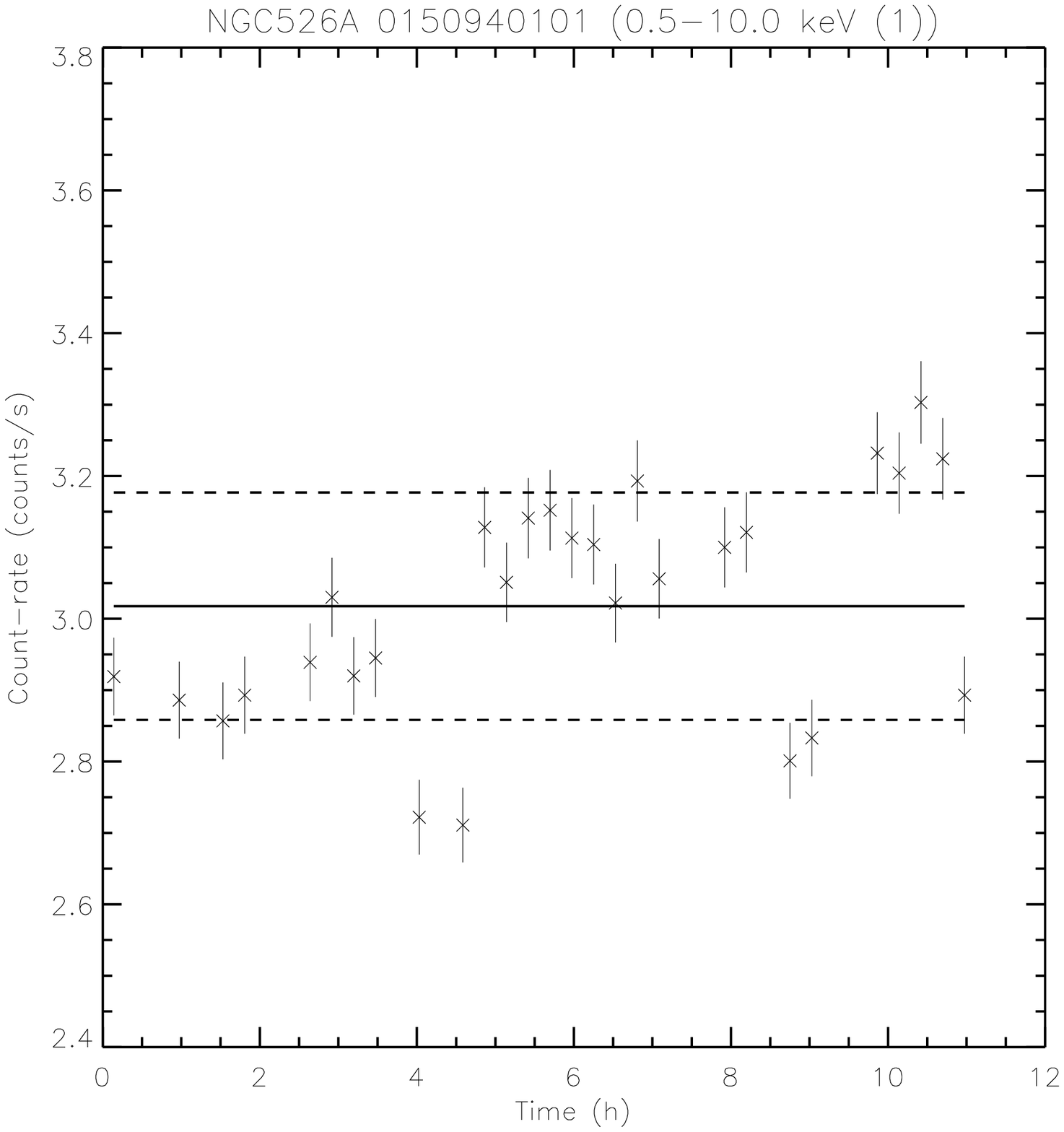}}

{\includegraphics[width=0.30\textwidth]{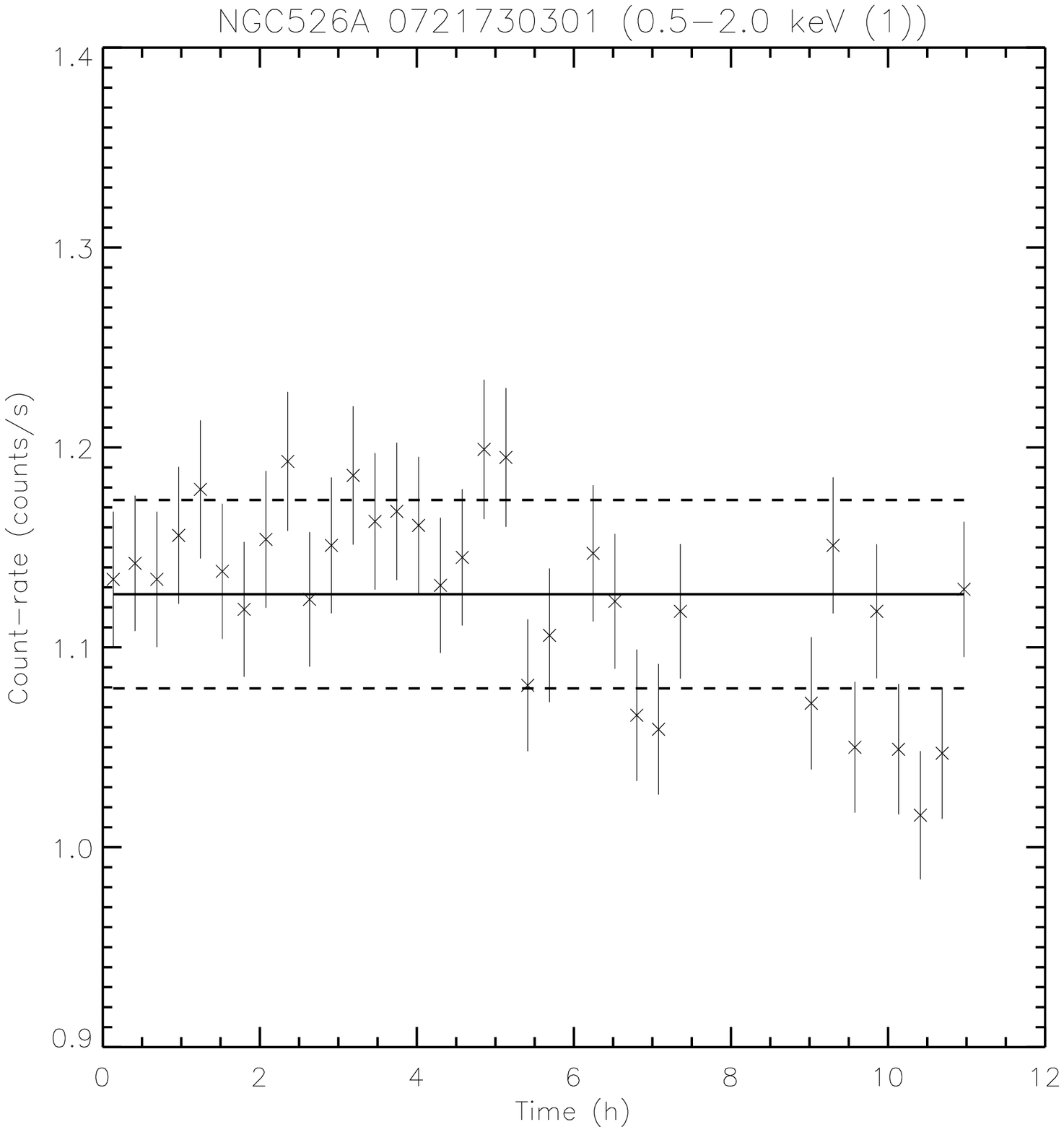}}
{\includegraphics[width=0.30\textwidth]{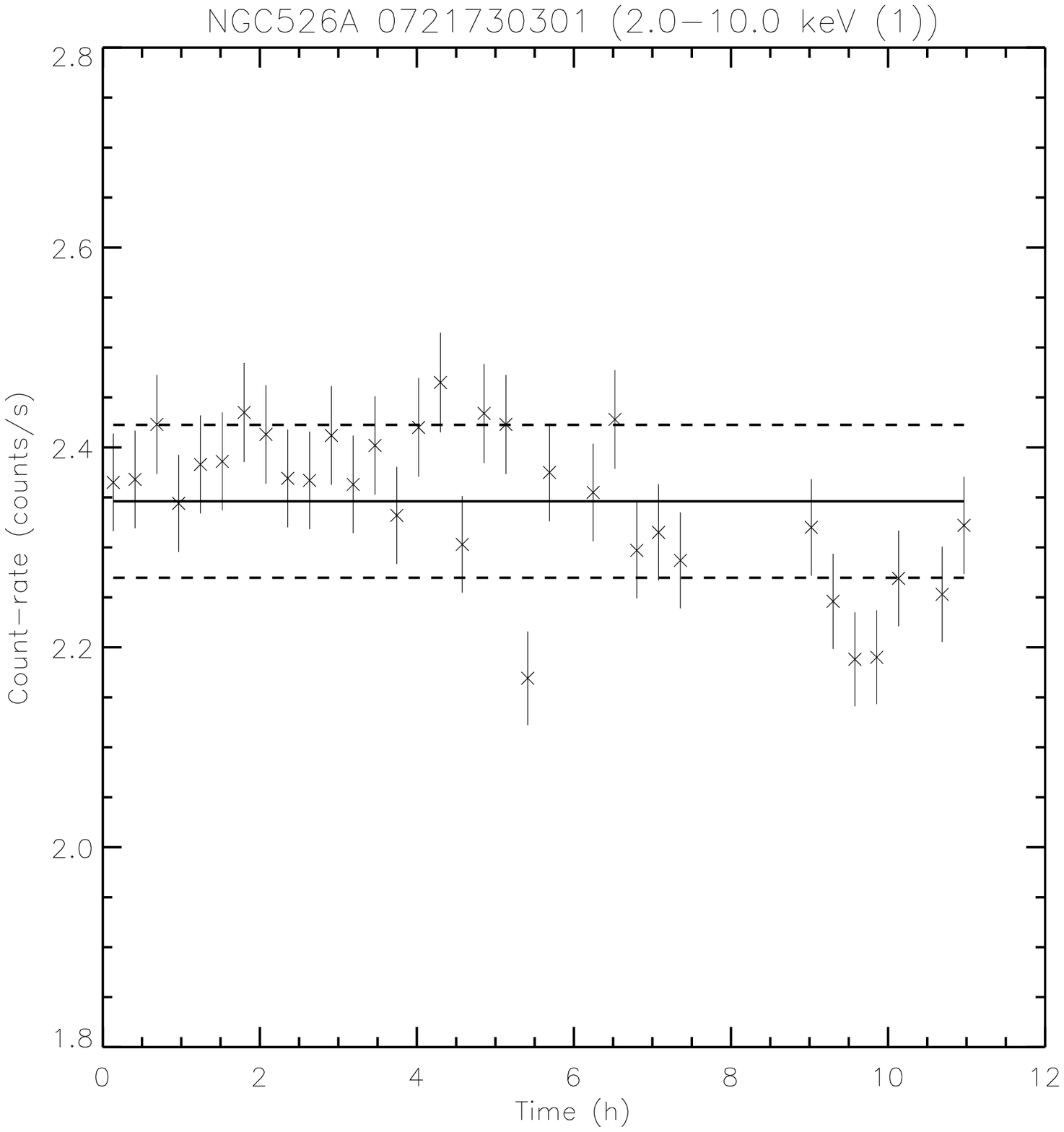}}
{\includegraphics[width=0.30\textwidth]{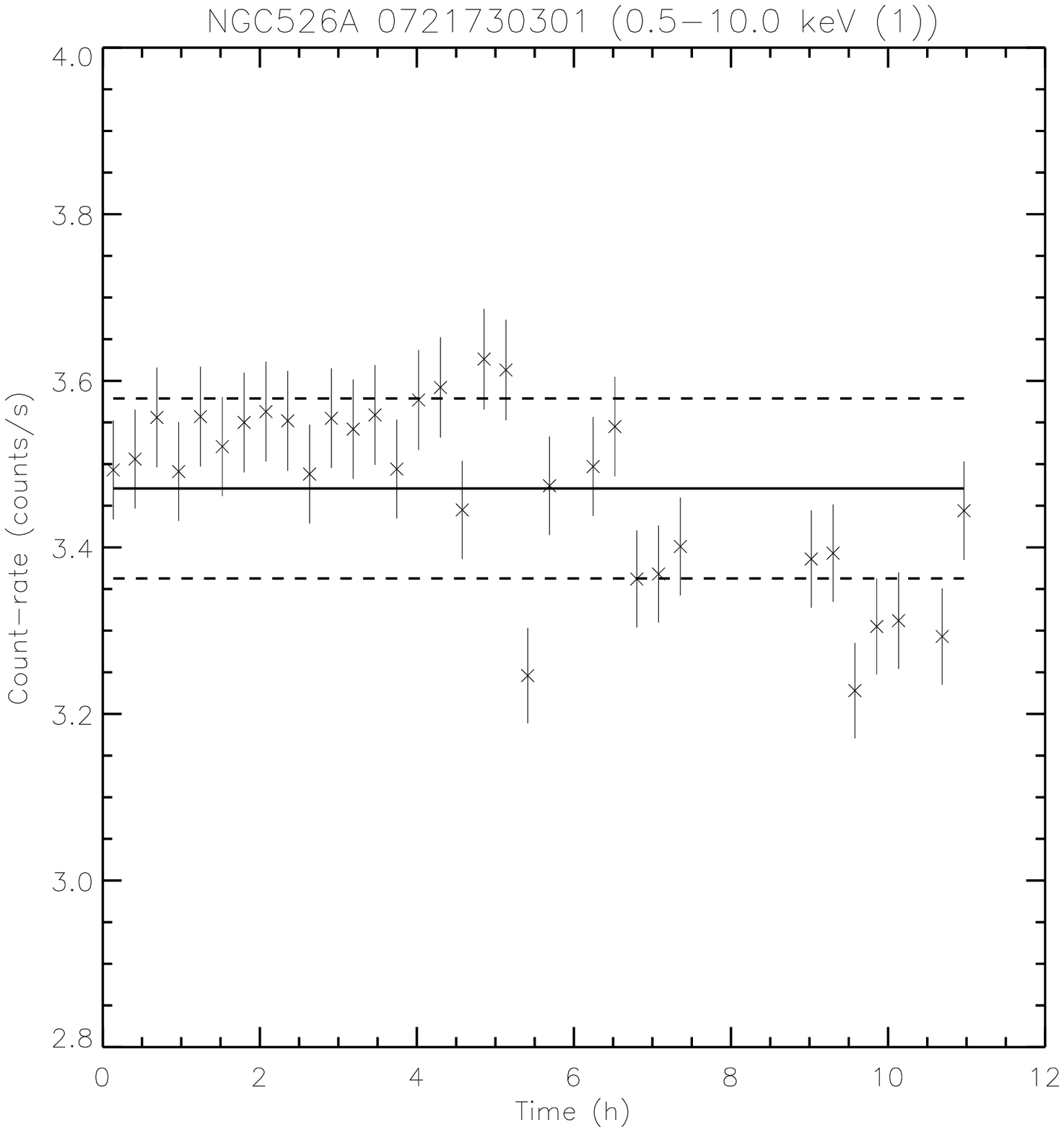}}

{\includegraphics[width=0.30\textwidth]{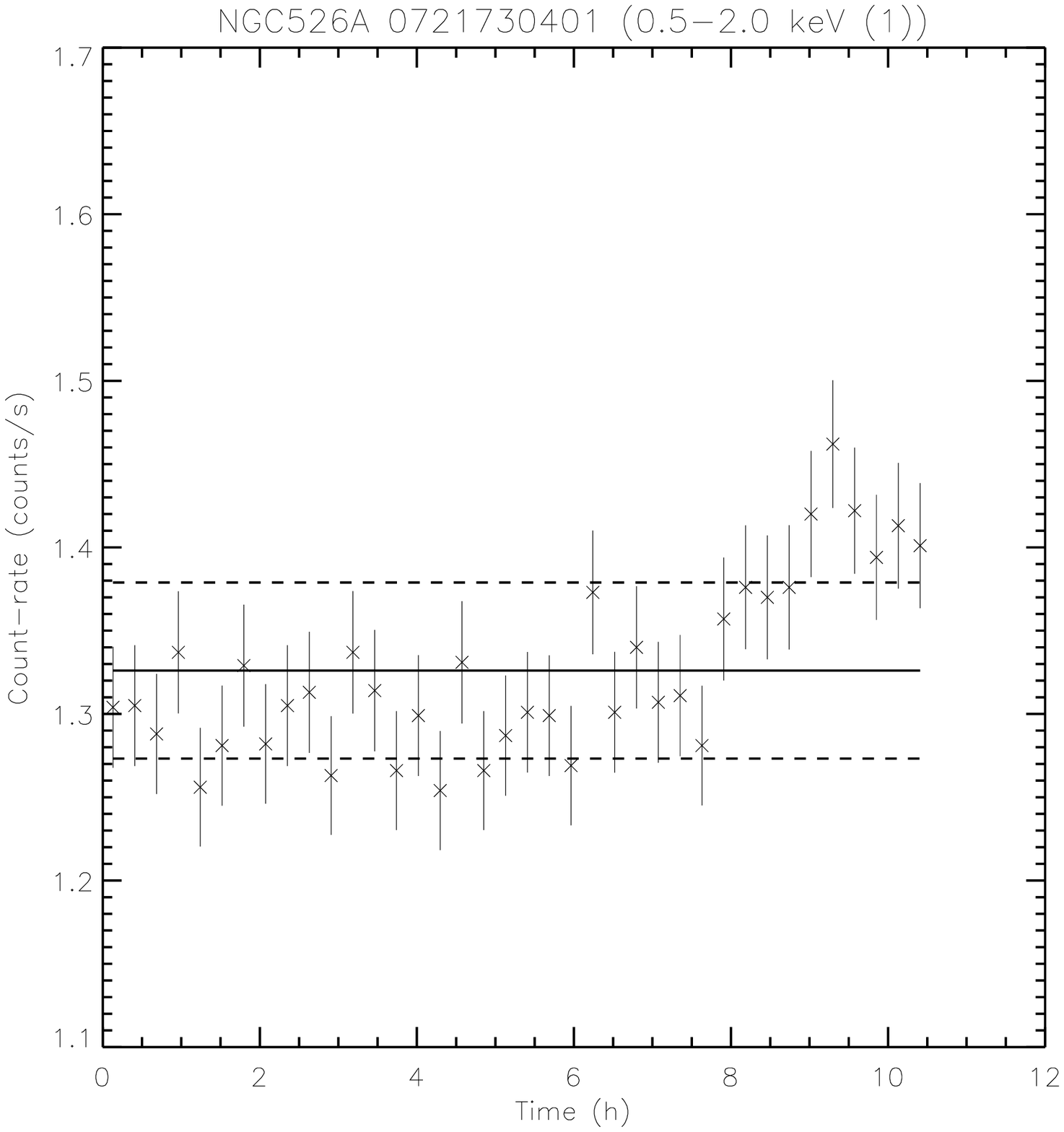}}
{\includegraphics[width=0.30\textwidth]{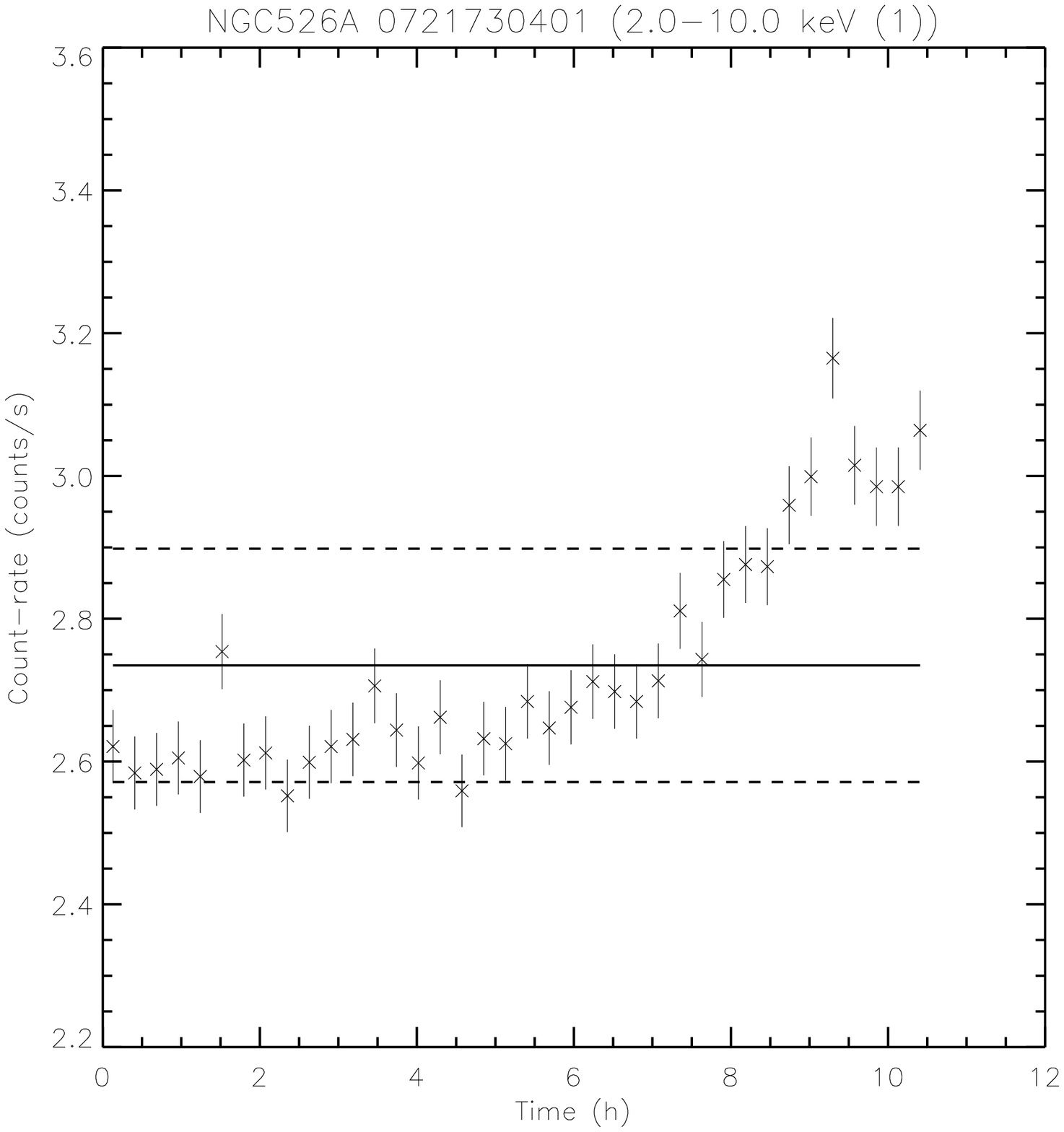}}
{\includegraphics[width=0.30\textwidth]{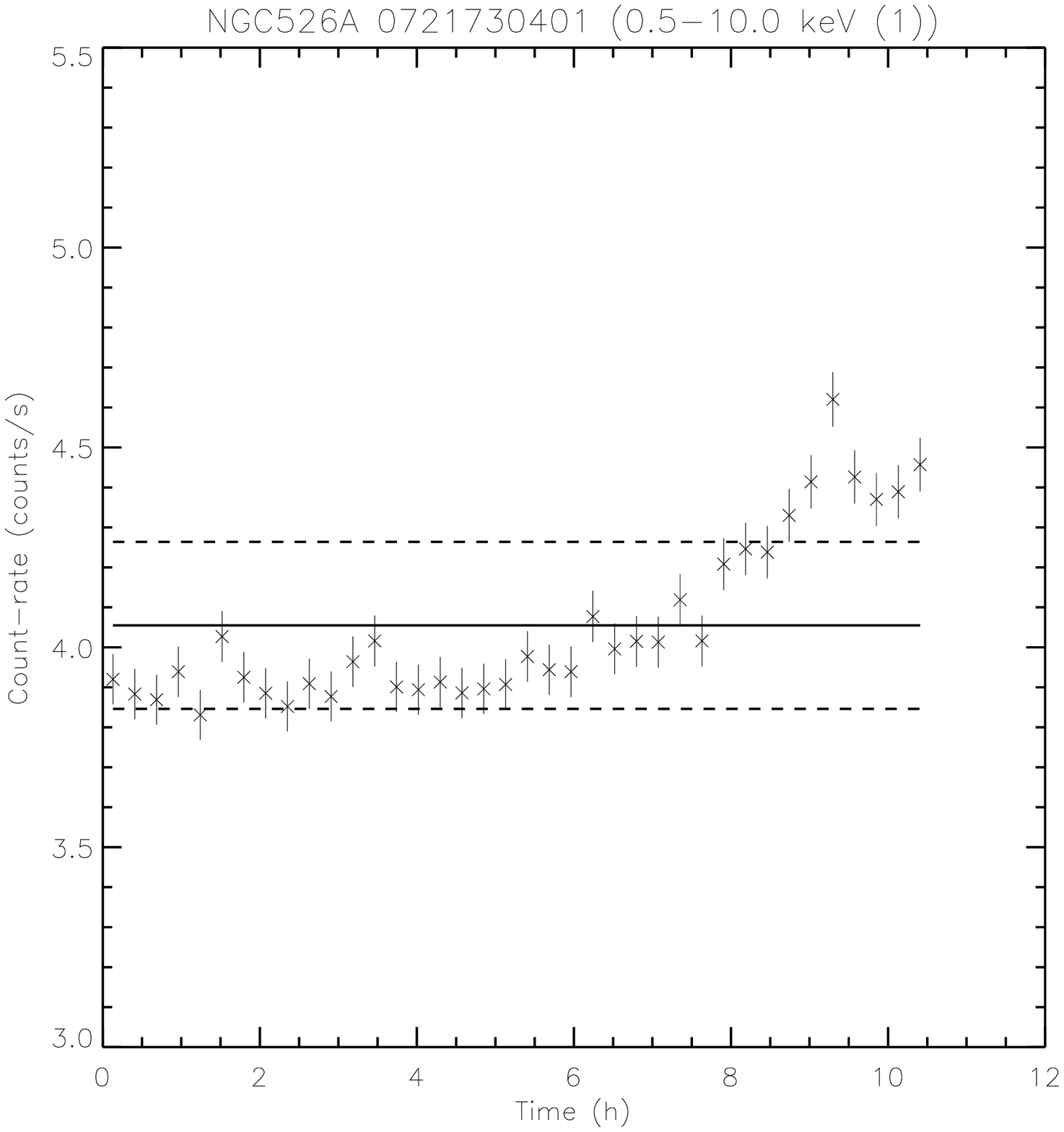}}
\caption{Light curves of NGC\,526A from \emph{XMM--Newton} data.}
\label{l526A}
\end{figure}

\begin{figure}
\centering
{\includegraphics[width=0.30\textwidth]{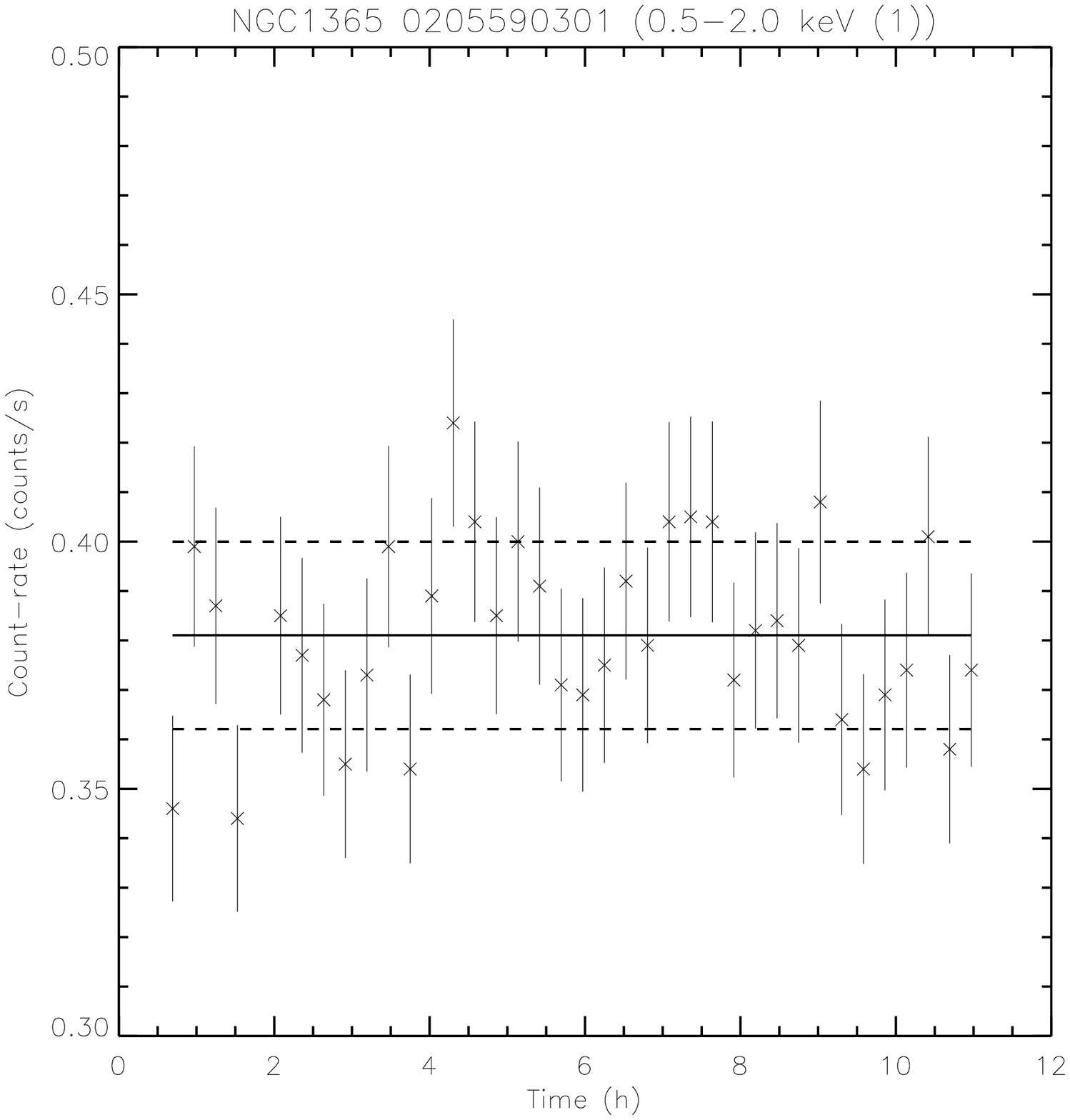}}
{\includegraphics[width=0.30\textwidth]{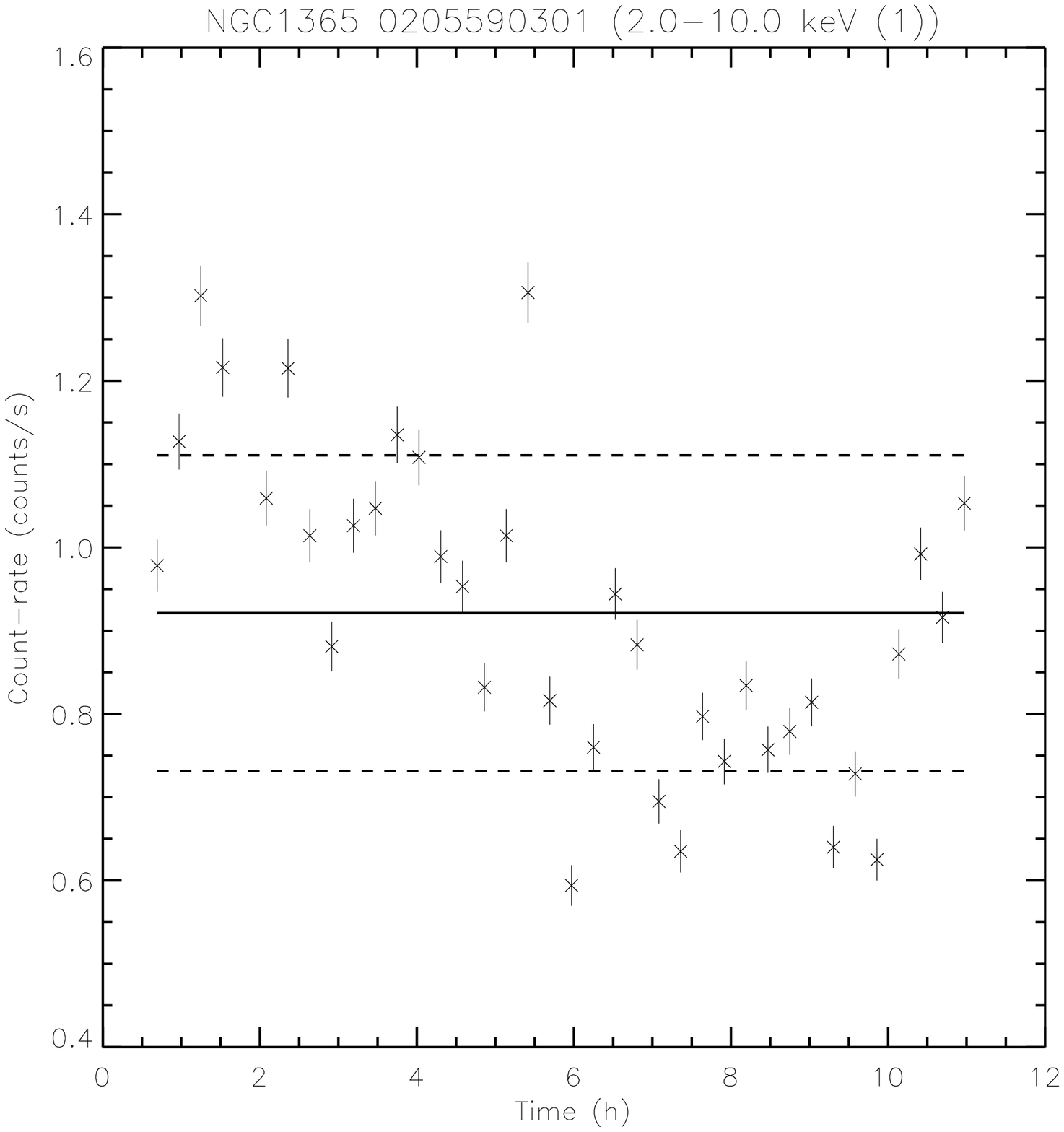}}
{\includegraphics[width=0.30\textwidth]{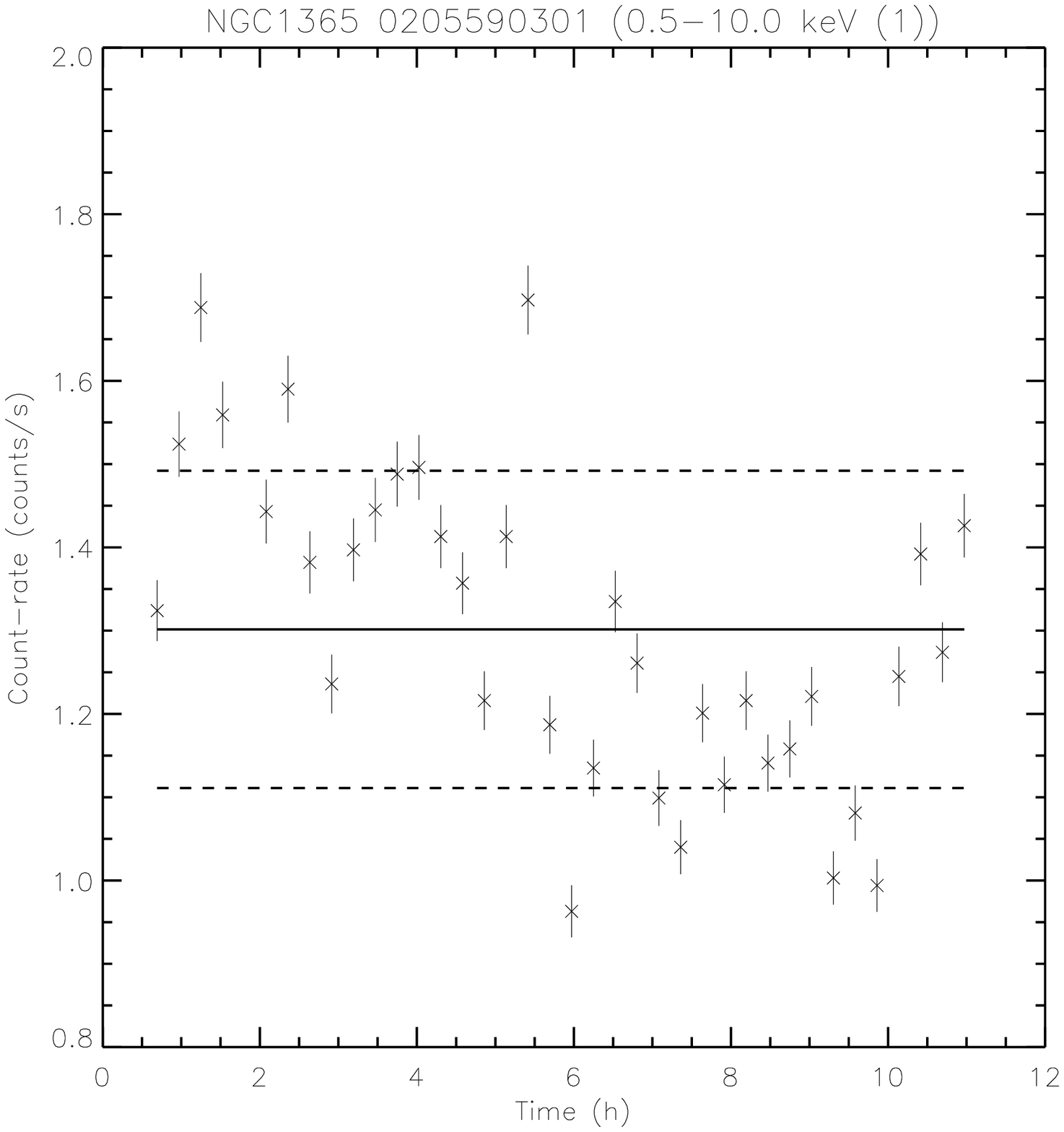}}

{\includegraphics[width=0.30\textwidth]{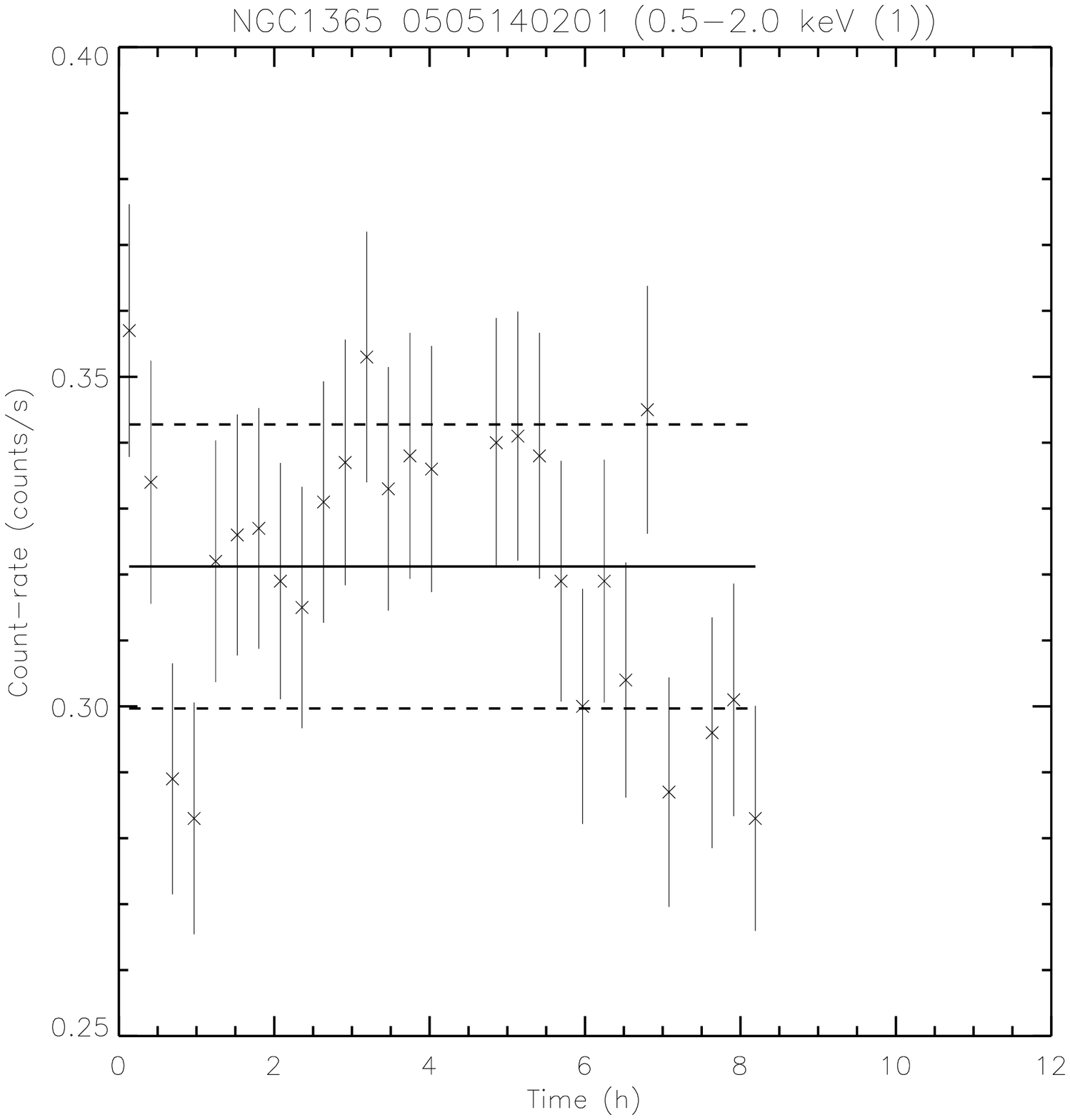}}
{\includegraphics[width=0.30\textwidth]{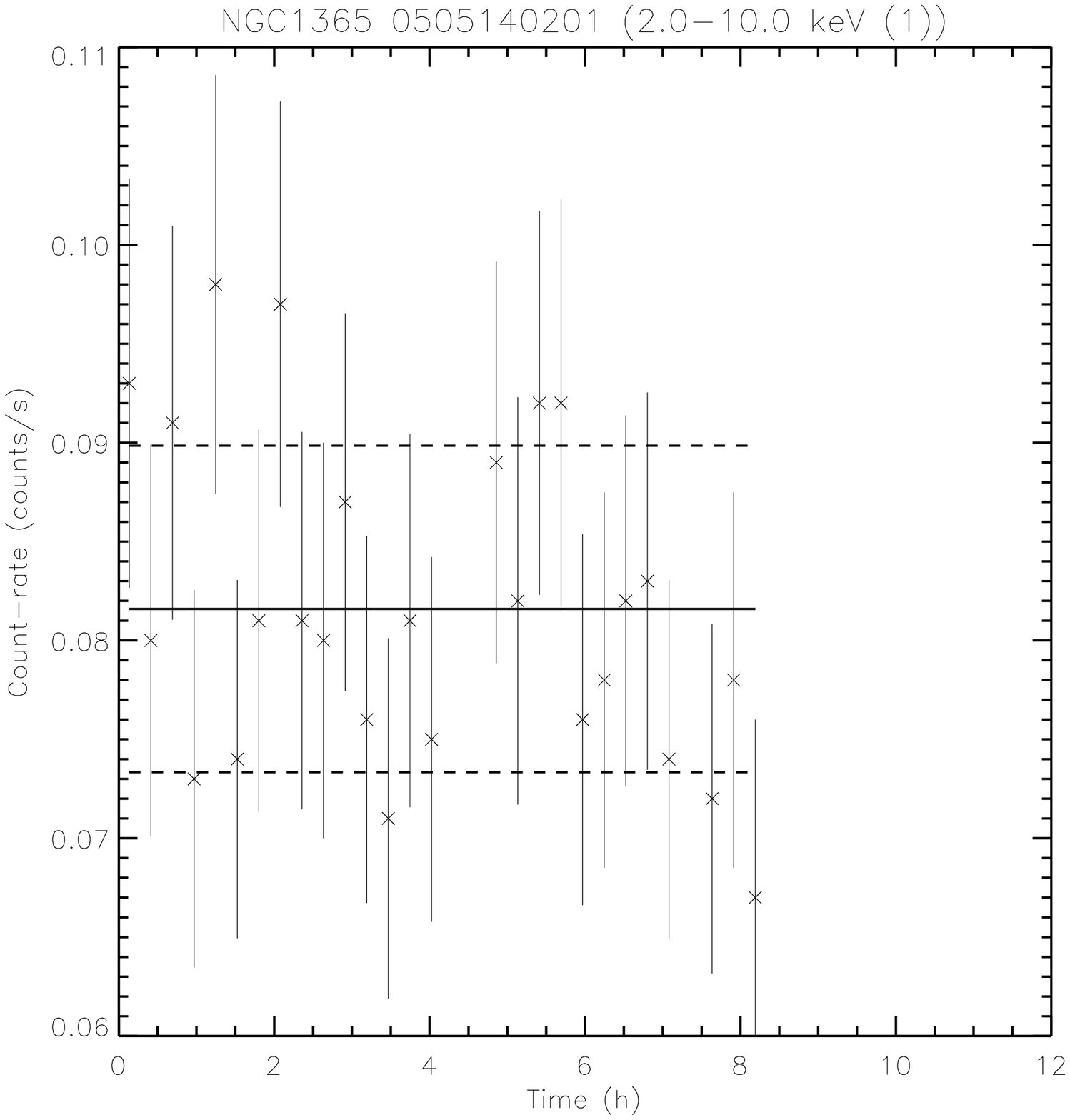}}
{\includegraphics[width=0.30\textwidth]{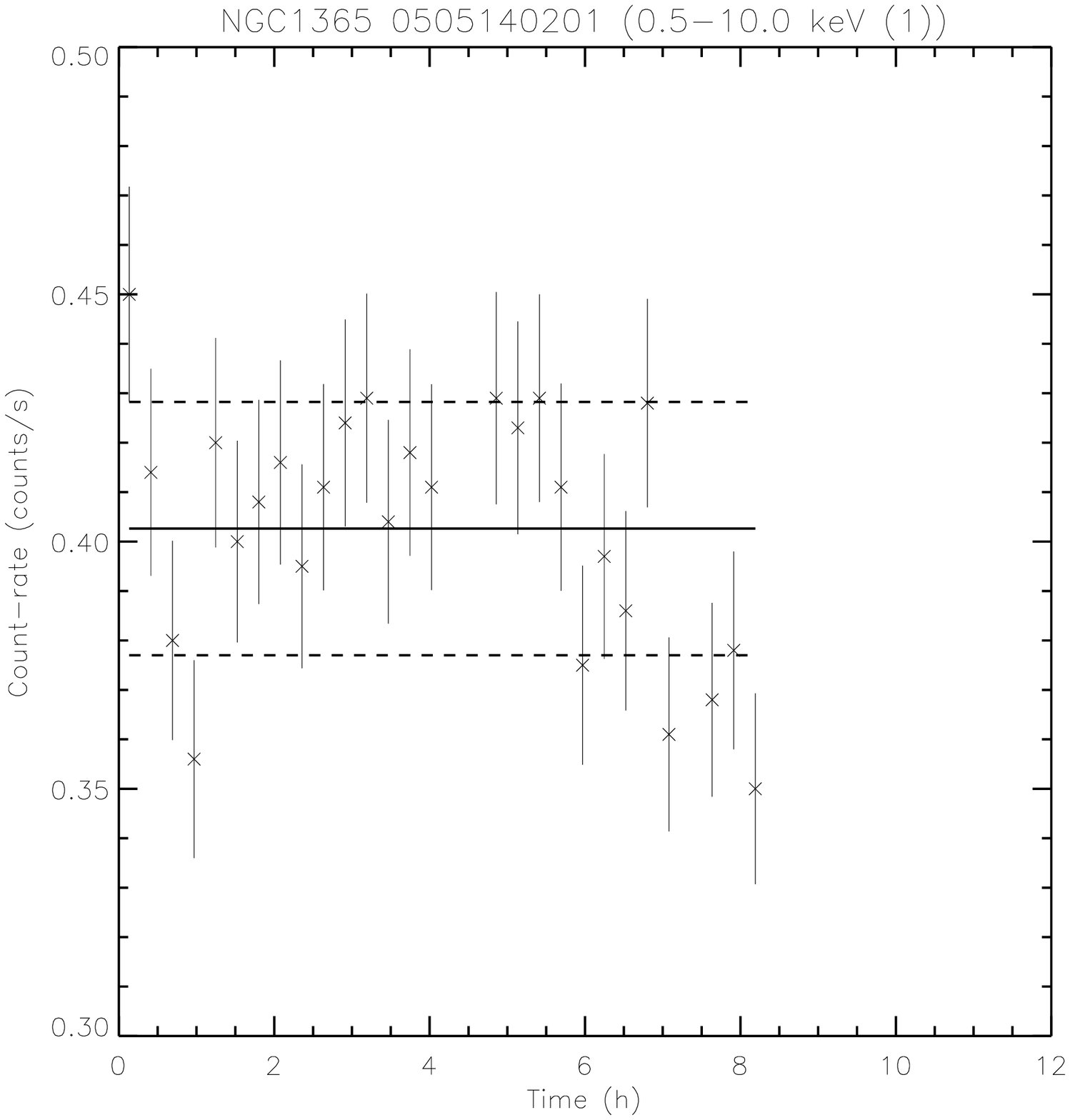}}

{\includegraphics[width=0.30\textwidth]{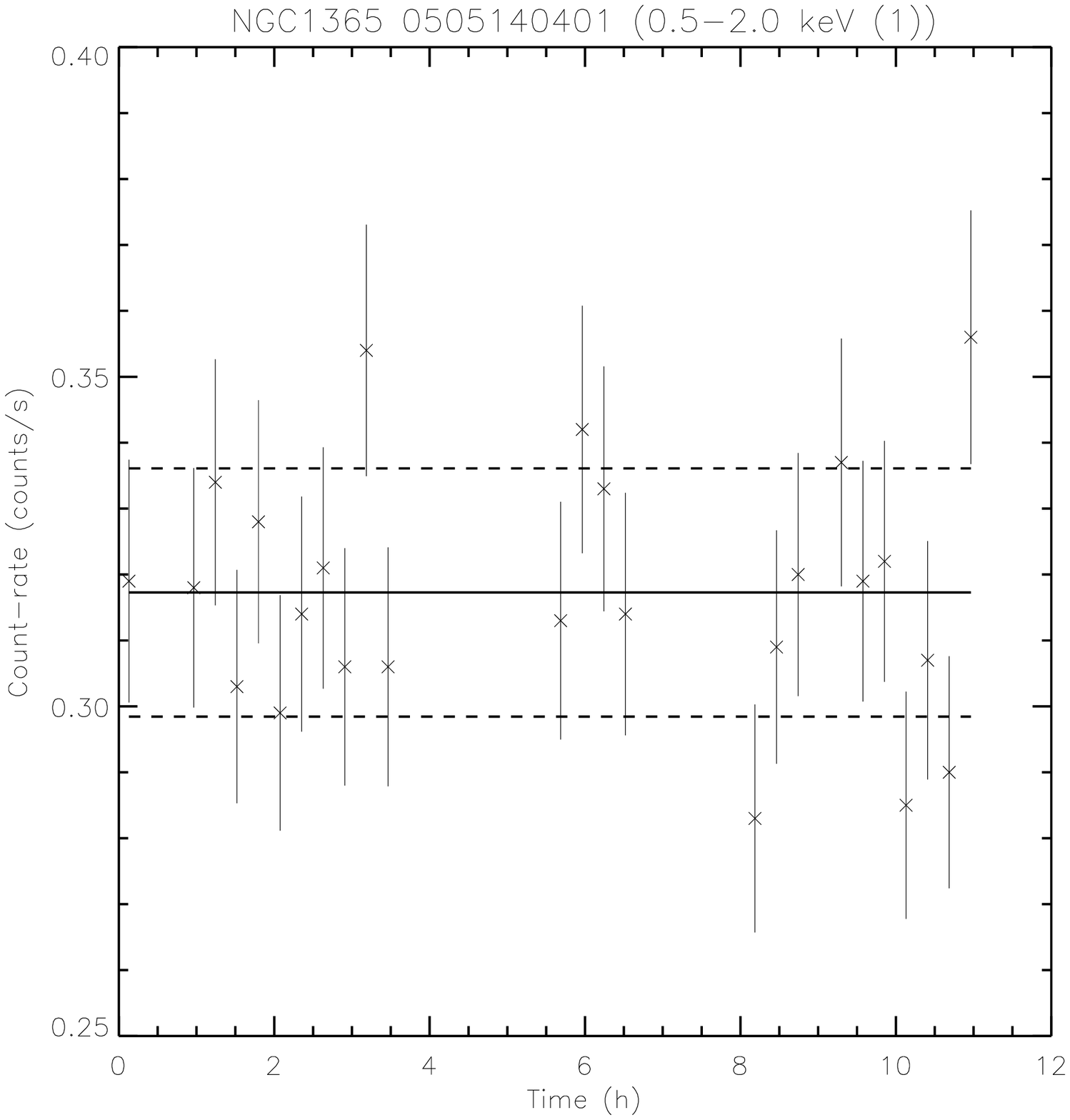}}
{\includegraphics[width=0.30\textwidth]{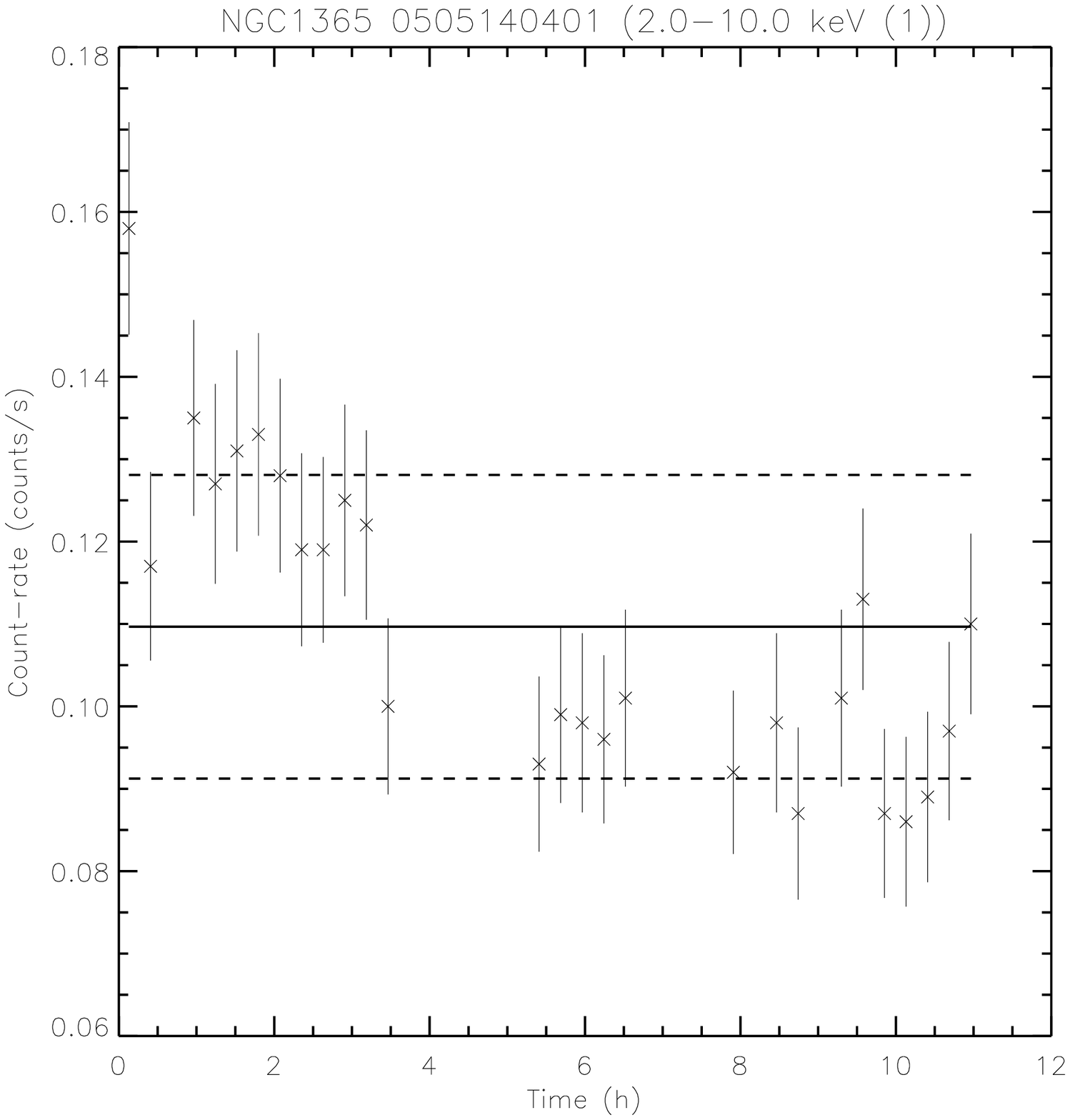}}
{\includegraphics[width=0.30\textwidth]{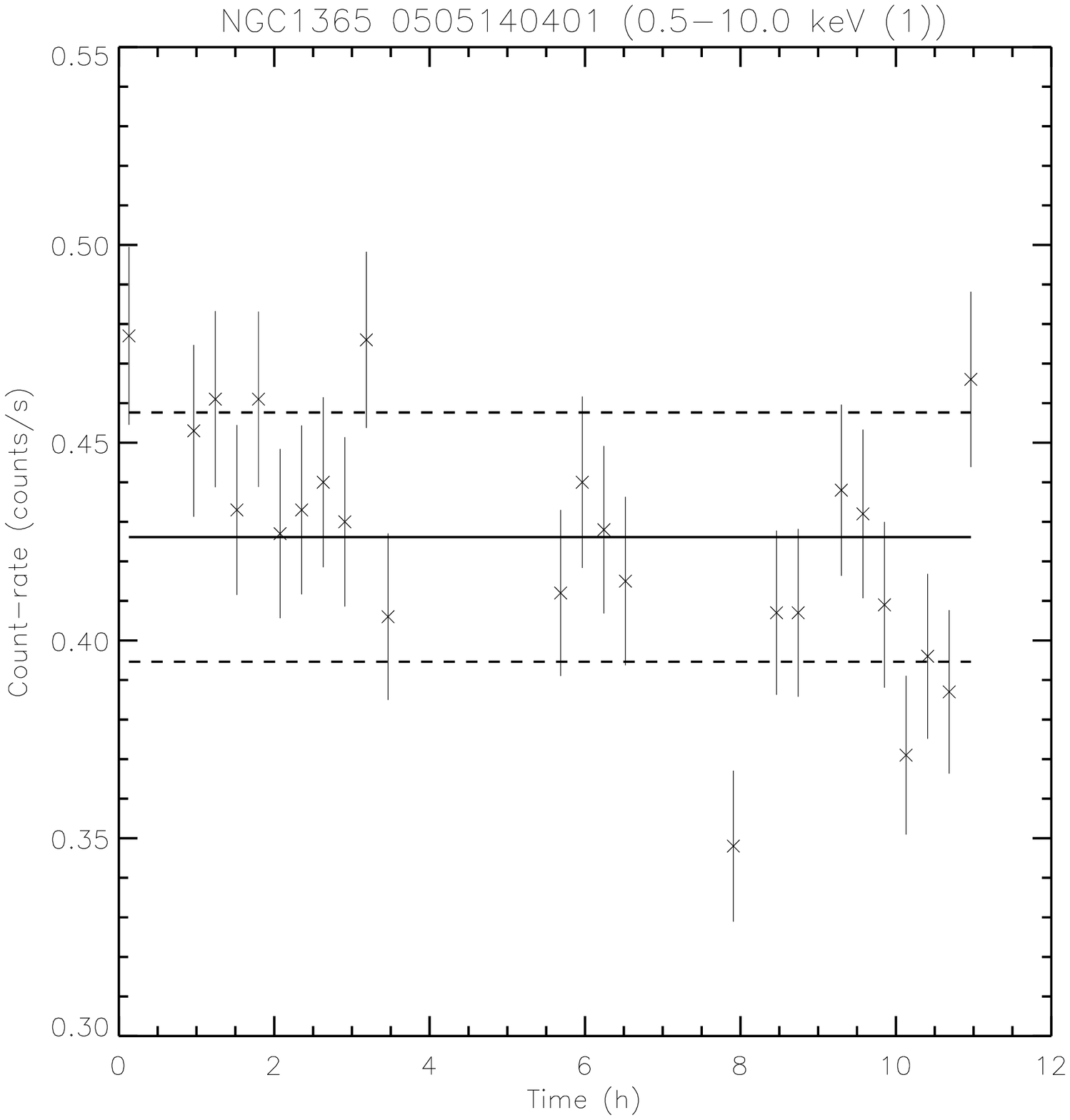}}

{\includegraphics[width=0.30\textwidth]{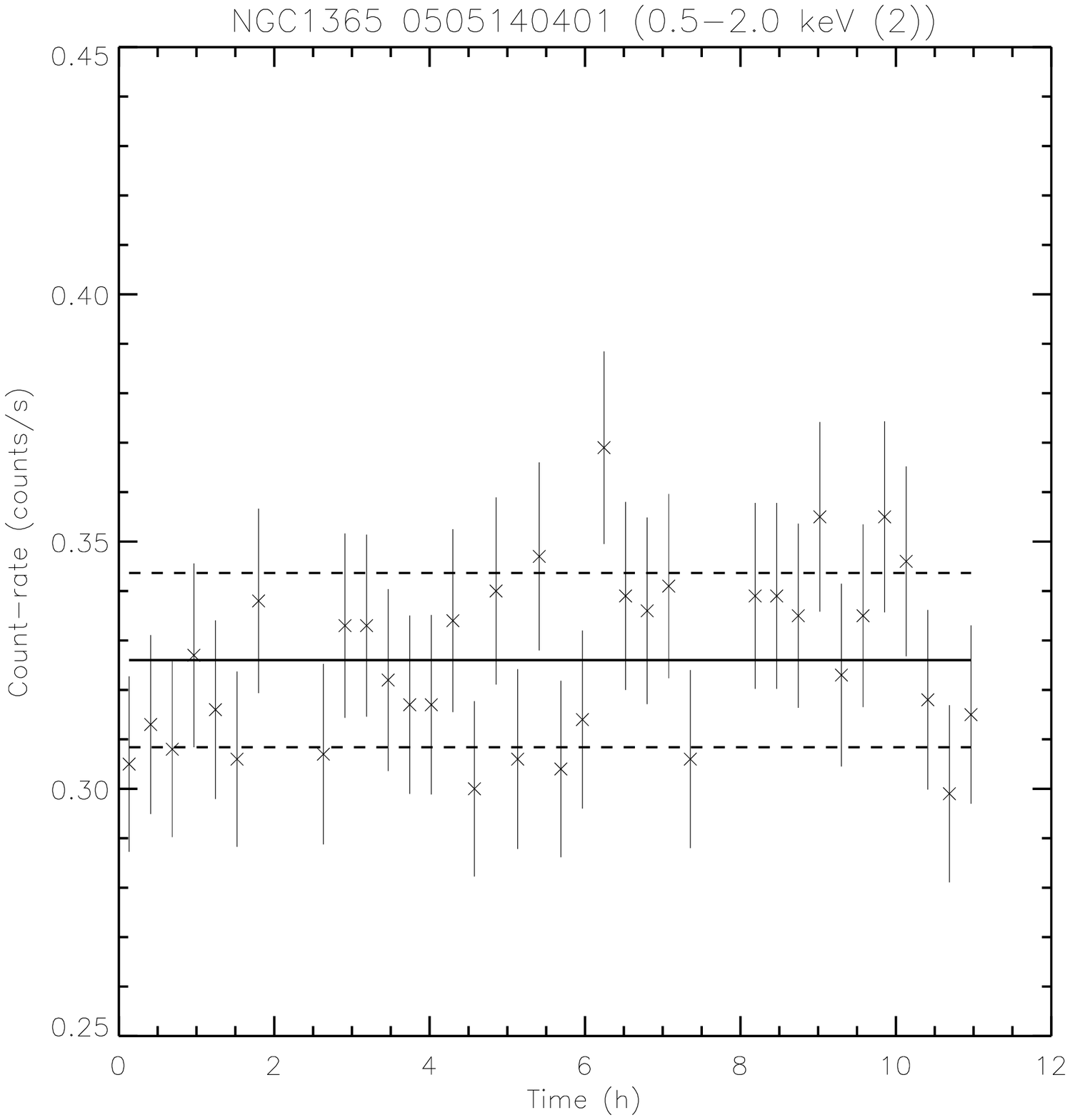}}
{\includegraphics[width=0.30\textwidth]{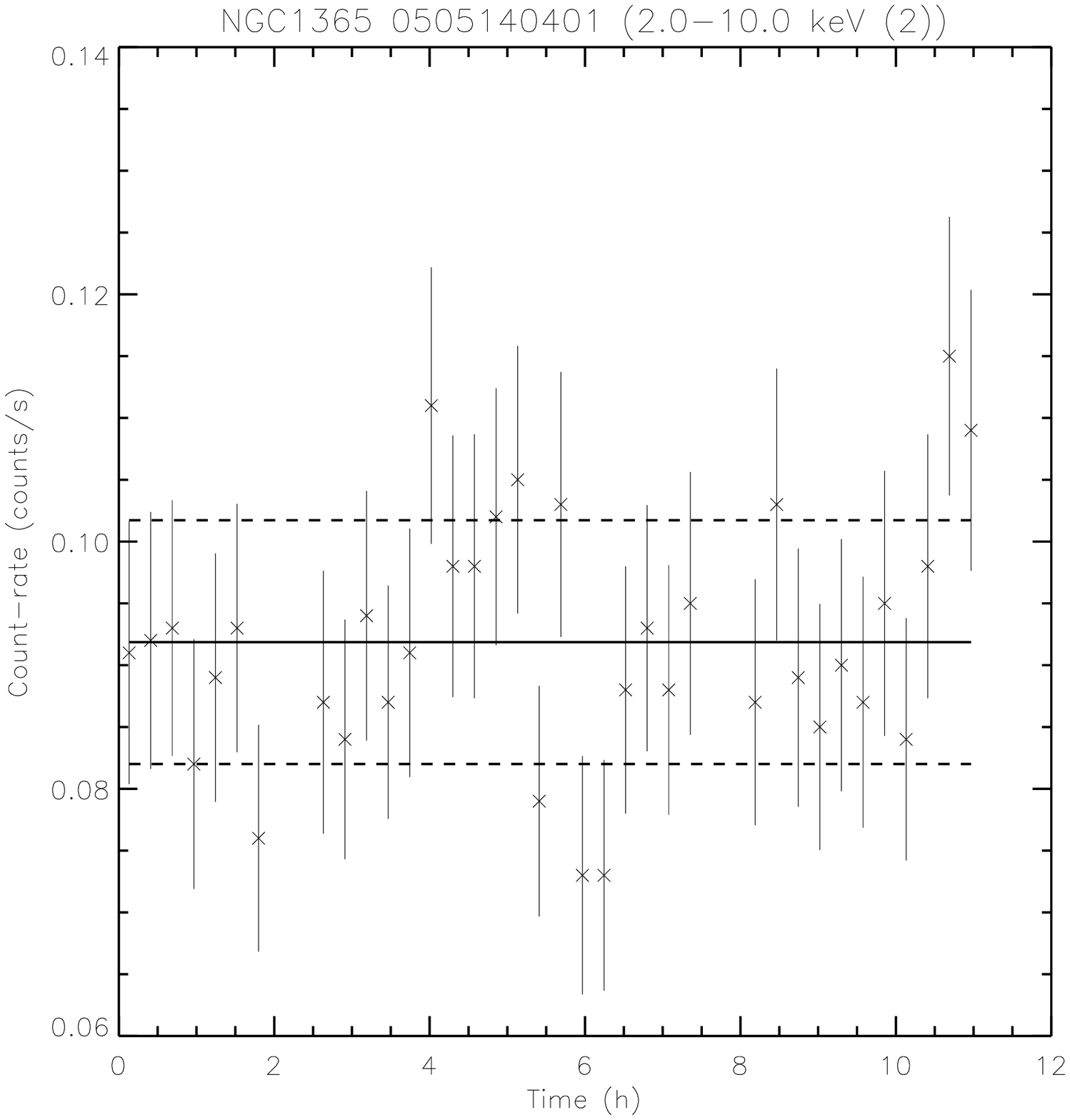}}
{\includegraphics[width=0.30\textwidth]{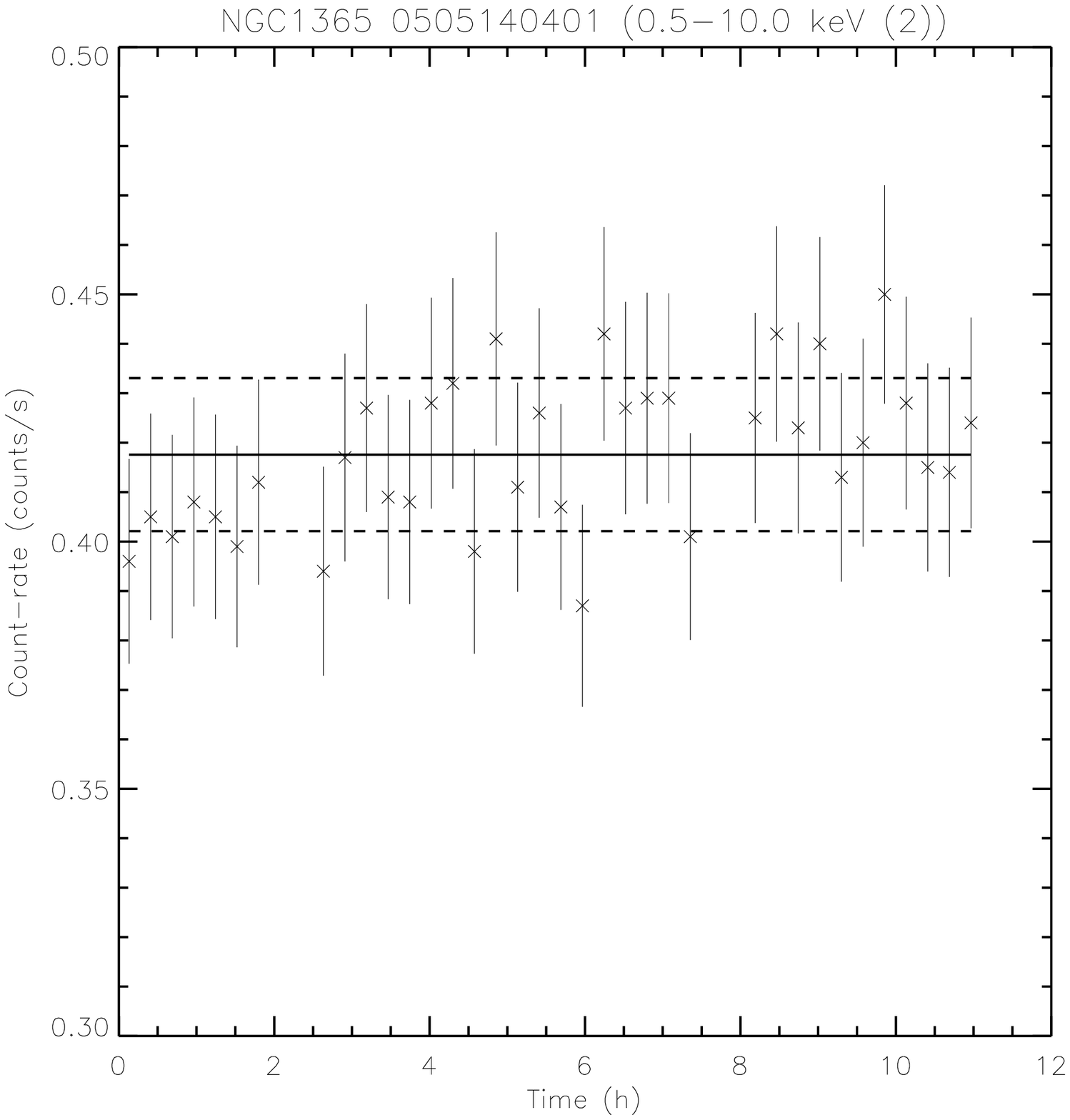}}

\caption{Light curves of NGC\,1365 from \emph{XMM--Newton} data.}
\label{l1365}
\end{figure}

\begin{figure}
\setcounter{figure}{2}
\centering
{\includegraphics[width=0.30\textwidth]{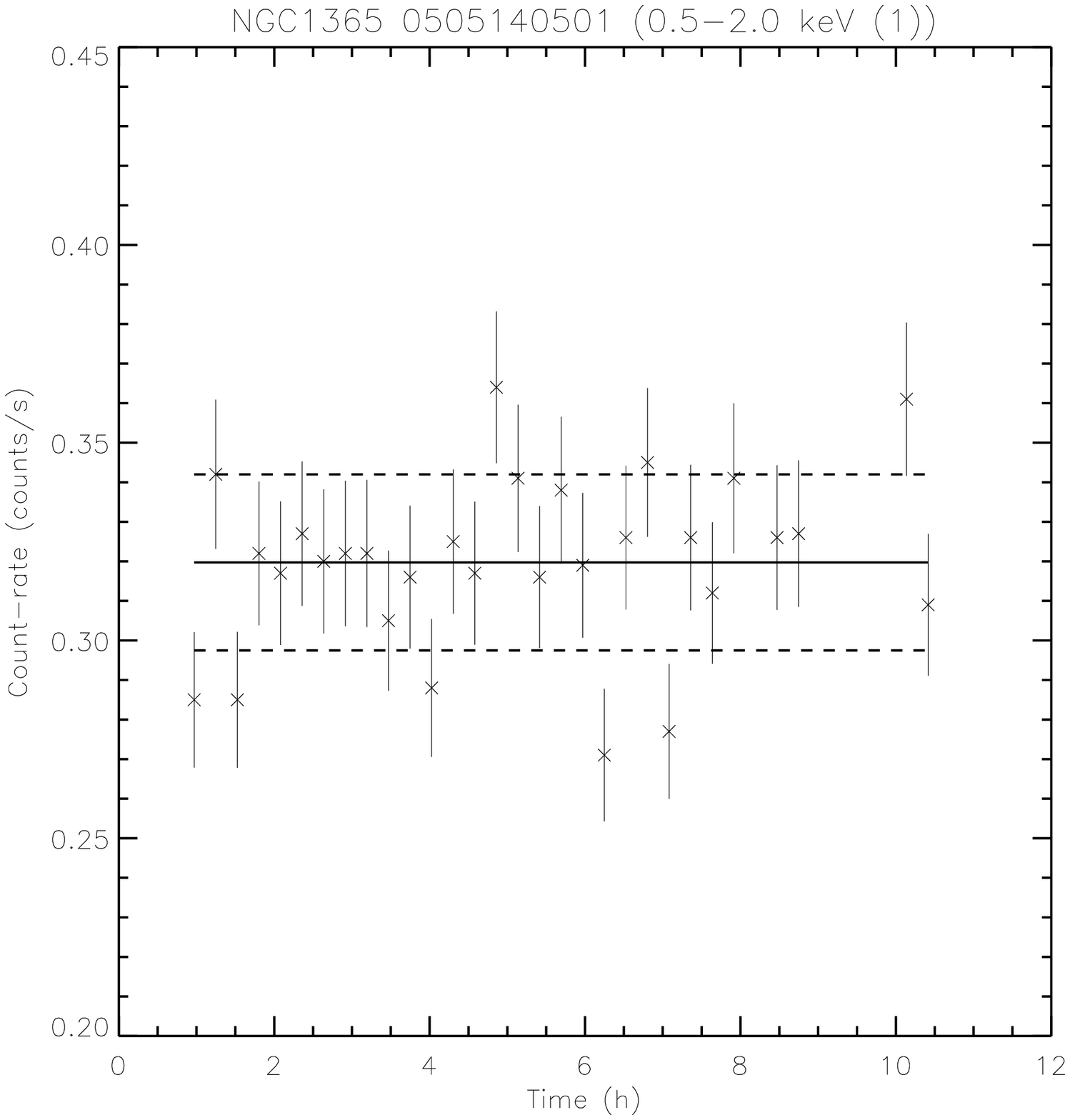}}
{\includegraphics[width=0.30\textwidth]{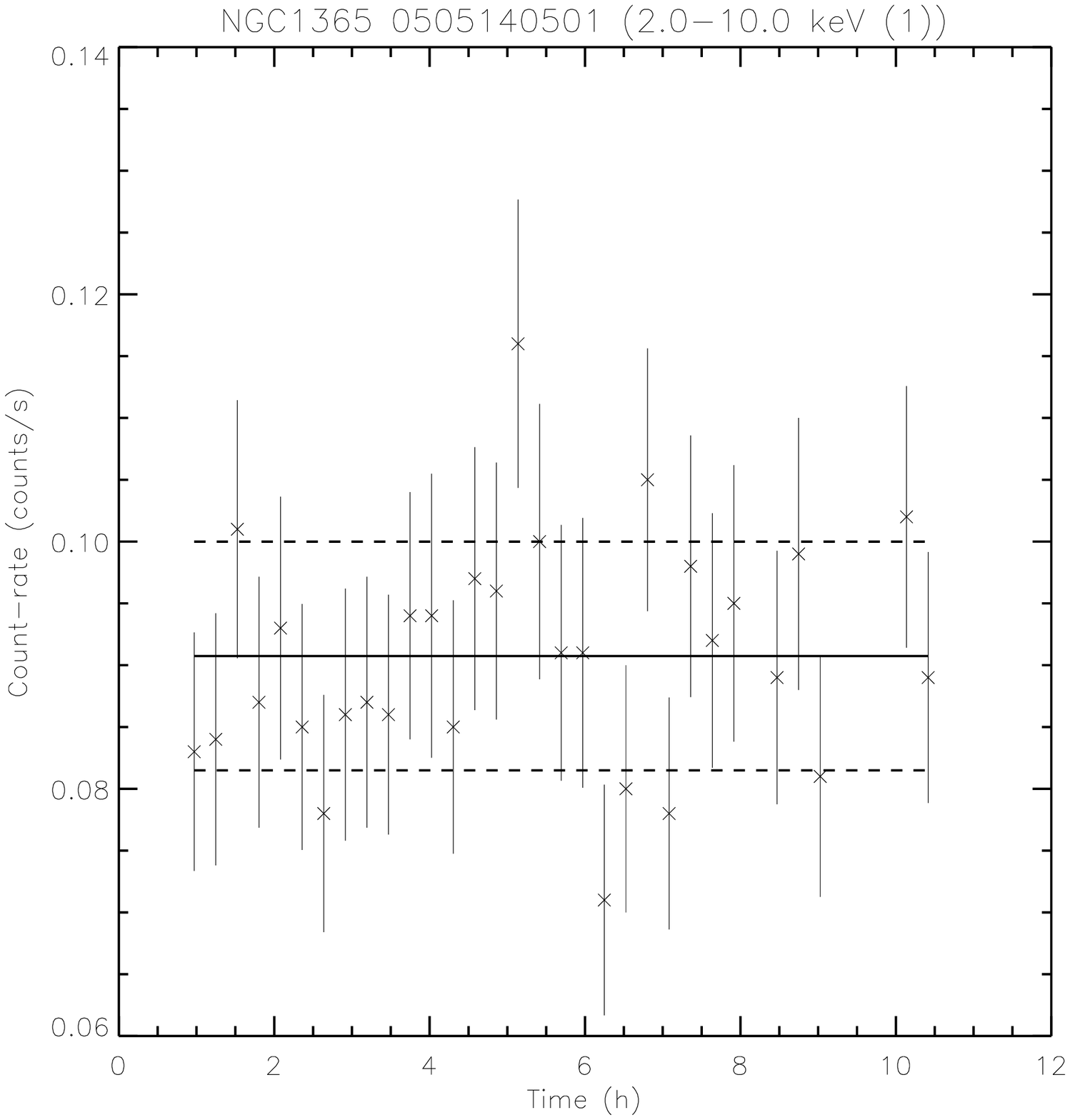}}
{\includegraphics[width=0.30\textwidth]{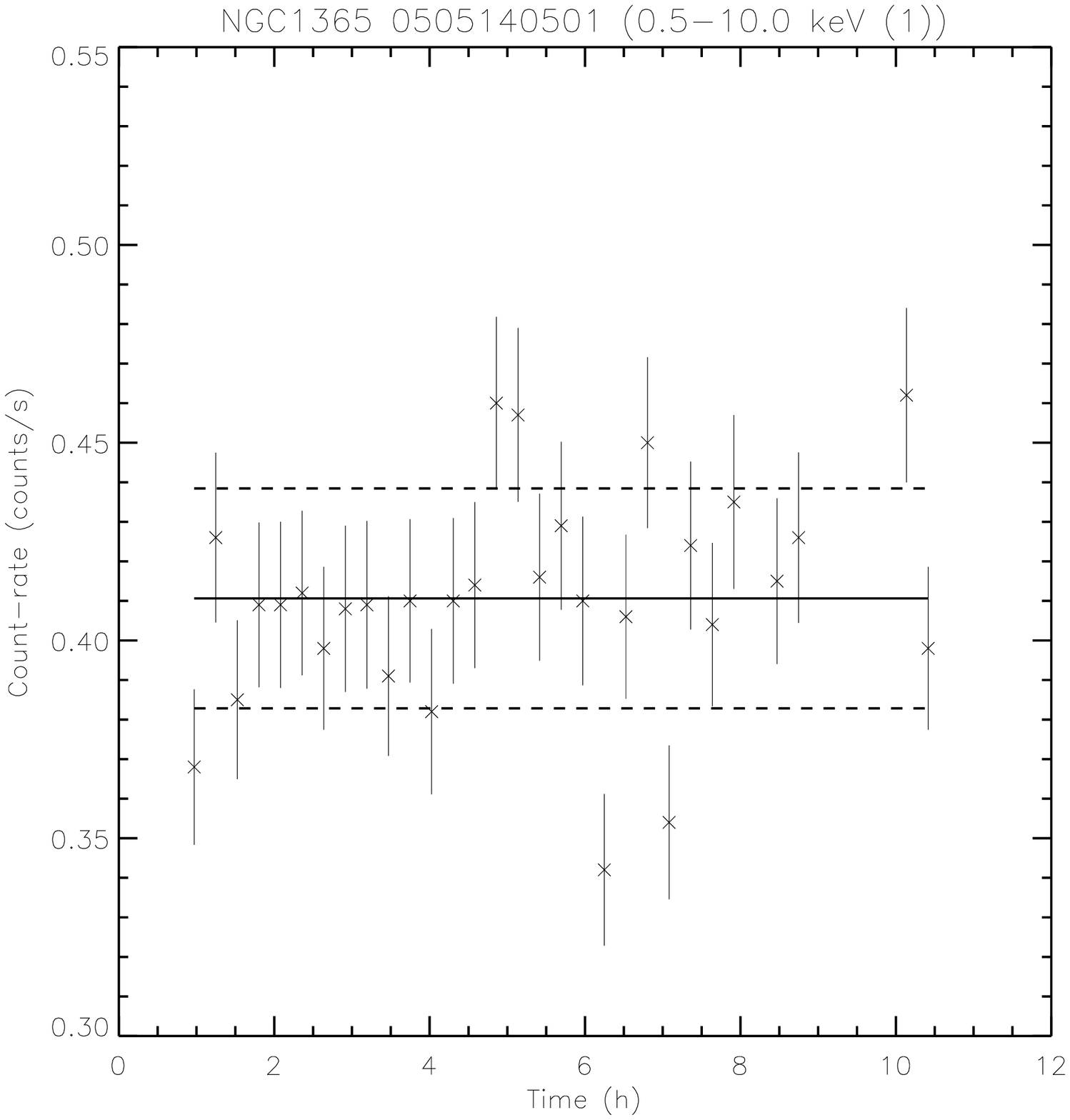}}

{\includegraphics[width=0.30\textwidth]{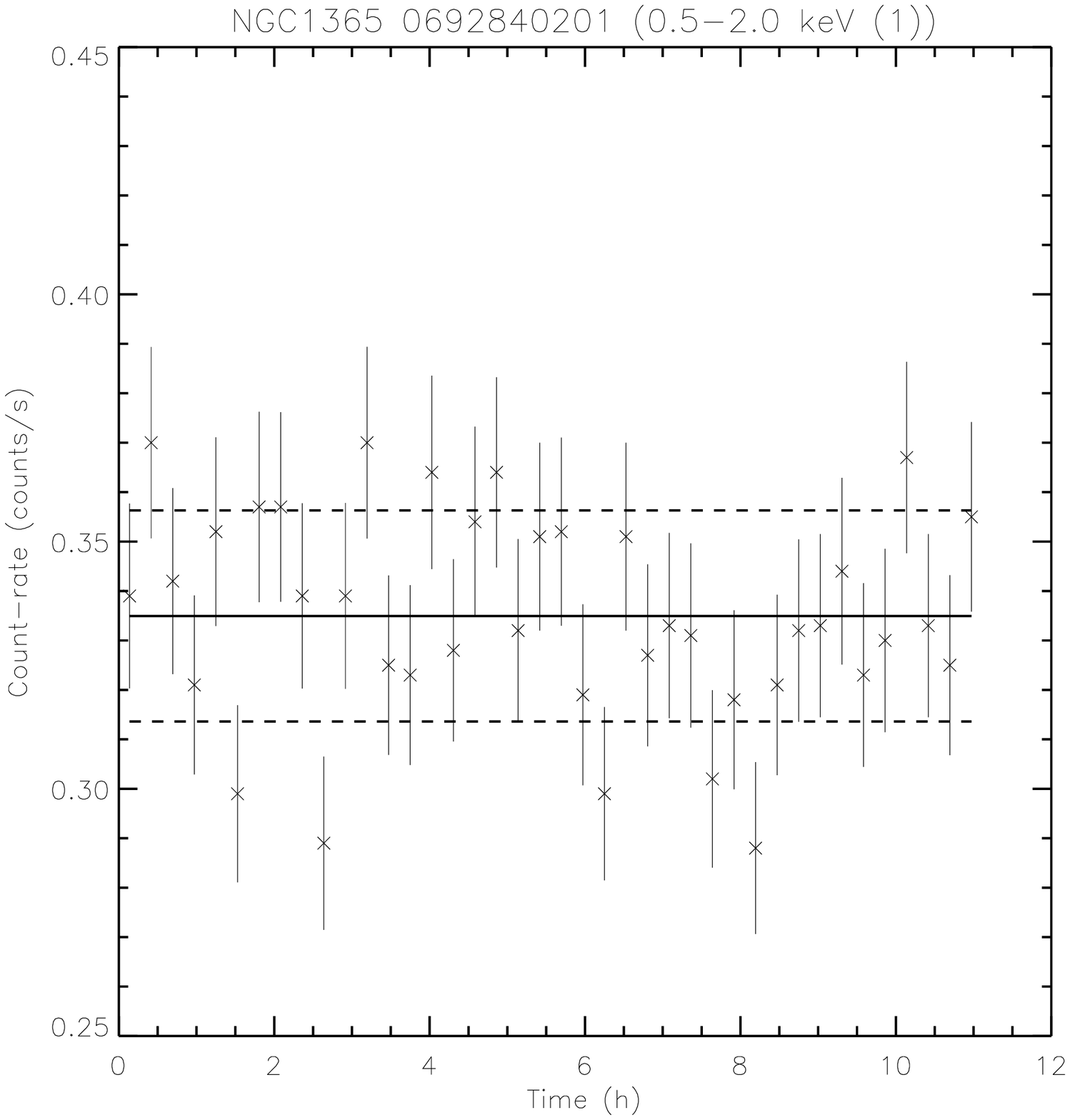}}
{\includegraphics[width=0.30\textwidth]{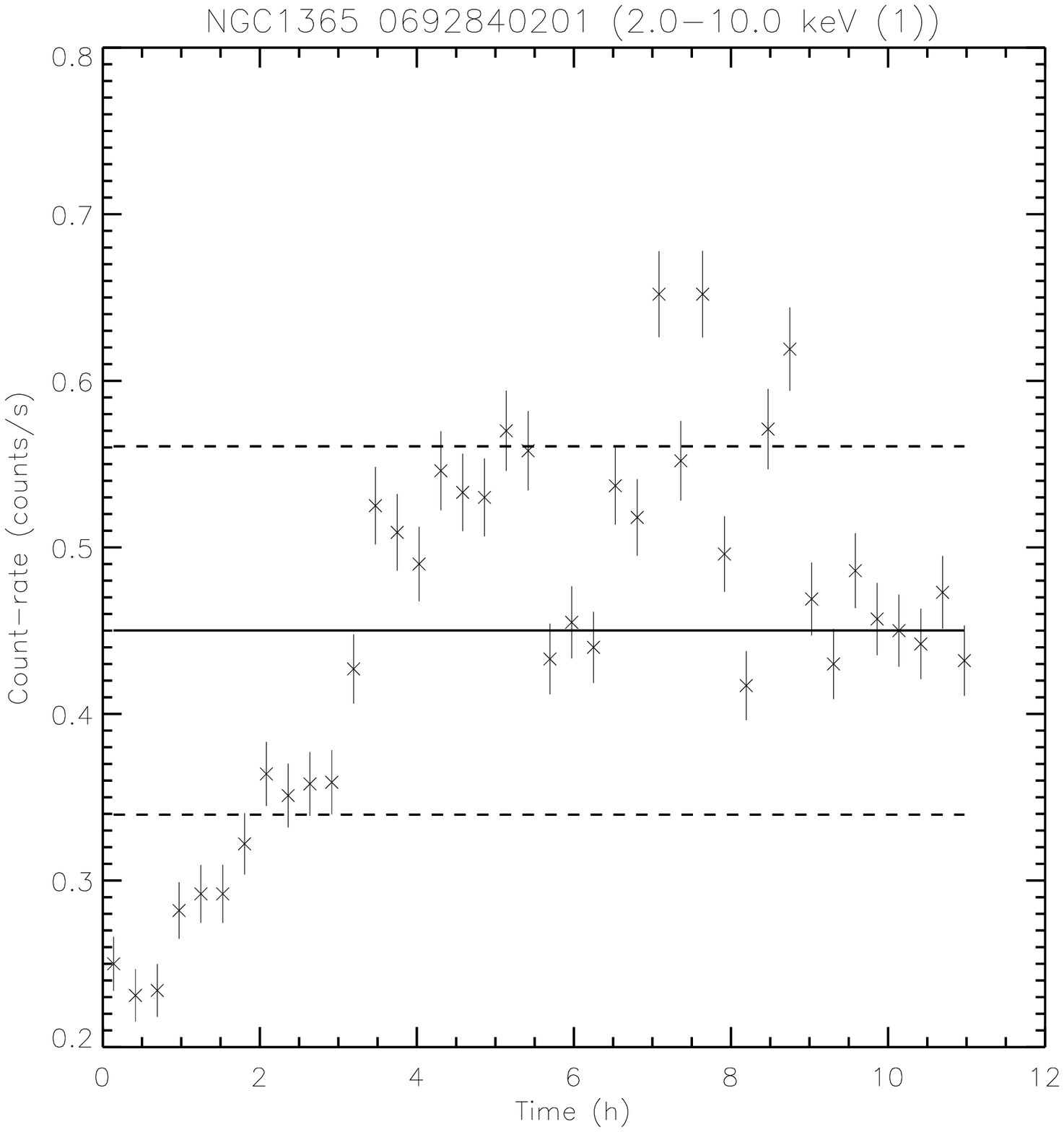}}
{\includegraphics[width=0.30\textwidth]{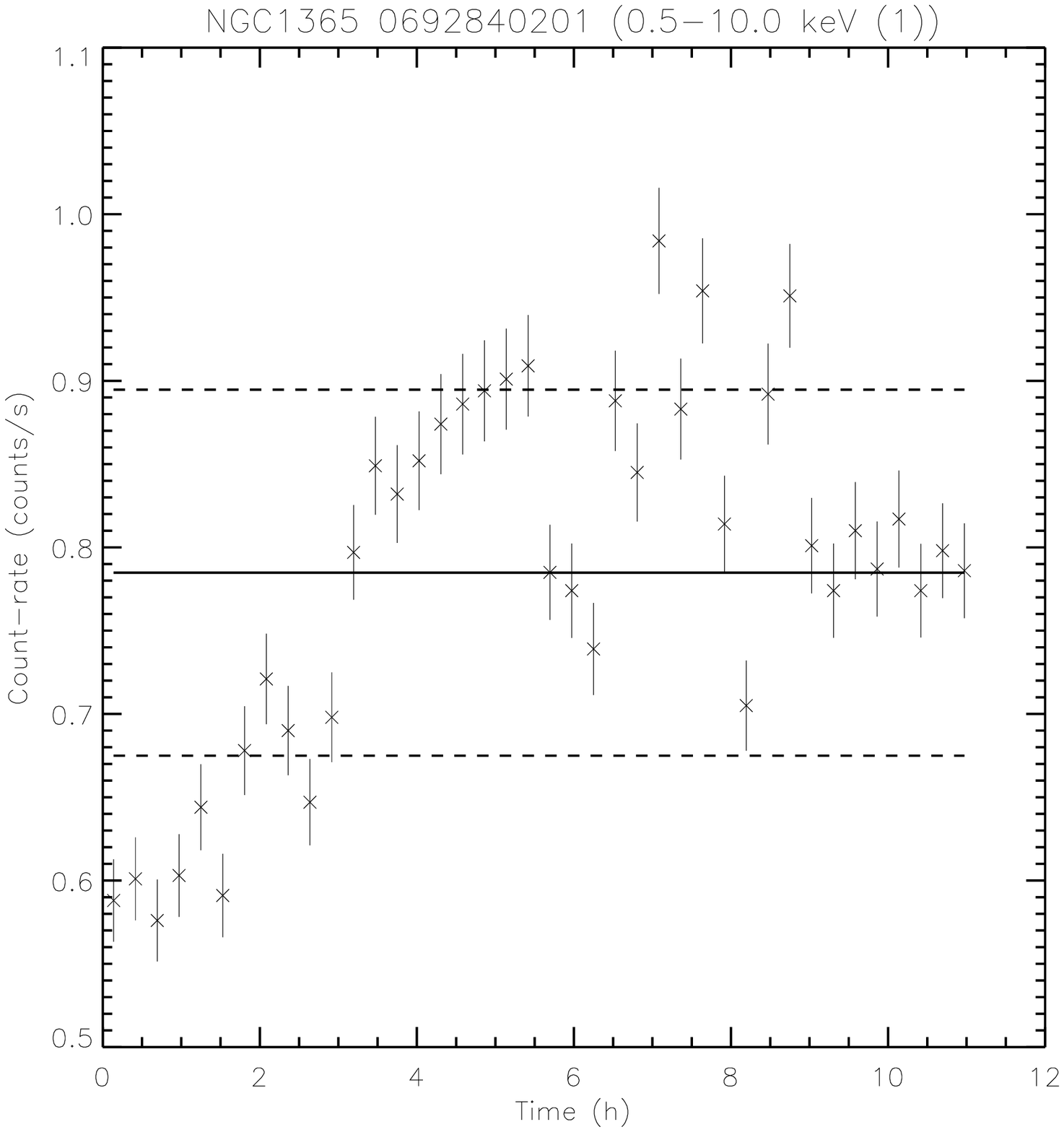}}

{\includegraphics[width=0.30\textwidth]{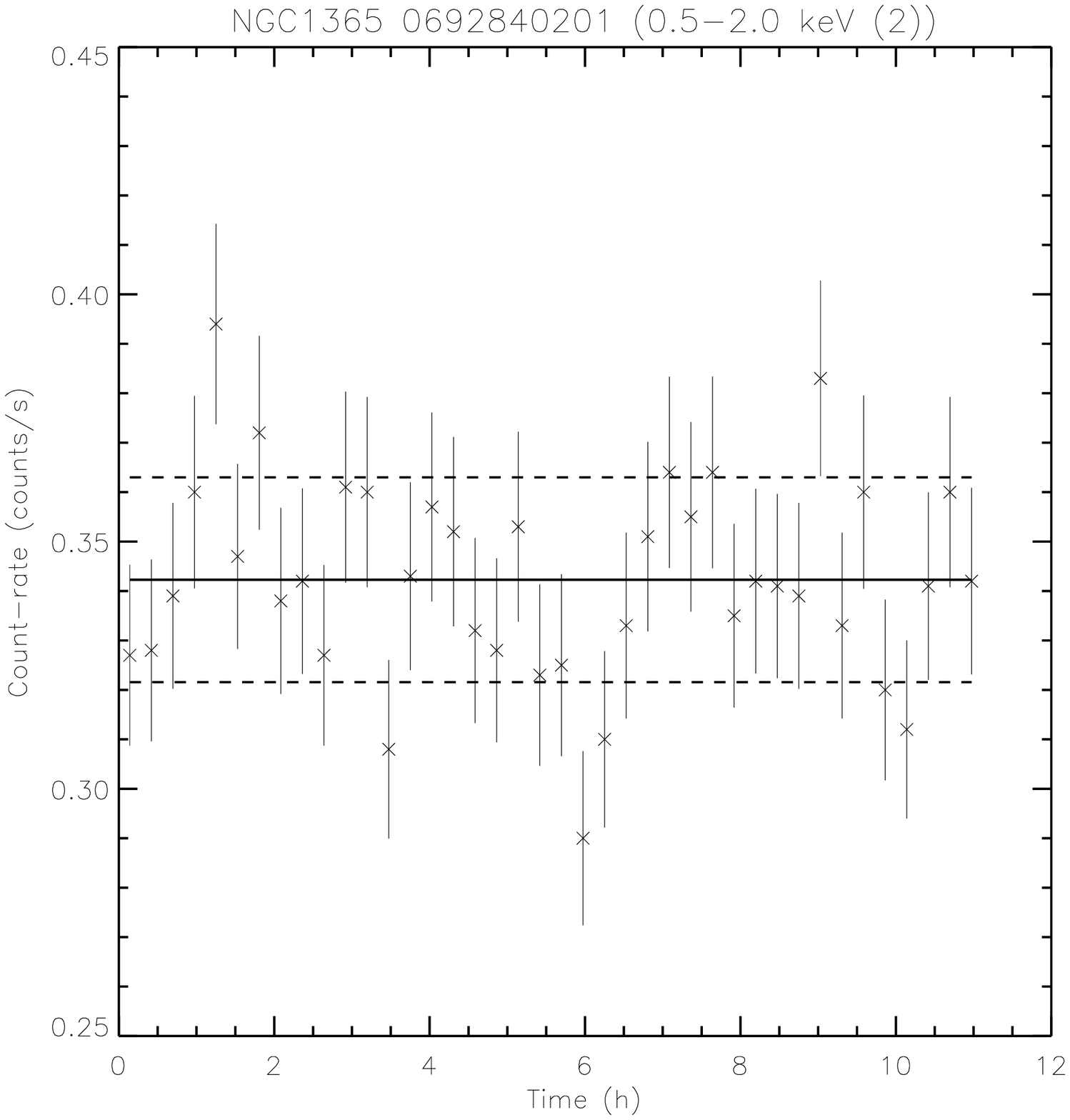}}
{\includegraphics[width=0.30\textwidth]{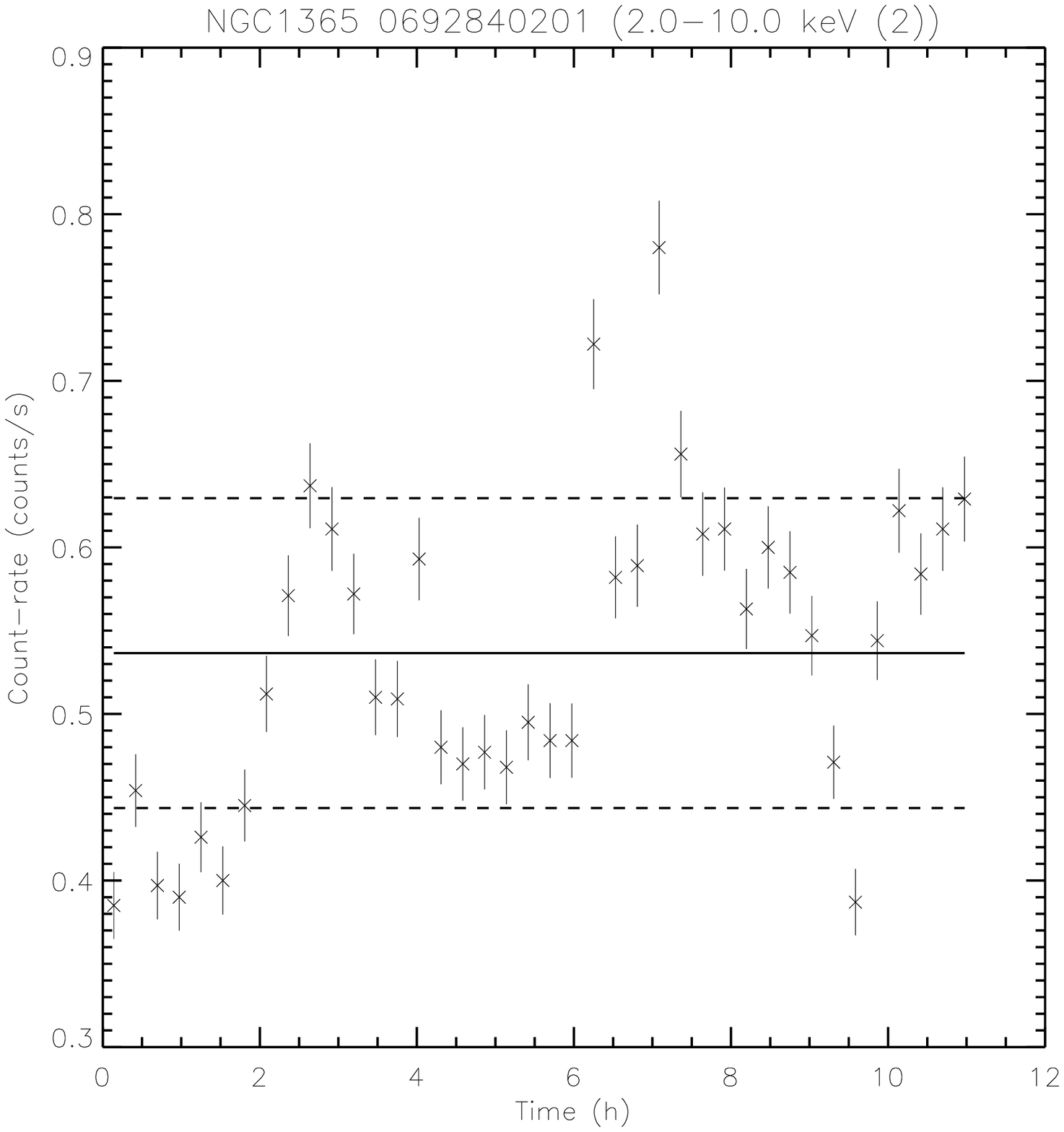}}
{\includegraphics[width=0.30\textwidth]{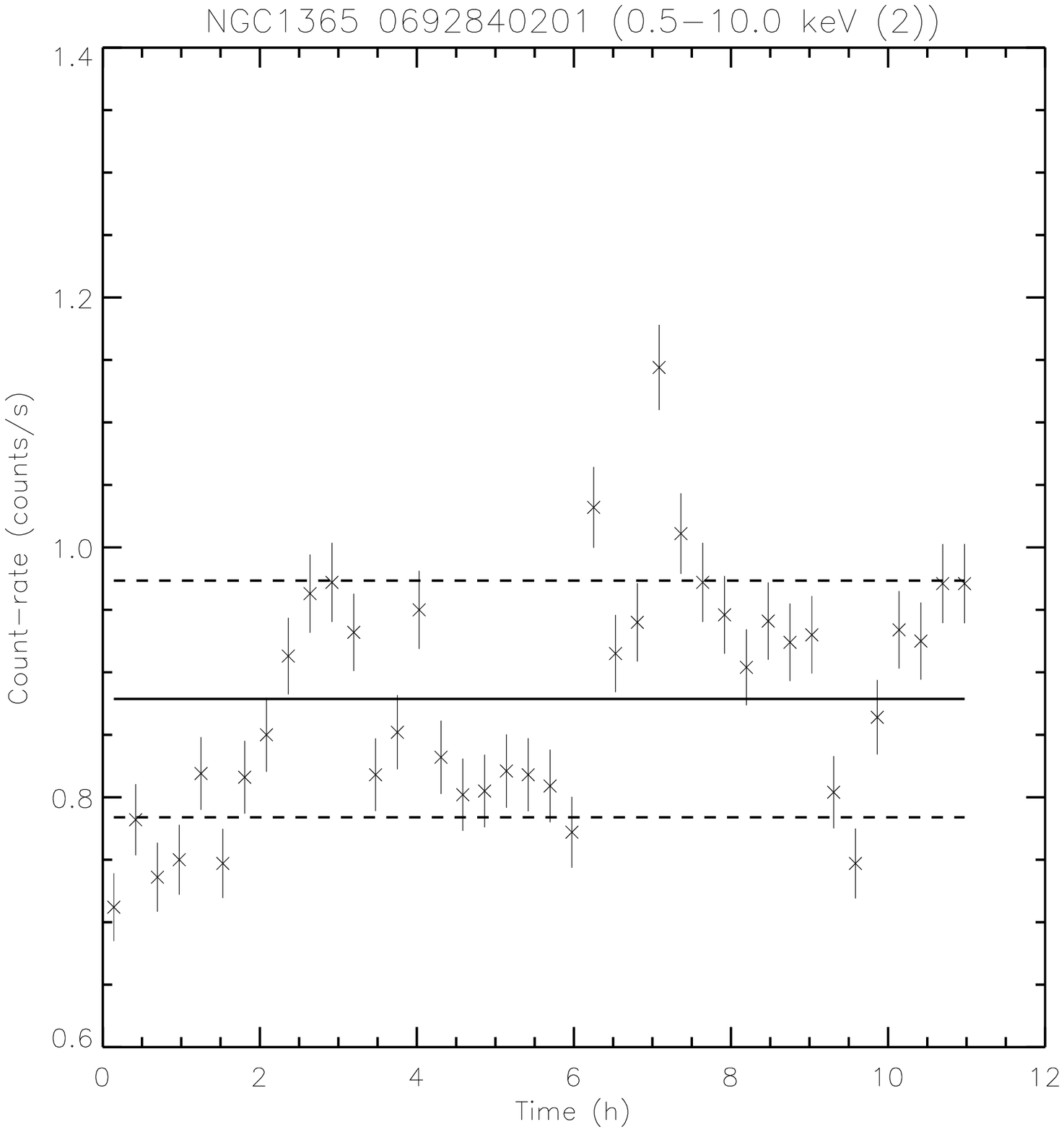}}

{\includegraphics[width=0.30\textwidth]{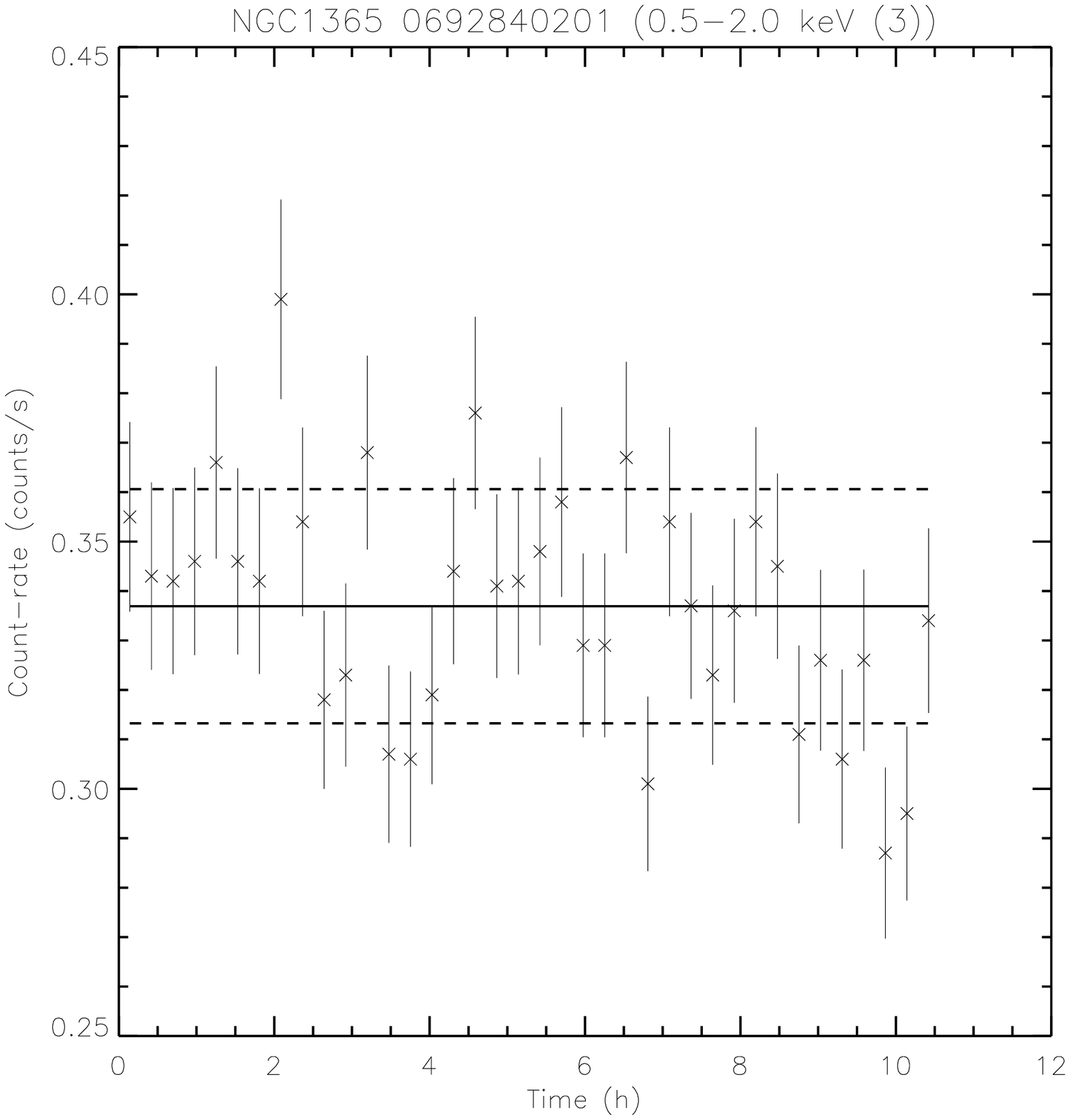}}
{\includegraphics[width=0.30\textwidth]{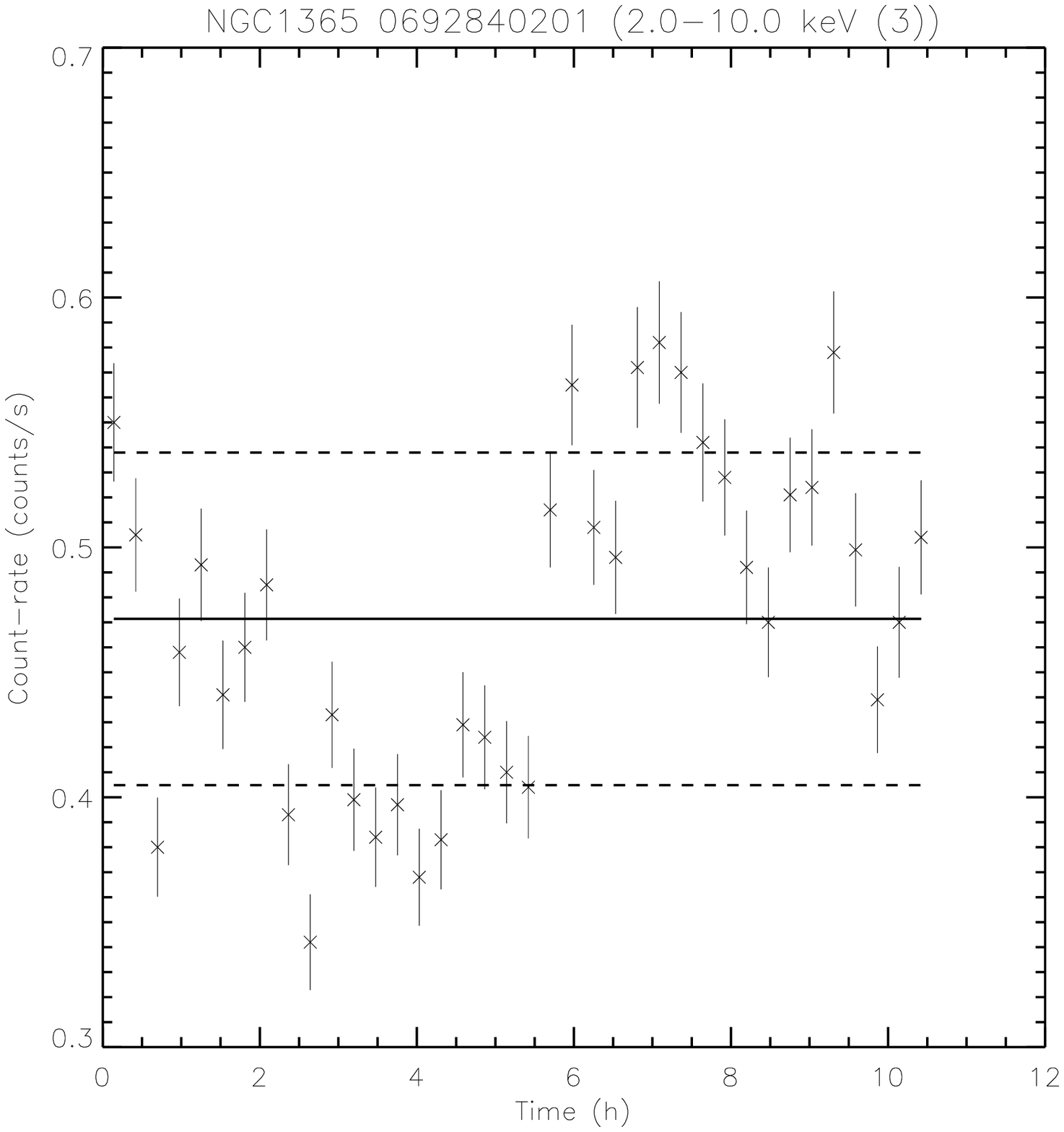}}
{\includegraphics[width=0.30\textwidth]{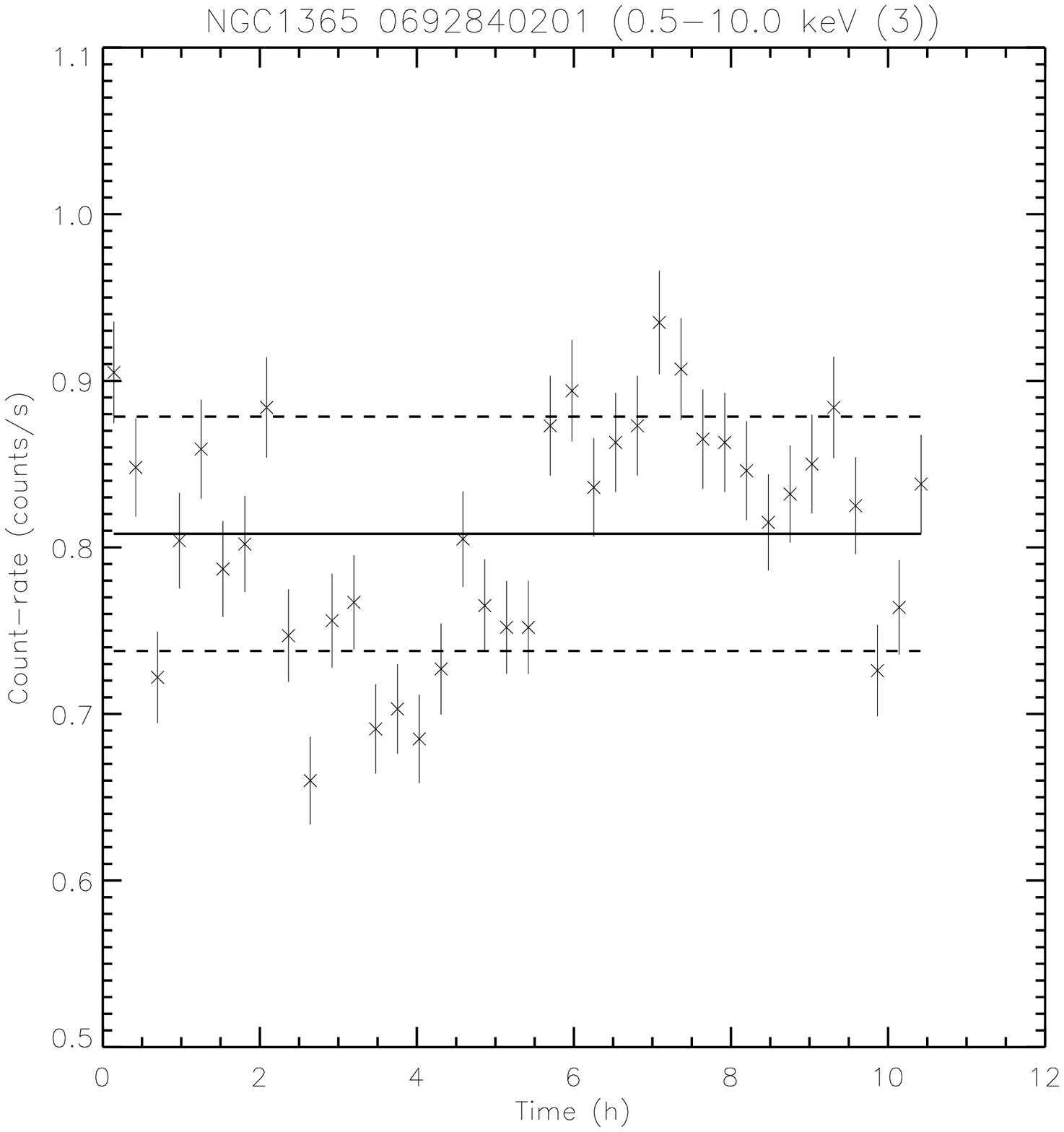}}
\caption{(Cont.)}
\end{figure}

\begin{figure}
\setcounter{figure}{2}
\centering
{\includegraphics[width=0.30\textwidth]{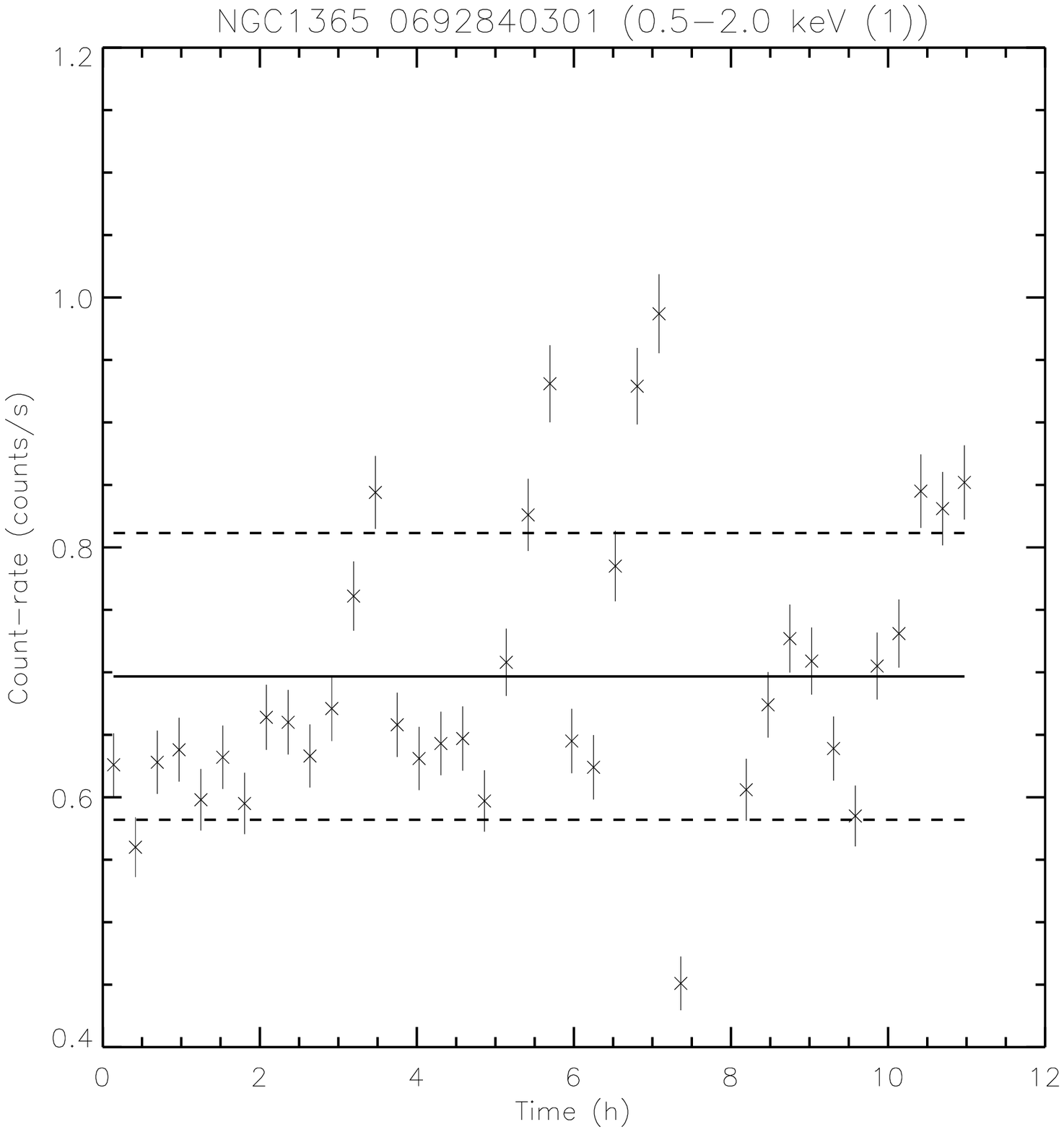}}
{\includegraphics[width=0.30\textwidth]{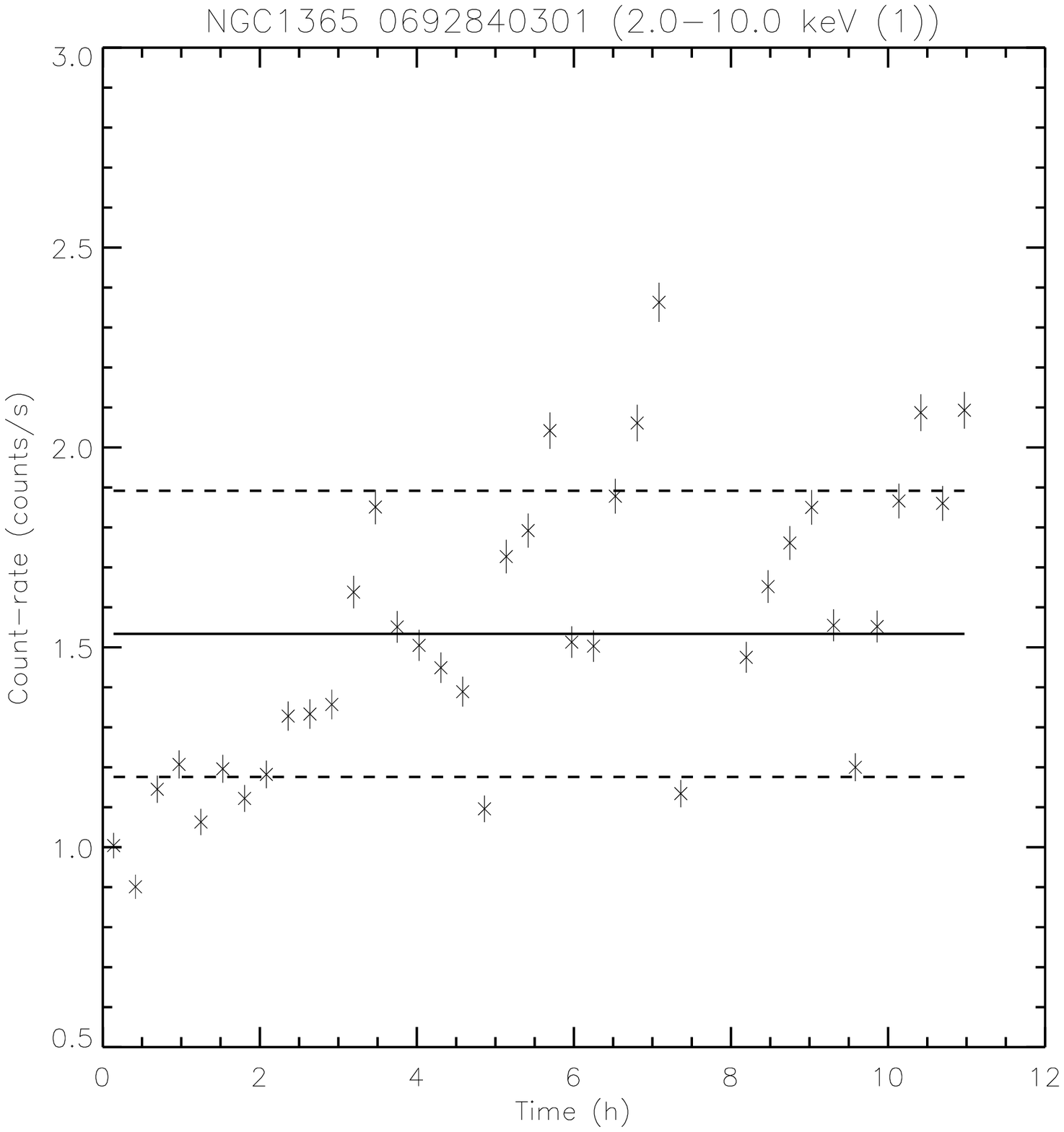}}
{\includegraphics[width=0.30\textwidth]{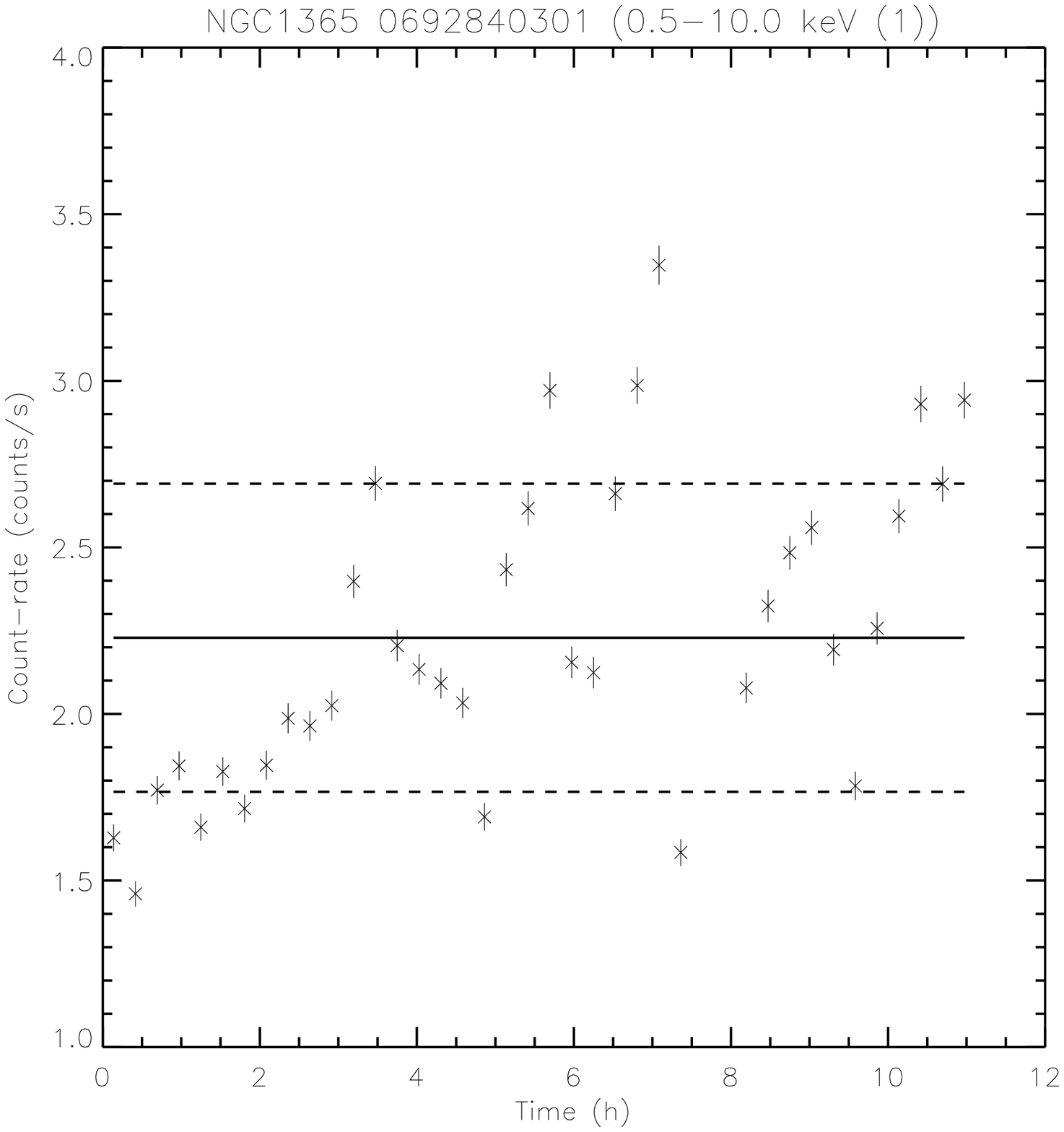}}

{\includegraphics[width=0.30\textwidth]{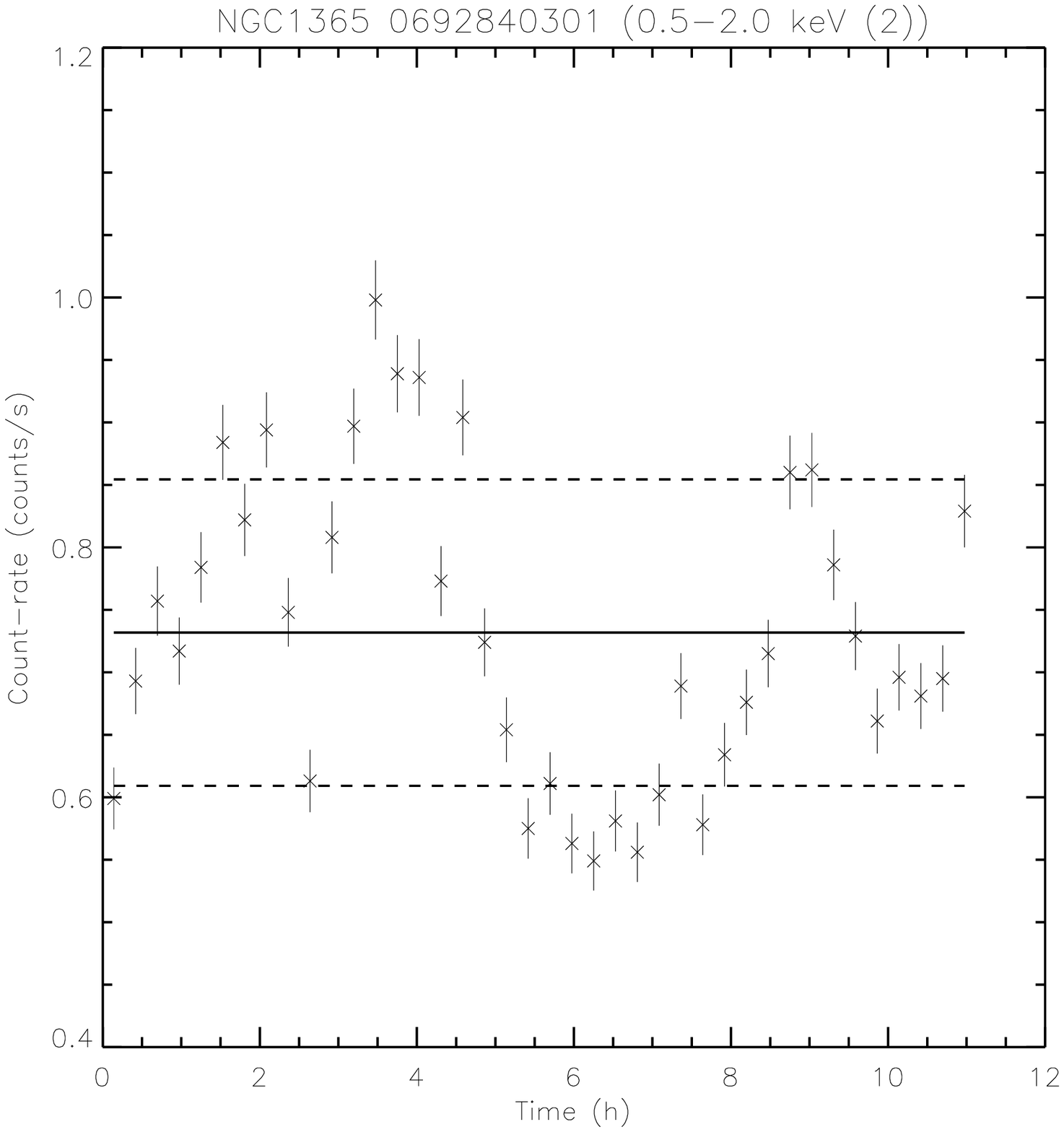}}
{\includegraphics[width=0.30\textwidth]{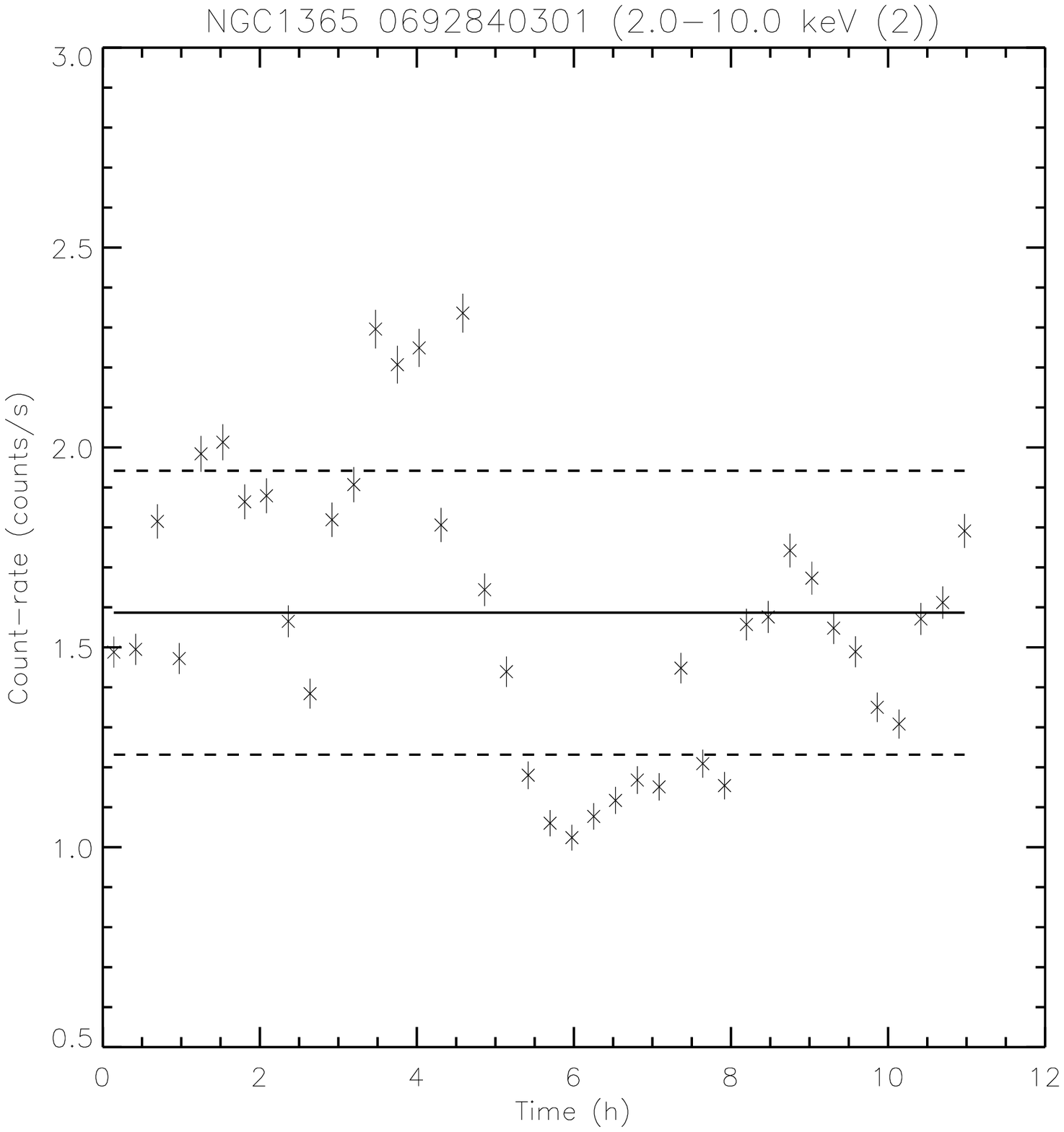}}
{\includegraphics[width=0.30\textwidth]{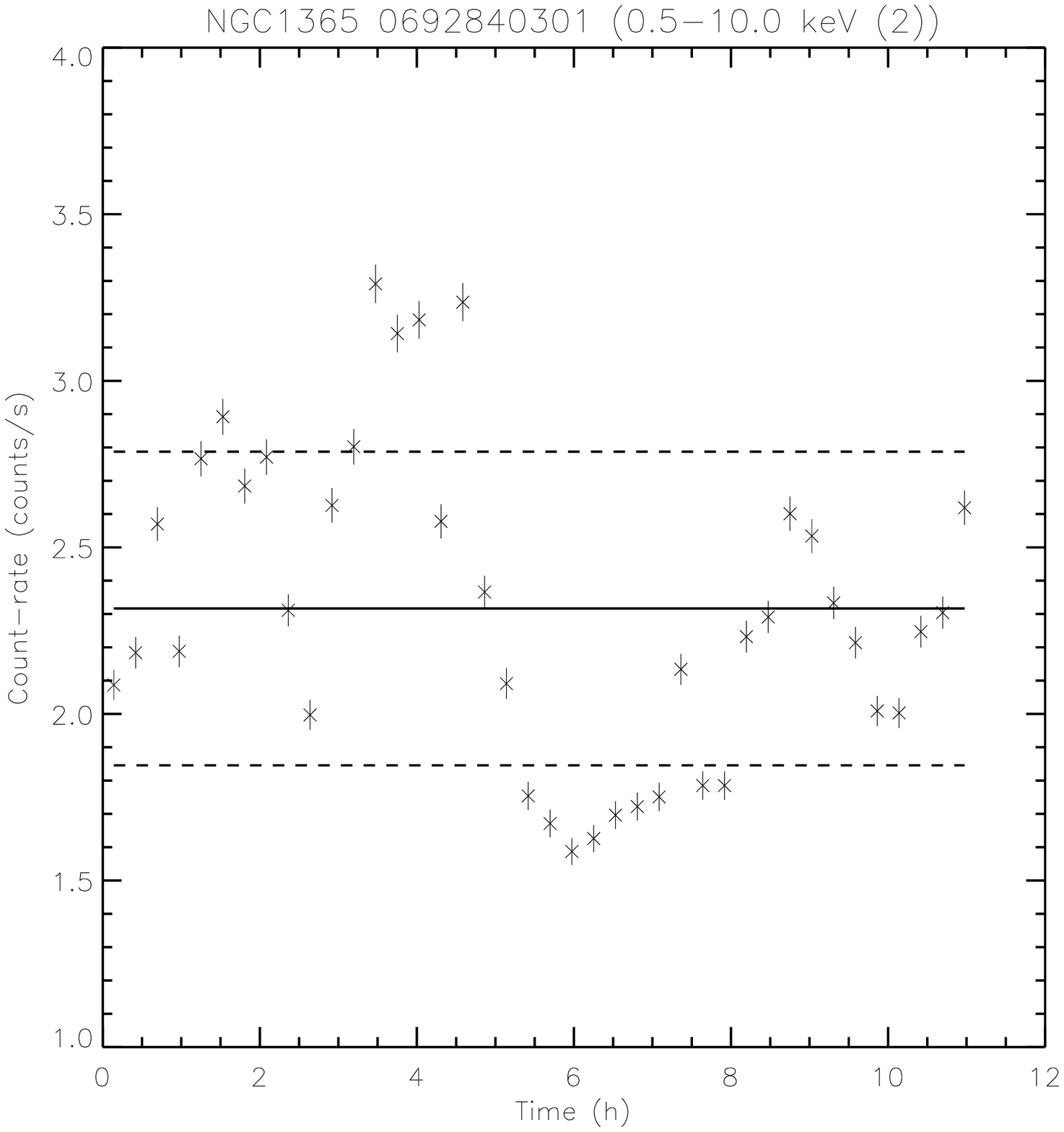}}

{\includegraphics[width=0.30\textwidth]{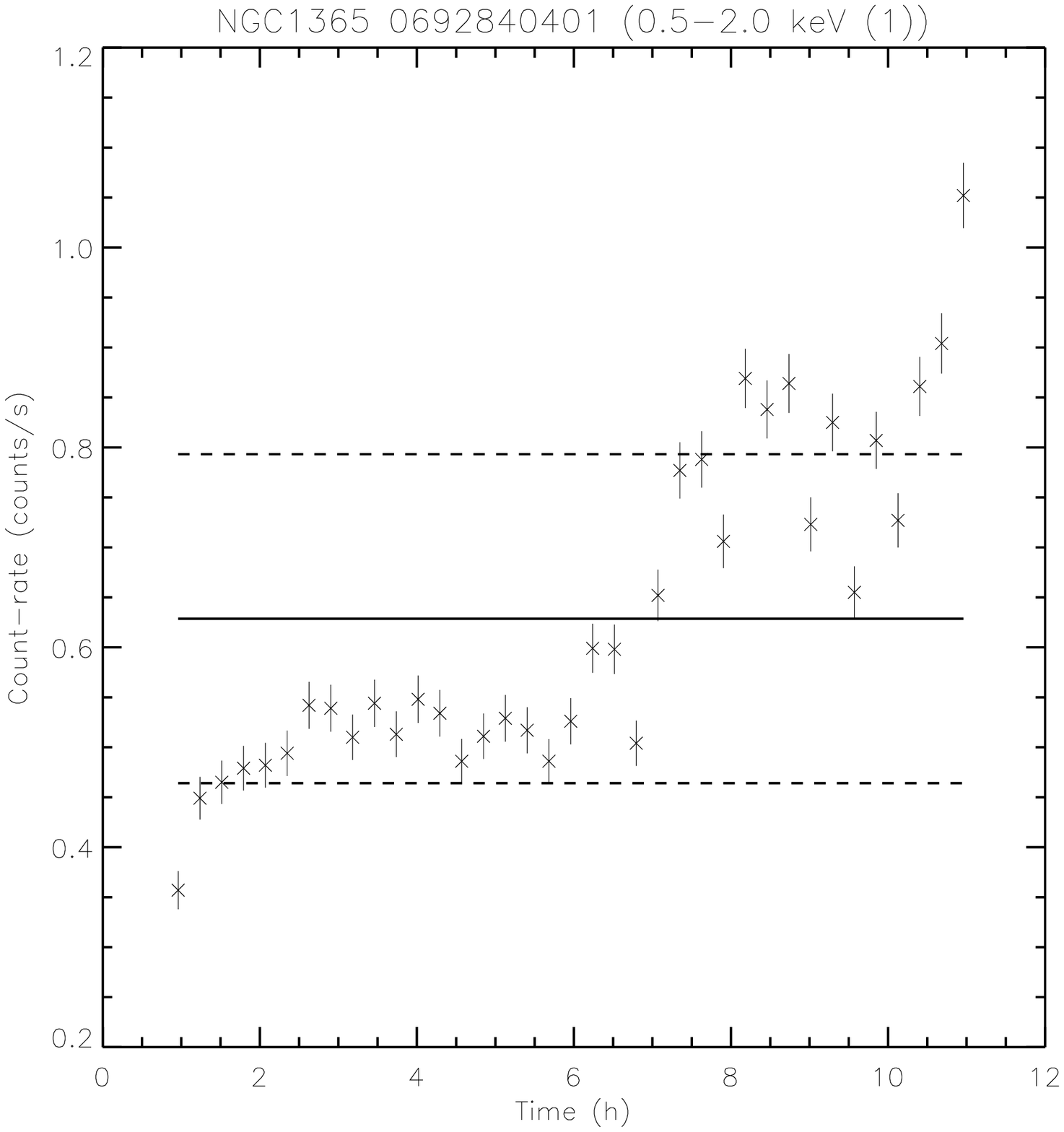}}
{\includegraphics[width=0.30\textwidth]{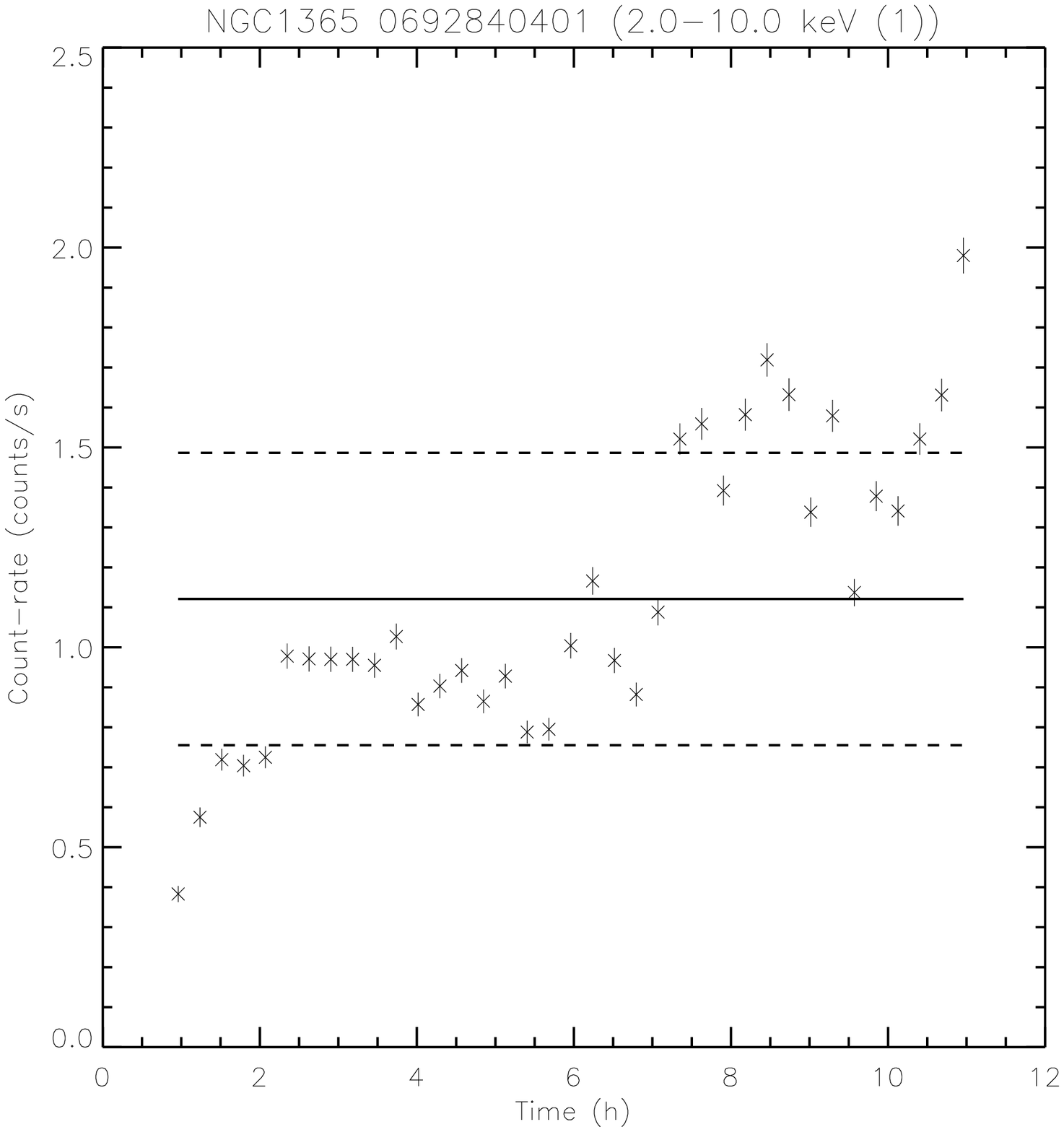}}
{\includegraphics[width=0.30\textwidth]{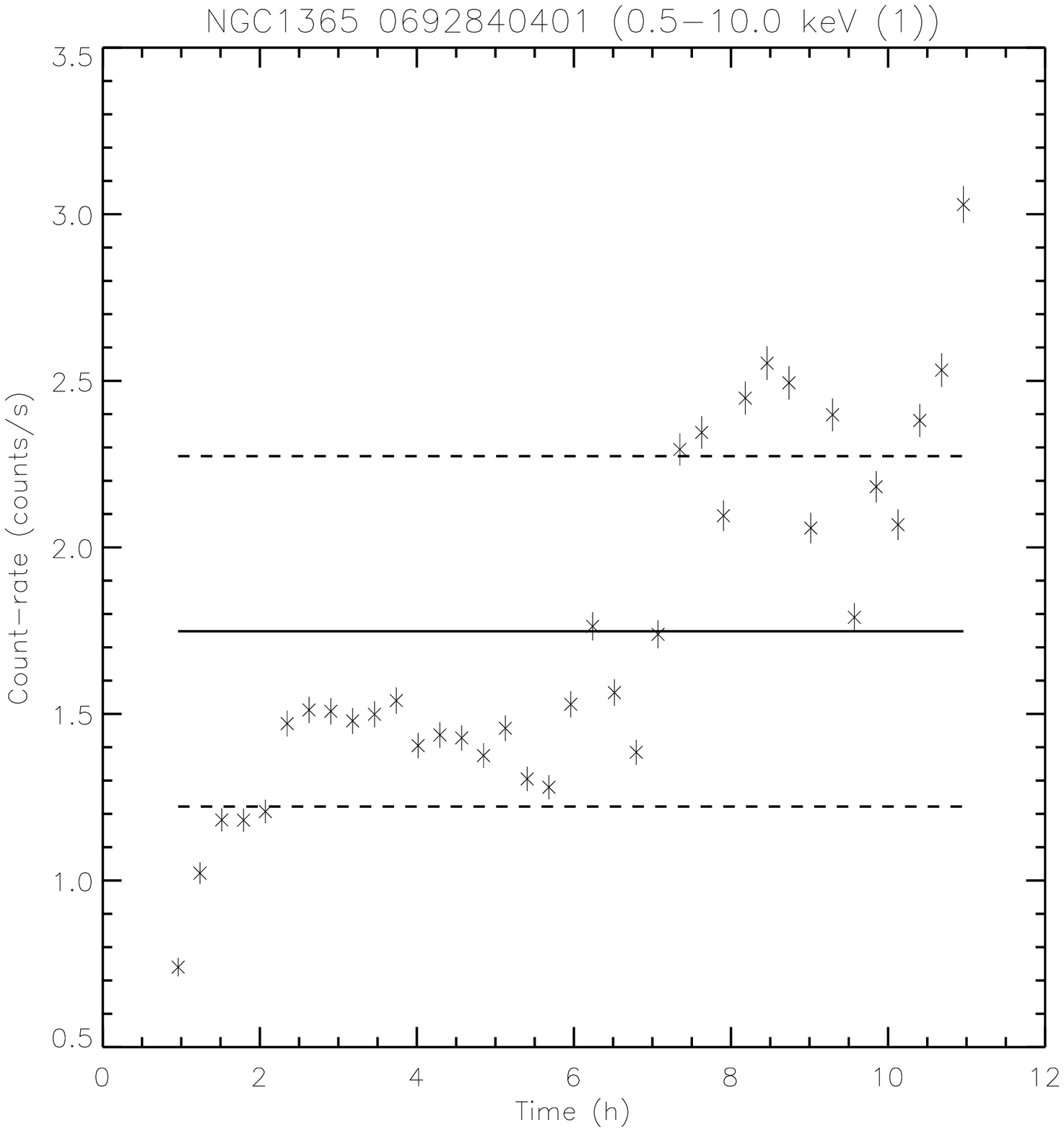}}

{\includegraphics[width=0.30\textwidth]{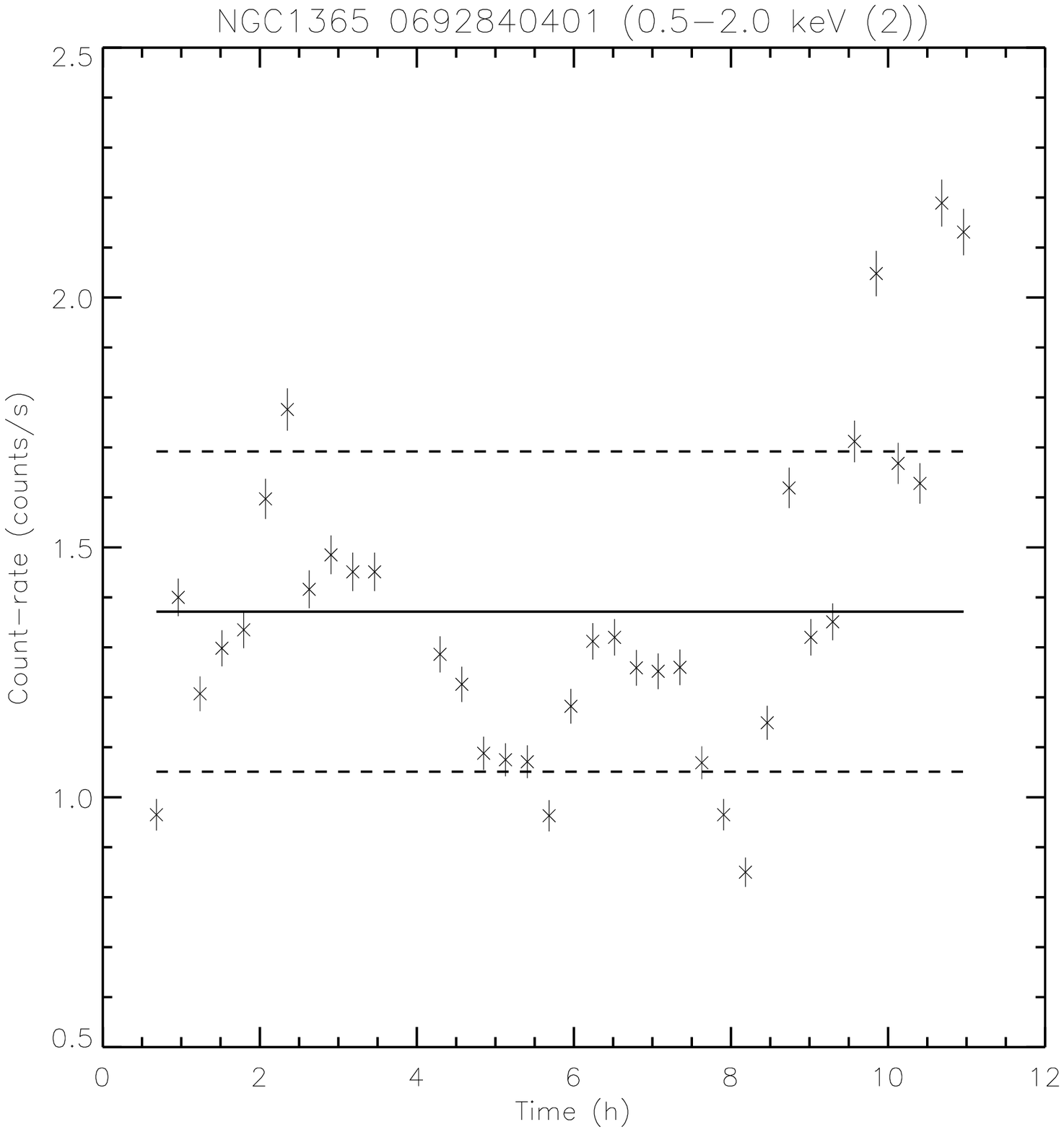}}
{\includegraphics[width=0.30\textwidth]{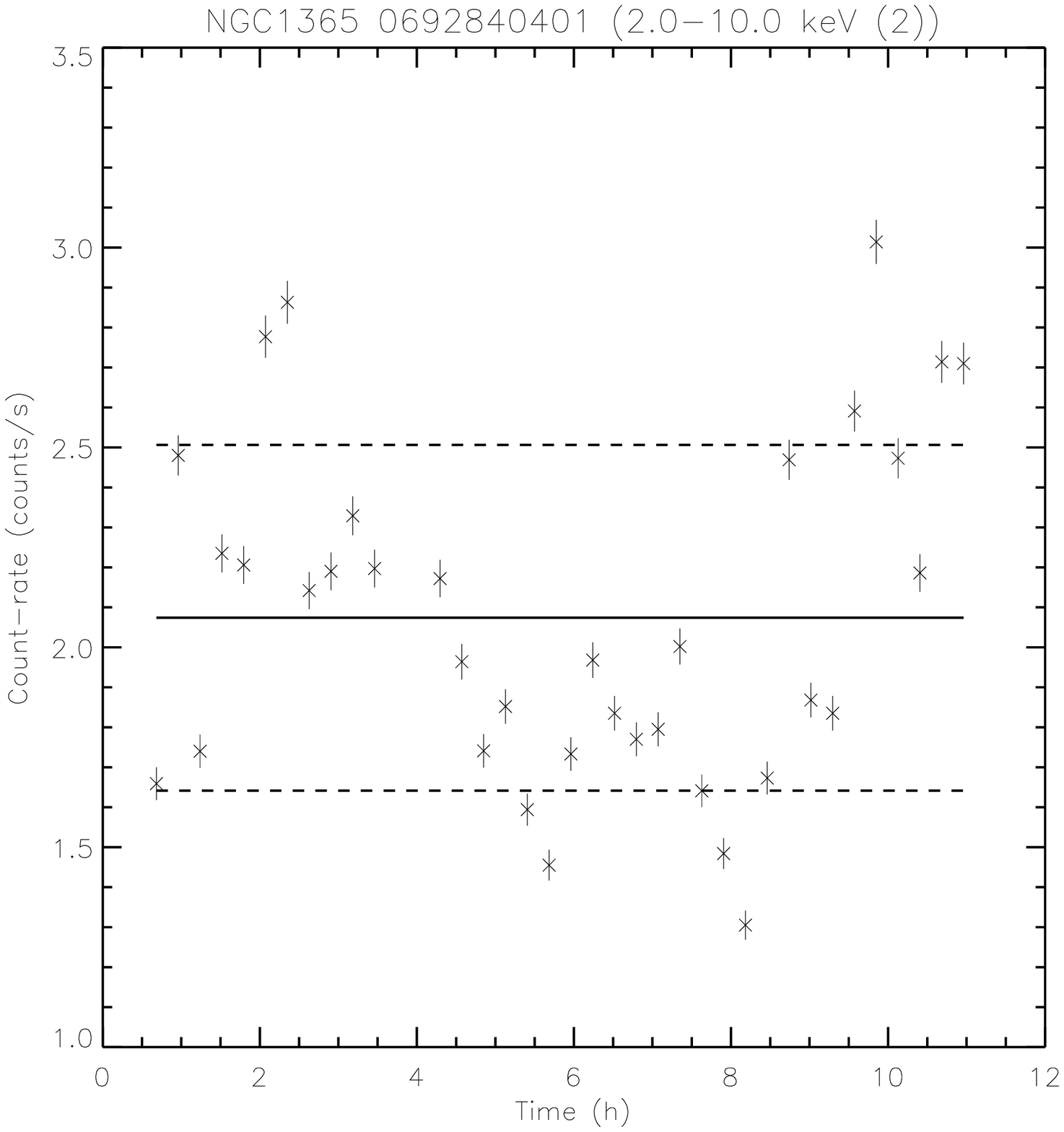}}
{\includegraphics[width=0.30\textwidth]{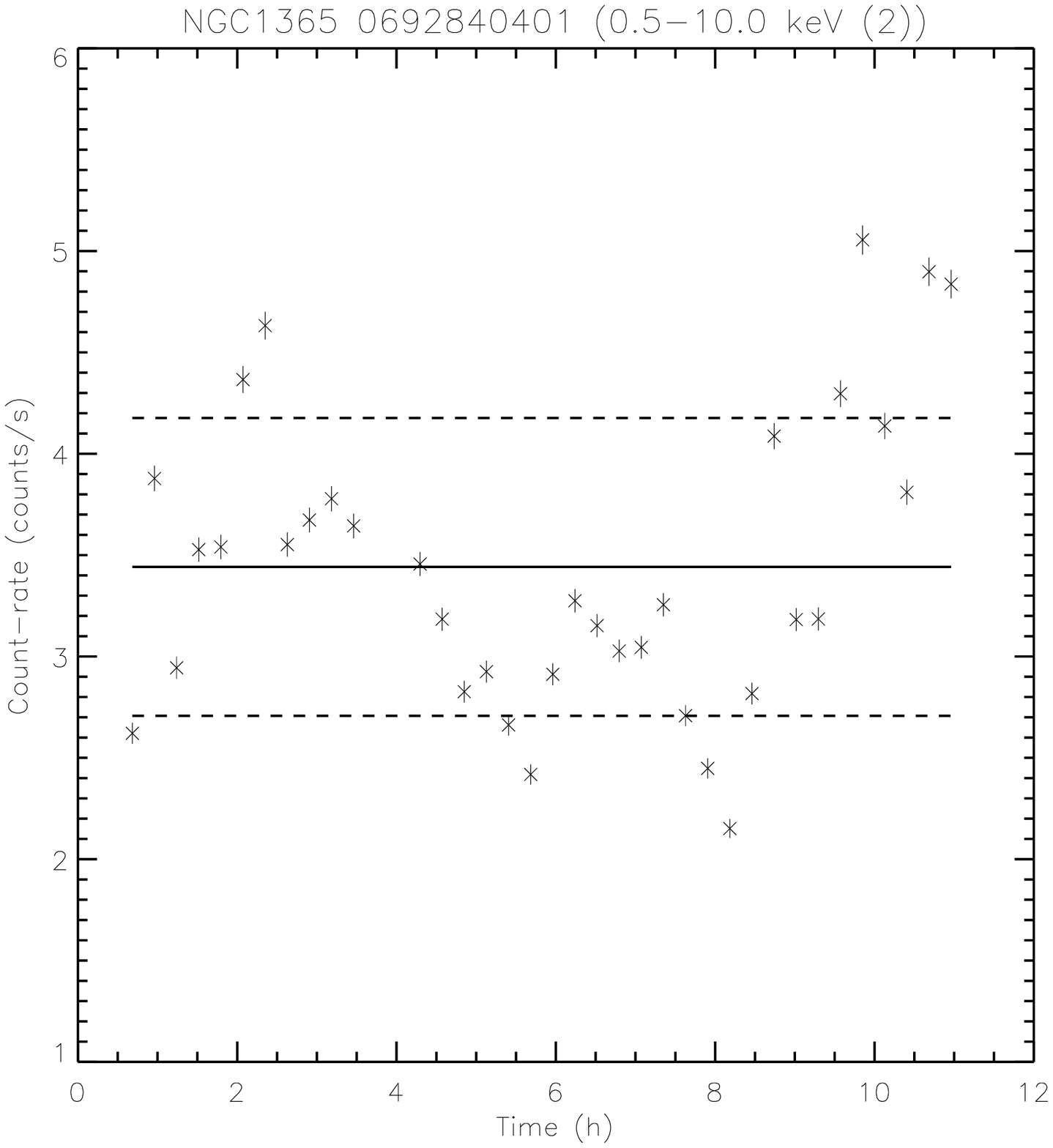}}
\caption{(Cont.)}
\end{figure}

\begin{figure}
\setcounter{figure}{2}
\centering
{\includegraphics[width=0.30\textwidth]{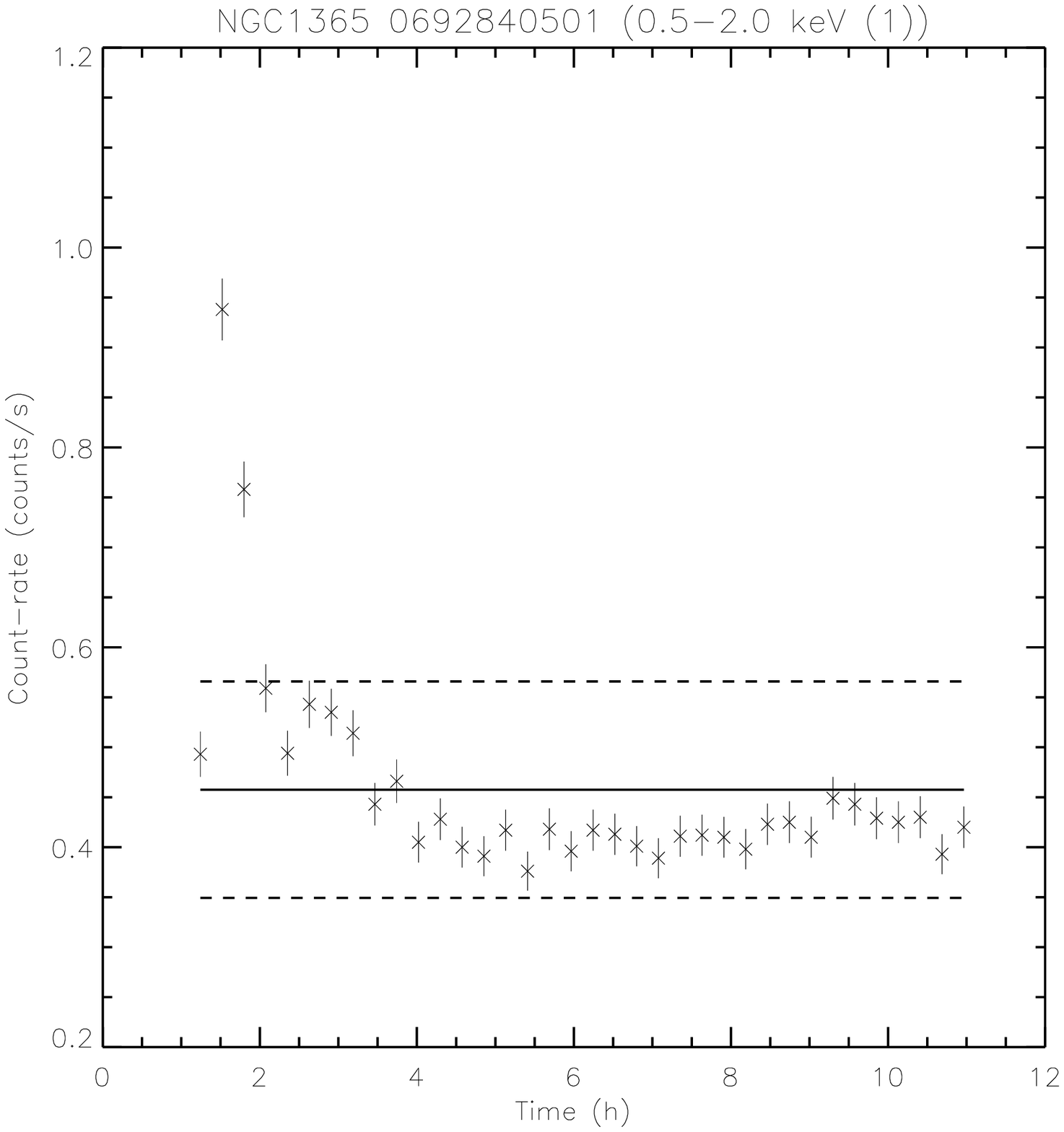}}
{\includegraphics[width=0.30\textwidth]{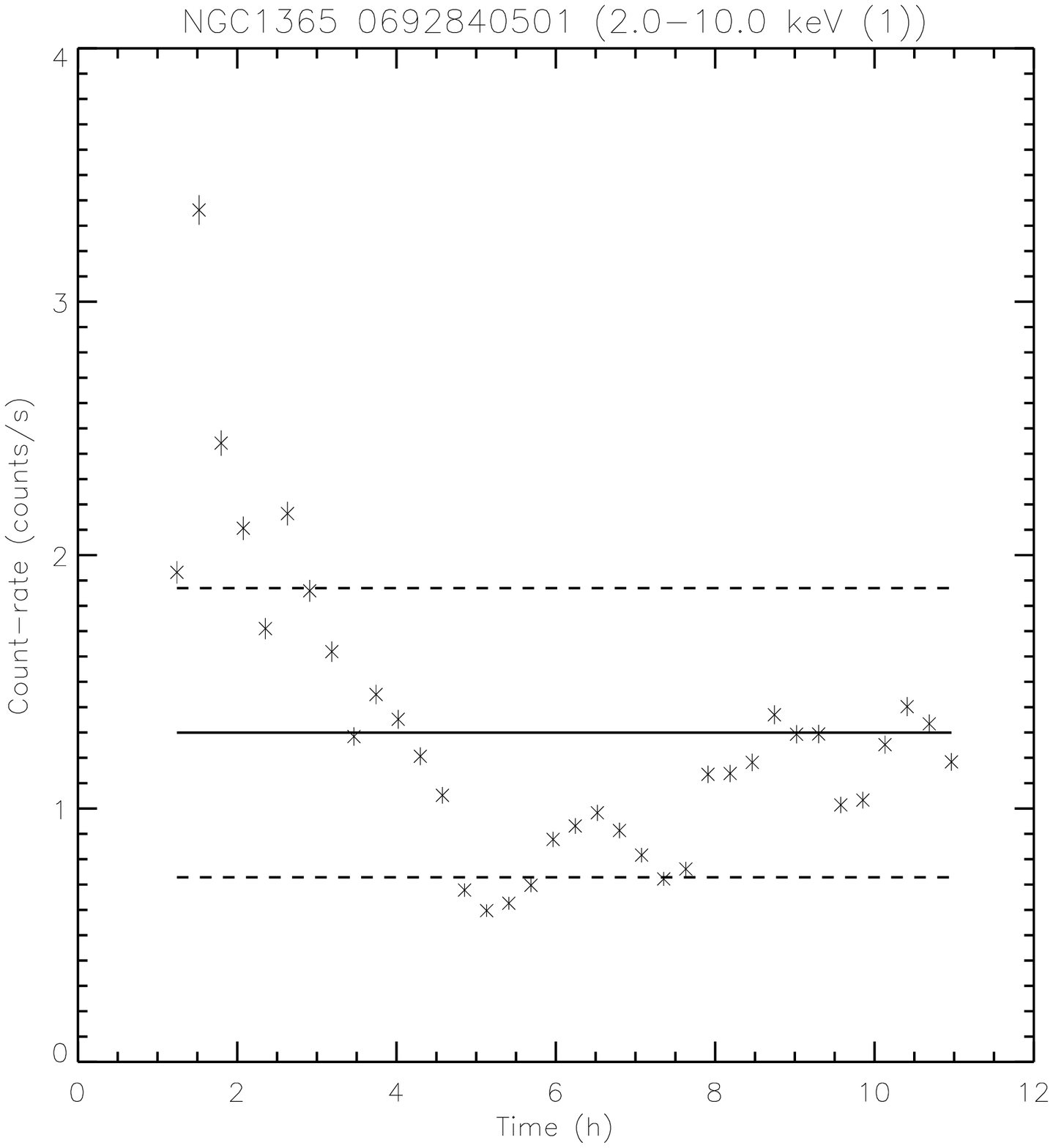}}
{\includegraphics[width=0.30\textwidth]{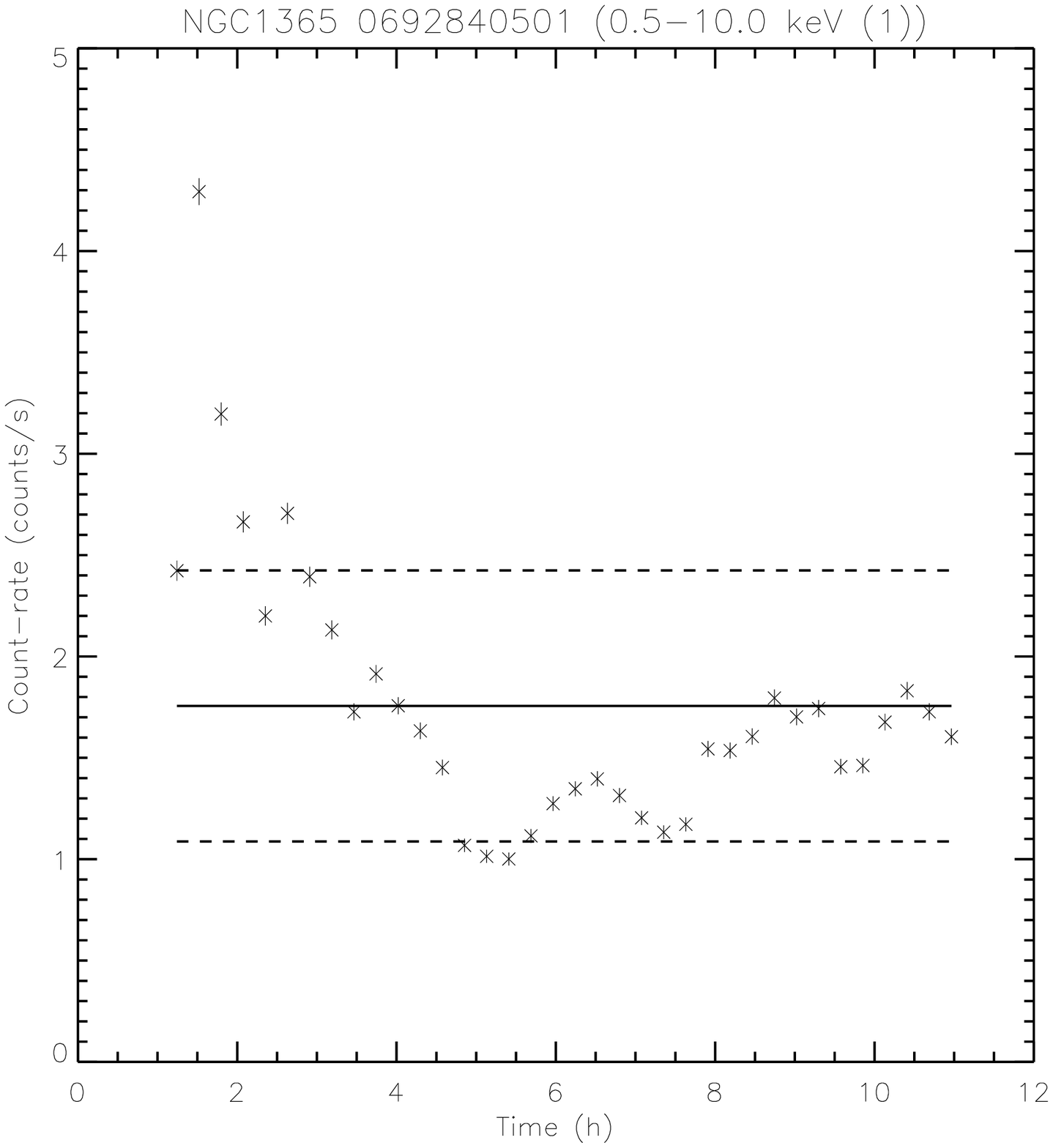}}

{\includegraphics[width=0.30\textwidth]{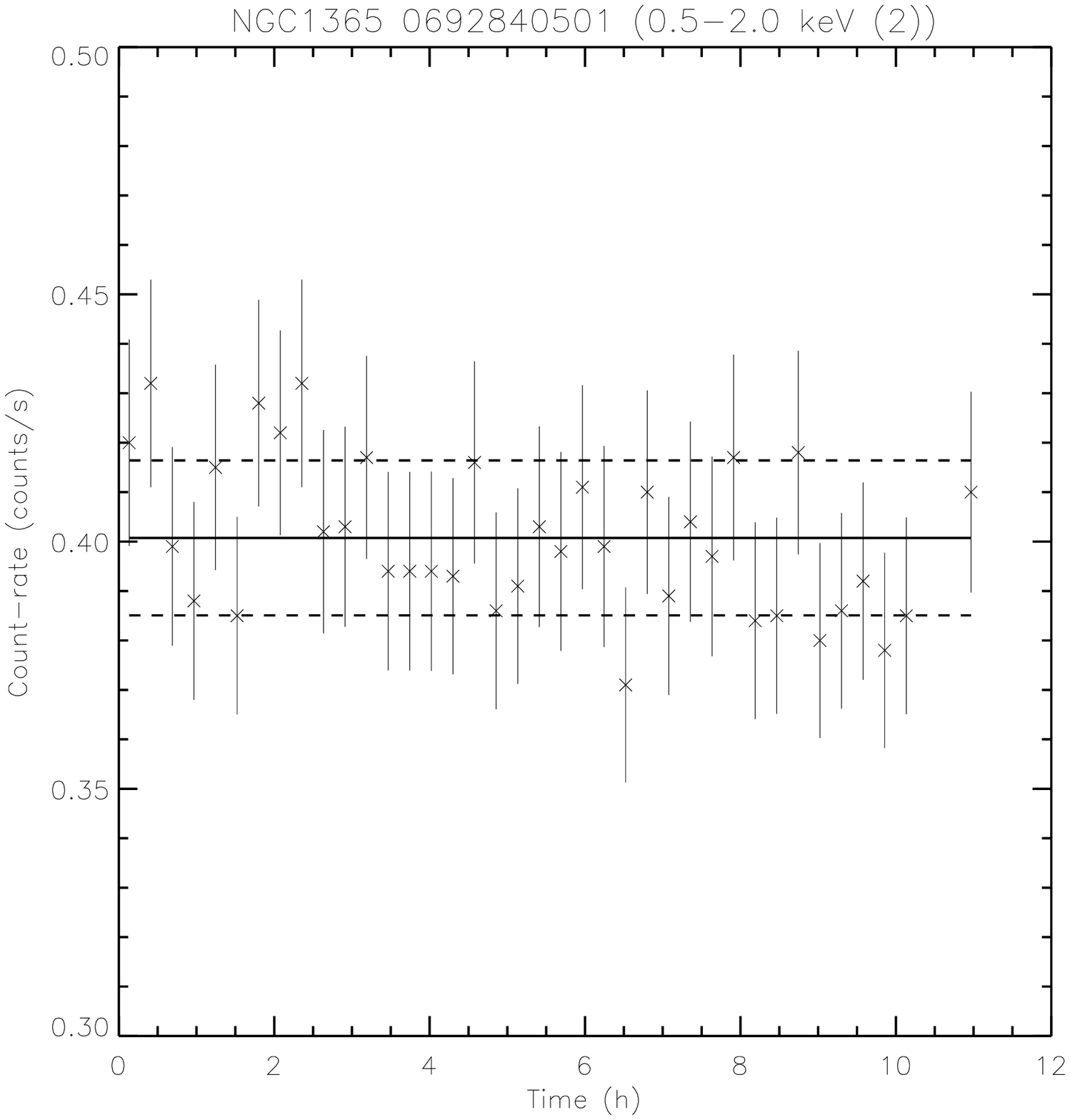}}
{\includegraphics[width=0.30\textwidth]{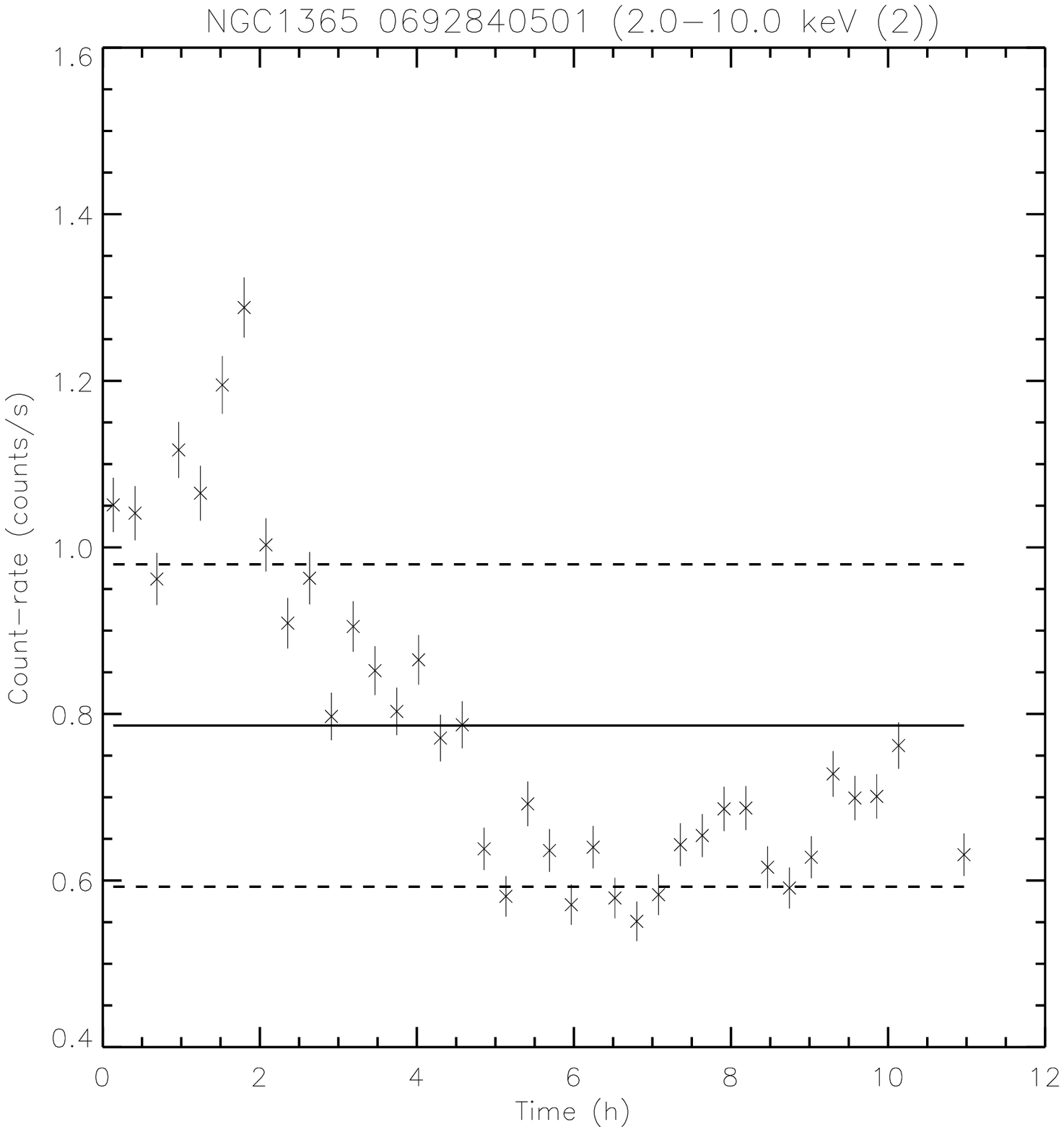}}
{\includegraphics[width=0.30\textwidth]{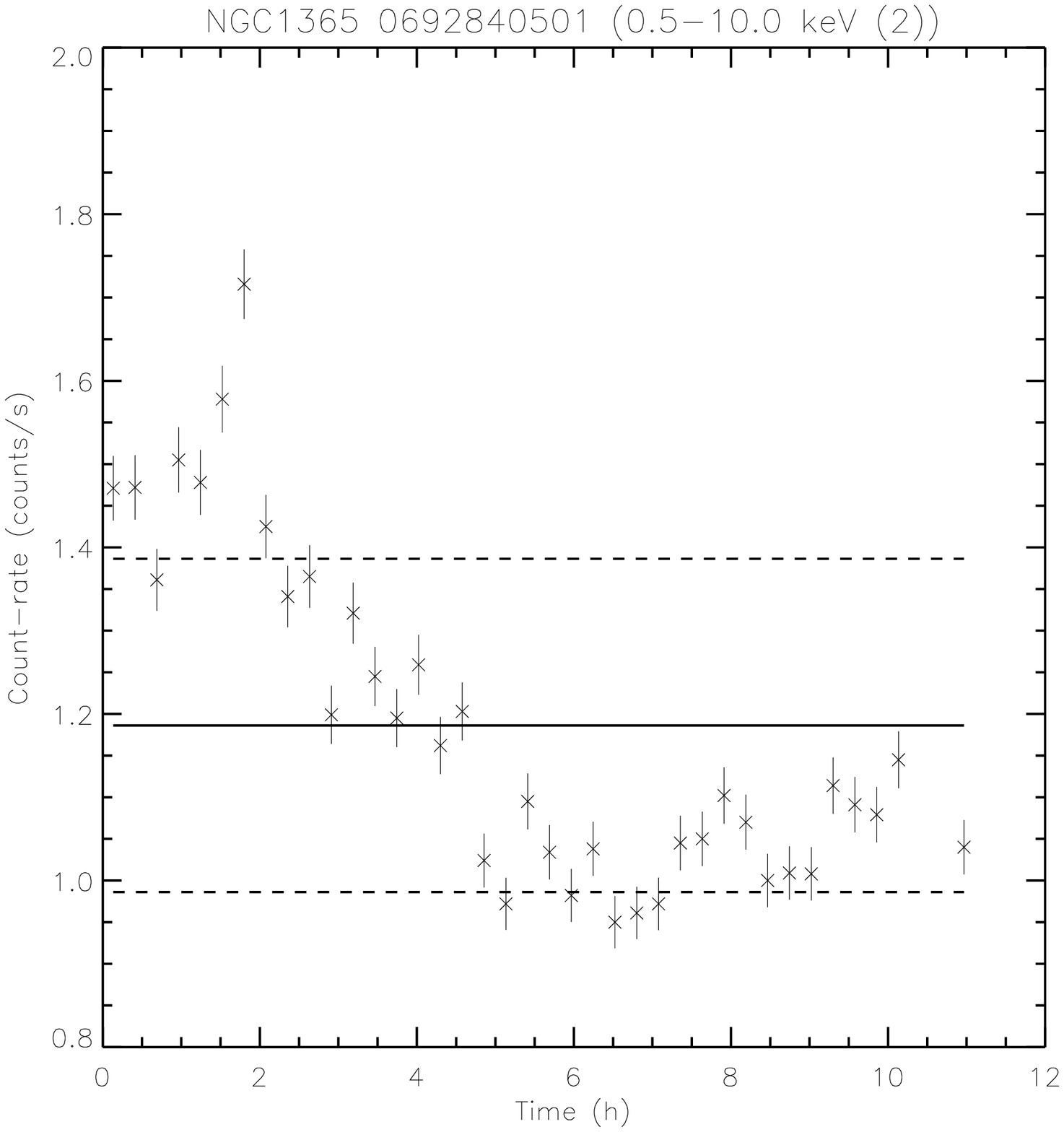}}

{\includegraphics[width=0.30\textwidth]{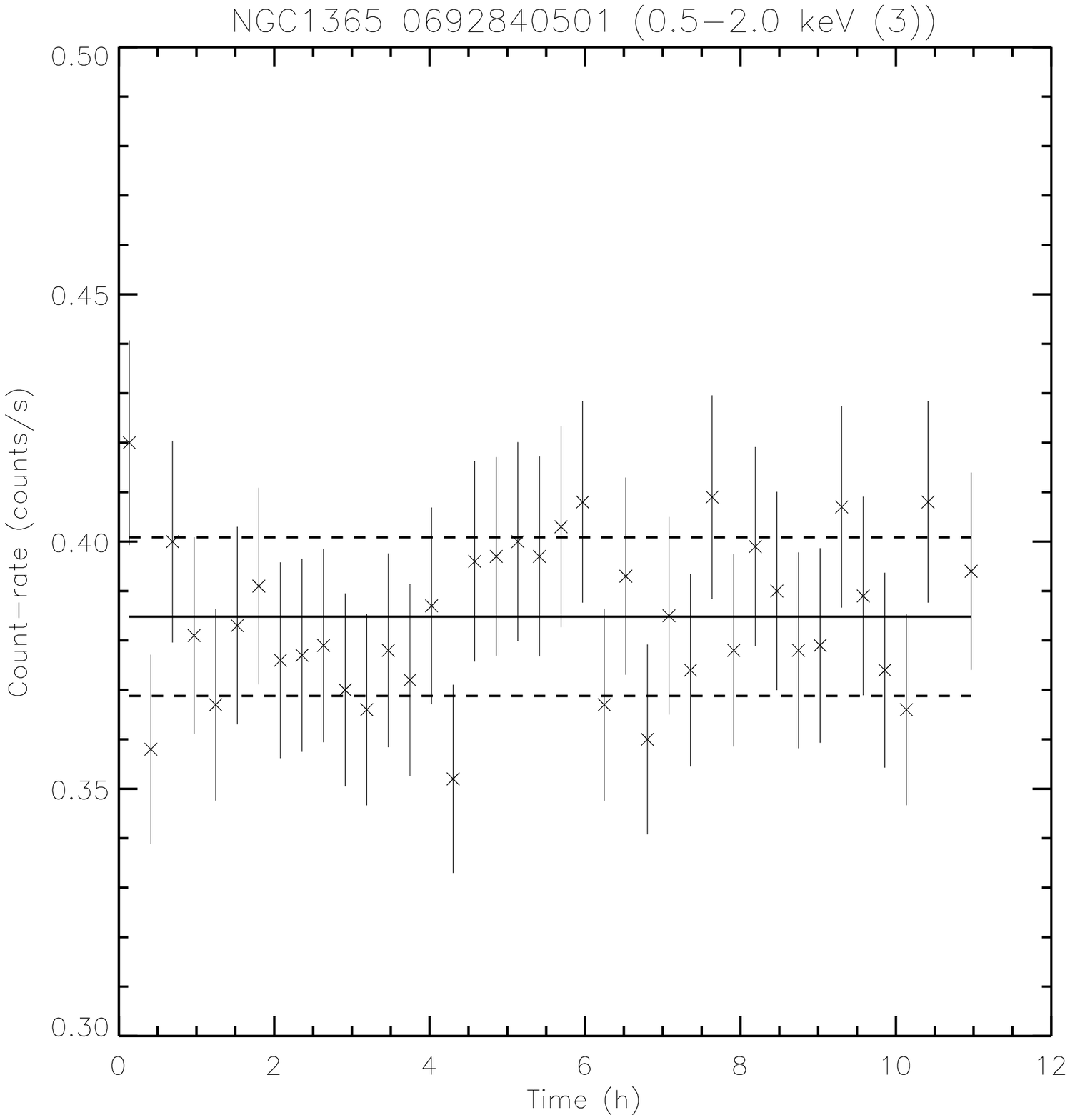}}
{\includegraphics[width=0.30\textwidth]{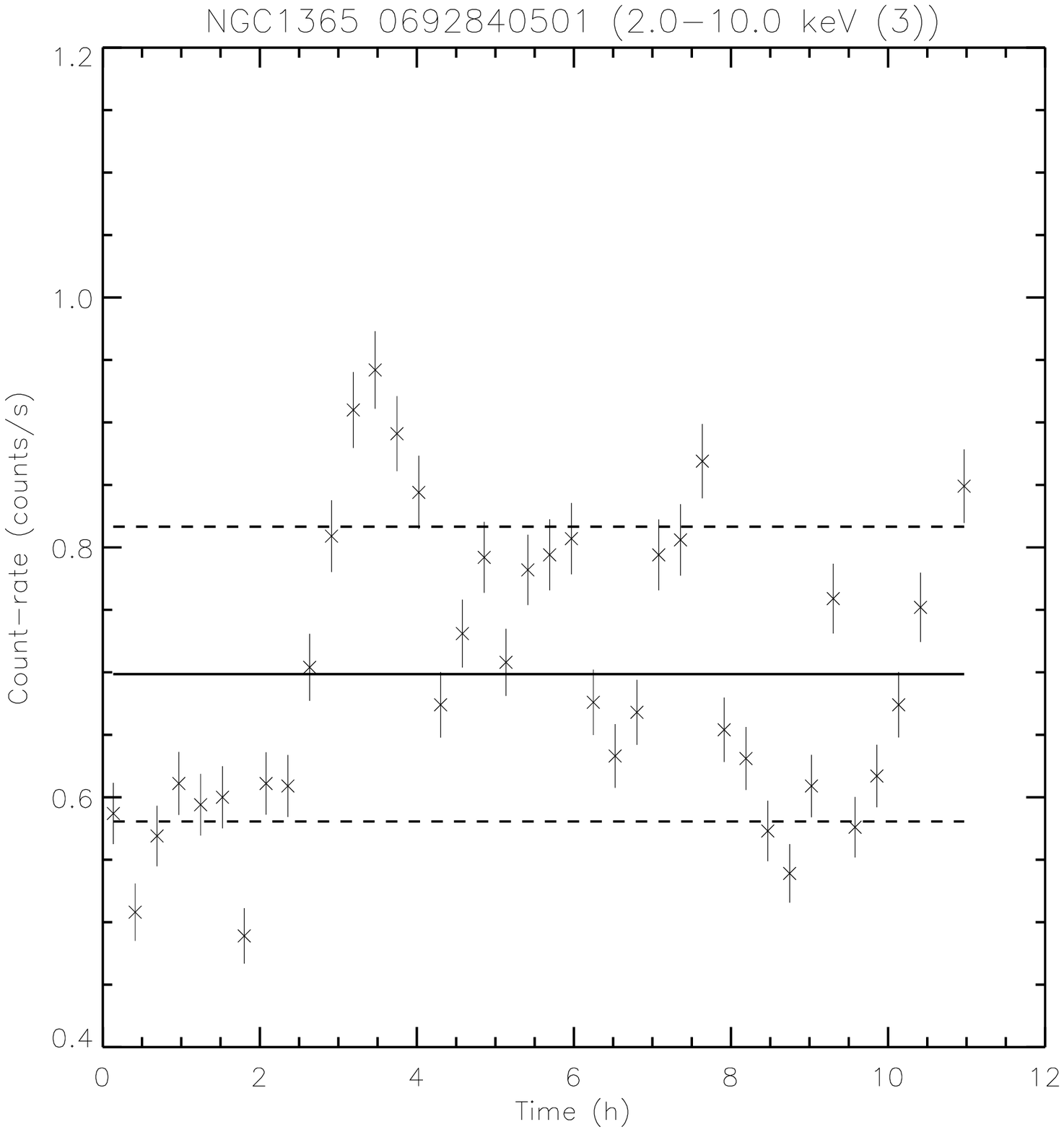}}
{\includegraphics[width=0.30\textwidth]{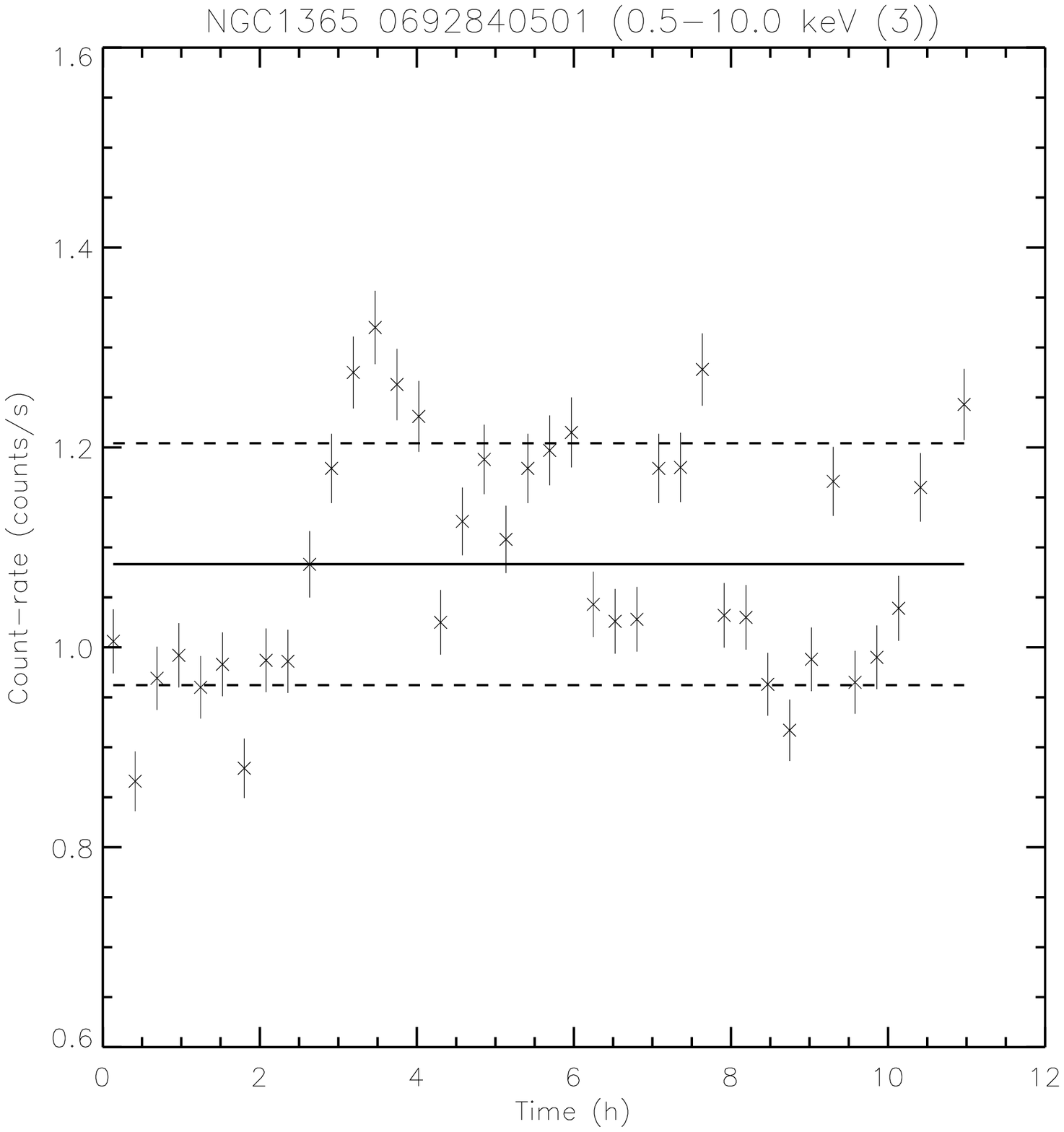}}
\caption{(Cont.)}
\end{figure}

\begin{figure}
\centering
{\includegraphics[width=0.30\textwidth]{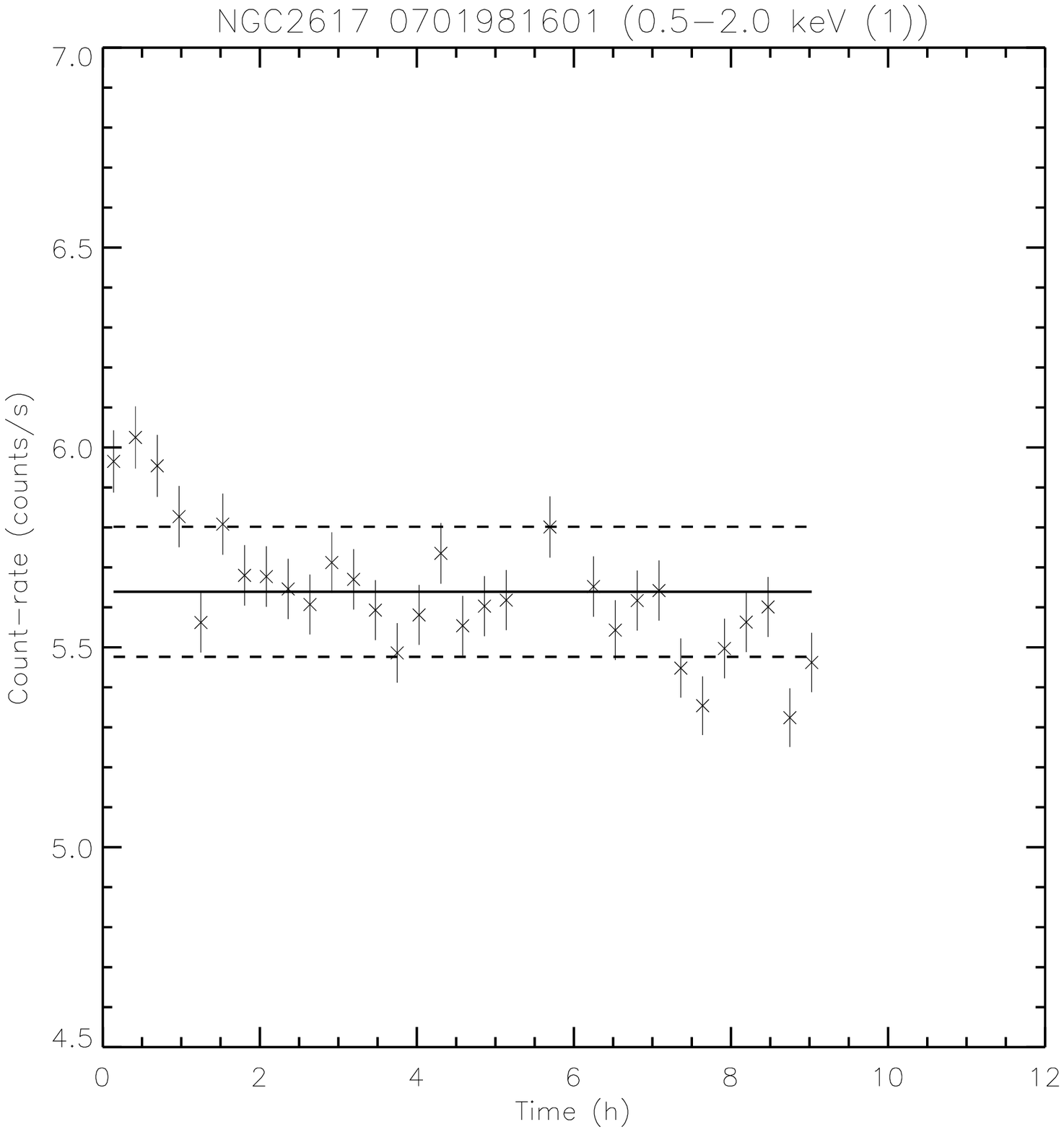}}
{\includegraphics[width=0.30\textwidth]{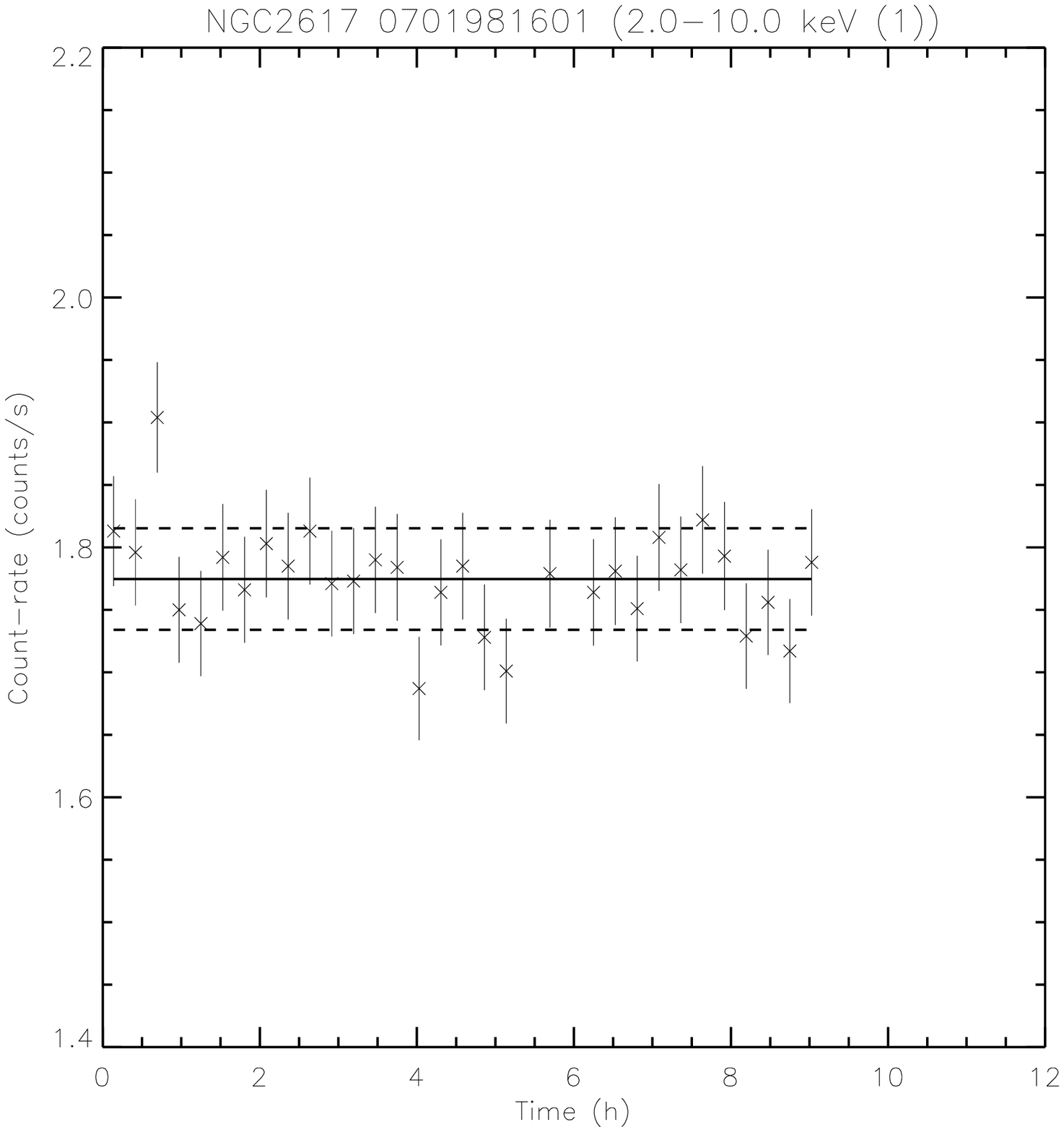}}
{\includegraphics[width=0.30\textwidth]{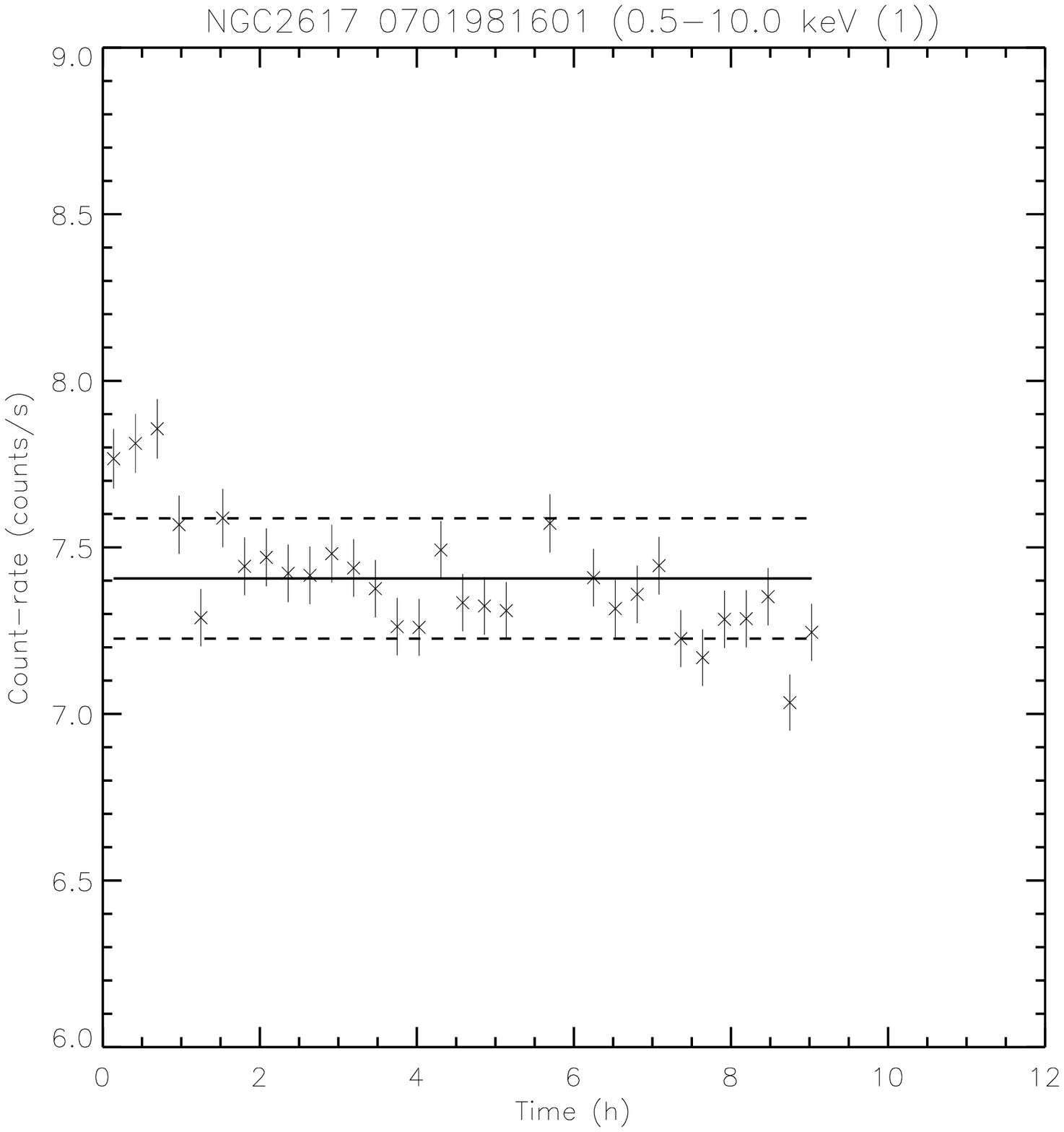}}
\caption{Light curves of NGC\,2617 from \emph{XMM--Newton} data.}
\label{l2617}
\end{figure}

\begin{figure}
\centering
{\includegraphics[width=0.30\textwidth]{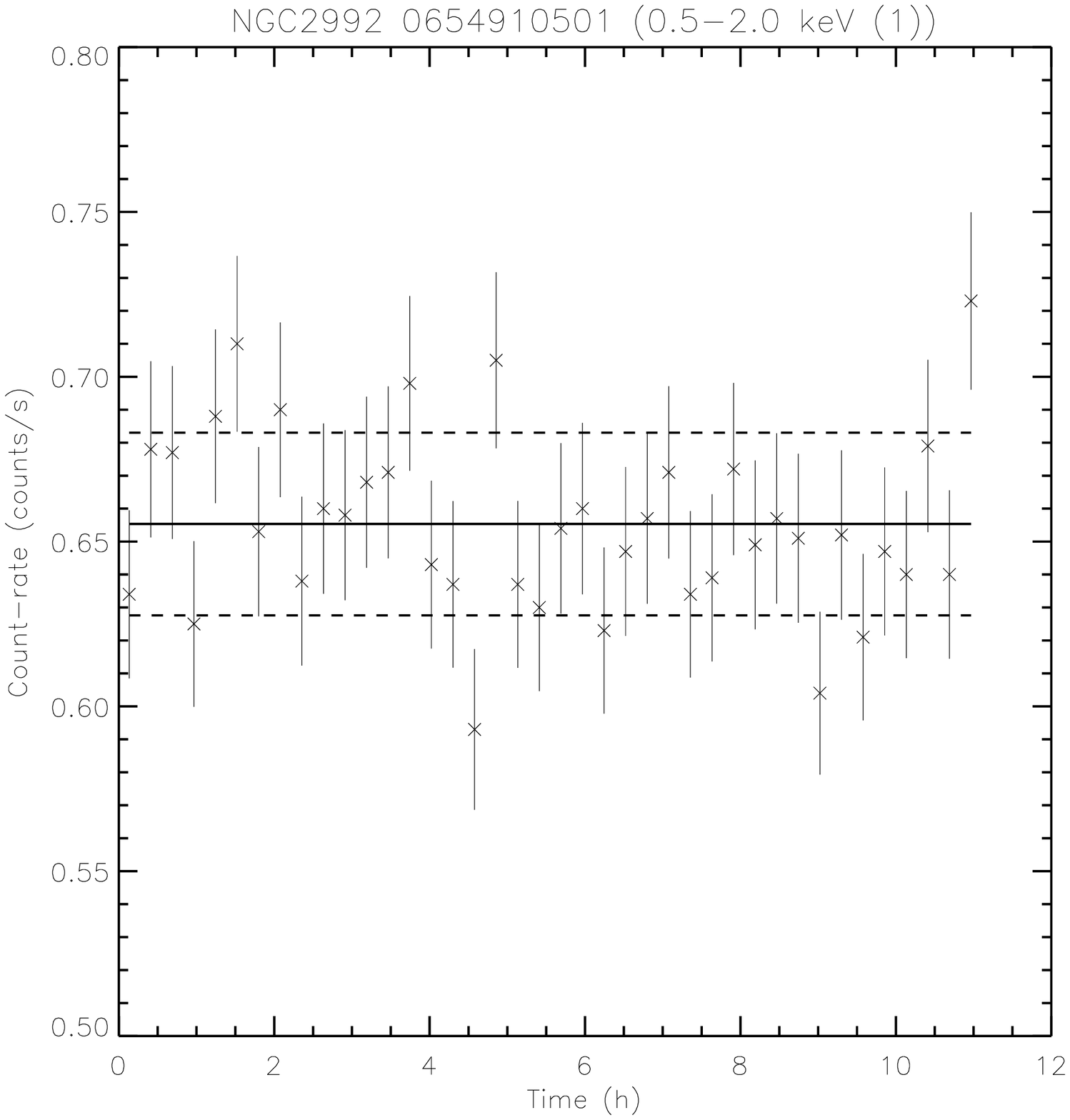}}
{\includegraphics[width=0.30\textwidth]{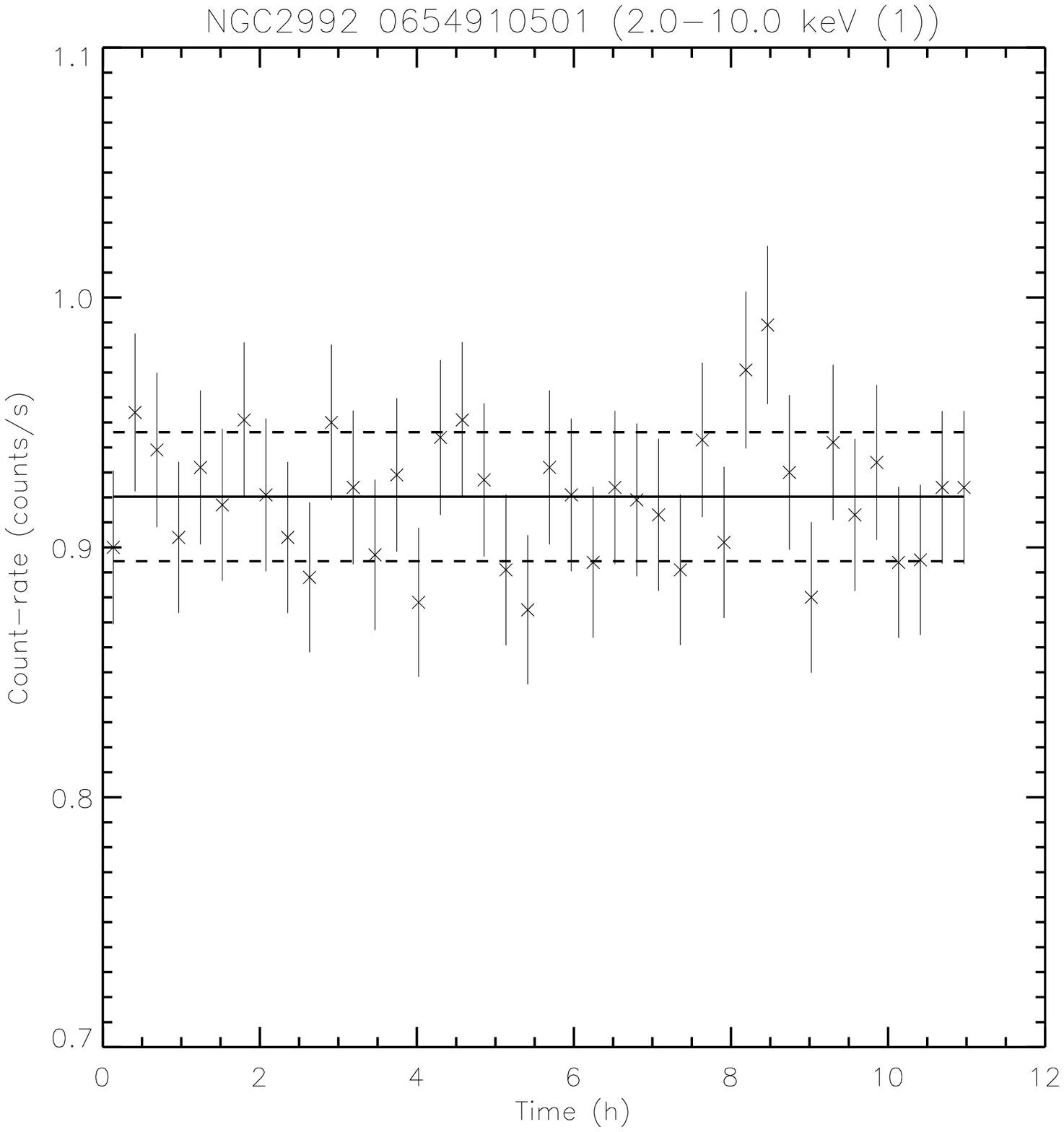}}
{\includegraphics[width=0.30\textwidth]{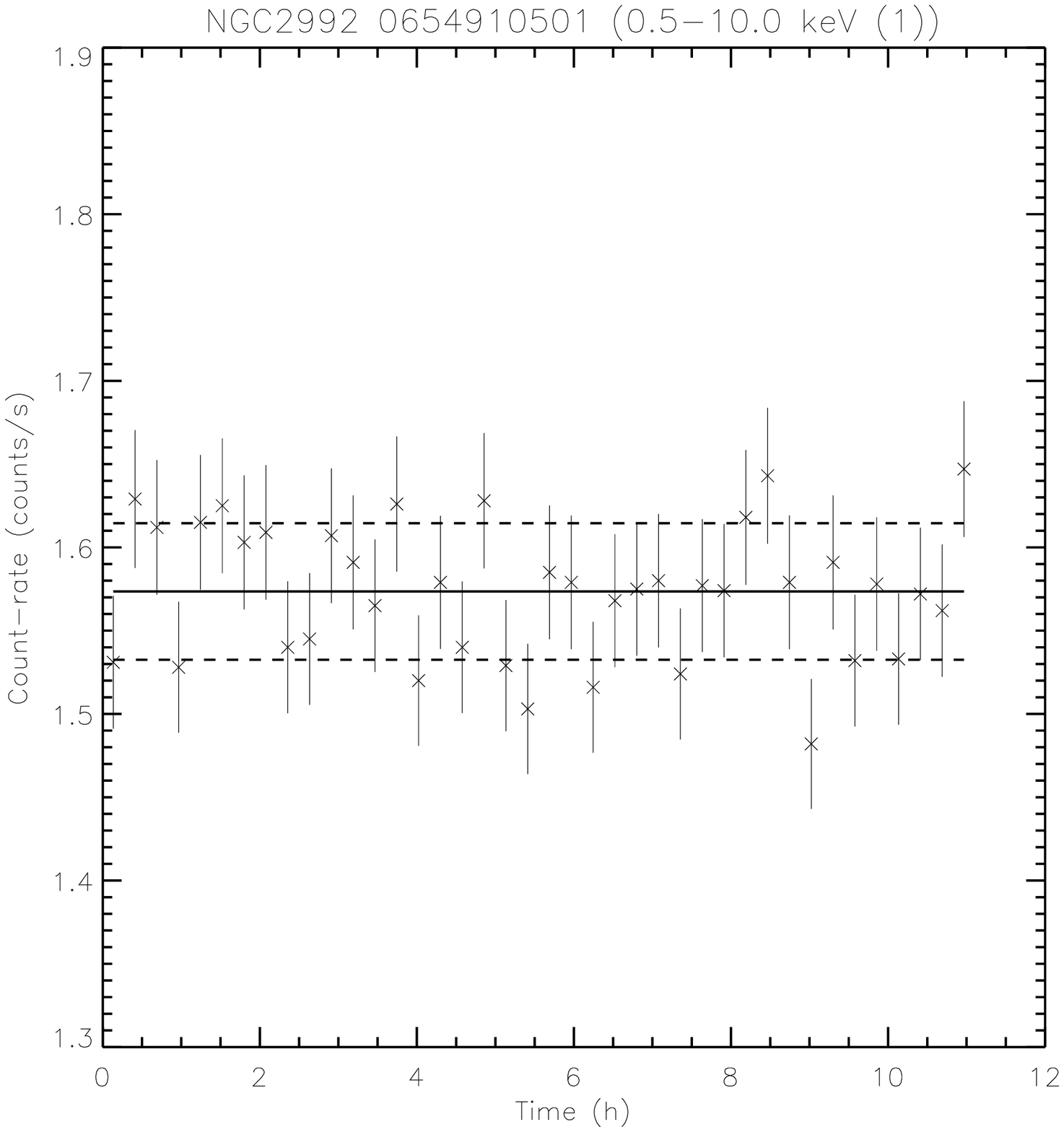}}

{\includegraphics[width=0.30\textwidth]{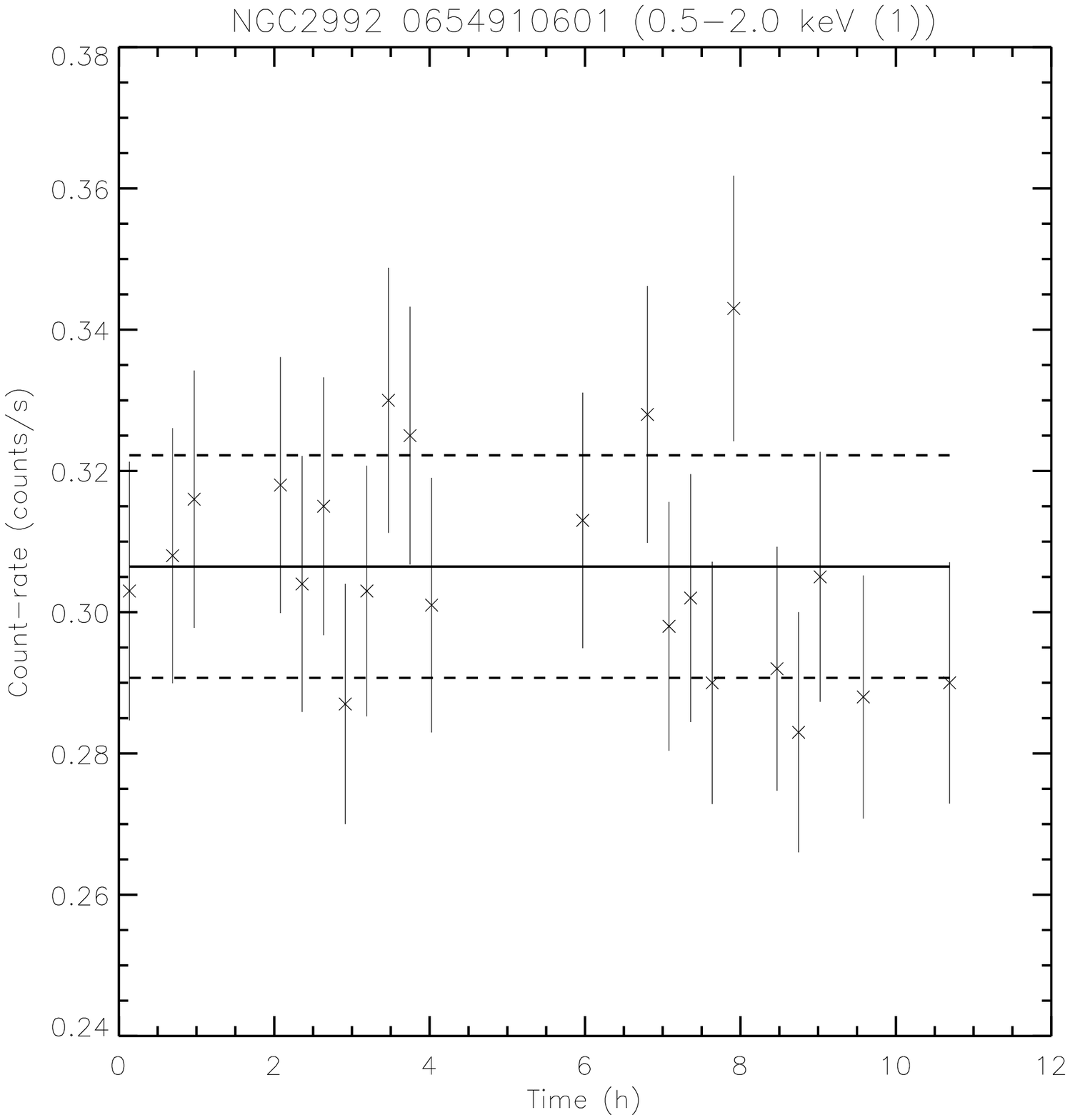}}
{\includegraphics[width=0.30\textwidth]{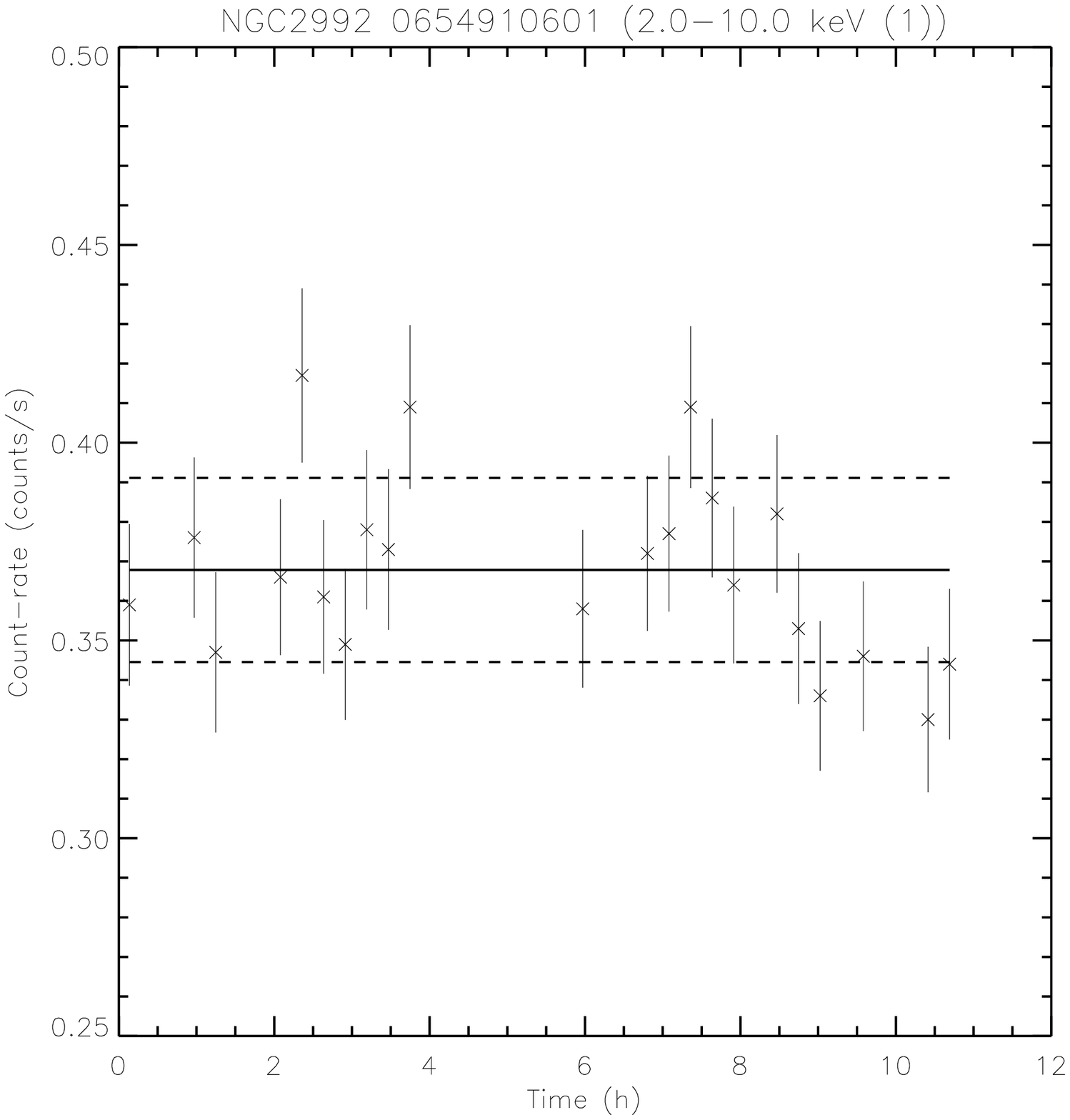}}
{\includegraphics[width=0.30\textwidth]{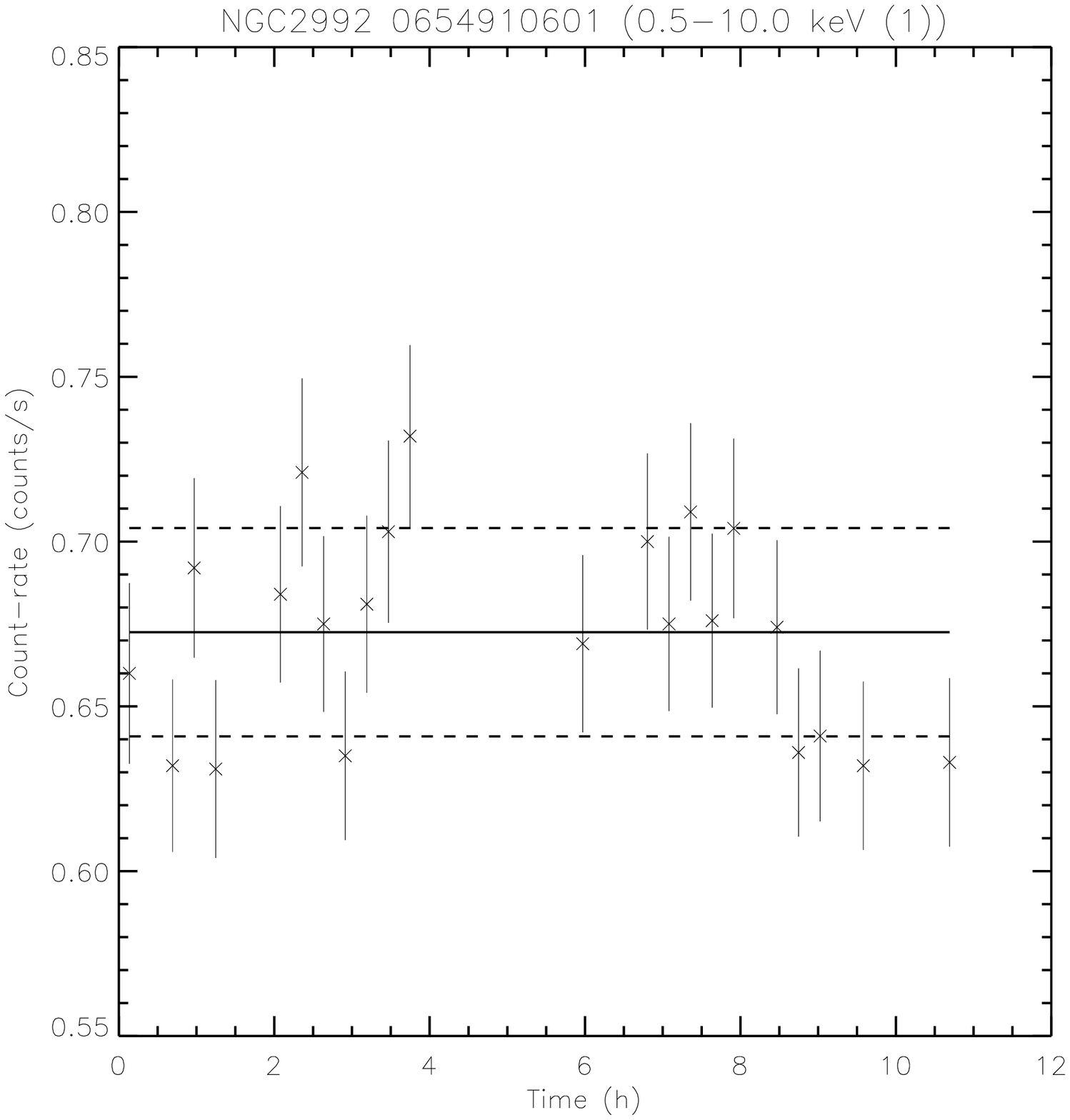}}

{\includegraphics[width=0.30\textwidth]{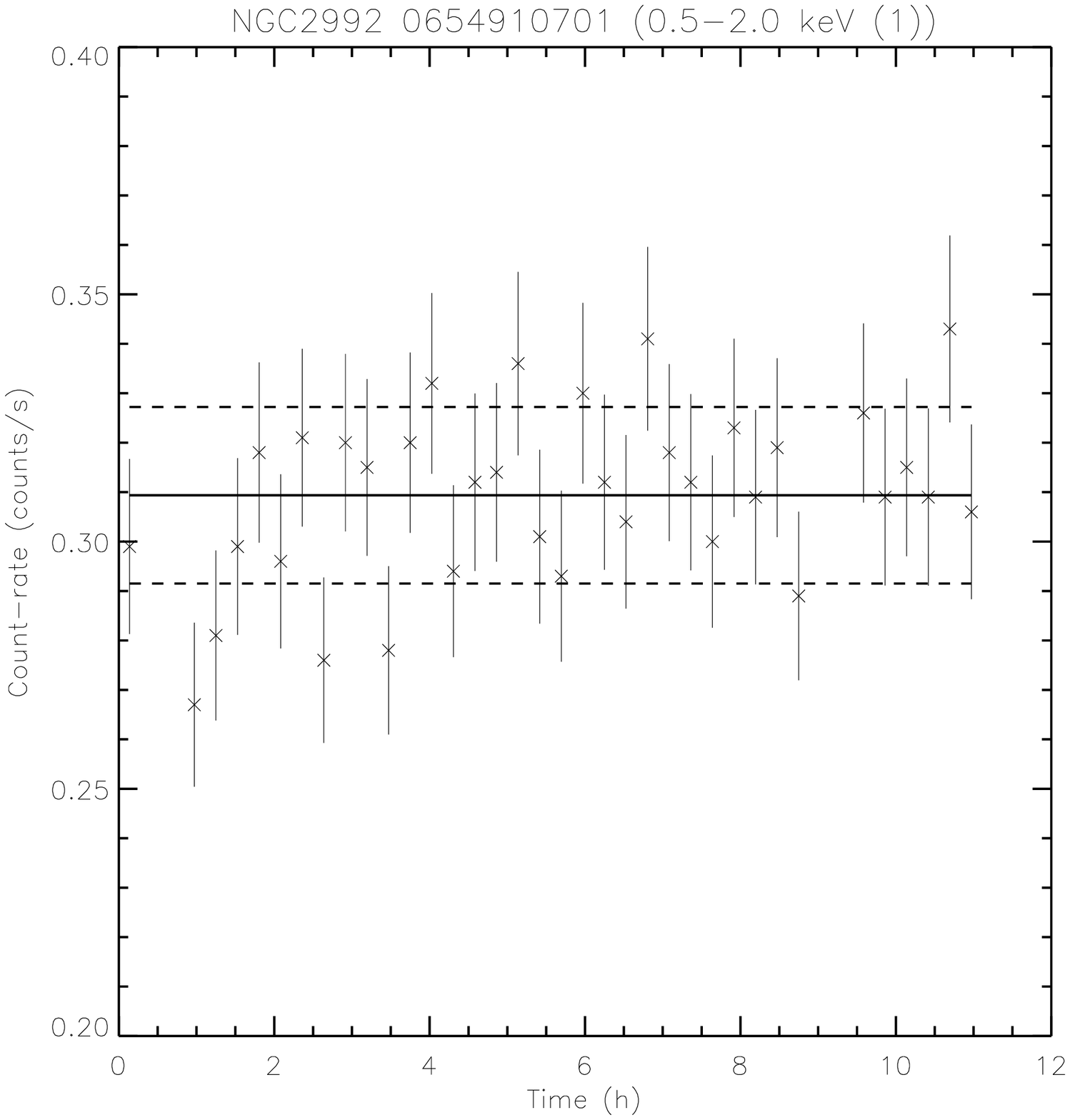}}
{\includegraphics[width=0.30\textwidth]{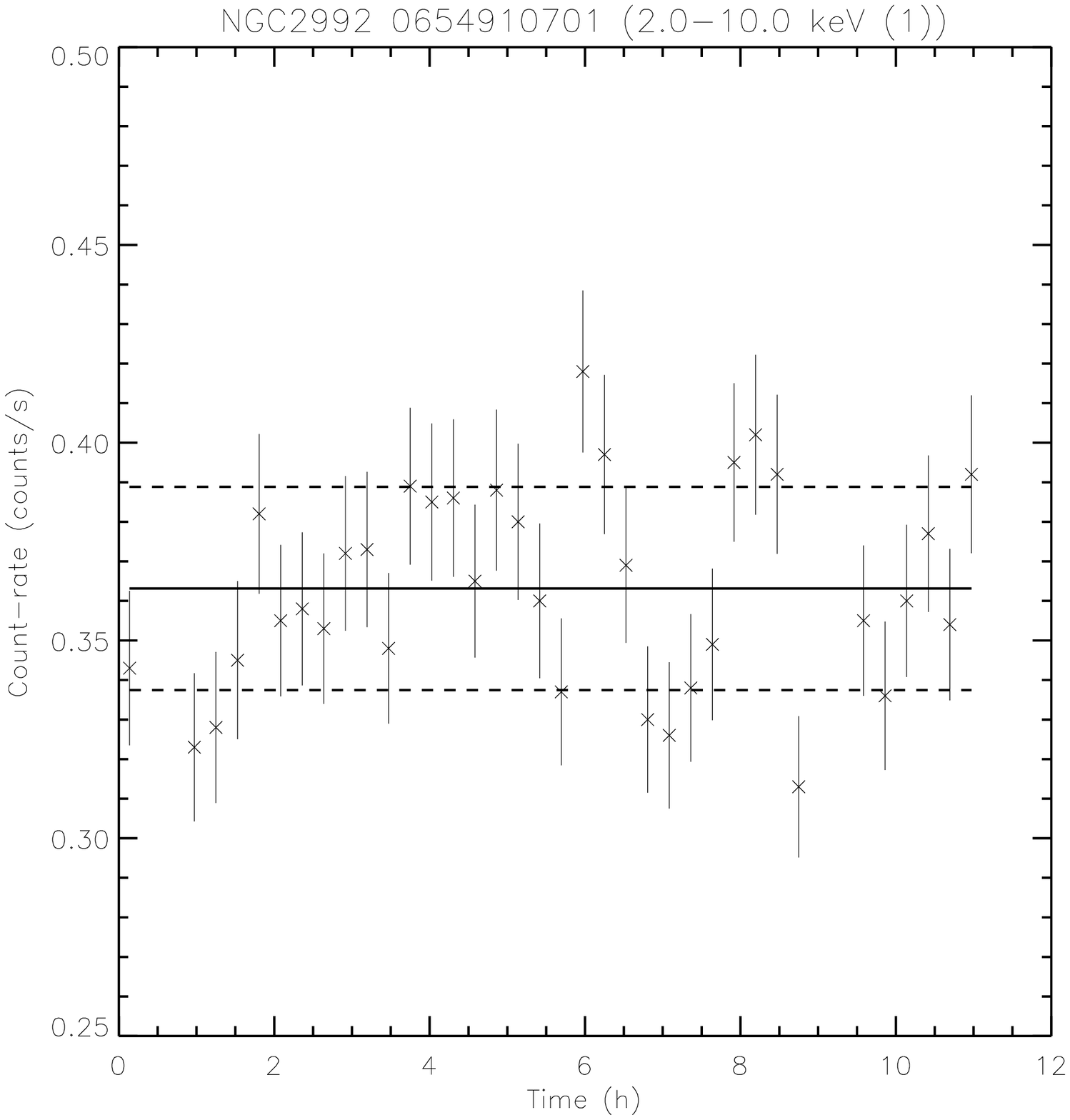}}
{\includegraphics[width=0.30\textwidth]{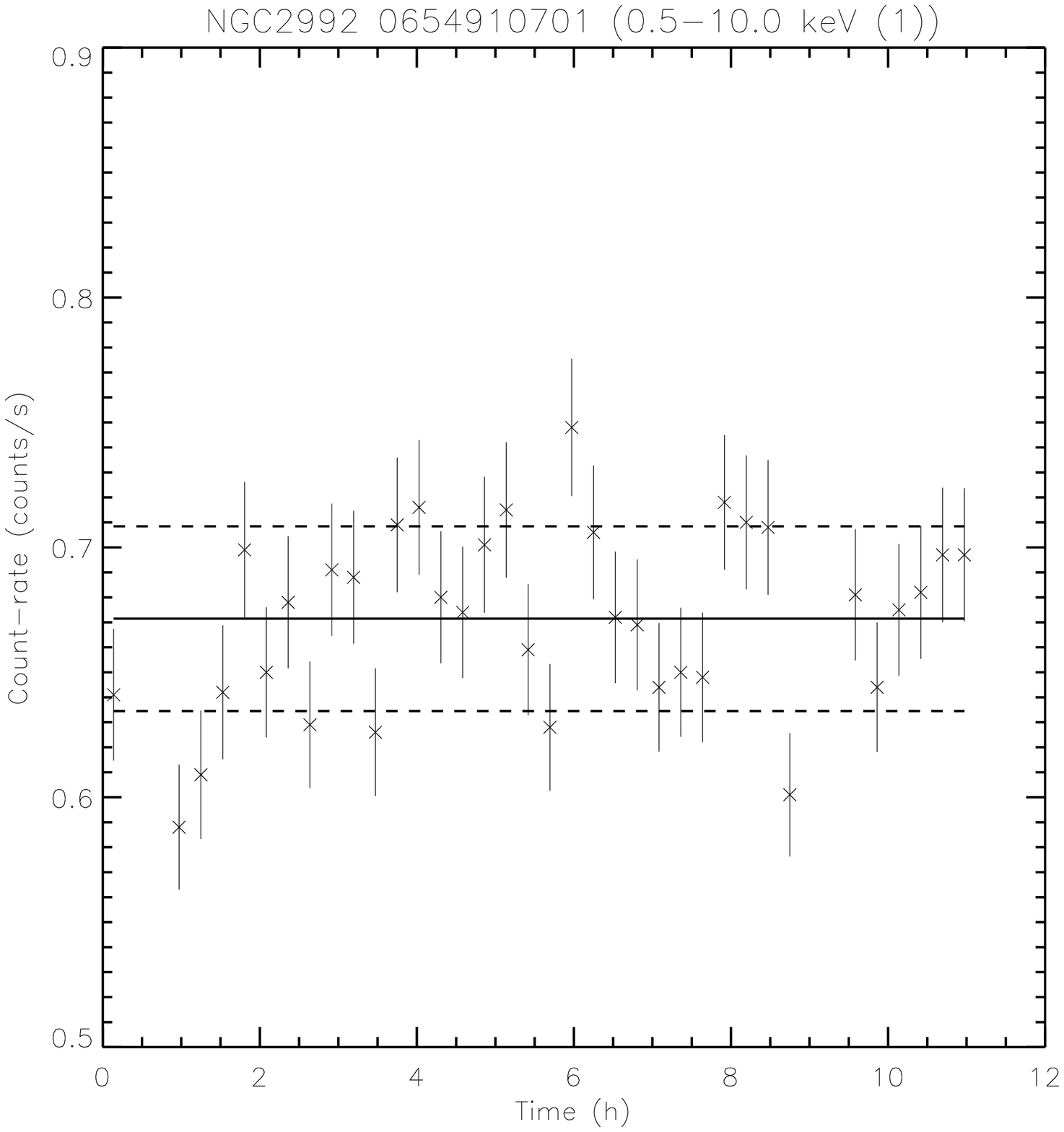}}

{\includegraphics[width=0.30\textwidth]{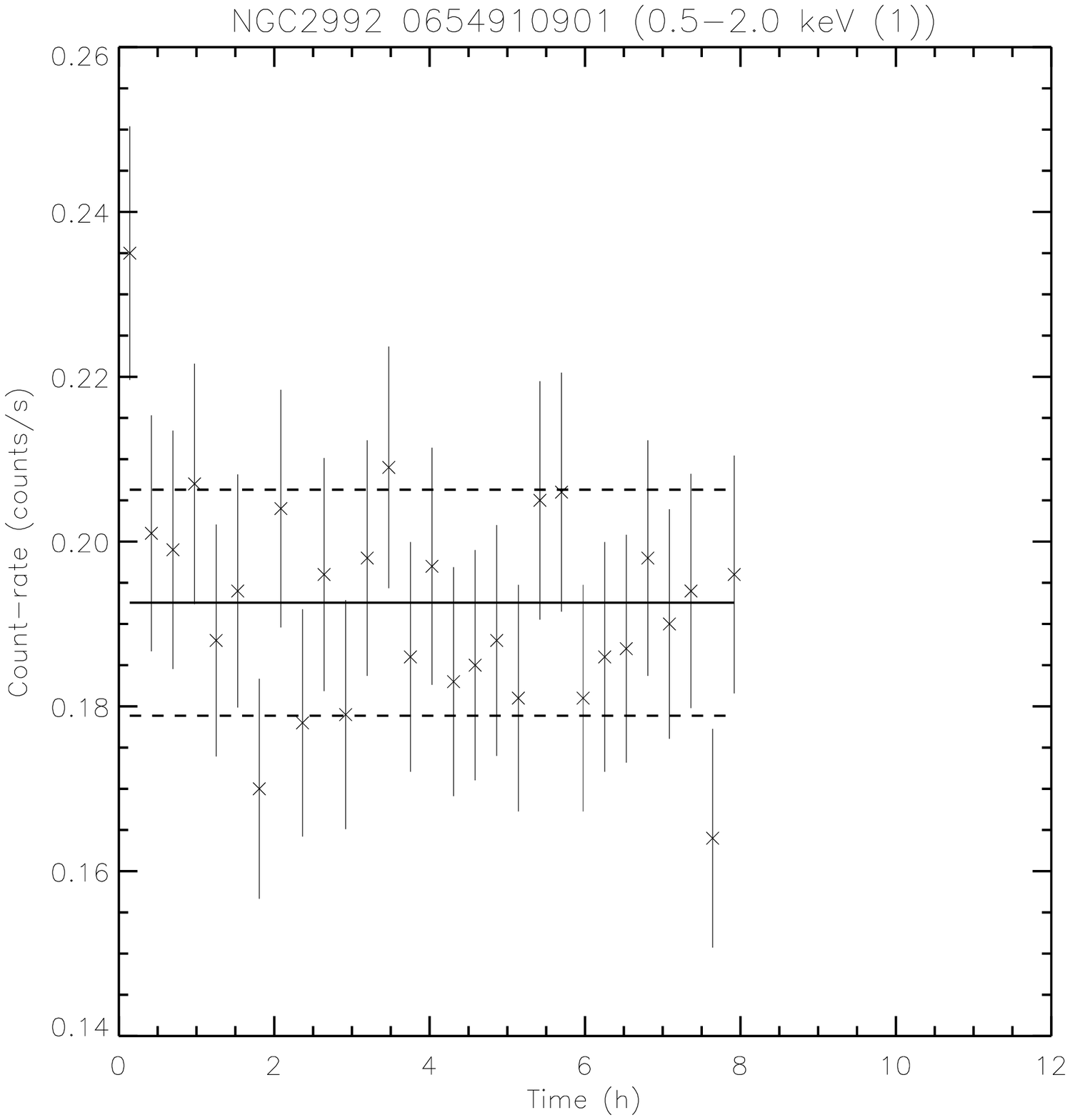}}
{\includegraphics[width=0.30\textwidth]{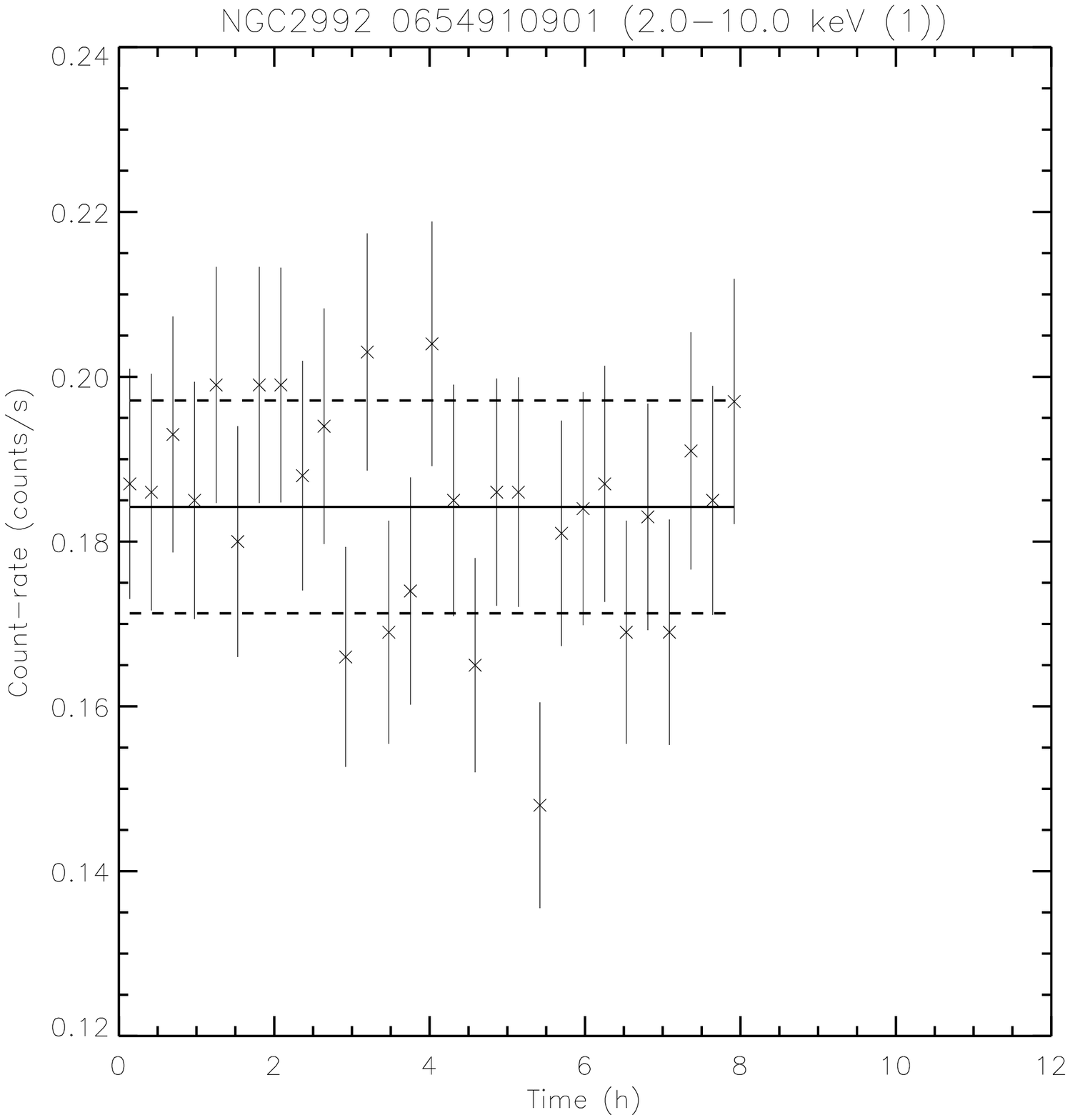}}
{\includegraphics[width=0.30\textwidth]{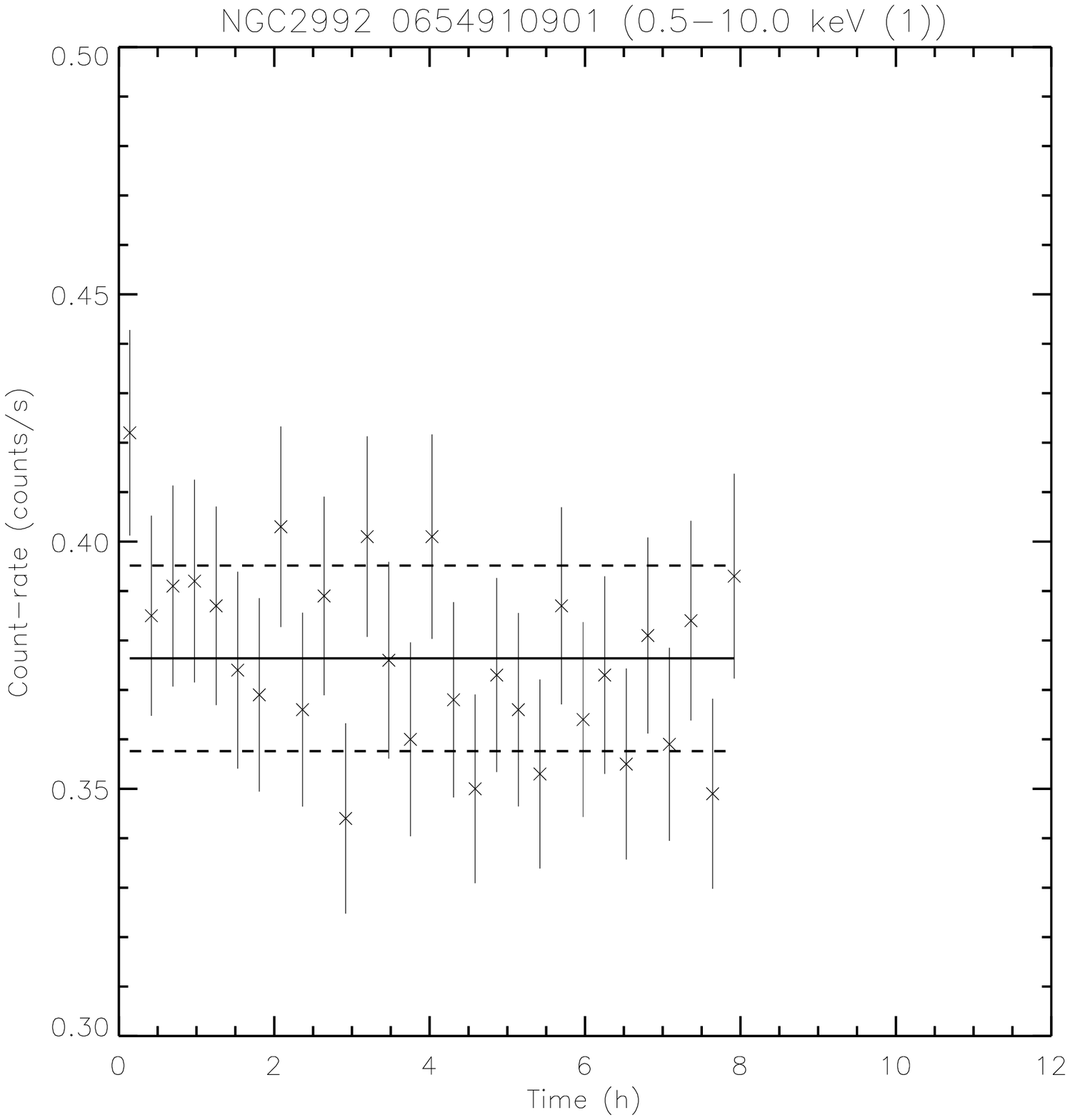}}
\caption{Light curves of NGC\,2992 from \emph{XMM--Newton} data.}
\label{l2992}
\end{figure}

\begin{figure}
\setcounter{figure}{4}
\centering
{\includegraphics[width=0.30\textwidth]{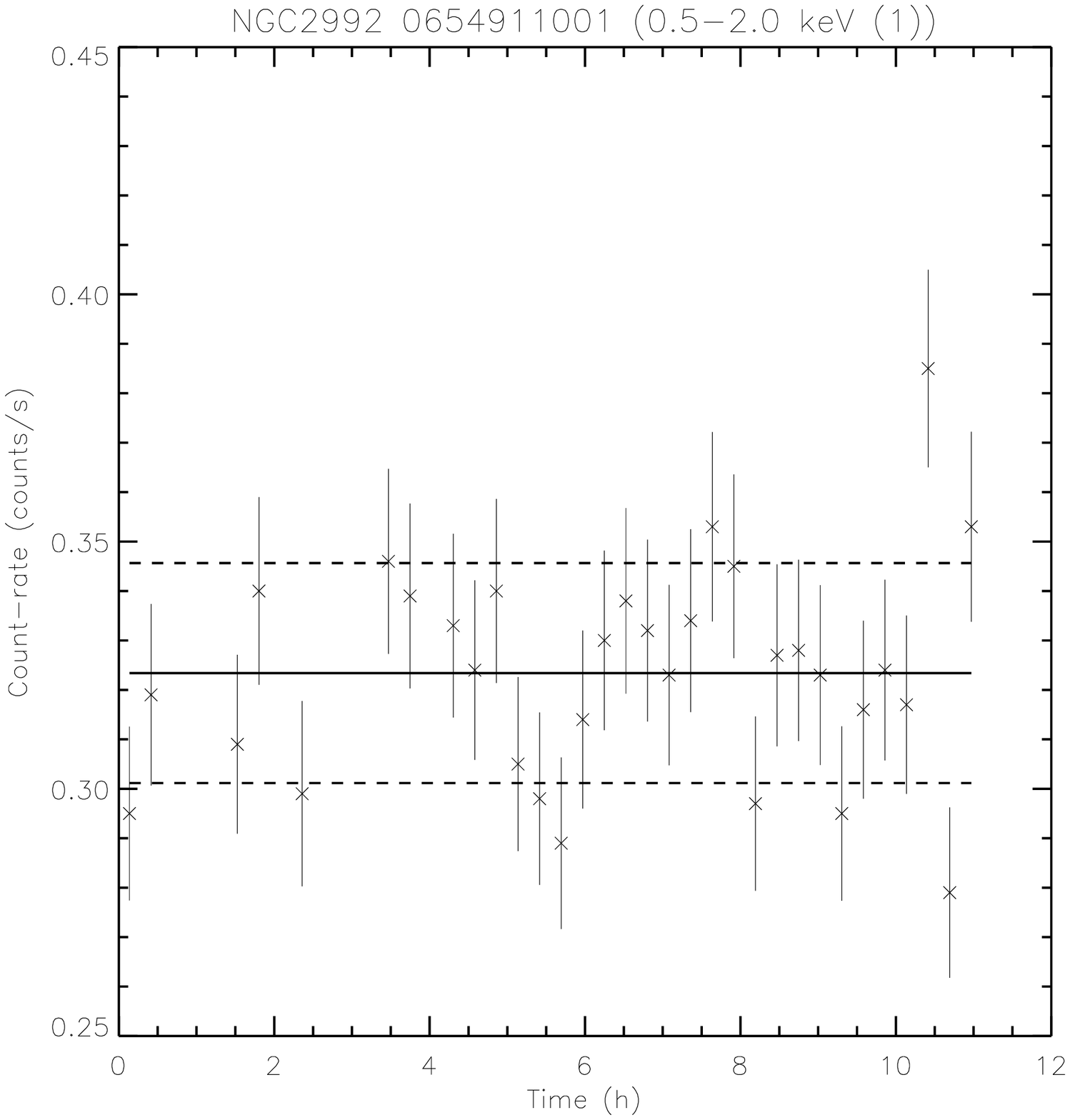}}
{\includegraphics[width=0.30\textwidth]{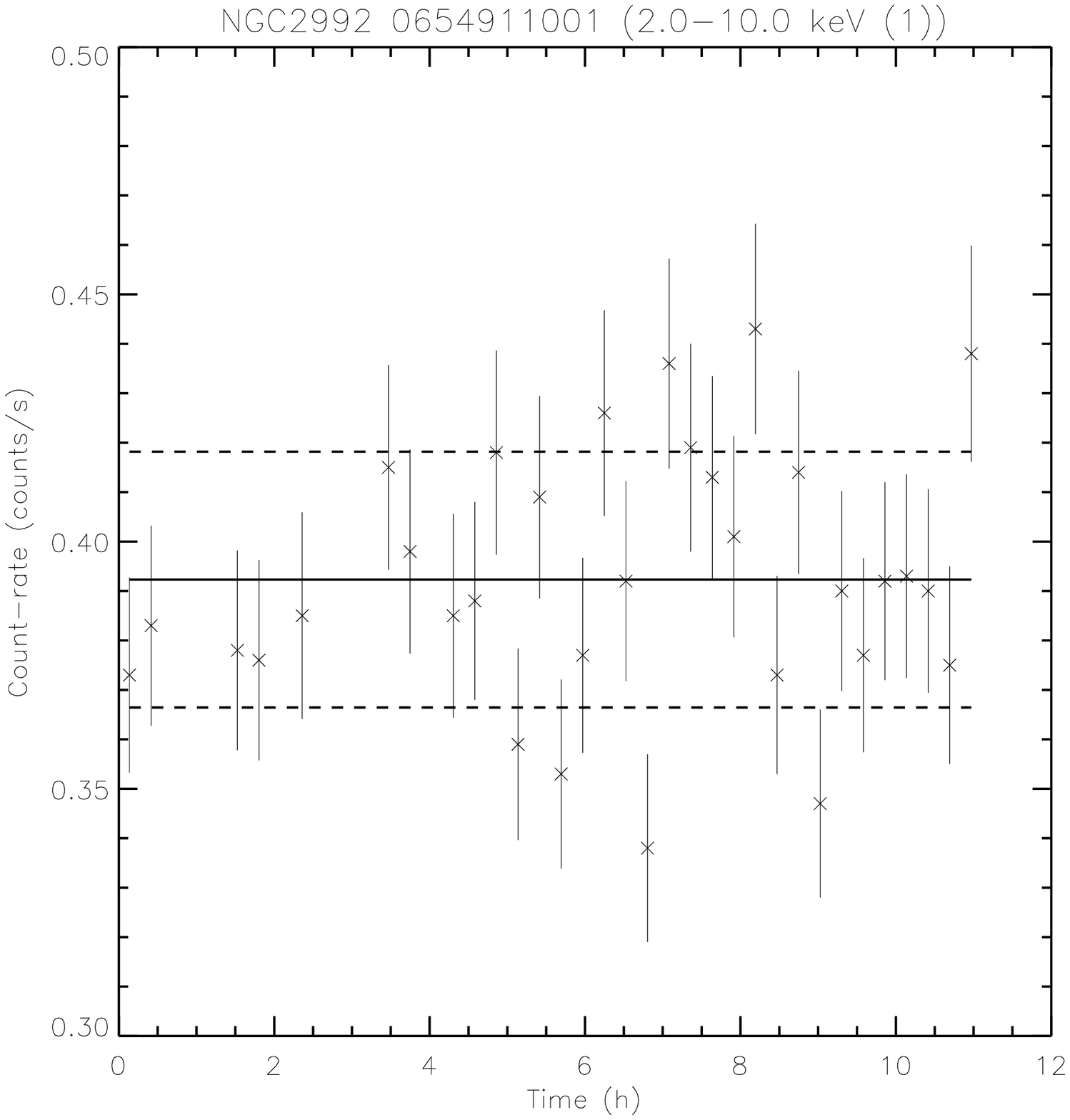}}
{\includegraphics[width=0.30\textwidth]{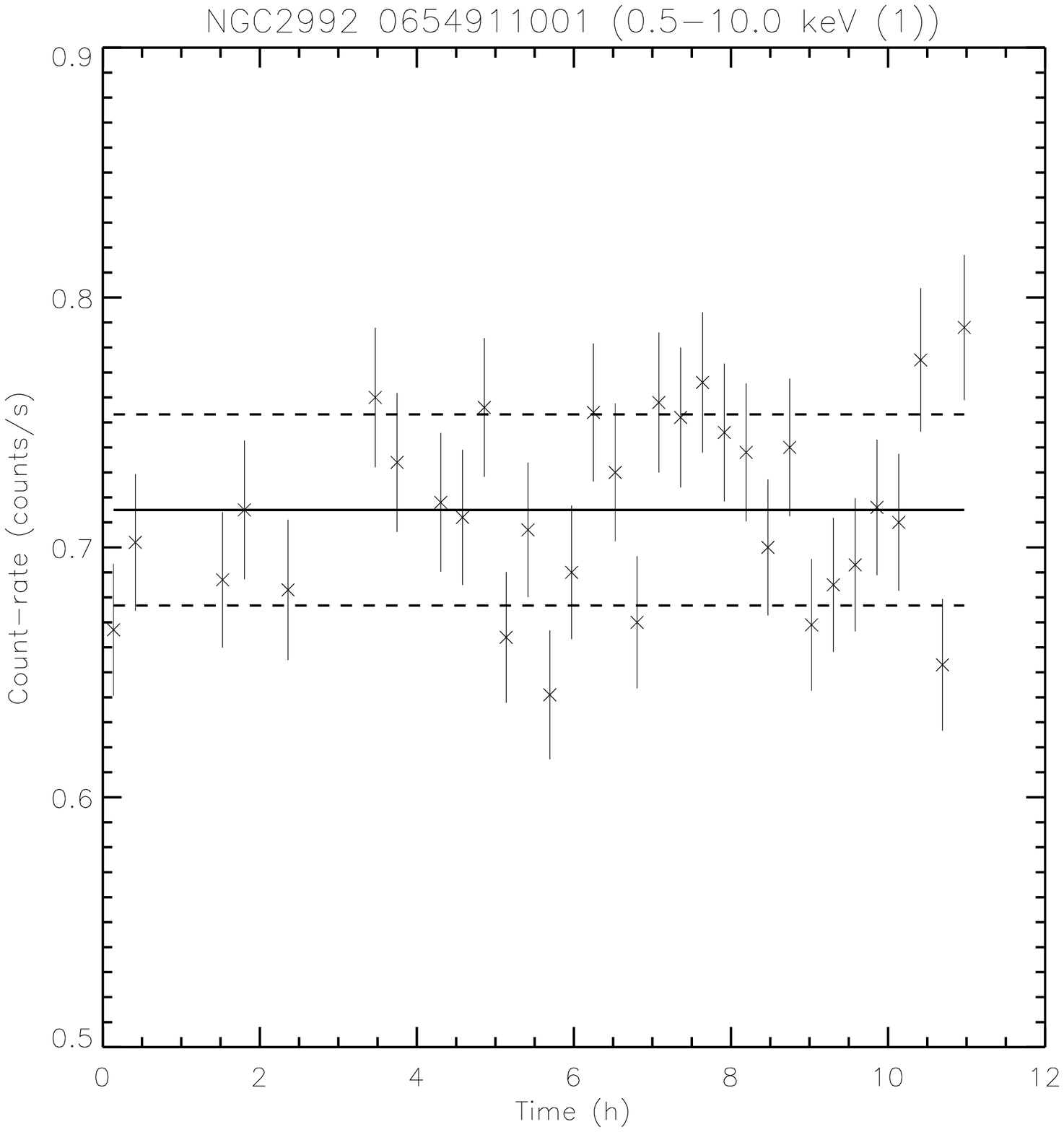}}
\caption{(Cont.)}
\end{figure}

\begin{figure}
\centering
{\includegraphics[width=0.30\textwidth]{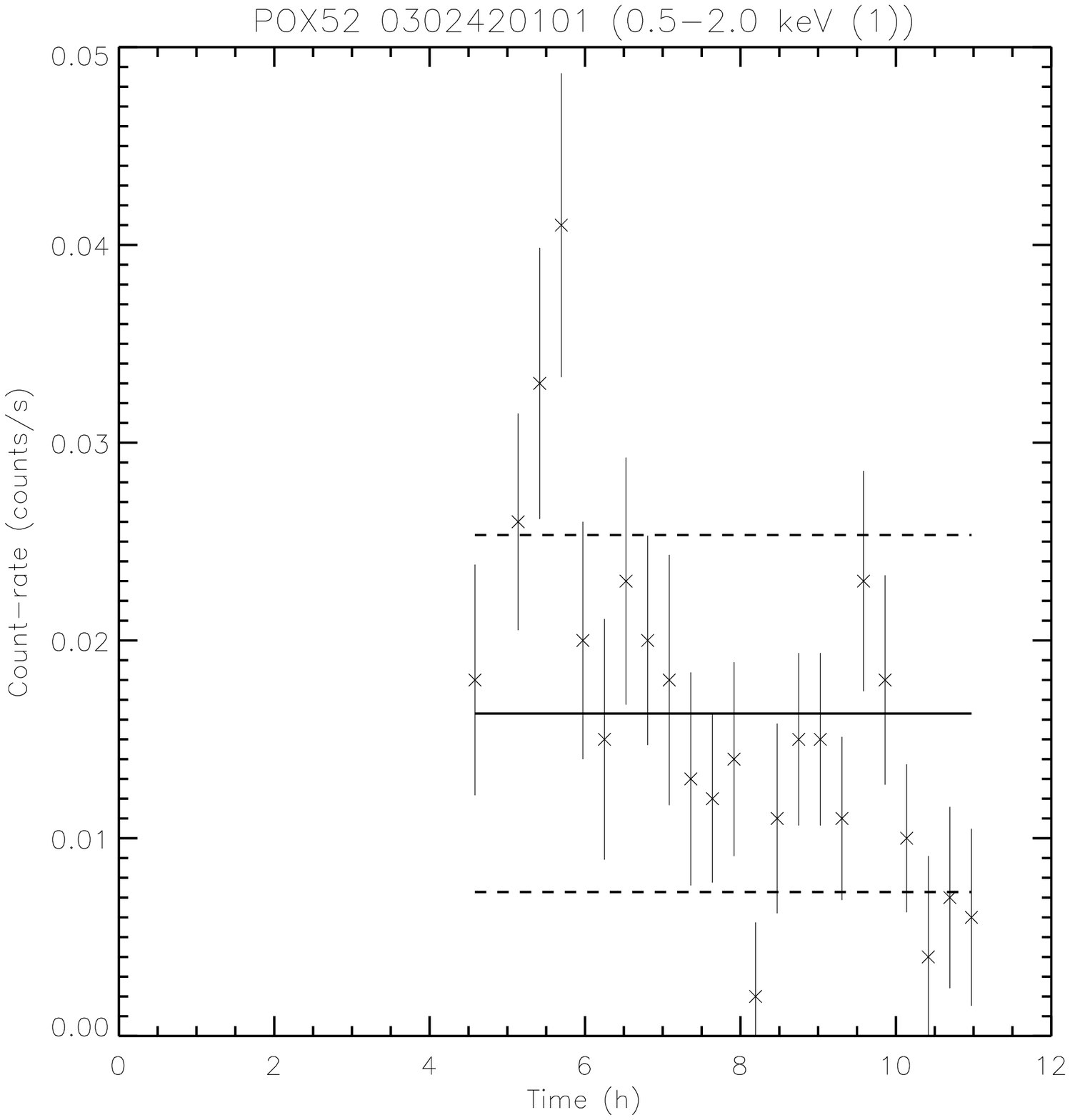}}
{\includegraphics[width=0.30\textwidth]{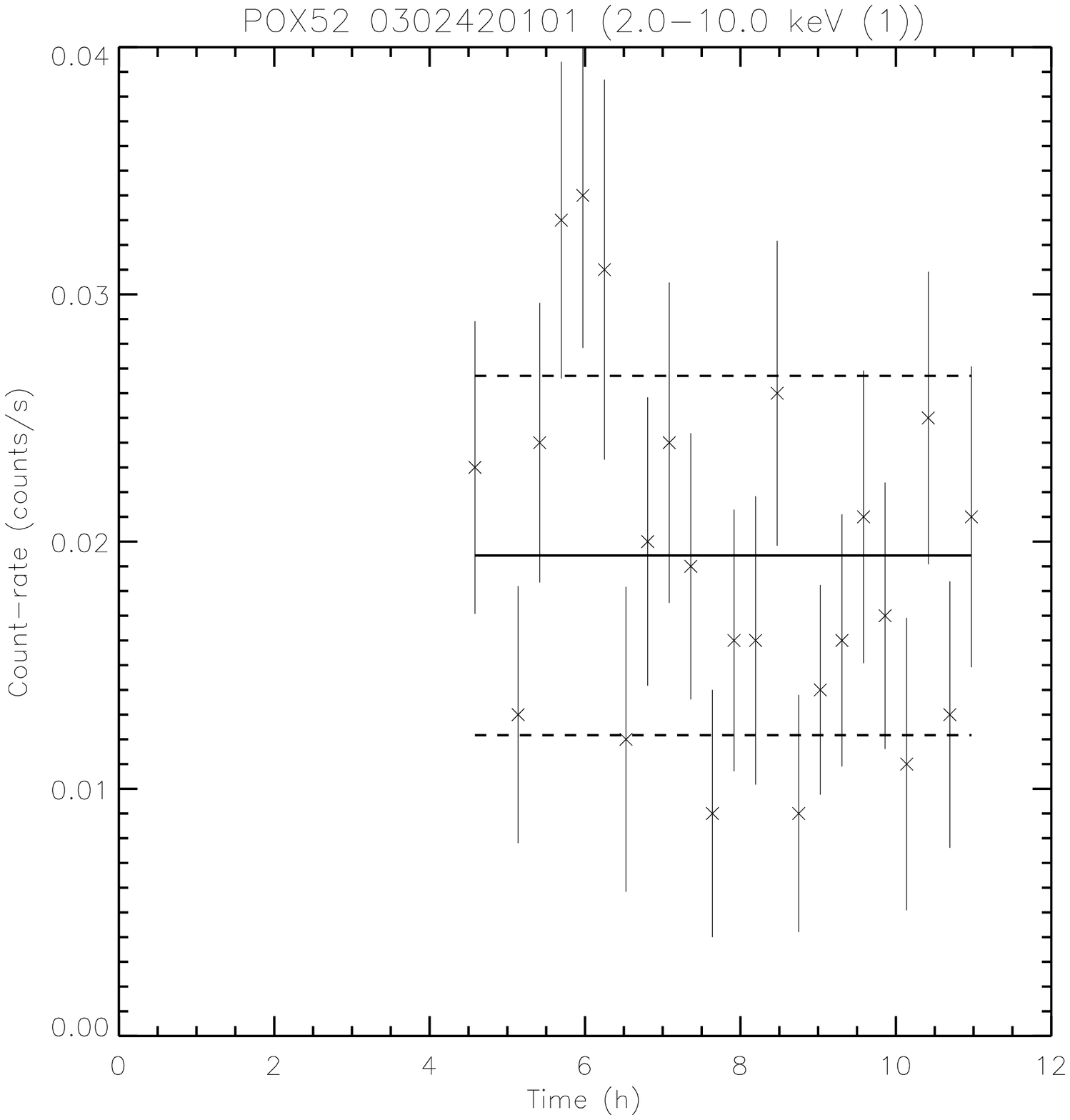}}
{\includegraphics[width=0.30\textwidth]{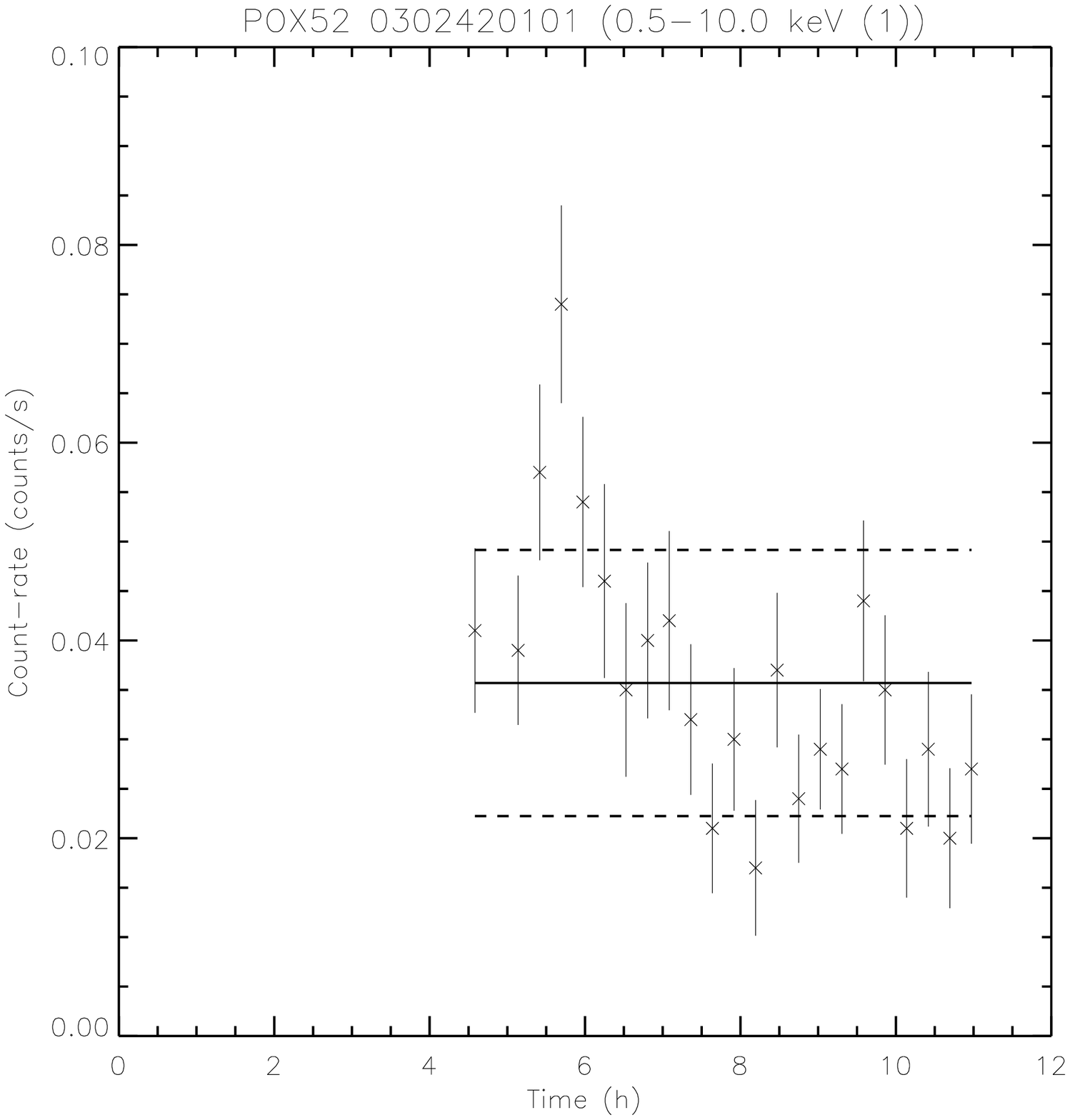}}

{\includegraphics[width=0.30\textwidth]{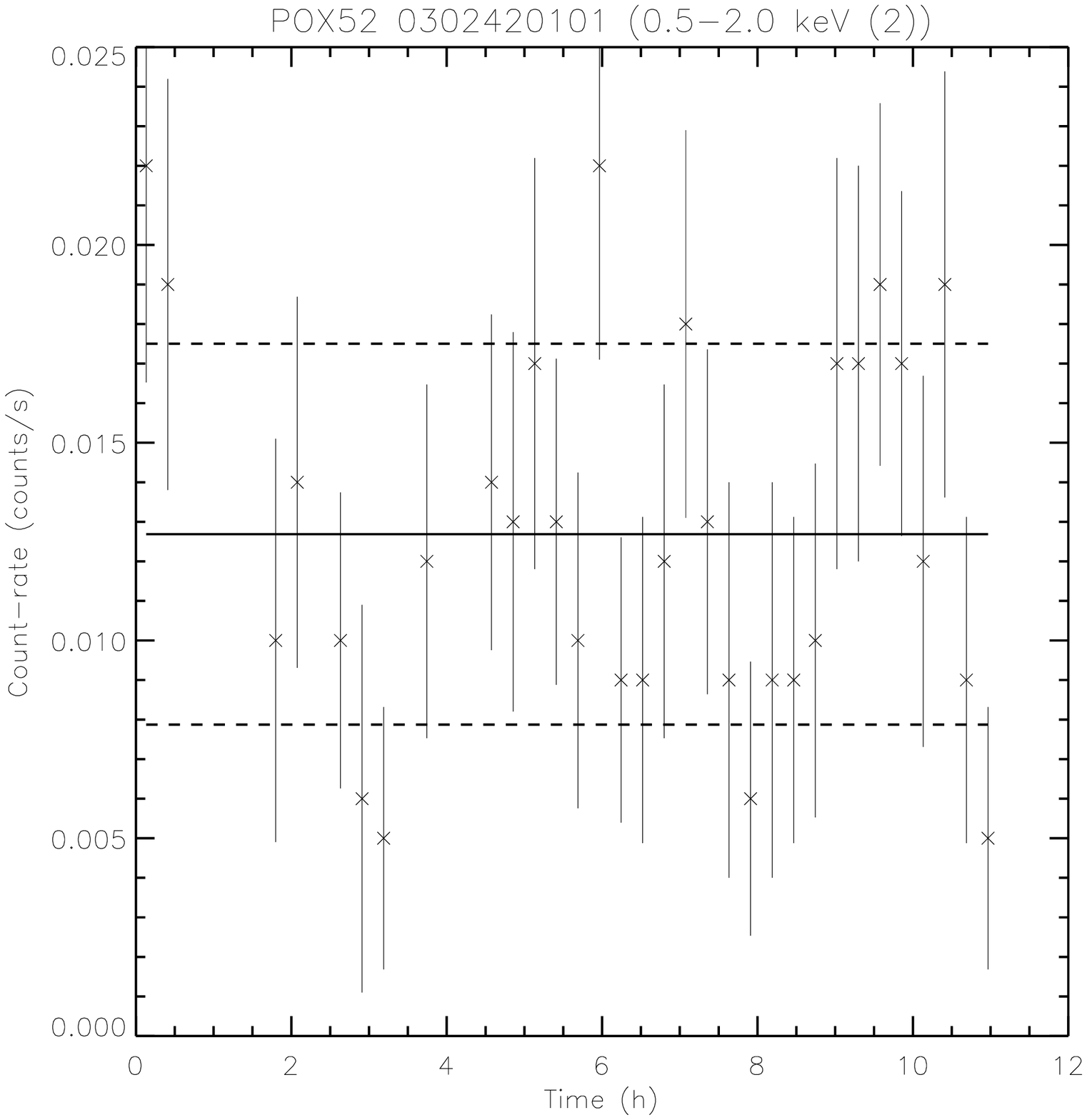}}
{\includegraphics[width=0.30\textwidth]{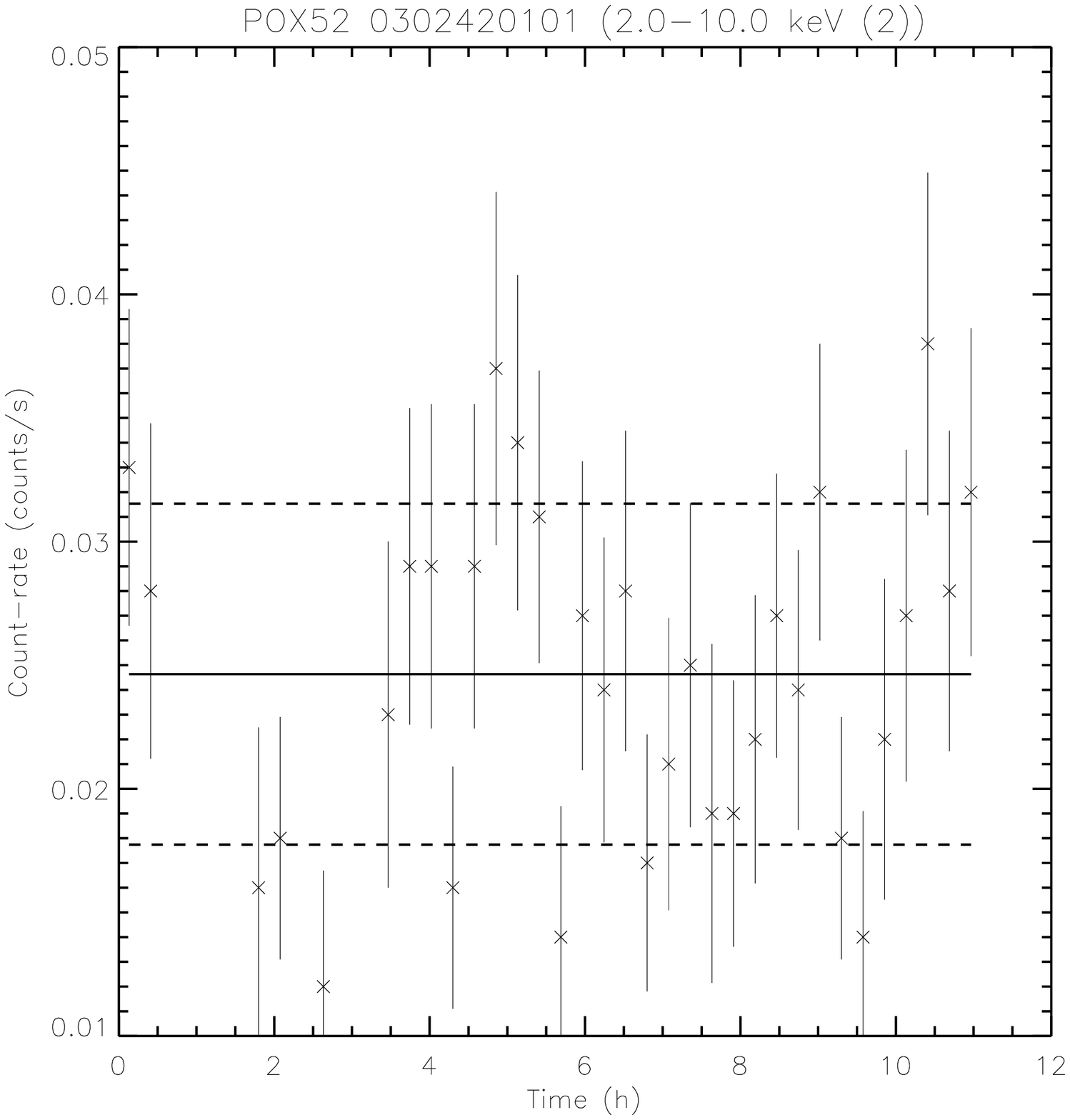}}
{\includegraphics[width=0.30\textwidth]{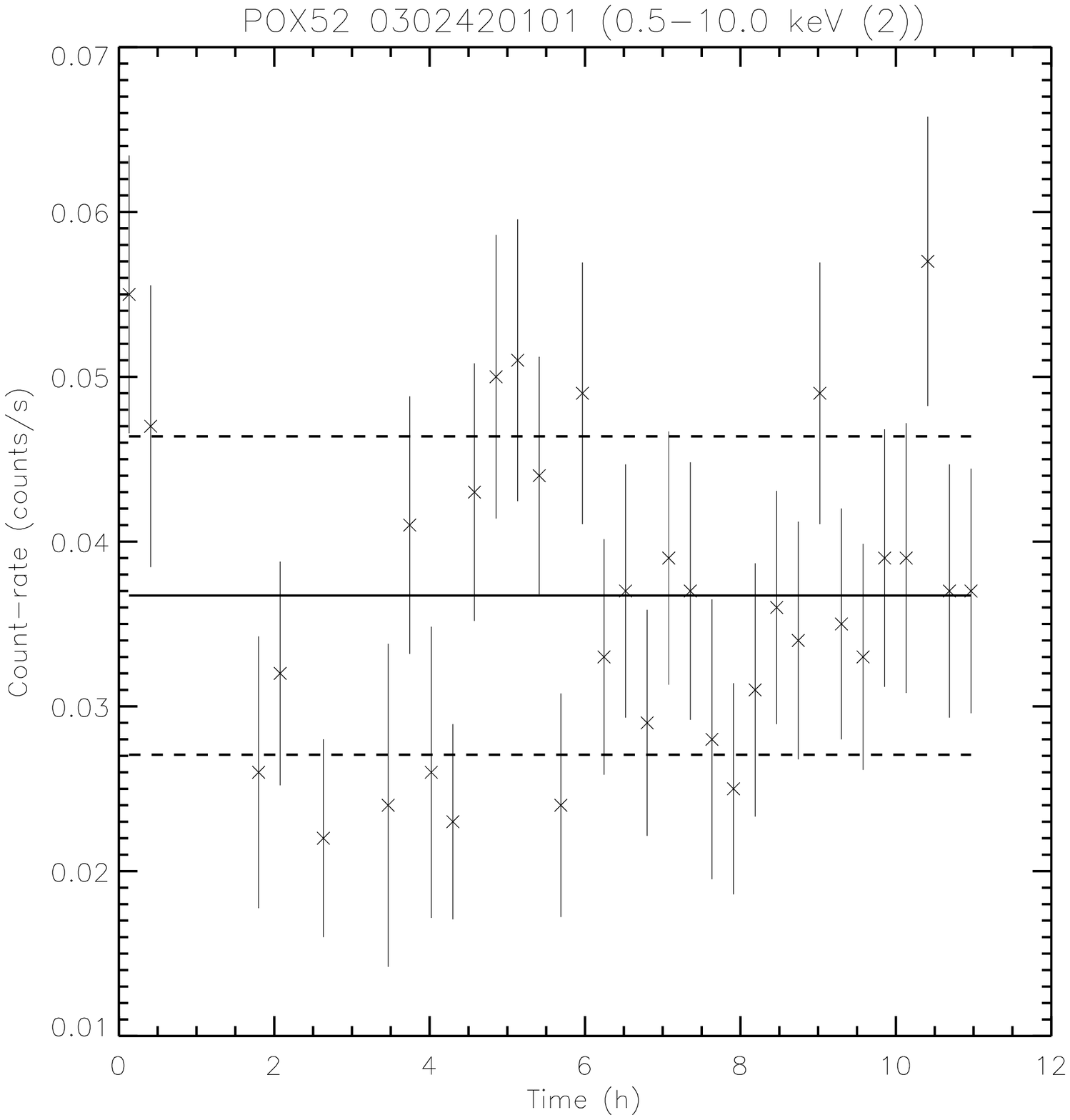}}
\caption{Light curves of POX\,52 from \emph{XMM--Newton} data.}
\label{lpox52}
\end{figure}

\begin{figure}
\centering
{\includegraphics[width=0.30\textwidth]{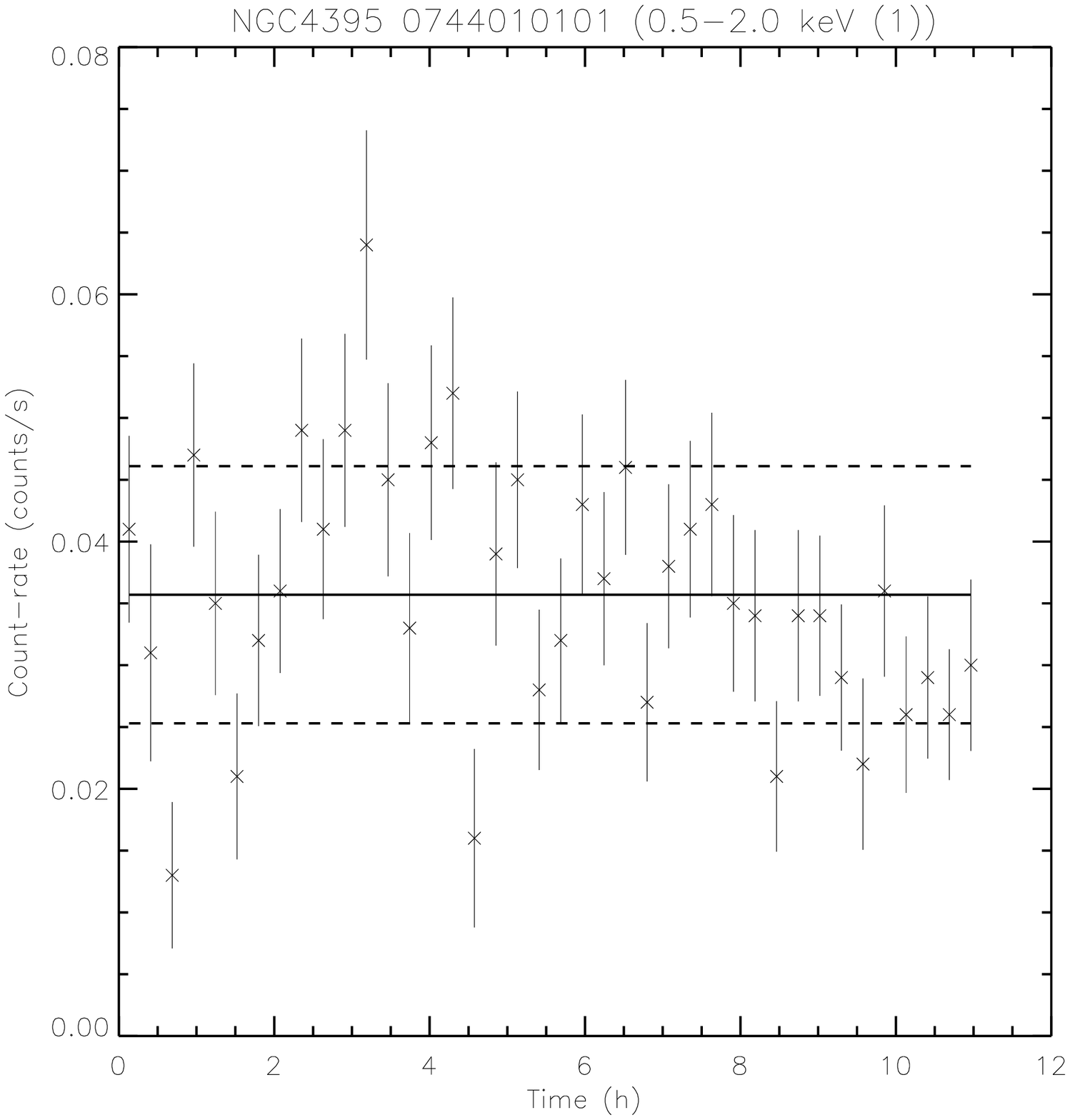}}
{\includegraphics[width=0.30\textwidth]{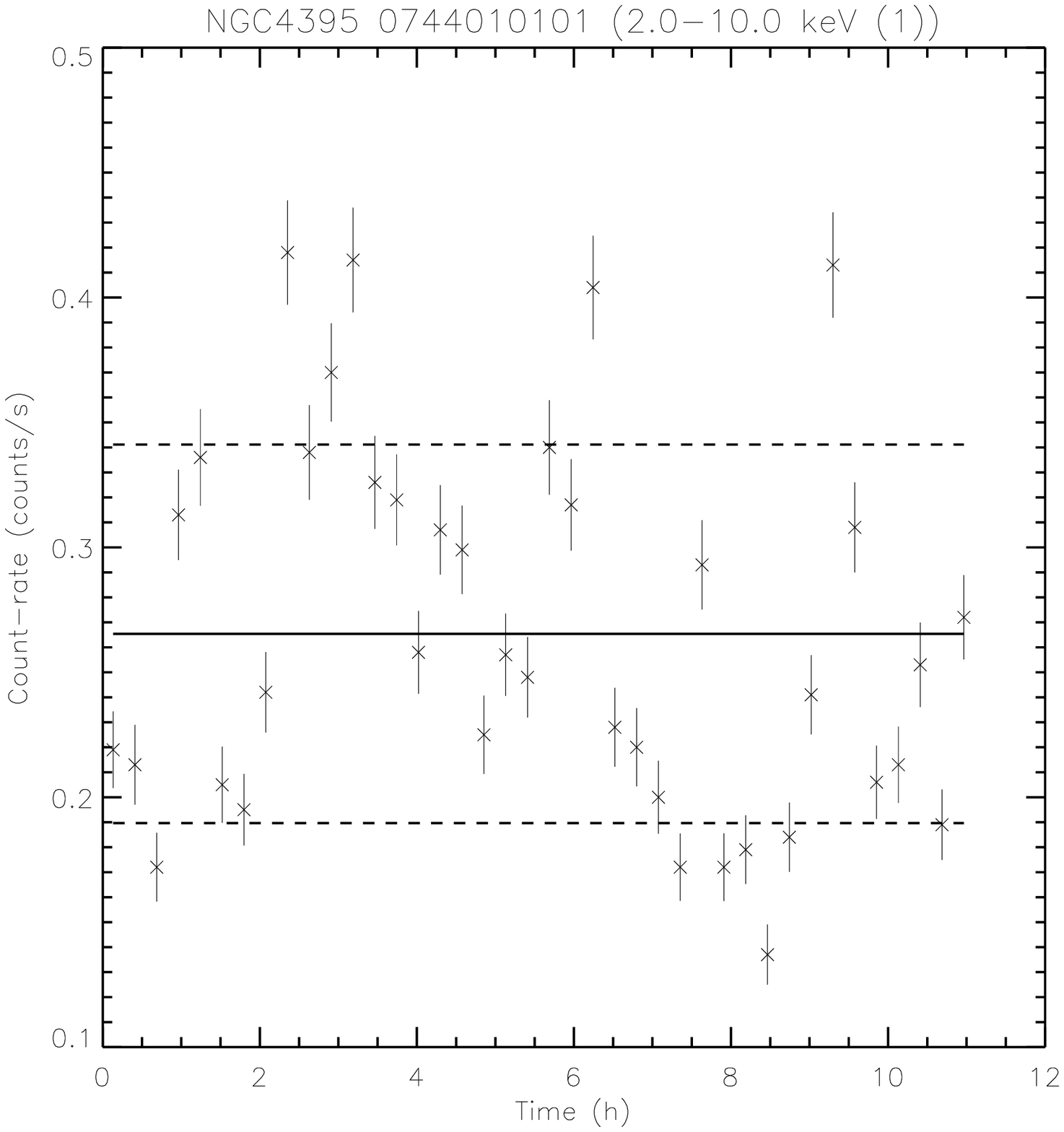}}
{\includegraphics[width=0.30\textwidth]{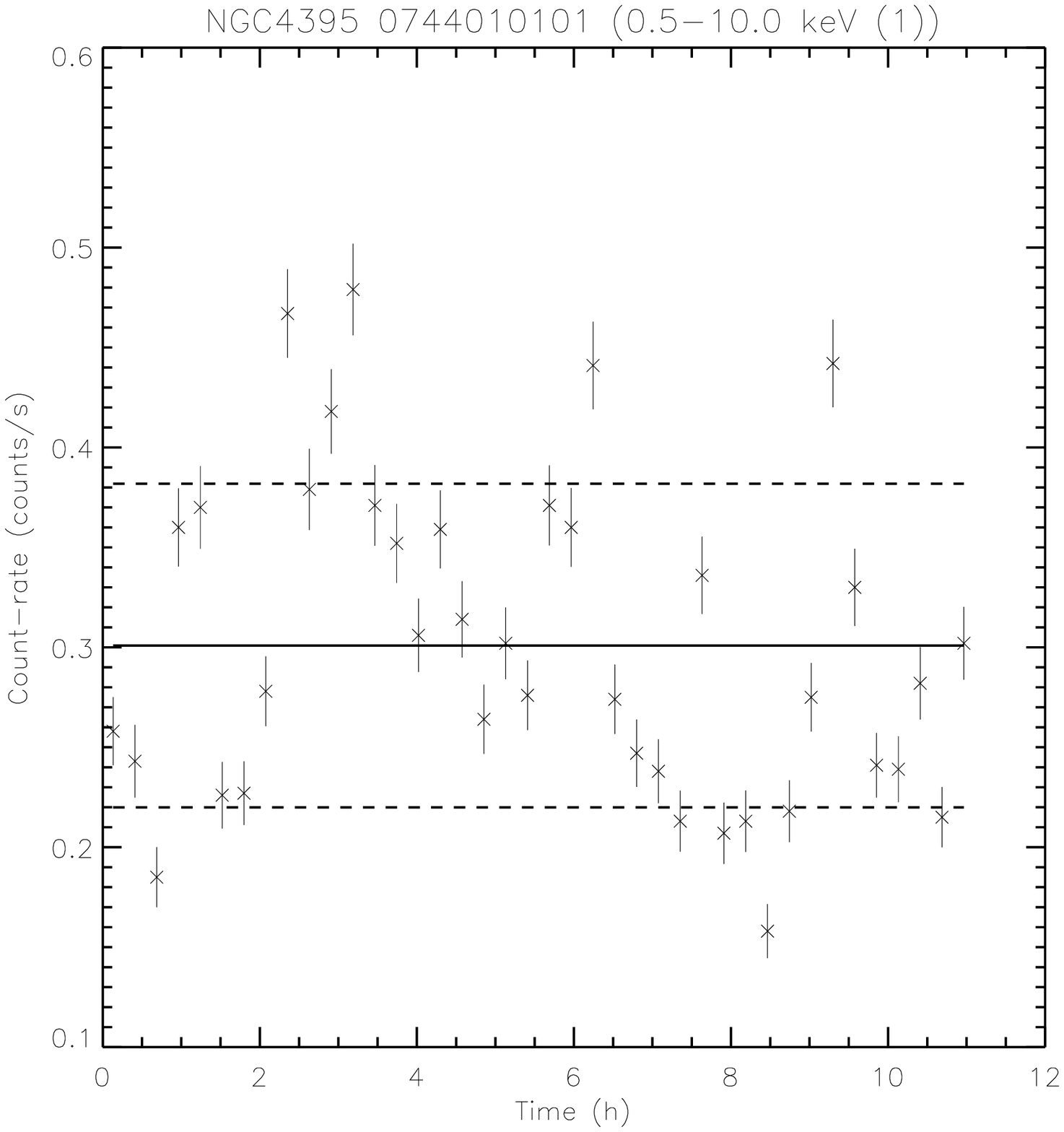}}
\caption{Light curves of NGC\,4395 from \emph{XMM--Newton} data.}
\label{l4395}
\end{figure}

\begin{figure}
\centering
{\includegraphics[width=0.30\textwidth]{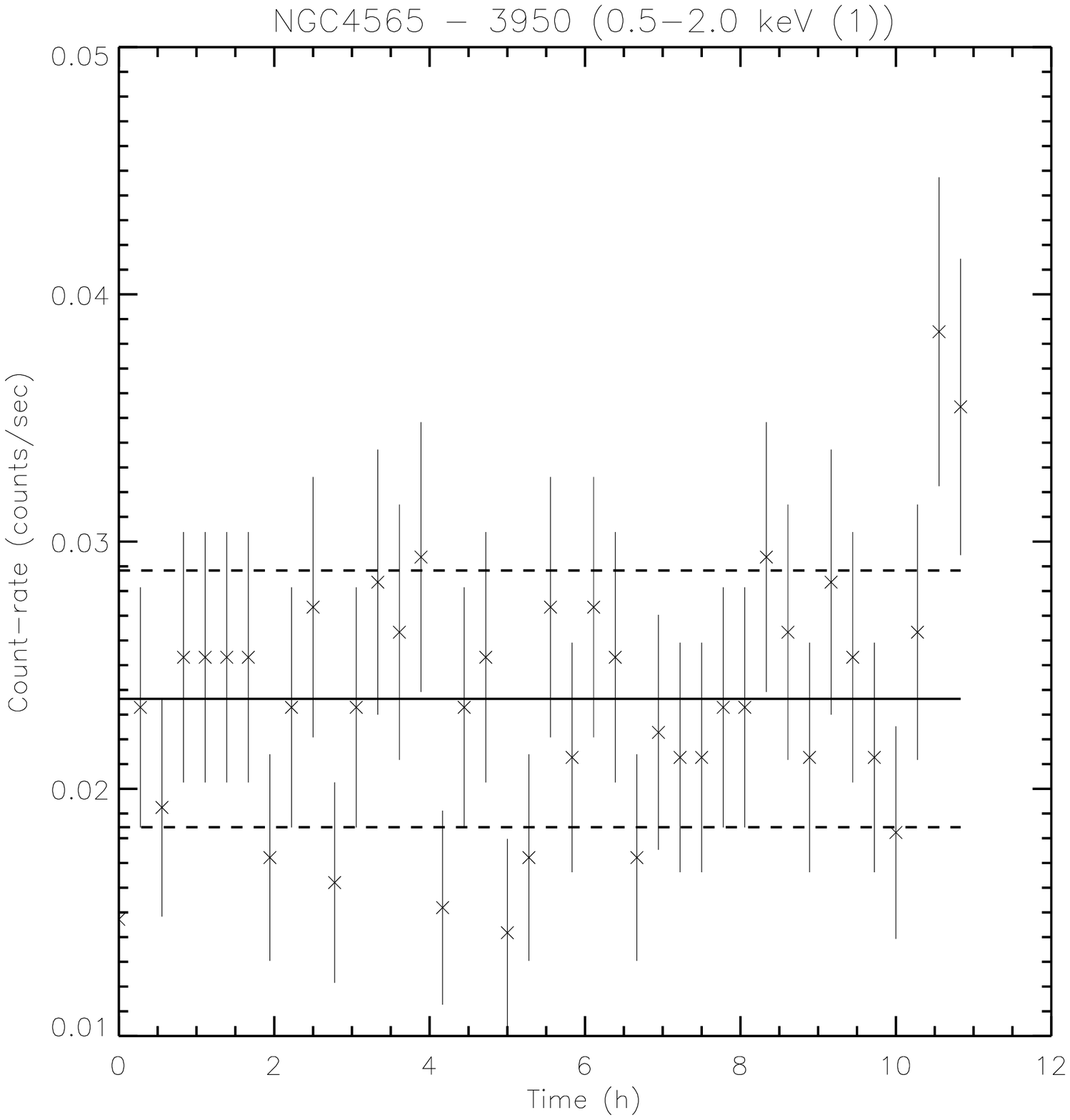}}
{\includegraphics[width=0.30\textwidth]{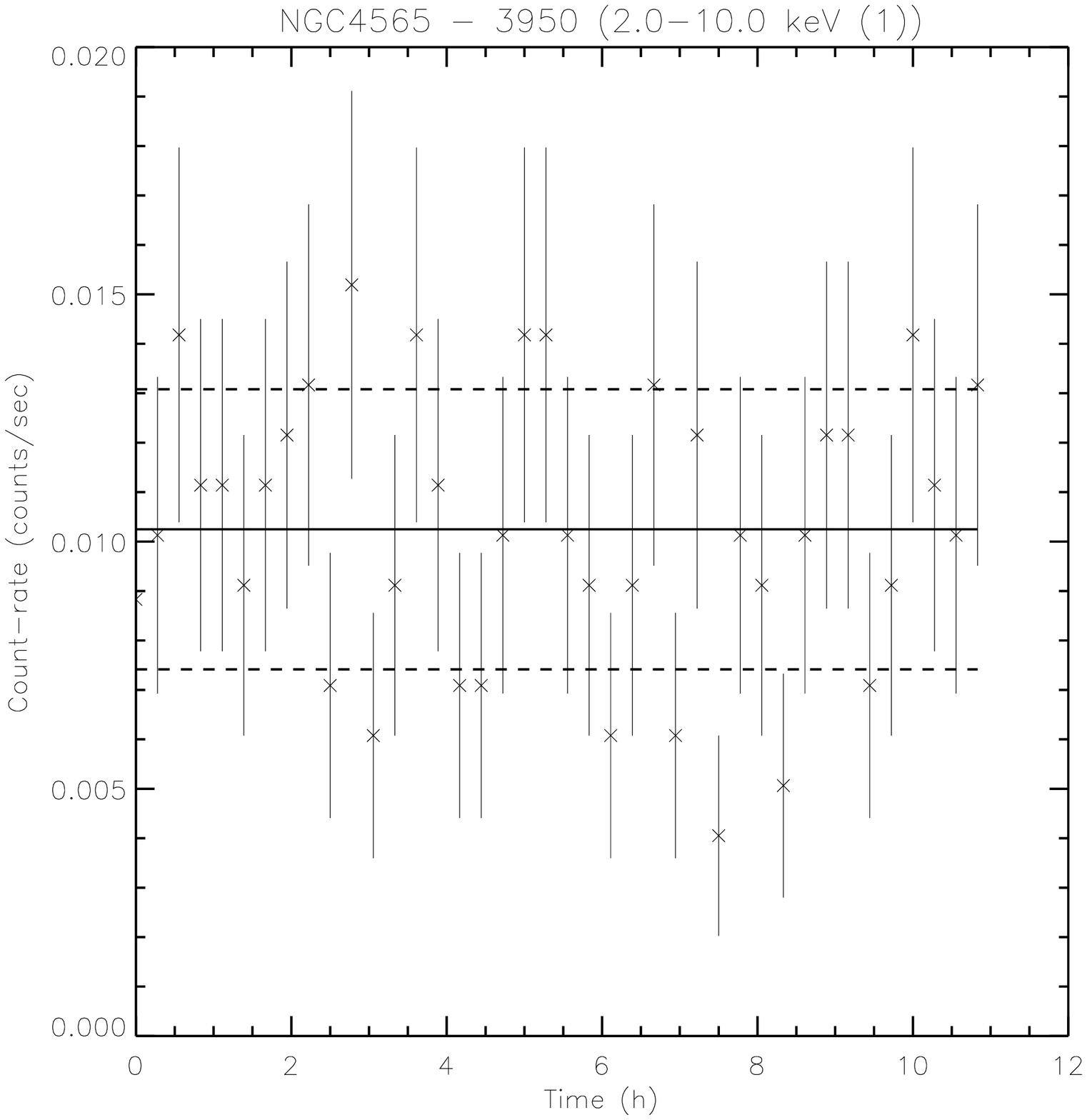}}
{\includegraphics[width=0.30\textwidth]{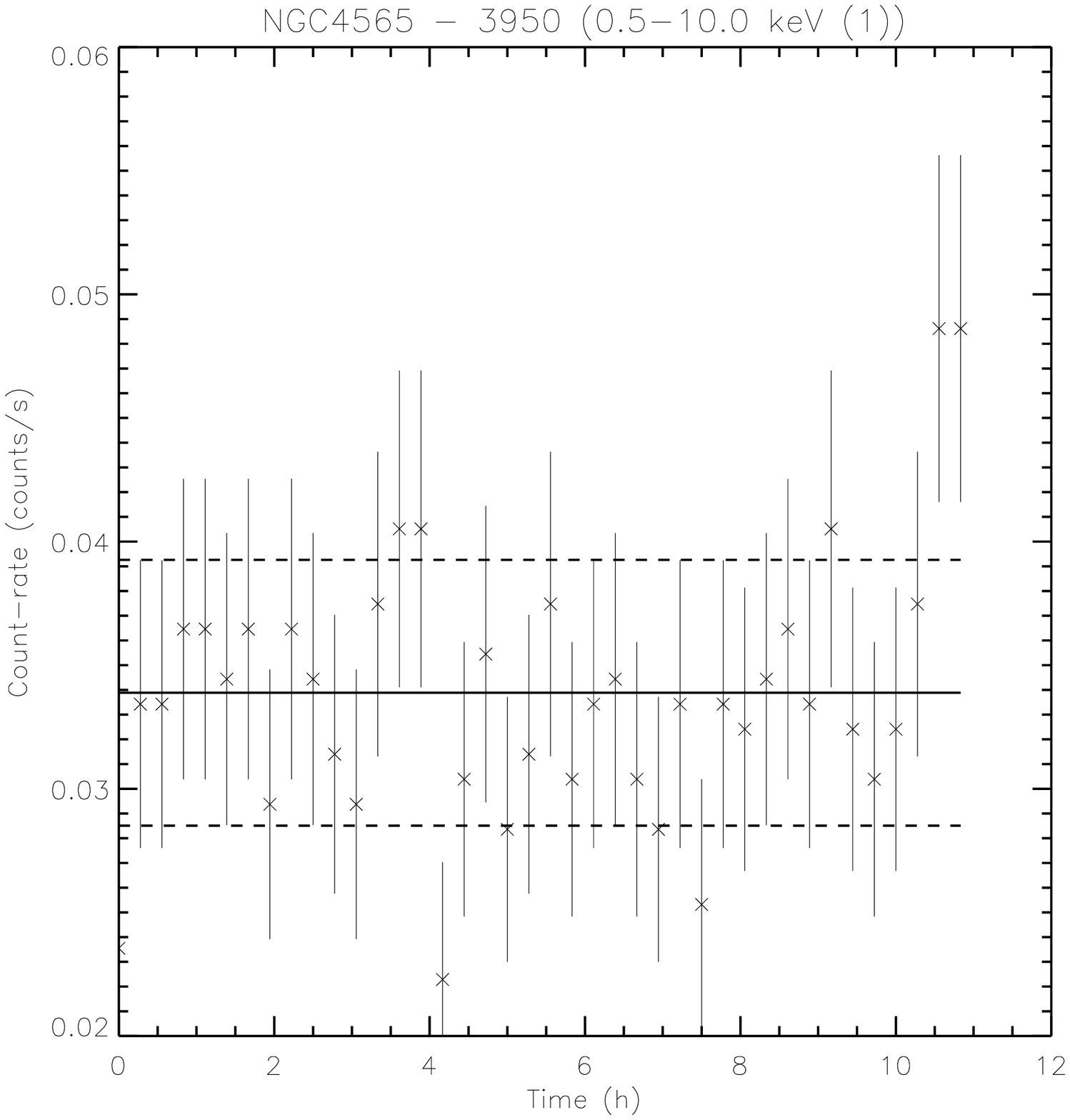}}

\caption{Light curves of NGC\,4565 from \emph{Chandra} data.}
\label{l4565}
\end{figure}

\end{document}